%% file: thesis.tex
\documentclass[10pt,b5paper,twoside,english]{latex_files/myphdthesis}

\usepackage{latex_files/ThALV} 

\usepackage[utf8]{inputenc} 
\usepackage{palatino} 
\usepackage{babel} 

\usepackage{lettrine}
\usepackage{epigraph} 

\makeatletter
\renewcommand{\@dotsep}{10000} 
\makeatother
\usepackage{tocbibind} 

\usepackage[figuresright]{rotating} 
\usepackage{subfigure}
\graphicspath{{./figure/}} 
\usepackage[small,bf]{caption} 

\usepackage{amsmath,amsfonts,amssymb,amsthm}  
\usepackage{latexsym}

\def\utw{\smash{\rlap{\lower5pt\hbox{$\sim$}}}}
\def\udtw{\smash{\rlap{\lower6pt\hbox{$\approx$}}}}

\usepackage[numbers,sort&compress]{natbib}
\usepackage{fancyhdr}
\renewcommand{\chaptermark}[1]{\markboth{#1}{}}
\renewcommand{\sectionmark}[1]{\markright{\thesection\ #1}}

\renewcommand{\headrulewidth}{0.2pt}
\renewcommand{\footrulewidth}{0pt}

\fancypagestyle{plain}{
\fancyhf{} 
\fancyfoot[C]{\bfseries \thepage} 
\renewcommand{\headrulewidth}{0pt}
\renewcommand{\footrulewidth}{0pt}}

\usepackage[binary-units=true]{siunitx}
\usepackage{isotope}
\usepackage{xcolor}
\usepackage{enumerate}
\usepackage{caption}
\usepackage[frozencache,cachedir=minted-files]{minted}
\usepackage{multirow}
\usepackage{booktabs}
\usepackage{quoting}
\usepackage{bm}
\usepackage{mathtools}
\usepackage[notoccite]{minitoc}

\let\minitocORIG\minitoc
\renewcommand{\minitoc}{\minitocORIG \vspace{1.5em}}

\newcommand{\myincludegraphics}[2][width=\textwidth]{\includegraphics[#1]{#2}}
\newcommand{\orderof}[1]{\ensuremath{\mathcal{O}(#1)}}
\newcommand{\abs}[1]{\ensuremath{\left\lvert #1 \right\rvert}}
\newcommand{\braces}[1]{\ensuremath{\left(#1\right)}}
\newcommand{\sbraces}[1]{\ensuremath{\left[#1\right]}}

\newcommand{\lrangles}[1]{\ensuremath{\left\langle#1\right\rangle}}
\newcommand{\bra}[1]{\ensuremath{\left\langle #1 \right\rvert}}
\newcommand{\ket}[1]{\ensuremath{\left\lvert #1 \right\rangle}}

\newcommand{\pdfflow}{\texttt{PDFFlow}~}
\newcommand{\madflow}{\texttt{MadFlow}~}
\newcommand{\vegasflow}{\texttt{VegasFlow}~}
\newcommand{\mgamc}{{\sc \small MG5\_aMC@NLO}~}

\dominitoc

\begin{document}

\frontmatter
\setcounter{page}{1}
\maketitle

\input{abstract/abstract}

\tableofcontents 
\thispagestyle{plain}


\listoffigures 
\renewcommand{\chaptermark}[1]{\markboth{#1}{}}
\renewcommand{\sectionmark}[1]{\markright{\thesection\ #1}}
\fancyhead[LE]{\bf \thepage}
\fancyhead[RO]{\bf \thepage}
\fancyhead[LO]{{\it List of Figures}}
\fancyhead[RE]{\it  List of Figures}
\renewcommand{\headrulewidth}{0.2pt}
\renewcommand{\footrulewidth}{0pt}

\listoftables 
\renewcommand{\chaptermark}[1]{\markboth{#1}{}}
\renewcommand{\sectionmark}[1]{\markright{\thesection\ #1}}
\fancyhead[RE,RO]{\bf \thepage}
\fancyhead[LO]{{\it List of Tables}}
\fancyhead[LE]{\it  List of Tables}
\renewcommand{\headrulewidth}{0.2pt}
\renewcommand{\footrulewidth}{0pt}

\renewcommand{\chaptermark}[1]{\markboth{#1}{}}
\renewcommand{\sectionmark}[1]{\markright{\thesection\ #1}}
\fancyhead[RO]{\bf \thepage}
\fancyhead[LE]{\bf \thepage}
\fancyhead[LO]{{\it Introduction}}
\fancyhead[RE]{\it Thesis overview}
\renewcommand{\headrulewidth}{0.2pt}
\renewcommand{\footrulewidth}{0pt}
\input{chapters/intro/Intro}
\thispagestyle{fancy}

\mainmatter

\renewcommand{\chaptermark}[1]{\markboth{#1}{}}
\renewcommand{\sectionmark}[1]{\markright{\thesection\ #1}}
\fancyhead[RO]{\bf \thepage}
\fancyhead[LE]{\bf \thepage}
\fancyhead[LO]{{\it \leftmark}}
\fancyhead[RE]{\it \rightmark}

\renewcommand{\headrulewidth}{0.2pt}
\renewcommand{\footrulewidth}{0pt}

\renewcommand{\theequation}{\arabic{chapter}.\arabic{equation}}

\input{chapters/partI/PartI}
\thispagestyle{empty}
\input{chapters/partI/chap1/chap1}

\input{chapters/partI/chap2/chap2}

\input{chapters/partII/PartII}
\thispagestyle{empty}
\input{chapters/partII/chap3/chap3}
\input{chapters/partII/chap4/chap4}
\vspace{0.2cm}

\input{chapters/partIII/PartIII}
\thispagestyle{empty}
\input{chapters/partIII/chap5/chap5}
\input{chapters/partIII/chap6/chap6}
\vspace{0.2cm}

\input{chapters/concl/Concl}
\vspace{0.2cm}


\makeatletter
\@addtoreset{equation}{section}  
\makeatother
\renewcommand{\theequation}
  {\thesection.\arabic{equation}}

\renewcommand{\theequation}{A.\arabic{equation}}

\renewcommand{\theequation}{\arabic{chapter}.\arabic{equation}}


\backmatter

\pagestyle{fancy}
\fancyhf{}
\renewcommand{\chaptermark}[1]{\markboth{#1}{}}
\renewcommand{\sectionmark}[1]{\markright{\thesection\ #1}}
\fancyhead[LE]{\bf \thepage}
\fancyhead[RO]{\bf \thepage}
\fancyhead[LO]{{\it Bibliography}}
\fancyhead[RE]{\it Bibliography}
\renewcommand{\headrulewidth}{0.2pt}
\renewcommand{\footrulewidth}{0pt}

\bibliography{bib/thesis}
\bibliographystyle{spphys}



\input{chapters/pubs/pubs}

\pagestyle{fancy}
\fancyhf{}
\renewcommand{\chaptermark}[1]{\markboth{#1}{}}
\renewcommand{\sectionmark}[1]{\markright{\thesection\ #1}}
\fancyhead[LE]{\bf \thepage}
\fancyhead[RO]{\bf \thepage}
\fancyhead[LO]{{\it List of Publications}}
\fancyhead[RE]{\it  List of Publications}
\renewcommand{\headrulewidth}{0.2pt}
\renewcommand{\footrulewidth}{0pt}

\input{chapters/ackno/Ackno}

\pagestyle{fancy}
\fancyhf{}
\renewcommand{\chaptermark}[1]{\markboth{#1}{}}
\renewcommand{\sectionmark}[1]{\markright{\thesection\ #1}}
\fancyhead[LE]{\bf \thepage}
\fancyhead[RO]{\bf \thepage}
\fancyhead[LO]{{\it Acknowledgments}}
\fancyhead[RE]{\it  Acknowledgments}
\renewcommand{\headrulewidth}{0.2pt}
\renewcommand{\footrulewidth}{0pt}

\end{document}

%% file: abstract/abstract.tex

\begin{center}\vspace{-0.8em}\small\bf Abstract\end{center}\quotation\small

    The thesis arises in the context of deep learning applications to particle
    physics. The dissertation follows two main parallel streams: the development
    of hardware-accelerated tools for event simulation in high-energy collider
    physics, and the optimization of deep learning models for reconstruction
    algorithms at neutrino detectors. These two topics are anticipated by a
    review of the literature concerning the recent advancements of artificial
    intelligence models in particle physics.
    
    Event generation is a central concept in high-energy physics phenomenology
    studies. The state-of-the-art software dedicated to Monte Carlo simulation is often
    written for general-purpose computing architectures (CPUs), which allow great
    flexibility but are not compatible with specialized accelerating devices,
    such as Graphics Processing Units (GPUs).
    
    The original tools presented in the thesis, PDFFlow and MadFlow,
    manage to combine these two aspects in Python and require no prior knowledge of
    specific programming languages for hardware accelerators. The former product,
    PDFFlow, is a Parton Distribution Functions (PDFs) interpolator, the latter,
    MadFlow, aims at building a complete tool suite to accelerate the whole event
    generation framework.

    The reconstruction pipeline at neutrino detectors is comprised of many
    different algorithms that work in synergy to extract a high-level
    representation of detector data. All the most important experiments in neutrino
    physics are developing software to automatically process and extract this
    information. This work describes the implementation of deep learning
    techniques to improve neutrino reconstruction efficiency at the ProtoDUNE-SP
    detector.
    
    Two original contributions are presented concerning raw data denoising and a
    hit-clustering procedure named "slicing". Both denoising and slicing involve
    the implementation and the training of novel neural network
    architectures, based on state-of-the-art models in machine learning,
    such as feed-forward, convolutional and graph neural networks. They represent
    a proof of concept that these models are indeed capable of providing an
    important impact on signal reconstruction at neutrino detectors.
\vspace{0.6em}\par\endquotation\normalsize\rm

%% file: chapters/intro/Intro.tex
\chapter*{Introduction}
\addstarredchapter{Introduction} 
\thispagestyle{plain}

\phantomsection
\section*{Motivation}
\addcontentsline{toc}{section}{Motivation} 

The present thesis arises in the context of particle physics and aims to
investigate novel deep learning solutions in the event simulation and reconstruction
pipeline. The key idea is that machine learning provides performance and accuracy
speed-ups which the physics community cannot waive to. We identify two main areas
of interest for our research: hardware acceleration for Monte Carlo simulation
and the implementation of new deep learning-based software for reconstruction
algorithms at neutrino detectors.

Hardware acceleration is defined as the process by which an application offloads
certain computing tasks onto specialized hardware components accessible by the
computing system, enabling greater software efficiency than would be possible
when running on a general-purpose CPU alone. Our techniques leverage mainly on
Graphics Processing Units (GPUs), but, in principle, other hardware might be
worth testing in the future, such as Tensor Processing Units (TPUs) and, perhaps,
Field Programmable Gate Arrays (FPGAs). 

Monte Carlo event simulators are software tools that play a central role in HEP
phenomenology. They exploit MC techniques to compute differential cross-sections
of particle physics processes and provide samples of weighted and unweighted events.
Simulation, indeed, is the first step into the production chain of events at hadron
colliders, followed by detector simulation and reconstruction. The quantity
of computational resources employed for running these tools is huge. Specifically,
it has been estimated~\citep{Albrecht:2017,Amoroso:2020} that the two major
LHC experiments, namely ATLAS and CMS, collect an order of $\mathcal{O}(10^{10})$
events for every year of data taking and, correspondingly, they further generate
a factor of $3$ more simulated events. Moreover, the quantity of data is bounded
to grow in the future during the High-Luminosity LHC phase, which will start towards
the end of 2027. In this context, one of the ideas to successfully face this
challenge is to optimize existing software to fully leverage modern hardware
architectures, such as Graphics Processing Units (GPUs).


Experimental setups in High Energy Physics (HEP), therefore, produce large \linebreak
datasets, containing huge amounts of information. As described in
section~\ref{sec:ai-review}, there have been high interest and numerous
published results about applying artificial intelligence (AI) models to particle
physics. The final goal is to be able to provide end-to-end data processing
solutions. The hope is that AI strategies and deep learning in particular will
eventually manage to scan the physics details hidden inside experimental data.
Although we do not know yet whether this will happen and, further, data
processing is often an obscure step towards obtaining the results, we cannot
leave without the automation and the speed at which these methods
manipulate data. Experiments, indeed, aim to build larger and more
complex machines, which put computing facilities under extreme stress.

We dedicate a considerable part of the present thesis to investigate deep learning
applications to the reconstruction workflow in the context of the Deep
Underground Neutrino Experiment (DUNE). In particular, we discuss two algorithms
and their solution employing neural network techniques.



\phantomsection
\section*{Thesis overview}
\addcontentsline{toc}{section}{Thesis overview} 

\subsection*{Main results}
\begin{list}{}{\leftmargin 15pt \itemsep 0pt \topsep 3pt}
\item {\bf Interpolating Parton Distribution Functions on GPUs: }
\pdfflow is the first Parton Distribution Functions access tool able to run on
Graphics Processing Units. We demonstrated the possibility to implement software
that can exploit hardware acceleration while maintaining the same level
of accuracy as the state-of-the-art methods in the field. The usage of the
software does not require specific knowledge or experience in low-level
programming languages dedicated to hardware accelerators. Coupling \pdfflow with
the existent \vegasflow program, we delivered the first particle physics
phenomenology calculation at next-to-leading order accuracy in perturbation
theory on GPU.
\item {\bf Automated Monte Carlo event generator on GPU:}
The \madflow tool-suite encapsulates the implementations of \vegasflow and
\pdfflow in a single software, that automates the computation particle physics
processes on hardware accelerators at leading order. The program takes advantage
of the \mgamc meta-programming capabilities to automatically produce code for GPU
matrix element calculations and phase space sampling for user-requested particle
physics processes. We fully test the performance and accuracy results of \madflow
for different physical processes, proving that the tool is able to handle
complex computations involving a high number of Feynman diagrams even despite
complying with the memory constraints of modern GPUs.
\item {\bf ProtoDUNE-SP raw data denoising with Graph Neural Networks:}
We investigate different strategies for denoising raw data at the ProtoDUNE-SP
experiment designing two novel Graph Neural Networks architectures. The work
addresses the first step of reconstruction for Liquid Argon Time Projecting Chamber
detectors. We train and test our neural networks on simulated datasets generated
with the LArSoft framework. We prove that our models' performance is
competitive and, in fact, outperforms the traditional methods currently employed
for denoising raw data by the latest neutrino experiments.
\item {\bf Slicing algorithm at ProtoDUNE-SP with deep neural networks:}
In the context of the neutrino event reconstruction, the slicing method aims to
cluster the detector hits into sets named slices, where all the hits in each
group belong to the same main interaction that happened within the detector. We
implement a neural network model to process the detector hits, to replace the
traditional slicing method included in the Pandora reconstruction
framework based on topological algorithms. We train our architecture on simulated
data and design a first test to compare the clustering performance of our model
against the state-of-the-art one at separate detector hits between test beam
and cosmic ray interactions. We discover that our approach outperforms the
traditional one implemented in Pandora.
\end{list}

\subsection*{Organizational note}

The present thesis consists of three Parts, for a total of six Chapters. Each
part is devoted to the study of an application of deep learning in the context
of particle physics. The three parts split the discussion into an introduction
of physical and computer science background, a presentation of our novel
contributions to simulation and a discussion of the techniques implemented for
detector reconstruction. Part I is composed of chapters~\ref{chap:introph}
and~\ref{chap:introcs} and give an overview of the physics and computer science
background, respectively. We discuss the Standard Model of particle physics
and give a more detailed insight into neutrino physics and Monte Carlo techniques.
On the computer science side, we introduce the fundamental concepts of
artificial intelligence and machine learning, providing also an overview of the
main state-of-the-art AI applications in the particle physics research field.
Part II comprehends chapters~\ref{chap:pdfflow} and~\ref{chap:madflow} and
presents the construction and characterization of efficient automated computation
methods for parton distribution functions interpolation and Monte Carlo
simulation of particle physics processes on hardware accelerators.
Part III encompasses chapter~\ref{chap:dunedn} and~\ref{chap:slicing} describing
the implementation of novel deep learning-based algorithms for event reconstruction
at the ProtoDUNE-SP detector. We discuss different strategies designed for
experimental raw data denoising and a clustering approach for detector hits
named slicing. Parts II and III have appeared in peer-reviewed publications in
scientific journals and international conference proceedings. Some variations
have been made in the presentation of previously published results, to maintain
consistency of style and content structure throughout the manuscript.

\begin{list}{\leftmargin 15pt \itemsep 0pt \topsep 3pt}
\item{\bf Chapter~\ref{chap:introph}: Physics background}
We give a high-level overview of the Standard Model of particle physics,
describing the model's particle content with a focus on the fermion sector.
We introduce the Higgs mechanism, the main technique used to include mass terms
for particle fields without explicitly breaking gauge symmetry. We discuss the
leptonic sector with a focus on neutrinos, investigating the neutrino oscillation
phenomenon and how to extend the Standard Model to include neutrino masses both
from a theoretical and experimental point of view. The last section of the
chapter is devoted to a presentation of numerical integration and Monte Carlo
techniques. We also describe the most widely used methods for sample generation
following given probability density functions: in this context, we focus also on
the physical phase space sampling techniques for collider events introducing
both a hierarchical and a democratic (RAMBO) approach.
\item{\bf Chapter \ref{chap:introcs}. Introduction to deep learning and its
physics applications} This chapter includes a review of the machine learning and
deep learning algorithms and a review of the most important applications
of these techniques to the particle physics research field available in the
literature. We define the concept of learning algorithms and neural networks,
giving examples of the main operations exploited in this field such as matrix
multiplications, image convolutions and the attention mechanism. We finally
present a study on stochastic optimization methods, used to train neural networks.
The last part of the chapter summarizes the main applications of such techniques
in particle physics, highlighting their great success and spread in many fields,
specifically we concentrate on reconstruction at colliders (jet physics and
particle tracking) and neutrino physics. Other applications, like fast event
simulation and detection of new physics, are mentioned in passing only.
\item{\bf Chapter \ref{chap:pdfflow}. Interpolating parton distribution
functions}
Our first novel contribution is in the field of parton distribution functions
(PDF): we design the \pdfflow \linebreak
software that can exploit hardware
acceleration to provide PDF values access. In this chapter, we briefly set the
stage by introducing the concept of PDFs and giving a detailed description of the
state-of-the-art tool for PDF access, namely LHAPDF6. We present the LHAPDF
framework concept from the historical point of view and its interpolating
algorithm. We leverage the TensorFlow library to demonstrate the performance gain
of our software with respect to the LHAPDF6 implementation while matching their
output accuracy. Finally, we benchmark our setup on particle physics experiments
both at leading and next to leading order in perturbation theory.
\item{\bf Chapter \ref{chap:madflow}. MadFlow: a Monte Carlo event generator on
GPUs}
The work on hardware accelerators continues with the implementation of a full
tool suite able to port a complete Monte Carlo event generator for particle
physics processes on Graphics Processing Units (GPUs). We exploit the same
concepts of \pdfflow to design a matrix element generator and a phase space
sampler that is compatible with GPUs. Further, we provide an algorithm to store the
generated events asynchronously to preserve the performance speed-up
delivered by our framework when coupled to hardware accelerators. As in the
\pdfflow chapter, the benchmarks are supported by particle physics processes
examples.
\item{\bf Chapter \ref{chap:dunedn}. Denoising ProtoDUNE-SP Raw Data with deep
learning}
This chapter starts the discussion about reconstruction algorithms for neutrino
LArTPC detectors, in particular, ProtoDUNE-SP. We introduce first the design of
the ProtoDUNE-SP detector and how Liquid Argon Time Projecting Chambers record
information about physics events. Our work focuses on the first stage of the
signal-processing pipeline: raw data denoising, that is taking the detector
signal and automatically mitigating the noise inherently introduced by the
electronics. The state-of-the-art technique that tackles this problem is
deconvolution. Instead, we investigate different deep learning approaches based
on graph neural networks.
\item{\bf Chapter \ref{chap:slicing}. Clustering at ProtoDUNE-SP with supervised
learning}
The final chapter of this thesis is devoted to the slicing algorithm. Slicing is
a step in the reconstruction chain at LArTPC detectors and is currently
managed by the Pandora framework. Thus, we first describe the Pandora concept
and how it implements event reconstruction. We present, then, our deep
learning-based approach to the slicing algorithm along with a benchmark comparison
of our method against the Pandora one.
\end{list}

%% file: chapters/partI/PartI.tex
\part*{Part I \vspace{0.5cm}\\Physics and computer science background}
\addstarredpart{\large Part I : Physics and computer science background}
\thispagestyle{empty}

%% file: chapters/partI/chap1/chap1.tex
\chapter{Physics background}
\label{chap:introph}
\thispagestyle{plain}

\adjustmtc[4]
\minitoc

In this chapter, we introduce the physics topics which the present thesis is based upon.
We discuss the Standard Model (SM) of particle physics, describing the model's
particle content and highlighting its gauge symmetries. Then, we restrict our
attention to the physics of neutrinos, given the central role of such particles
in part~\ref{part:dl4nu}. In particular, we focus on the neutrino oscillation
mechanism, which implies a non-zero mass of such particles and requires extending
the SM to include lagrangian mass terms for neutrinos. We give both theoretical
and experimental insights into this research field.

We conclude the chapter with a dissertation on Monte Carlo (MC) techniques, the
fundamental tools for phenomenology studies. We present the mathematical
formalism behind these methods to show how MC methods are suited to sample
from high-dimensional probability distributions to generate events for particle
physics processes.

\input{chapters/partI/chap1/sm.tex}

\input{chapters/partI/chap1/nu.tex}

\input{chapters/partI/chap1/mc.tex}

%% file: chapters/partI/chap1/sm.tex
\section{The standard model of particle physics}

In this section, we set the stage by giving a brief introduction of the standard
model~\cite{Novaes:1999} of particle physics. The SM describes our current
understanding of three out of four fundamental interactions in nature: the strong,
weak and electromagnetic forces. The last fundamental force, gravity, is not
included in the theory, given that its mathematical properties, e.g.
non-renormalizability, make it incompatible with the model.
The SM lagrangian, containing only 19 free parameters, effectively yields
accurate predictions of physical processes validated by many experimental
results~\citep{Butterworth:2016}. Despite its success, the model still contains
open questions~\cite{Paudel:2021}, such as the lack of a consistent description of
neutrinos, whose discussion will be presented in section~\ref{sec:nu_phys}. Here,
we present a descriptive overview of the SM, leaving the interested reader to the
detailed dissertation of specialized quantum field theory (QFT) books, such
as~\cite{Peskin:1995}.

Formally, the SM is a Yang Mills theory with $SU(3)_C {\times} SU(2)_L {\times} U(1)_Y$
gauge symmetry group, where the subscripts $C$, $L$ and $Y$ stand respectively for
the color group of Quantum Chromodynamics (QCD), the fact that the weak force
couples only to left chiral components of fermion fields and the hypercharge.
In such theories, there is always a one-to-one correspondence between the generators
of the symmetry group and the bosonic fields representing force carriers.
The physical fields for the gluons correspond to the $8$ generators of $SU(3)_C$.
The $3+1$ generators of $SU(2)_L {\times} U(1)_Y$, instead, are linked with the electroweak
physical bosons, namely $\mathrm{W^\pm{/}Z{/}\gamma}$, through a spontaneous
symmetry breaking (SSB) mechanism~\cite{Higgs:1964}.
In QFT, including a mass term for gauge bosons in the lagrangian would explicitly
break the local symmetry. Instead, the SSB is a process that allows gauge bosons
to acquire mass coupling with a scalar field, which gains a non-zero vacuum
expectation value (VEV). In the SM, this scalar field is called the Higgs boson
$\phi$ and it is realized as a doublet of fields:
\begin{equation}
    \label{eqn:higgs-field}
    \phi =
    \begin{pmatrix}
        \phi^+\\ \phi^0
    \end{pmatrix}
    \qquad
    \left\langle \phi\right\rangle =
    \begin{pmatrix}
        0\\ \frac{v}{\sqrt{2}}
    \end{pmatrix}
\end{equation}
where upper and lower components of $\phi$ are two complex scalar fields
and the angular parentheses represent the VEV operator. When $\phi^0$ acquires a
non-zero VEV $v\sqrt{2}$, it is responsible to give mass to the SM particles.

\begin{figure}
    \centering
    \includegraphics[width=0.5\textwidth]{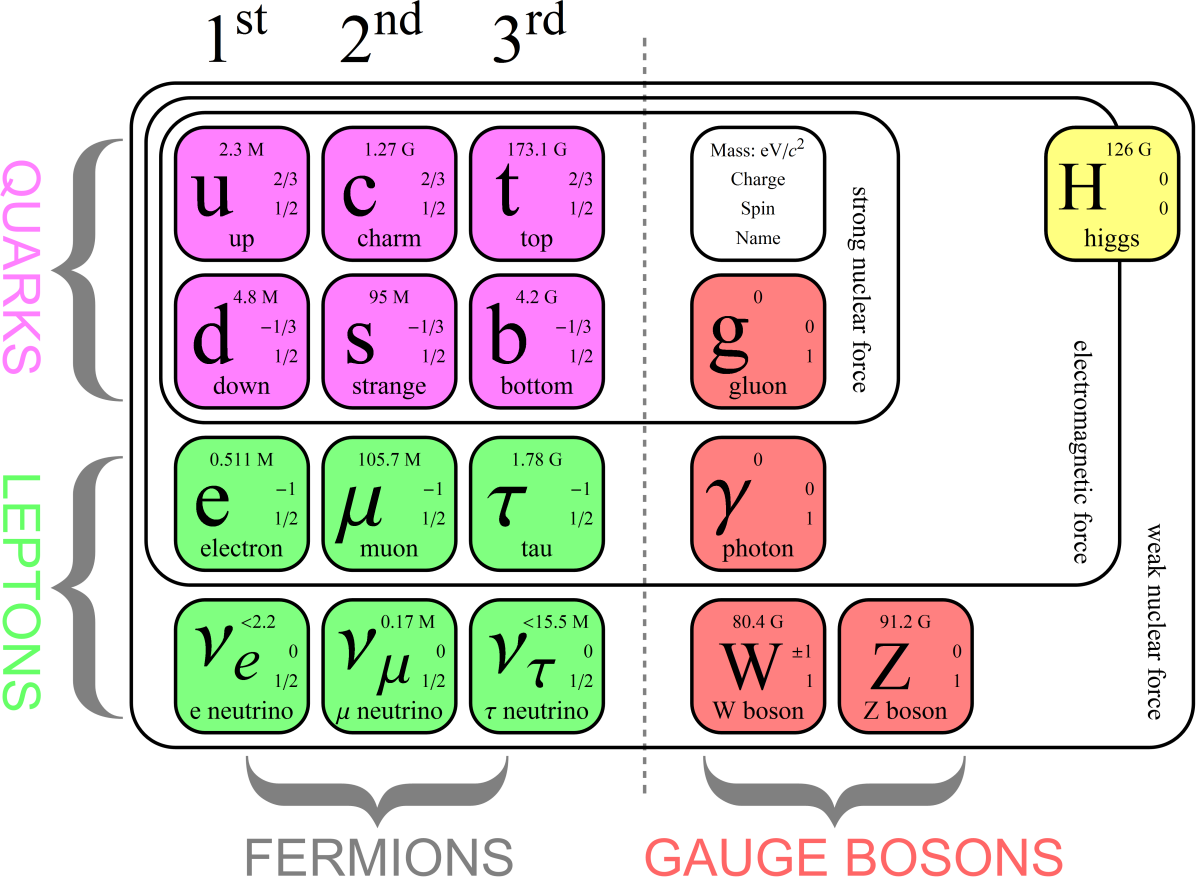}
    \caption{The standard model of particle physics. The figure is taken
    from~\cite{Matic:sm}.}
    \label{fig:sm}
\end{figure}

The fermionic sector of the SM is organized in three families, containing
quark doublets with one up ($\mathrm{u{/}c{/}t}$) and one down
($\mathrm{d{/}s{/}b}$) quark type each and the leptons, gathering the charged
leptons ($\mathrm{e{/}\mu{/}\tau}$) and their associated neutrinos
($\mathrm{\nu_e{/}\nu_\mu{/}\nu_\tau}$). Leptons interact with the electroweak
force only. In particular, the SM neutrinos are massless and carry no
electromagnetic charge and, therefore, cannot be easily experimentally revealed.
Figure~\ref{fig:sm} schematically shows the elementary particles predicted by the SM.
In particular, quarks in the SM lagrangian acquire masses through Yukawa
interactions with the Higgs boson fields. The lagrangian terms for such
interactions are:
\begin{equation}
    \label{eqn:quark-mass-lagrangian}
    -\mathcal{L}_{Y,\mathrm{quark}} =
    Y_{ij}^d \bar{Q}_{Li}^I \phi d_{Rj}^I
    +Y_{ij}^u \bar{Q}_{Li}^I \epsilon \phi^* u_{Rj}^I
    + \mathrm{h.c.}
\end{equation}
where the $Y^{u,d}$ are complex $3\times3$ matrices, $i,j$ are the generation
labels ($u,c,t$ in the case of up-type quarks), $\epsilon$ is the $2\times2$
anti-symmetric tensor. $Q_L^I$, $d_R^I$ and $u_R^I$ are the quark fields in the
weak eigenstate basis, namely $Q_L^I$ are left-handed quark doublets and the
others represent the quark right-handed singlets.

When the Higgs field $\phi$ acquires a non-zero VEV, the
equation~\ref{eqn:quark-mass-lagrangian} can be expanded around that value,
yielding the explicit quark mass terms. However, the physical states are obtained
by diagonalizing the $Y^{u,d}$ matrices by the following bi-unitary
transformations:
\begin{equation}
    Y^f \to M_{\mathrm{diag}}^f = V_L^f Y^f V_R^{\dagger\,f}
\end{equation}
for $f=u,d$ and all the $V$ are unitary matrices. This is possible of course only
if the quark fields transform accordingly. In this new basis, i.e. the mass
basis, the equation~\ref{eqn:quark-mass-lagrangian} is properly diagonalized,
while the charged current weak interactions $W^\pm$ are modified as:
\begin{equation}
    \label{eqn:cmk-cclagrangian}
    \frac{-g}{\sqrt{2}}
    \braces{\bar{u}_L, \bar{c}_L, \bar{t}_L}
    \gamma^\mu W_\mu^+ V_{\mathrm{CKM}}
    \begin{pmatrix}
        d_L \\ s_L \\ b_L
    \end{pmatrix}
    + \mathrm{h.c}
    ,
    \qquad
    V_{\mathrm{CKM}} = V_L^{u}V_L^{\dagger\,d} =
    \begin{pmatrix}
        V_{ud} & V_{us} & V_{ub} \\
        V_{cd} & V_{cs} & V_{cb} \\
        V_{td} & V_{ts} & V_{tb} \\ 
    \end{pmatrix}
\end{equation}

The $V_{\mathrm{CKM}}$ is the Cabibbo-Kobayashi-Maskawa (CKM)
matrix~\cite{Gilman:2000}, which mixes the weak-flavor eigenstates to obtain the
mass ones. The standard parametrization of the CKM matrix~\citep{Chau:1984}
involves three angles $\braces{\theta_{12}, \theta_{13}, \theta_{23}}$ and a
CP-violating phase $\delta$~\cite{Kobayashi:1973}:
\begin{equation}
    \label{eqn:ckm-standard}
    V_{\mathrm{CKM}} =
    \begin{pmatrix}
        1 & 0 & 0 \\
        0 & c_{23} & s_{23} \\
        0 & -s_{23} & c_{23} \\
    \end{pmatrix}
    \begin{pmatrix}
        c_{13} & 0 & s_{13}e^{-i\delta} \\
        0 & 1 & 0 \\
        -s_{13}e^{i\delta} & 0 & c_{13} \\
    \end{pmatrix}
    \begin{pmatrix}
        c_{12} & s_{12} & 0 \\
        -s_{12} & c_{12} & 0 \\
        0 & 0 & 1 \\
    \end{pmatrix}
\end{equation}
where $s_{ij} = \sin\theta_{ij}$ and $c_{ij}=\cos\theta_{ij}$. A review on the
latest experimental measures of the values of the CKM matrix can be found
at~\cite{Workman:2022ckm}.


%% file: chapters/partI/chap1/nu.tex
\section{Neutrino physics}
\label{sec:nu_phys}

\subsection{SM neutrino}

The standard model neutrinos are part of the lepton doublets of $SU(2)_L$:
\begin{equation}
    L_{L\alpha} =
    \begin{pmatrix}
        \nu_\alpha \\ l_{\alpha}
    \end{pmatrix}_L
\end{equation}
where the subscripts $\alpha$ and $L$ indicate the $\alpha{-}$th lepton family the
left-handed projection of the fields obtained acting with the projector
$P_L =\frac{1}{2}(1-\gamma_5)$. We note that within the SM the right-handed
charged leptons are singlets of weak interactions and no right-handed
neutrinos exist. Further, neutrinos do not carry any strong or electromagnetic
charge, but are subject to the following charged-current (CC) and
neutral-current (NC) weak interactions:
\begin{align}
    -\mathcal{L}_{CC} &=
    \frac{g}{\sqrt{2}} \sum_\alpha \bar{\nu}_{L\alpha} \gamma^\mu l_{L\alpha}^- W_\mu^+
    + \mathrm{h.c.}
    \label{eqn:nu-cc} \\
    -\mathcal{L}_{NC} &=
    \frac{g}{2\cos\theta_W} \sum_\alpha \bar{\nu}_{L\alpha} \gamma^\mu \nu_{L\alpha}Z_\mu^0
\end{align}
where $g$ is the weak coupling constant and $\cos\theta_W$ is the Weinberg angle.
In particular, the NC interaction influences the decay width of the
$Z^0$ boson in the SM, which, in turn, is related to the number of measured
active light neutrinos $N_\nu$.

In general, the SM neutrinos are dubbed active neutrinos since they couple
with the SM interactions as seen above. There exist, however, Beyond Standard
Model (BSM) theories that include the possibility to have other neutrinos that
are singlets of the SM gauge group and therefore are called sterile. Both light
(with mass $m_\nu \leq M_Z/2$) and heavy sterile neutrino hypotheses are present
in BSM theories. Experiments at LEP on the measurement of the decay width of
the $Z^0$ boson constrained the $N_\nu$ number~\citep{Schael:2006}:
\begin{equation}
    N_\nu =
    \frac{\Gamma_Z^{\mathrm{invisible}}}{\Gamma(Z\to\nu\bar{\nu})_{\mathrm{SM}}}
    = 2.9840 \pm 0.0082
\end{equation}
This measurement excluded the fourth active light neutrino option.

A neutrino mass term cannot be included in the SM because it would break the
symmetries of the model. In fact, the SM exhibits an accidental global symmetry
besides the gauge $SU(3)_C {\times} SU(2)_L {\times} U(1)_Y$ one:
\begin{equation}
    \mathrm{G^{global}_{SM}} =
    U(1)_B {\times} U(1)_{L_e} {\times} U(1)_{L_\mu} {\times} U(1)_{L_\tau}
\end{equation}
which reflects the conservation of the baryon number $B$ and the leptonic flavor
numbers $L_{e,\mu,\tau}$. Moreover, also the total lepton number
$L=L_e + L_\mu + L_\tau$ is conserved, being the quantum number associated with
the sub-group of $\mathrm{G^{global}_{SM}}$. We show that these symmetries forbid
any neutrino mass term in the SM.

The SSB mechanism allows all the SM fermions to acquire a mass coupling with the
Higgs boson field like in equation~\ref{eqn:quark-mass-lagrangian}. Charged
leptons admit a similar contribution:
\begin{equation}
    \label{eqn:lepton-mass-lagrangian}
    -\mathcal{L}_{Y,\mathrm{lep}} =
    Y^l_{\alpha\beta}\bar{L}_{L\alpha}\phi E_{R\beta}
    + \mathrm{h.c.}
\end{equation}
where $E_{Ri}$ are the right-handed singlets leptons of $SU(2)_L$ and $\phi$
is the Higgs boson field. The $Y_{\alpha\beta}^l$ is the $3\times3$ Yukawa matrix
for leptons analogous to the quark one. Expanding the Higgs field around is VEV
$v$ one obtains terms such as $Y^l_{\alpha\beta} \frac{v}{\sqrt{2}}\bar{l}_{L\alpha}l_{R\beta}$,
yielding mass $m_{\alpha\beta}^l = Y^l_{\alpha\beta} \frac{v}{\sqrt{2}}$ to the leptons. Such a
lagrangian term cannot be written for neutrinos, since the right-handed field is
not available.

In principle, a mass term like $\bar{L}_L L_L^c$ might arise at loop level in
the SM, where $L^c = C\bar{L}^T$ is the charge conjugated field and $C$ is the
charge conjugation matrix satisfying $C\gamma_\mu C^{-1} = -\gamma_\mu^T$.
However, this would break the total lepton number conservation accidental
symmetry of the model, since the term would produce interactions with
$\abs{\Delta L} = 2$. In light of these considerations, the SM neutrinos are, in
fact, massless, opening the problem of consistently describing the neutrino
oscillation mechanism (see the next section~\ref{subsec:nu-oscillation}) while
preserving gauge invariance and renormalizability of the theory.

\subsection{Neutrino oscillation}
\label{subsec:nu-oscillation}

The first intuition of the neutrino oscillation dates back to 1957 by
Pontecorvo~\citep{Pontecorvo:1957cp,Pontecorvo:1957qd}, which advocated a
mechanism similar to the kaon anti-kaon mixing for the mesonium $(\mu^+e^-)$ and
anti-mesonium $(\mu^-e^+)$ oscillation. This matter-antimatter oscillation was
possible only by violating the leptonic number conservation. Some years later, the
first quantitative neutrino mixing model was proposed in~\cite{Maki:1962} and
further refined in~\cite{Pontecorvo:1967,Gribov:1968}.

The idea is that the left-handed flavor $\nu_{(e,\mu,\tau)L}$ and mass
$\nu_{(1,2,3)L}$ eigenstates of the active neutrinos are linked by a unitary
transformation, represented by the so-called Pontecorvo-Maki-Nakagawa-Sakata
(PMNS) matrix $U$:
\begin{equation}
    \begin{pmatrix}
        \nu_e \\ \nu_\mu \\ \nu_\tau
    \end{pmatrix}_L = 
    \begin{pmatrix}
        U_{e1} & U_{e2} & U_{e3} \\
        U_{\mu1} & U_{\mu2} & U_{\mu3} \\
        U_{\tau1} & U_{\tau2} & U_{\tau3} \\
    \end{pmatrix}
    \begin{pmatrix}
        \nu_1 \\ \nu_2 \\ \nu_3
    \end{pmatrix}_L
\end{equation}
Employing greek indices for the flavor states and latin ones for the mass states,
the equation above can be written in short form:
\begin{equation}
    \nu_{\alpha L}(x) = \sum_i U_{\alpha i} \nu_{iL}(x)
\end{equation}
The matrix $U$ satisfies the unitary conditions:
\begin{equation}
    (UU^\dagger)_{\alpha \beta}
    = \sum_i U_{\alpha i} U^*_{\beta i}
    = \delta_{\alpha\beta},
    \qquad
    (U^\dagger U)_{ij}
    = \sum_\alpha U^*_{\alpha i} U_{\alpha j}
    = \delta_{ij},
\end{equation}

A $3\times3$ unitary matrix can be parametrized by $3$ angles and $6$ phases.
However, it is possible to transform the neutrino and charged lepton fields with
an unobservable global phase to absorb $5$ out of the $6$ PMNS phases. The
aforementioned transformation can be made manifest considering the lagrangian
term of the CC interaction, equation~\ref{eqn:nu-cc}, which is left invariant
by the following:
\begin{equation}
    \label{eqn:phase-transformation}
    l_{\alpha}(x) \to e^{i\phi_\alpha}l_\alpha(x), \quad
    \nu_i(x) \to e^{i\phi_i}\nu_i(x), \quad
    U_{\alpha i} \to e^{i\phi_{\alpha i}} U_{\alpha i}
\end{equation}
where $\phi_{\alpha i} = \phi_\alpha - \phi_i$.
There exist indeed only $5$ independent phase differences $\phi_{e1}$,
$\phi_{e2}$, $\phi_{e3}$, $\phi_{\mu1}$, $\phi_{\tau1}$ among the $9$ possible
ones.

This is only true when the neutrinos are described by Dirac fields, while it is
not the case if neutrinos are Majorana. Indeed the phase transformation introduced
by equation~\ref{eqn:phase-transformation} would yield a complex mass for the
Majorana fields:
\begin{equation}
    \frac{1}{2}m_i \nu^T_{iL} C \nu_{iL} \to
    \frac{1}{2}m_i e^{2i\phi_i} \, \nu^T_{iL} C \nu_{iL}
\end{equation}
Therefore, in this case, it is not possible to redefine the $\nu_i$ fields to
absorb the phases and we have, in fact, $3$ phases left to describe the PMNS
matrix. We report a summary of the number of independent parameters needed to
describe the PMNS matrix for the general case of $N$ lepton flavors in
table~\ref{tab:pmns-count}.
\begingroup
\renewcommand{\arraystretch}{1.2}
\begin{table}
    \centering
    \scriptsize
    \begin{tabular}{|c|c|c|}
        \hline\noalign{\smallskip}
        Count & Majorana & Dirac \\
        \noalign{\smallskip}\hline\noalign{\smallskip}
        Rescalable fields & $N$ & $2N-1$\\
        Angles $\theta_i$ & $\frac{1}{2}N(N-1)$ & $\frac{1}{2}N(N-1)$ \\ 
        Phases $\delta_i$ & $\frac{1}{2}N(N-1)$ & $\frac{1}{2}(N-1)(N-1)$ \\ 
        Total free parameters & $N(N-1)$ & $(N-1)^2$\\
        \noalign{\smallskip}\hline
    \end{tabular}
    \caption{The number of mixing angles and phases needed to parametrize the
    general unitary $N\times N$ PMNS matrix both in the case of Dirac and
    Majorana neutrino fields.}
    \label{tab:pmns-count}
\end{table}
\endgroup

The standard parametrization of the PMNS matrix for $3$ flavor neutrinos is
similar to the CKM one, see equation~\ref{eqn:ckm-standard}, except for the
presence of an extra $U(1)$ matrix $P$ in the Majorana case:
\begin{equation}
    \label{eqn:pmns}
    U =
    \begin{pmatrix}
        1 & 0 & 0 \\
        0 & c_{23} & s_{23} \\
        0 & -s_{23} & c_{23} \\
    \end{pmatrix}
    \begin{pmatrix}
        c_{13} & 0 & s_{13}e^{-i\delta_{CP}} \\
        0 & 1 & 0 \\
        -s_{13}e^{i\delta_{CP}} & 0 & c_{13} \\
    \end{pmatrix}
    \begin{pmatrix}
        c_{12} & s_{12} & 0 \\
        -s_{12} & c_{12} & 0 \\
        0 & 0 & 1 \\
    \end{pmatrix}
    P
\end{equation}
where $c_{ij} = \cos\theta_{ij}$, $s_{ij} = \cos\theta_{ij}$ with the mixing
angles $\theta_{ij}$ lying in the first quadrant $[0,\pi/2]$ and
$\delta_{CP} \in [0,2\pi)$ is the Dirac phase responsible for the violation of
the CP symmetry.
Multiple representations of the $P$ matrix are present in the literature and all
depend on two Majorana phases, which can be taken in the interval $[0,\pi)$:
\begin{equation}
    P =
    \begin{pmatrix}
        e^{i\alpha_1} & 0 & 0 \\
        0 & e^{i\alpha_2} & 0 \\
        0 & 0 & 1\\
    \end{pmatrix}, \quad
    \begin{pmatrix}
        1 & 0 & 0\\
        0 & e^{i\phi_2} & 0 \\
        0 & 0 & e^{i(\phi_3+\delta_{CP})} \\
    \end{pmatrix}, \quad
    \begin{pmatrix}
        e^{i\rho} & 0 & 0 \\
        0 & 1 & 0\\
        0 & 0 & e^{i\sigma}\\
    \end{pmatrix}
\end{equation}
Majorana masses do not have any influence on the neutrino oscillation process we
are going to discuss, but enter in the calculation of lepton violating processes
such as the neutrinoless double beta decay
$0\nu\beta\beta$~\cite{Vergados:2012,DellOro:2016}.

In the following, we introduce the description of the neutrino oscillation \linebreak
phenomenon, which will be described by the transition probability
$P(\nu_\alpha\to\nu_\beta)$ of a certain neutrino with flavor $\alpha$ to become
a neutrino with flavor $\beta$. We will show that the oscillation mechanism 
requires massive neutrinos to be observed. Therefore we will consider these
particles to have a small mass $m_i$ for $i=1,2,3$ and always move at
ultra-relativistic speed as it happens in all realistic experimental conditions.
We will present the main techniques to design neutrino mass terms for BSM
lagrangians in the next section~\ref{subsec:nu-masses}.

The ideal neutrino oscillation experiment in vacuum involves a three-step process:
the production of a neutrino of pure flavor state $\alpha$, the propagation of
the particle along a distance $L$ and the detection of the final flavor $\beta$
at the end of the apparatus setup. Let us consider, then, an initial state
produced by some CC process, (e.g charged pion decay $\pi^+ \to \mu^+ \nu_\mu$):
\begin{equation}
    \ket{\nu(t=0)} = \ket{\nu_\alpha} = \sum_i U^*_{\alpha i} \ket{\nu_i}
\end{equation}
The propagation through the vacuum is described by the evolution operator
obtained by the exponentiation of the Hamiltonian and leads to the following
state at time $t$:
\begin{equation}
    \ket{v(t)} = \sum_i U^*_{\alpha i} e^{-iE_i t} \ket{v_i} =
    \sum_i U^*_{\alpha i} e^{-iE_i t} \sum_\beta U_{\beta i} \ket{v_\beta}
\end{equation}
The final probability amplitude is obtained by acting with the state
$\bra{v_\beta}$ and the corresponding probability by applying the modulus square
to the bra-ket notation:
\begin{equation}
    \label{eqn:nu-oscillation0}
    P(\nu_\alpha \to \nu_\beta) = \abs{\langle\nu_\beta \ket{v(t)} }^2 =
    \abs{\sum_i U_{\beta i} U^*_{\alpha i} e^{-iE_i t}}^2 = 
    \sum_{ij} (J_{\alpha\beta})_{ij} e^{-i(E_i-E_j)t}
\end{equation}
where we have defined the self-adjoint matrix
$(J_{\alpha\beta})_{ij} = U_{\alpha_i} U^*_{\beta i} U^*_{\alpha j} U_{\beta j}$.
It is clear from this definition that the $J_{\alpha\beta}$ matrices do not depend
on the Majorana phases, since the $P$ factor of equation~\ref{eqn:pmns} cancel
out in the product.

The energy eigenvalues $E_i$ in the ultra-relativistic limit are dominated by
the particle kinetic energy and can be expanded for $m_i^2/\mathbf{p}^2 \ll 1$:
\begin{equation}
    E_i = \sqrt{\mathbf{p}^2 + m_i^2} = \abs{\mathbf{p}} + \frac{m_i^2}{2E},
    \qquad \mathrm{with} \; E \simeq \abs{\mathbf{p}}
\end{equation}
Inserting this approximation into equation~\ref{eqn:nu-oscillation0} yields:
\begin{equation}
    P(\nu_\alpha \to \nu_\beta) =
    \sum_{ij} (J_{\alpha\beta})_{ij}
    \exp\braces{i\frac{\Delta m^2_{ji}L}{2E}}
\end{equation}
where $L = ct$ and $\Delta m^2_{ji} = m^2_j - m^2_i$ is the squared mass
difference between the neutrino mass eigenstates. After some algebraic
manipulations, we obtain:
\begin{equation}
    \label{eqn:nu-oscillation}
    P(\nu_\alpha \to \nu_\beta) =
    \delta_{\alpha\beta}
    -4 \sum_{i<j}\Re(J_{\alpha\beta})_{ij}
        \sin^2\braces{\frac{\Delta m^2_{ij}L}{4E}}
    +2 \sum_{i<j}\Im(J_{\alpha\beta})_{ij}
        \sin\braces{\frac{\Delta m^2_{ij}L}{2E}}
\end{equation}
This result makes evident that in order to have neutrino oscillation, we should
not have a degenerate mass spectrum and a trivial flavor mixing matrix, namely
$\Delta m^2_{ij} \neq 0$ and $U \neq \mathbb{I} $, respectively. The oscillation
probability $P(\nu_\alpha \to \nu_\beta)$ depends on the four parameters of the
PMNS matrix ($\theta_{12}$, $\theta_{13}$, $\theta_{23}$ and $\delta_{CP}$)
and the two independent squared mass differences ($\Delta m^2_{21}$,
$\Delta m^2_{31}$).

\newlength{\arrow}
\settowidth{\arrow}{\scriptsize$\;CPT\;$}
\newcommand*{\myrightarrow}[1]{\xrightarrow{\mathmakebox[\arrow]{#1}}}
The violation of the CP symmetry is given by the fact that the probabilities
for the particles are not the same as for the anti-particles. We define the
CP asymmetry parameter as\footnote{
    The C, P, T discrete symmetries transform the $P(\nu_\alpha \to \nu_\beta)$
    as follows:
    \begin{equation*}
        \begin{split}
            P(\nu_\alpha \to \nu_\beta) &\,\myrightarrow{CP}\,
            P(\bar\nu_\alpha \to \bar\nu_\beta) \\
            P(\nu_\alpha \to \nu_\beta) &\,\myrightarrow{T}\,
            P(\nu_\beta \to \nu_\alpha) \\
            P(\nu_\alpha \to \nu_\beta) &\,\myrightarrow{CPT}\,
            P(\bar\nu_\beta \to \bar\nu_\alpha)\\
        \end{split}
    \end{equation*}
    Note that the exchange from neutrinos to  anti-neutrinos is achieved by CP
    rather than C, because the latter would transform $\nu_L\to \bar\nu_L$
    which, in fact, do not exist in the SM.
}:
\begin{equation}
    \Delta P_{\alpha\beta} =
    P(\nu_\alpha \to \nu_\beta)
    - P(\bar{\nu}_\alpha \to \bar{\nu}_\beta)
\end{equation}
As a consequence of equation~\ref{eqn:nu-oscillation}, the investigation of the
CP violation should be done in the appearance channels ($\beta\neq\alpha$), rather
than in disappearance ones ($\beta = \alpha$). The latter case, indeed, yields
the same probability for neutrino and anti-neutrinos since the PMNS matrix appear
in the formula only in the combination $\abs{U_{\alpha i} U^*_{\alpha i}}$, which
is real. Constraints on the CP violation in neutrino oscillation can be derived
expressing the asymmetry parameter in the following
form~\citep{Bilenky:1980,Barger:1980}:
\begin{equation}
    \Delta P_{\alpha\beta} =
        \pm 16 J
        \sin\braces{\frac{\Delta m^2_{21}L}{4E}}
        \sin\braces{\frac{\Delta m^2_{31}L}{4E}}
        \sin\braces{\frac{\Delta m^2_{32}L}{4E}}
\end{equation}
where the sign of the right-hand side is $+$ for even permutations of
$(e,\mu,\tau)$, $-$ for the odd ones instead, and the quantity $J$ is called
the Jarlskog invariant and equals to:
\begin{equation}
    J = \Im (J_{e\mu})_{12} =
    \frac{1}{8} \cos\theta_{13}
    \sin2\theta_{12} \sin2\theta_{13} \sin2\theta_{23}
    \sin\delta_\mathrm{CP}
\end{equation}
Interestingly, the form of the asymmetry parameter tells that it is universal
and do not depend on the particular observation channel:
$\Delta P_{e\mu} = \Delta P_{\tau e} = \Delta P_{\mu\tau}$ .
The asymmetry parameter product is non-zero only if all of its factors are
non-zero. Therefore, to observe CP violation, the mass spectrum should be
non-degenerate (as it should have been to observe the oscillation) and, further,
the parameters of the PMNS matrix should obey:
\begin{equation}
    \theta_{ij} \neq 0, \qquad \delta_\mathrm{CP} \neq 0,\pi
\end{equation}

Equation~\ref{eqn:nu-oscillation} is valid in general for all numbers of neutrino
flavors $N$. In fact, in many situations, such as in the oscillation of solar
and atmospheric neutrinos, it is possible to neglect subleading terms and CP
violation effects and consider just a two-flavor approximation. In these cases,
the PMNS matrix reduces to a $2\times2$ rotation by an angle $\theta$ and there 
is just one squared mass difference $\Delta m^2$ to take into account. The
appearance probability becomes:
\begin{equation}
    P(\nu_\alpha \to \nu_\beta) =
    P(\bar{\nu}_\alpha \to \bar{\nu}_\beta) =
    \sin^2 \theta \sin^2 \braces{\frac{\Delta m^2_{ij}L}{4E}}
\end{equation}

The oscillation probability for neutrino appearance,
equation~\ref{eqn:nu-oscillation}, describes the \linebreak probability that a neutrino
$\nu_\alpha$ produced by some source A is revealed after propagation in the
vacuum for a distance L by a detector B as $\nu\beta$. This ideal scheme, however,
represents only an approximation of the real case. First, source A never
produces a beam of monoenergetic neutrinos, but rather a flux of particles with
an energy spectrum $\Phi(E)$. Second, the propagation over long distances does
not happen in vacuum, therefore, especially in the case of layers of matter with
high density, a proper description of the propagation should be implemented
(see~\cite{Giganti:2017,Workman:2022nu} for a detailed description of this topic).
Third, detector B does not have infinite energy resolution but rather
presents an efficiency $\epsilon(E)$ to reveal a certain process with
cross-section $\sigma(E)$.

In this context, the final result for the appearance probability, given in
equation~\ref{eqn:nu-oscillation}, is an expected value given by the convolution
of the theoretical probability $P_{\alpha\beta}(E)$ with the detection
ingredients named above:
\begin{equation}
    \begin{split}
        \lrangles{P_{\alpha\beta}} &= 
        \frac{
            \int d\Phi(E)\,\sigma(E) P_{\alpha\beta}(E) \epsilon(E)
        }{
                \int d\Phi(E) \,\sigma_\mathrm{tot}(E) \epsilon(E)
        } \\
        &= \delta_{\alpha\beta}
        -4 \sum_{i<j}\Re(J_{\alpha\beta})_{ij}
            \lrangles{\sin^2X_{ij}}
        +2 \sum_{i<j}\Im(J_{\alpha\beta})_{ij}
            \lrangles{\sin\braces2 X_{ij}}
    \end{split}
\end{equation}
where the $X_{ij}$ quantity can be expressed by restoring the SI units of
measure:
\begin{equation}
    X_{ij}
    = \frac{\Delta m^2_{ij}L}{4E}
    = 1.267 \, \frac{\Delta m^2_{ij}}{\SI{}{\eV^2}} \frac{L/E}{\SI{}{\m/\MeV}}
\end{equation}

The above equations highlight that the probability oscillates along the distance
traveled by the neutrinos with period: $4\pi E / \abs{\Delta m^2_{ij}}$. Also,
the $\sin$ functions are averaged, which brings two alternative behaviors
discriminated by the value of the ratio between energy and distance: if
$(E/L) \ll \abs{\Delta m^2_{ij}}$, the argument of the oscillating functions is
small and the probability goes to zero, instead if
$(E/L) \geq \abs{\Delta m^2_{ij}}$, the function is rapidly oscillating and it
gets averaged out to $\lrangles{\sin^2 X_{ij}} = 1/2$. As a consequence, the
experiments looking for measuring the neutrino mass difference will be only
sensible to some mass intervals based on their experimental setup.
Table~\ref{tab:nu-mass-exp-ranges} collects the neutrino energy and the baseline
distance parameters for the experiments in neutrino oscillation. Each of these
kinds of experiments will be sensible to just a portion of the whole mass
difference spectrum.
\begin{table}
    \scriptsize
    \centering
    \begin{tabular}{llccc}
\hline\noalign{\smallskip}
Experiment & Baseline & $L (\SI{}{\m})$ & $E (\SI{}{\MeV})$ & $\abs{\Delta m^2_{ij}} (\SI{}{\eV})$ \\
\noalign{\smallskip}\hline\noalign{\smallskip}
Solar & & $10^{10}$ & 1 & $10^{-10}$ \\
\noalign{\smallskip}\hline\noalign{\smallskip}
Atmospheric & & $10^4-10^7$ &  $10^2-10^5$ & $10^{-1}-10^{-4}$ \\
\noalign{\smallskip}\hline\noalign{\smallskip}
\multirow{2}*{Reactor} & VSBL-SBL-MBL & $10-10^3$ & \multirow{2}*{1} & $1-10^{-3}$ \\
                       & LBL & $10^4-10^5$ & & $10^{-4}-10^{-3}$ \\
\noalign{\smallskip}\hline\noalign{\smallskip}
\multirow{2}*{Accelerator} & SBL & $10^2$ & \multirow{2}*{$10^3-10^4$} & $>0.1$ \\
                           & LBL & $10^5-10^6$ & & $10^{-2}-10^{-3}$ \\
\noalign{\smallskip}\hline
    \end{tabular}
    \caption{The list of experiments that are sensitive to certain neutrino
    mass difference ranges: characteristic values for $L$ and $E$ and relative
    ranges of $\abs{\Delta m^2_{ij}}$. The experiments are categorized as Very
    Short Baseline (VBSL), Short Baseline (SBL), Medium Baseline (MBL) and Long
    Baseline (LBL) depending on the neutrino source-detection distance. All the
    numbers neglect matter effects on the oscillation probability equation.
    The table is taken from~\cite{Workman:2022nu}.}
    \label{tab:nu-mass-exp-ranges}
\end{table}

So far, most of the PMNS matrix~\ref{eqn:pmns} parameters have been measured at
$3\sigma$ precision by several analyses. The better-known parameters are, indeed,
$\theta_{12}$ and $\theta_{13}$, while the remaining $\theta_{23}$ and the CP
violating phase $\delta_\mathrm{CP}$ involve major uncertainties. In particular,
the octant of the $\theta_{23}$ angle has yet to be discovered with enough
statistical precision, namely greater than $3\sigma$.
Furthermore, CP conservation is still a possible scenario, given that the
hypothesis $\delta_\mathrm{CP} = \pi$ is discouraged, but cannot be completely
rejected with a high enough confidence level.

Regarding the other parameters governing the neutrino oscillation, the square
mass difference $\Delta m^2_{21}$ is known and is positive, yielding $m_2 > m_1$.
On the other hand, the known precision measurement of the $\Delta m^2_{32}$
quantity concerns its absolute value only. Nothing has been discovered yet
about the sign of $\Delta m^2_{32}$. This leads to two mass schemes, named
the normal ordering (NO):
\begin{equation}
    m_1 < m_2 < m_3
\end{equation}
and inverted ordering (IO):
\begin{equation}
    m_3 < m_1 < m_2
\end{equation}

Observations signal less agreement with experimental data of the inverted mass
ordering compared to the normal ordering. This is measured quantitatively
with  an excess in the $\chi^2$ test for the IO best fit:
$\Delta\chi^2 = \chi^2_\mathrm{IO} - \chi^2_\mathrm{NO} > 0$.
We collect the best-known measurements of the oscillation parameters in the $3$
neutrino scheme in table~\ref{tab:pmns-params}.
\begingroup
\setlength{\tabcolsep}{0.1pt}
\begin{table}
    \caption{Global fits for the $3\nu$ parameters. The two mass-ordering
    schemes are presented separately. The excess in the $\chi^2$ by the IO is
    given by the $\Delta\chi^2$ quantity for each analysis. All the results
    include the contribution by the Super-Kamiokande experiment~\cite{Abe:2017},
    which provided only the tabulated $\chi^2$ test values for their measurements.
    The table is taken from~\cite{Workman:2022nu}.
    }
    \label{tab:pmns-params}
    \scriptsize
    \centering
    \begin{tabular}{lcccccc}
\hline\noalign{\smallskip}
& \multicolumn{2}{c}{Ref~\cite{nufit:2021}} & \multicolumn{2}{c}{Ref~\cite{Capozzi:2018}} & \multicolumn{2}{c}{Ref~\cite{deSalas:2017}}\\
Param & bfp $\pm1\sigma$ & $3\sigma$ range & bfp $\pm1\sigma$ & $3\sigma$ range & bfp $\pm1\sigma$ & $3\sigma$ range \\
\cmidrule(r){1-1}\cmidrule(lr){2-3}\cmidrule(lr){4-5}\cmidrule(l){6-7}
NO & \multicolumn{2}{c}{Best Fit Ordering} & \multicolumn{2}{c}{Best Fit Ordering} & \multicolumn{2}{c}{Best Fit Ordering} \\
\noalign{\smallskip}\hline\noalign{\smallskip}
$\frac{sin^2\theta_{12}}{10^{-1}}$            & $3.10^{+0.13}_{-0.12}$     & $2.75\to3.50$   & $3.04^{+0.14}_{-0.13}$     & $2.65 \to3.46 $ & $3.20^{+0.20}_{-0.16}$   & $2.73 \to3.79 $ \\
$\theta_{12}/^\circ$                          & $33.82^{+0.78}_{-0.76}$    & $31.61\to36.27$ & $33.46^{+0.87}_{-0.88}$    & $30.98\to36.03$ & $34.5^{+1.2}_{-1.0}$     & $31.5 \to38.0 $ \\
$\frac{sin^2\theta_{23}}{10^{-1}}$            & $5.63^{+0.18}_{-0.24}$     & $4.33\to6.09$   & $5.51^{+0.19}_{-0.80}$     & $4.30 \to6.02 $ & $5.47^{+0.20}_{-0.30}$   & $4.45 \to5.99 $ \\
$\theta_{23}/^\circ$                          & $48.6 ^{+1.0 }_{-1.4 }$    & $41.1 \to51.3 $ & $47.9^{+1.1 }_{-4.0 }$     & $41.0 \to50.9 $ & $47.7^{+1.2}_{-1.7}$     & $41.8 \to50.7 $ \\
$\frac{sin^2\theta_{13}}{10^{-2}}$            & $2.237^{+0.066}_{-0.065}$  & $2.044\to2.435$ & $2.14^{+0.09}_{-0.07}$     & $1.90 \to2.39 $ & $2.160^{+0.083}_{-0.69}$ & $1.96 \to2.41 $ \\
$\theta_{13}/^\circ$                          & $8.60^{+0.13}_{-0.13}$     & $8.22 \to8.98 $ & $8.41^{+0.18}_{-0.14}$     & $7.9  \to8.9  $ & $8.45^{+0.16}_{-0.14}$   & $8.0  \to8.9  $ \\
$\delta_\mathrm{CP}/^\circ$                   & $221^{+39   }_{-28   }$    & $144  \to357  $ & $238^{+41   }_{-33   }$    & $149  \to358  $ & $218^{+38  }_{-27  }$    & $157  \to349  $ \\
$\frac{\Delta m^2_{21}}{10^{-5}\SI{}{\eV^2}}$ & $7.39 ^{+0.21 }_{-0.20 }$  & $6.79 \to8.01 $ & $7.34^{+0.17 }_{-0.14 }$   & $6.92 \to7.91 $ & $7.55^{+0.20}_{-0.16}$   & $7.05 \to8.24 $ \\
\noalign{\smallskip}
$\frac{\Delta m^2_{32}}{10^{-3}\SI{}{\eV^2}}$ & $2.454^{+0.029}_{-0.031}$  & $2.362\to2.544$ & $2.419^{+0.035}_{-0.032}$  & $2.319\to2.521$ & $2.424^{+0.03}_{-0.03}$  & $2.334\to2.524$ \\
\noalign{\smallskip}\hline\noalign{\smallskip}
IO & \multicolumn{2}{c}{$\Delta \chi^2 = 10.4$} & \multicolumn{2}{c}{$\Delta \chi^2 = 9.5$} & \multicolumn{2}{c}{$\Delta \chi^2 = 11.7$} \\
\noalign{\smallskip}\hline\noalign{\smallskip}
$\frac{sin^2\theta_{12}}{10^{-1}}$            & $3.10^{+0.13}_{-0.12}$     & $2.75\to3.50$     & $3.03^{+0.14}_{-0.13}$      & $2.65 \to3.45 $   & $3.20^{+0.20}_{-0.16}$    & $2.73 \to3.79 $ \\
$\theta_{12}/^\circ$                          & $33.82^{+0.78}_{-0.75}$    & $31.62\to36.27$   & $33.40^{+0.87}_{-0.81}$     & $30.92\to35.97$   & $34.5^{+1.2}_{-1.0}$      & $31.5 \to38.0 $ \\
$\frac{sin^2\theta_{23}}{10^{-1}}$            & $5.65^{+0.17}_{-0.22}$     & $4.36\to6.10$     & $5.57^{+0.17}_{-0.24}$      & $4.44 \to6.03 $   & $5.51^{+0.18}_{-0.30}$    & $4.53 \to5.98 $ \\
$\theta_{23}/^\circ$                          & $48.8 ^{+1.0 }_{-1.2 }$    & $41.4 \to51.3 $   & $48.2^{+1.0 }_{-1.4 }$      & $41.8 \to50.9 $   & $47.9^{+1.0}_{-1.7}$      & $42.3 \to50.7 $ \\
$\frac{sin^2\theta_{13}}{10^{-2}}$            & $2.259^{+0.065}_{-0.065}$  & $2.064\to2.457$   & $2.18^{+0.08}_{-0.07}$      & $1.95 \to2.43 $   & $2.220^{+0.074}_{-0.076}$ & $1.99 \to2.44 $ \\
$\theta_{13}/^\circ$                          & $8.64^{+0.12}_{-0.13}$     & $8.26 \to9.02 $   & $8.49^{+0.15}_{-0.14}$      & $8.0  \to9.0  $   & $8.53^{+0.14}_{-0.15}$    & $8.1  \to8.9  $ \\
$\delta_\mathrm{CP}/^\circ$                   & $282^{+23}_{-25}$          & $205  \to348  $   & $247^{+26   }_{-27   }$     & $193  \to346  $   & $281^{+23  }_{-27  }$     & $202  \to349  $ \\
$\frac{\Delta m^2_{21}}{10^{-5}\SI{}{\eV^2}}$ & $7.39 ^{+0.21 }_{-0.20 }$  & $6.79 \to8.01 $   & $7.34^{+0.17 }_{-0.14 }$    & $6.92 \to7.91 $   & $7.55^{+0.20}_{-0.16}$    & $7.05 \to8.24 $ \\
\noalign{\smallskip}
$\frac{\Delta m^2_{32}}{10^{-3}\SI{}{\eV^2}}$ & $-2.510^{+0.030}_{-0.031}$ & $-2.601\to-2.419$ & $-2.478^{+0.035}_{-0.033}$  & $-2.577\to-2.375$ & $-2.50^{+0.04}_{-0.03}$   & $-2.59\to-2.39$ \\
\noalign{\smallskip}\hline
    \end{tabular}
\end{table}
\endgroup

\subsection{Neutrino masses}
\label{subsec:nu-masses}

In this paragraph, we present the main techniques exploited to extend the SM
and give mass to neutrinos. The crucial point is to introduce new particles that
behave like the neutrinos but are singlets of the SM interactions, these are
called sterile neutrinos $\nu_{si}$ for $i=1,\dots,m$. These new particles do
not break the fundamental properties of the SM, namely, gauge symmetry
and renormalizability. It is possible, then, to extend the SM building two kinds
of mass terms:
\begin{equation}
    \label{eqn:nu-bsm-terms}
    -\mathcal{L}_{M_\nu} =
    (M_D)_{ij} \bar\nu_{si} \nu_{Lj}
    + \frac{1}{2}(M_N)_{ij}\bar\nu_{si}\nu^c_{sj}
    + \mathrm{h.c.}
\end{equation}
where the $\nu^c_{sj}$ field is the charge conjugate field for the $j{-}$th
sterile neutrino and $\nu_{Lj}$ is the $i{-}$th SM left-handed neutrino. We
designed two complex mass matrices: $M_D$ of dimension $m\times3$ and $M_N$ of
dimension $m\times m$. The gauge symmetry of the SM is indeed respected, since
the sterile neutrino couples only with the active neutrinos and do not contribute
to any interaction with the SM gauge bosons.

The two lagrangian contributions just written are a Dirac-like term that can arise
from an SSB of a Yukawa interaction with the Higgs boson, such as
$Y_{ij}^\nu \bar\nu_{si}\phi^\dagger L_{Lj}$, and a Majorana term that violates
the total lepton number conservation but is allowed since the $\nu_{Si}$ do not
carry any additional conserved charge.

The equation~\ref{eqn:nu-bsm-terms} can be formatted in a compact shape
introducing a vector of $3+m$ neutrinos ${\bm\nu} = ({\bm\nu}_L, {\bm\nu}_s^c)$:
\begin{equation}
    \label{eqn:nu-bsm-compact}
    -\mathcal{L}_{M_\nu} = \frac{1}{2}
    \begin{pmatrix}
        \bar{\bm\nu}_L^c & \bar{\bm\nu}_s
    \end{pmatrix}
    \begin{pmatrix}
        0 & M_D^T \\
        M_D & M_N
    \end{pmatrix}
    \begin{pmatrix}
        {\bm\nu}_L \\ {\bm\nu}_s^c
    \end{pmatrix}
    + \mathrm{h.c.}
    =  \bar{\bm\nu}^c M_\nu {\bm\nu} + \mathrm{h.c.}
\end{equation}
The resulting matrix $M_\nu$ is symmetric and can be diagonalized to retrieve
the mass eigenvalues $(m_1,m_2,m_3,\dots,m_{3+m})$ with the transformation
induced by a unitary mixing matrix $V^\nu$:
\begin{equation}
    {\bm\nu} \to {\bm\nu}_{\mathrm{mass}} = (V^\nu)^\dagger {\bm\nu},
    \qquad
    M_\nu \to \mathrm{diag}(m_k) = (V^\nu)^T M_\nu V^\nu
    \quad
    \mathrm{for}\;k=1,\dots 3+m
\end{equation}
The neat effect on equation~\ref{eqn:nu-bsm-compact} is:
\begin{equation}
    \label{eqn:nu-bsm}
    \begin{split}
        -\mathcal{L}_{M_\nu} &=
        \frac{1}{2} \sum_{k=1}^{3+m} m_k \sbraces{
            (\bar{\bm\nu}^c_\mathrm{mass})_k ({\bm\nu}_\mathrm{mass})_k
            + (\bar{\bm\nu}_\mathrm{mass})_k ({\bm\nu}^c_\mathrm{mass})_k
        } \\
        &= \frac{1}{2}\sum_{k=1}^{3+m} m_k (\bar{\bm\nu}_M)_k ({\bm\nu}_M)_k
    \end{split}
\end{equation}
where we have defined\footnote{The charge conjugation matrix acts on spinors like:
\begin{equation*}
    \begin{split}
        \Psi^c \coloneqq C^{-1}\Psi C  = C \bar{\Psi}^T\\
        \bar{\Psi}^c \coloneqq C^{-1}\bar{\Psi} C  = \Psi^T C
    \end{split}
\end{equation*}
Hence the bilinears $(\bar{\bm\nu}_\mathrm{mass})_k ({\bm\nu}_\mathrm{mass})_k
+ (\bar{\bm\nu}^c_\mathrm{mass})_k ({\bm\nu}^c_\mathrm{mass})_k$ cancel in
equation~\ref{eqn:nu-bsm}.
}:
\begin{equation}
    \label{eqn:nu-majorana}
    {\bm\nu}_M
    = {\bm\nu}_\mathrm{mass} + {\bm\nu}^c_\mathrm{mass}
    = V^{\nu\dagger}{\bm\nu} + (V^{\nu\dagger}{\bm\nu})^c
\end{equation}
It is easy to check that the fields in ${\bm\nu}_M$ are Majorana neutrinos since
they satisfy the condition ${\bm\nu}_M = {\bm\nu}_M^c$. The Majorana condition
imposes a further constraint on the spinor representation of the fields and acts
halving the number of independent complex components of the spinor from four
(Dirac spinors) to two (Majorana spinors). The situation is analogous to the
scalar case, where a two real-component complex field $\phi$ reduces to a single
real scalar field imposing the reality condition $\phi = \phi^*$.

Equation~\ref{eqn:nu-majorana} can be inverted to obtain the $3$ left-handed
active neutrinos of the SM:
\begin{equation}
    \nu_{Li} = P_L \sum_{j=1}^{3+m} V^\nu_{ij} ({\bm\nu}_M)_j
    \quad \mathrm{for}\;i = 1,2,3
\end{equation}
where $P_L$ is the left chiral projector as usual.

We presented the generic workflow that allows extending the SM to include
neutrino masses. We see that the neutrino mass terms arise from the inclusion of
a number $m$ of sterile neutrinos. In this context, the SM left-handed neutrinos
are retrieved with a superposition of $3+m$ Majorana neutrino fields by a mixing
matrix $V^\nu$ of rank $3+m$. In the following paragraphs, we will present some
examples of practical realizations of this general framework.

\paragraph{Dirac neutrinos}

Setting $M_N = 0$ in equation~\ref{eqn:nu-bsm-compact} equals to exclude 
Majorana mass terms from the model lagrangian~\ref{eqn:nu-bsm-terms}, namely
to impose lepton number conservation. In the particular case of $m=3$, we have a
sterile partner for each neutrino in the SM that can be seen as the missing
right-handed component of a four-spinor Dirac field for the neutrino.

The mass matrix $M_\nu$ can be diagonalized with the following transformations:
\begin{equation}
    \nu_{Li} \to (V_L^\nu)_{ij} \nu_{Lj}, \qquad 
    \nu_{si} \to (V_R^\nu)_{ij} \nu_{sj}, \qquad
    M_D \to \mathrm{diag}(m_1,m_2,m_3) = V_R^{\nu\dagger}M_D V_L^\nu
\end{equation}
where $V_L^\nu$ and $V_R^\nu$ are two $3\times3$ unitary matrices.

Then, the lagrangian mass term becomes:
\begin{equation}
    -\mathcal{L}_{M_\nu} = \sum_{k=1}^3 m_k (\bar{\bm\nu}_D)_k ({\bm\nu}_D)_k
\end{equation}
which is obtained by defining:
\begin{equation}
    {\bm\nu}_D = V_L^\nu {\bm\nu}_L + V_R^{\nu*} {\bm\nu}_s^c
\end{equation}
Hence, the weak doublet of the SM neutrino fields is retrieved inverting the
previous formula acting with the left-handed chiral projector $P_L$:
\begin{equation}
    {\bm\nu}_L = P_L V_L^\nu {\bm\nu}_D
\end{equation}

This extension of the SM introduces $3$ sterile neutrinos that provide the
missing right-handed degrees of freedom to SM neutrinos. The theory presents a
low-energy matter content different from the original SM, which cannot be seen
as a low-energy effective field theory of the model. Moreover, the model does not
explain why the neutrinos are much lighter than the charged leptons. This sounds
strange since both leptons and quarks acquire masses with the same Yukawa
mechanism: up and down-type quarks in the same doublet have comparable masses,
while this would not be true for the leptons and the neutrinos.

\paragraph{The see-saw mechanism}

The presence of a non-zero Majorana mass matrix $M_N$ in
equation~\ref{eqn:nu-bsm-compact} opens the possibility to explain the
appearance of just $3$ active light neutrinos observed so far, with the
introduction of the see-saw mechanism. This model predicts that the $3$ active
light neutrinos are accompanied by $m$ heavy neutrinos with mass much higher
then the scale of the electroweak symmetry breaking $v$.

The mechanism can be understood by inspecting a $2\times2$ matrix:
\begin{equation}
    A = 
    \begin{pmatrix}
        0 & M \\
        M & N
    \end{pmatrix}
\end{equation}
with $M\ll N$. The eigenvalues of the matrix are given by:
\begin{equation}
    \begin{split}
        \lambda_{(+)} = \frac{N + \sqrt{N^2+4M^2}}{2} \to N
        \lambda_{(-)} = \frac{N - \sqrt{N^2+4M^2}}{2} \to -\frac{M^2}{N}
    \end{split}
\end{equation}
One of the two eigenvalues is suppressed by the magnitude of $N$, while
the other is amplified by it, hence the name see-saw.

In the general case, the matrix $M_D$ from equation~\ref{eqn:nu-bsm-compact} can
be diagonalized through the adjoint action of a unitary matrix $V^\nu$, obtaining
the following neutrino mass lagrangian terms:
\begin{equation}
    -\mathcal{L}_{M_\nu} =
    \frac{1}{2} \bar{\bm\nu}_l M^l {\bm\nu}_l
    + \frac{1}{2}\bar{\bm N} M^h {\bm N}
\end{equation}
where the two mass matrices $M^l$ and $M^h$ are given by:
\begin{equation}
    M^l \simeq - V^T_l M^T_D M^{-1}_L M_D V_l, \qquad
    M^h \simeq V^T_h M_N V_h
\end{equation}
and
\begin{equation}
    V^\nu \simeq
    \begin{bmatrix}
        \braces{1-\frac{1}{2}M_D^\dagger M_N^{*-1}M_N^{-1}M_D}V_l &
        M_D^\dagger M_N^{*-1}V_h \\
        -M_N^{-1} M_D V_l &
        \braces{1-\frac{1}{2}M_N^{-1}M_D M_D^\dagger M_N^{*-1}} V_h
    \end{bmatrix}
\end{equation}
where $V_l$ and $V_h$ are $3\times3$ and $m\times m$ unitary matrices,
respectively. Such as in the $2{-}$dimensional toy case, the $M^l$ is
proportional to the inverse matrix $M_N^{-1}$, while the heavier mass matrix
contains $M_N$. In this case, the SM is a good low-energy effective field theory
and, further, both the light and heavy neutrinos are Majorana particles.

\paragraph{Neutrino masses from generic new physics}

In the previous examples, we extended the standard model introducing new
lagrangian terms that respected the fundamental properties of the SM, namely
gauge invariance and renormalizability. In general, new physics (NP) beyond the
standard model might arise at energy scales $\Lambda_{\mathrm{NP}}$ much higher
than those currently probed by the current experiments. Then, like in the first
Fermi's theory of weak interactions, it should possible to include
non-renormalizable terms in the lagrangian that are suppressed by powers
of the characteristic scales of the new theory $\Lambda_{\mathrm{NP}}$: the
theory in this case manifests as non-renormalizable interactions at scales
lower than $\Lambda_{\mathrm{NP}}$, but is, in fact, renormalizable at higher
energies.

The least suppressed NP effects come from dimension $5$ operators. Considering
the SM model fields and imposing the gauge symmetry, we can build the following
term:
\begin{equation}
    \mathcal{O}_5 = \frac{Z^\nu_{ij}}{\Lambda_\mathrm{NP}}
    (\bar{L}_{Li}\phi)(\phi^T\bar{L}_{Li}^C) + \mathrm{h.c.}
\end{equation} 
where the $Z^\nu$ is a $n\times n$ matrix of coefficients, which leads after SSB
to the Majorana mass term for the left-handed neutrinos:
\begin{equation}
    -\mathcal{L}_{M_\nu} =
    \frac{Z^\nu_{ij}}{2}\frac{v^2}{\Lambda_\mathrm{NP}} \bar\nu_{Li} \nu^c_{Lj}
    + \mathrm{h.c.}
\end{equation}
Then, in this case the mass matrix of the neutrino fields is given by:
\begin{equation}
    (M_\nu)_{ij} = Z^\nu_{ij}\frac{v^2}{\Lambda_\mathrm{NP}}
\end{equation}
which contains the suppressed factor $v^2/\Lambda_\mathrm{NP}$, that would
explain why the neutrinos have such low masses compared to the other particles
in the theory (i.e. charged leptons).

The lagrangian term breaks both the total lepton number and the flavor
symmetry $U(1)_e\times U(1)_\mu\times U(1)_\tau$ and, in absence of further
symmetries at the level of the $Z\nu_{ij}$ coefficients, we expect lepton flavor
mixing and CP violation, as discussed in the previous section on neutrino
oscillation~\ref{subsec:nu-oscillation}. Moreover, we stress that the see-saw
mechanism might be seen as a particular model of NP theory: if we consider
$Z^\nu$ to be a $(3+m)\times(3+m)$ matrix and $M_N = \Lambda_\mathrm{NP}$, the
mass coefficients can be arranged to contain $3$ active light neutrinos with
mass suppressed by the NP theory scale plus $m$ heavy sterile neutrinos.










%% file: chapters/partI/chap1/mc.tex
\section{Monte Carlo techniques}
\label{sec:mc_gen}
\renewcommand{\myincludegraphics}[2][width=\textwidth]{
    \includegraphics[#1]{chapters/partI/chap1/plots/#2}
}

\subsection{Numerical Integration}
\label{subsec:num-int}

Both theoretical and experimental particle physics need to evaluate complex \linebreak
high-dimensional integrals whose analytical solution is not always known. It is
of \linebreak paramount importance to be able to compute them numerically and several
methods are known for centuries. These algorithms in general require to sample
points in the integration domain and evaluate the integrand function in them. The
simplest classical formulae have been developed for $1{-}$dimensional integration
and are usually split between Newton-Cotes type and Gaussian quadrature rules.
The former evaluates the integrand at equally spaced abscissas points, as
opposed to the latter. In this discussion, we focus on the Newton-Cotes type
rules only, leaving the reader to specialized articles for further
details~\cite{Atkinson:1989}.

\subsubsection{Newton-Cotes type formulae}
\label{ssubsec:newton-cotes}
The Newton-Cotes type rules compute an integral over a finite interval, weighting
the integrand function evaluated at equally spaced abscissas with suitable
coefficients. The following equation shows the simplest rule, namely the
trapezoidal rule:
\begin{equation}
    \int\displaylimits_{x_0}^{x_0 + \Delta x} dx\,f(x) =
    \frac{\Delta x}{2} [f(x_0)
    + f(x_0 + \Delta x)]
    - \frac{(\Delta x )^3}{12} f^{\prime\prime}(\xi)
\end{equation}
where $\xi \in [x_0, x_0 + \Delta x] \subset \mathbb{R}$. This procedure can be
repeated to cover macroscopic intervals $[x_0, x_n]$, containing a set of $n+1$
abscissas $\{x_i | x_i = x_0 + i \cdot \Delta x\}_{i=0}^n$:
\begin{equation}
    \label{eqn:trapezoids}
    \mathrm{I} = \int\displaylimits_{x_0}^{x_n} dx\,f(x) =
    \frac{x_n-x_0}{n} \sum_{i=0}^n w_i\,f(x_i)
    - \frac{1}{12} \frac{(x_n - x_0)^3}{n^2}
    \tilde{f}^{\prime\prime}
\end{equation}
with $w_0 = w_n = 1/2$ and $w_i = 1$ for $1 \leq i \leq n - 1$. Also
$\tilde{f}^{\prime\prime}$ is the average of all the $f^{\prime\prime}(\xi_i)$
values in the sub-intervals:
\begin{equation}
    \tilde{f}^{\prime\prime} =
    \frac{1}{n} \sum_{i=0}^{n} f^{\prime\prime}(\xi_i)
\end{equation}
for some $\xi_i \in [x_i, x_{i+1}]$. Equation~\ref{eqn:trapezoids} is an exact
formula for the integral $\mathrm{I}$. However, the term with the second
derivative of the integrand function should be evaluated in a point dependent on
$\mathrm{I}$ itself and, being unknown a priori, it is usually dropped. This
approximation results in an error of order $\orderof{n^{-2}}$ at the price of
$\orderof{n}$ function evaluations.

Another formula that leads to a better approximation is Simpson's rule
evaluating the integrand function in three points:
\begin{equation}
    \int\displaylimits_{x_0}^{x_2} dx\,f(x) =
    \frac{\Delta x}{3}
    [f(x_0) + 4 f(x_1) + f(x_2)]
    - \frac{(\Delta x )^5}{90} f^{(4)}(\xi)
\end{equation}
which gives:
\begin{equation}
    \label{eqn:simpson}
    \mathrm{I} = \int\displaylimits_{x_0}^{x_n} dx\,f(x) =
    \frac{x_n-x_0}{n} \sum_{i=0}^n w_i\,f(x_i)
    - \frac{1}{180} \frac{(x_n - x_0)^5}{n^4}
    \tilde{f}^{(4)}
\end{equation}
where this time $n$ should be an even number, $w_0 = w_n = 1/3$ and for $1 \leq
i < n$, $w_i = 4/3$ if $i$ is odd, $w_i = 2/3$ if $i$ is even. Again, the second
term is usually dropped, but this time it goes to zero way faster than the
trapezoidal rule one, namely as $\orderof{n^{-4}}$.

For the sake of completeness, we give here the prescriptions more refined
approximations rules. The Newton's $3/8$ rule, which employs $4$ integrand
evaluations for each sub-interval, but does not improve the final approximation
error:
\begin{equation}
    \int\displaylimits_{x_0}^{x_3} dx\,f(x) = \frac{3\Delta x}{8} [f(x_0)
    + 3 f(x_1) + 3 f(x_2) + f(x_3)] - \frac{3(\Delta x )^5}{80} f^{(4)}(\xi)
\end{equation}
and the five-points Boole's rule:
\begin{equation}
    \int\displaylimits_{x_0}^{x_4} dx\,f(x) =
    \frac{2\Delta x}{45}
    [7f(x_0) + 32 f(x_1) + 12 f(x_2) + 32 f(x_3) + 7 f(x_4)]
    - \frac{8(\Delta x )^7}{945} f^{(6)}(\xi)
\end{equation}
which has an error that scales as $\orderof{n^{-6}}$.

In general, if the number of derivatives for the error term is $k+1$, the rule
is said to be of degree $k$ and the approximation error behaves like
$\orderof{n^{-k}}$. It can be shown that an even $2k{-}$starting point rule does
not improve the degree of the odd $(2k-1){-}$rule: one has to resort to the
$(2k+1){-}$point formula to gain better precision. Although higher degree
formulae provide errors that decay faster, it can be verified that the $w_i$
coefficients start growing large in magnitude and with opposite signs, leading
to numerical cancellations that do not improve consistently the
approximations. Moreover, $k{-}$degree Newton-Cotes rules assume that the
integrand function is differentiable at least $k$ times, with continuous
derivative: applications of the rule to functions not satisfying this requirement
provide a wrong error estimate. Therefore, higher-order rules are not used in
practice.

\subsubsection{Multi-dimensional integration}
\label{ssubsec:multidim-int}
\newcommand{\mum}{\ensuremath{\mu(\mathcal{M})}}
\newcommand{\mumn}[1]{\ensuremath{\mu^{#1}(\mathcal{M})}}

The results presented in the previous paragraph refer to $1{-}$dimensional
integration only. Here, we focus on extending the discussion to multi-dimensional
integrals. Let $f$ be a function defined over a $d{-}$dimensional volume
$\mathcal{M} \subseteq \mathbb{R}^{d}$ and consider the integral:
\begin{equation}
    \label{eqn:multidim-int}
    I = \int_\mathcal{M} d\mu(x) \, f(x_1, \dots, x_d)
\end{equation}
where $\mu(x)$ is some measure function on the integration domain, such that
$\mum$ is the volume of $\mathcal{M}$ itself. The trapezoidal
formula~\ref{eqn:trapezoids} can be extended to the multi-dimensional case
sampling $n$ points along each of the $d$ the $1{-}$dimensional axes projections
of the function $f$. That is, building a $d{-}$dimensional histogram covering the
integration volume:
\begin{equation}
    I =
    \frac{\mum}{n^d} \sum_{i_1=0}^{n}
    \dots
    \sum_{i_d = 0}^{n} w_{i_1}
    \dots w_{i_d} f(x_{i_1},\dots,x_{i_d}) + \orderof{n^{-2}}
\end{equation}
where the integrand
function is evaluated in the $nd$ sampled points ${x_{i_k}}$ for
$i = 0, \dots, n$ and $k= 1, \dots, d$.
The total number of points used to cover the $d{-}$dimensional set is
$N = (n+1)^d \approx n^d$, which leads to an approximation error of order
$\orderof{N^{-2/d}}$. The same argument holds for the other Newton-Cotes type
formulae, showing how their scaling behavior is optimum in the $1{-}$dimensional
case, but rather poor when the dimensionality increases. This problem can be
regarded as the \textit{curse of dimensionality}, namely the issue afflicting
integration and histogram methods for which the number of points required to
cover a multi-dimensional space grows exponentially with the number of
dimensions. The next paragraph introduces a technique suited for integration in
a high number of dimensions: Monte Carlo integration.

\subsection{Monte Carlo Integration}
\label{ssubsec:mc_int}

Monte Carlo integration relies on a probabilistic approach to the measure
of sets: inside an event space $\Omega$, it could be possible to measure the
volume of a subset $\mathcal{M} \subseteq \Omega$, drawing a sequence of random
events $\{\omega_i\}_{i=0}^N$ and keeping track of the number of events that
fall inside the subset $\mathcal{M}$, $N_{in}$. The measure $\mu$ of the subset
$\mathcal{M}$, in the limit of large N, is approximated by the ratio:
\begin{equation}
    \mum = \lim_{N\to\infty} \mu(\Omega) \, \frac{N_{in}}{N}
\end{equation}

This idea can be transposed to find the value $I$ of the integral of a
multi-dimensional function $f$, defined as in equation~\ref{eqn:multidim-int}.
Further, we assume that $f$ is square integrable. In the following, we normalize
to unity the volume of the integration domain; the general case can be retrieved
using dimensional analysis, multiplying quantities appearing in equations by a
suitable power of $\mum$. The Monte Carlo estimate $E_N$ of the integral $I$ is
given by the following formula:
\begin{equation}
    \label{eqn:mc_est}
    E_N = \frac{1}{N} \sum_{n=1}^{N} \, f(x_n) \xrightarrow[N\to\infty]{} I 
\end{equation}
where $\{x_n\}_{n=1}^N$ represent random points where the function $f$ should be
evaluated. In the large $N$ limit, the MC estimate equals the integral.

An estimation of the error for approximating the integral $I$ with a finite $N$
computation of the MC estimate $E_N$ should be discussed defining the variance
$\sigma^2[f]$ of the function $f$:
\begin{equation}
    \label{eqn:var-f}
    \sigma^2[f] := \int_{\mathcal{M}} d\mu(x)\,(f(x) - I)^2
\end{equation}
The variance for the MC estimate can be found by integrating over each of the
randomly drawn points:
\begin{equation}
    \label{eqn:var-en}
    \sigma^2 [E_N] =
    \int_{\mathcal{M}} d\mu(x_1) \dots d\mu(x_N) \,(E_N - I)^2
    = \frac{\sigma^2[f]}{N}
\end{equation}
which can be derived first proving that $\int d\mu(x) (f(x)-I) = 0$ and then
working out the square in equation~\ref{eqn:var-en}, followed by the integration
on the drawn points $\{x_n\in\mathcal{M}\}_{n=1}^N$. As opposed to the exact
remainders associated with the Newton-Cotes type formulae (see
paragraph~\ref{ssubsec:newton-cotes}) this error has a probabilistic
interpretation: evaluating the integrand function on a finite sample of $N$
points, the average error that one commits approximating $I \approx E_N$ is
$\frac{\sigma[f]}{\sqrt{N}}$. Equivalently, due to Central Limit Theorem,
equation~\ref{eqn:var-en} implies that the sequence of random variables
$\{(E_n - I)\}_{n=1}^{N}$ converges in probability to a centered gaussian with
standard deviation of $\sigma[f]/\sqrt{N}$, where $\sigma[f]$ is the integrand
function standard deviation. In practice, the exact value for the variance of
the MC estimation $\sigma^2[E_N]$ is unknown and an approximated formula is
employed:
\begin{equation}
    \label{eqn:var-en-est}
    S_N^2 = \frac{1}{N-1} \sum_{n=1}^N \braces{f(x_n) - E_N}^2 = 
    \frac{1}{N} \sum_{n=1}^N (f(x_n))^2 - E_N^2
\end{equation}

An immediate consequence of equations~\ref{eqn:var-en} and~\ref{eqn:var-en-est}
is that the approximation error of the MC estimation scales as
$\orderof{N^{-1/2}}$ with the number of function evaluations, which unfortunately
is rather poor compared to the $1{-}$dimensional trapezoidal or Simpson's rules.
However, by construction, the error does not depend on the dimensionality $d$ of
the integration domain $\mathcal{M}$. Hence, the MC integration is particularly
suited to evaluate high dimensional integrals. We report here the master formula
for MC integration:
\begin{equation}
    \label{eqn:mc_formula}
    I \approx E_N \pm S_N = \frac{1}{N} \sum_{n=1}^N f(x_i) \pm
    \sbraces{\frac{1}{N-1} \sum_{n=1}^N \braces{f(x_n)-E_N}^2}^{1/2}
\end{equation}

\subsubsection{Variance Reducing Techniques}

The approximated formula of the variance of the MC integral,
equation~\ref{eqn:var-en-est}, is the ratio of the variance of the integrand
function $\sigma^2[f]$ and the number of function evaluations $N$. It follows
that if it is possible to somehow reduce the numerator, the poor
$\orderof{N^{-1/2}}$ behavior can be improved dramatically. Several
variance-reducing techniques have been designed for this purpose: we present the
ideas behind the main strategies in this paragraph.

\paragraph{Stratified sampling}
Stratified sampling consists in dividing the integration domain into smaller
sub-domains, perform an MC integration in each of them separately and, finally,
gather all the contributions to provide the final MC estimate of the integral.
This technique is supported by the following property of the Riemann integral:
let $\{\mathcal{M}_i\}_{i=0}^k$ a numerable family of disjoint subsets of the
integration domain $\mathcal{M}$, such that
$\cup_{i=0}^k \mathcal{M}_i = \mathcal{M}$, then the following equation holds:
\begin{equation}
    I =
    \int_{\mathcal{M}} d\mu(x)\,f(x)
    = \sum_{i=0}^k \int_{\mathcal{M}_i} d\mu(x)\,f(x)
\end{equation}
Therefore, the MC estimate of the integral I becomes:
\begin{equation}
    E_N = \sum_{i=1}^k \frac{\mu(\mathcal{M}_i)}{N_i}\sum_{n=1}^{N_i}f(x_{in})
\end{equation}
where we have drawn $N_i$ random points for each subset $\mathcal{M}_i$.

The variances of the function in each subset add up together to retrieve the
variance of the MC estimate. Note that the $\mu(\mathcal{M_i})$ is not unity
anymore and must be included explicitly in the formula (dimensional analysis
helps infer the correct power of the integration volume in each equation):
\begin{equation}
    \label{eqn:var-strat}
    \sigma^2[E_N] = \sum_{i=0}^k \frac{\mu^2(\mathcal{M}_i)}{N_i} \,
    \sigma^2[f]\Big\rvert_{\mathcal{M}_i}
\end{equation}
with:
\begin{equation}
    \begin{split}
        \sigma^2[f]\Big\rvert_{\mathcal{M}_i} &=
        \frac{1}{\mu({\mathcal{M}_i})} \int_{\mathcal{M}_i} d\mu(x)\,
        \braces{f(x) - \frac{I}{\mu({\mathcal{M}_i})}}^2 \\
        &= \frac{1}{\mu({\mathcal{M}_i})} \int_{\mathcal{M}_i} d\mu(x) f(x)^2
        - \braces{
            \frac{1}{\mu({\mathcal{M}_i})} \int_{\mathcal{M}_i} d\mu(x) f(x)
        }^2
    \end{split}
\end{equation}
It can be shown that the value of the MC variance in equation~\ref{eqn:var-strat}
can be minimized by an appropriate choice of the points sampled for each subset
$N_i$. Here, we consider the case with $K=2$ and infer the general rule. If
there are only two subsets $a$ and $b$, we parametrize the number of points where
to evaluate the integrand function in the two subsets with $N_a$ and $N_b = N- N_a$
respectively, where $N$ stands for the total number of sampled points
$\{x_{in}\}$. Taking the derivative with respect to $N_a$, the formula for
$\sigma^2[E_N]$ can be minimized:
\begin{equation}
    \frac{\partial\,\sigma^2[E_N]}{\partial N_a}  =
    - \braces{\frac{\mu(\mathcal{M}_a)}{N_a} \, \sigma_a}^2
    + \braces{\frac{\mu(\mathcal{M}_b)}{N - N_a} \, \sigma_b}^2
    \overset{!}{=} 0
\end{equation}
where $\sigma_i$ is a short-hand for $\sigma[f] \Big\rvert_{\mathcal{M}_i}$. So,
solving for $N_a$:
\begin{equation}
    \frac{N_a}{N} = \frac{\mu(\mathcal{M}_a) \sigma_a}{\mu(\mathcal{M}_a)\sigma_a
    + \mu(\mathcal{M}_b)\sigma_b}
\end{equation}
This result shows how to minimize the variance of the MC integration splitting
the integration domain into $2$ parts.
We can generalize it by claiming that the least possible variance for an MC
integration can be obtained drawing in each subset $\mathcal{M}_i$ a fraction of
points equal to the relative product of the volume of the subset
$\mu(\mathcal{M}_i)$ times the variance of the integrand function evaluated in it.
This means that evaluations of $f(x)$ must be concentrated in subsets where the
potential error given by high variance of the integrand function is the highest,
namely where the function is both large or rapidly changing.

\paragraph{Importance sampling}

The importance sampling technique refers to a change of variables in the
integration, resulting in a new integrand function. We show in this paragraph
that carefully tweaking this transformation leads to more efficient
computation. Formally, we have:
\begin{equation}
    \label{eqn:importance-sampling}
    I = \int_{\mathcal{M}_i} d\mu(x) \,f(x)
      = \int_{\mathcal{M}} d\mu(x) p(x)\, \frac{f(x)}{p(x)}
\end{equation}
where the function $p(x)$ can be chosen to be positive and normalized to unity,
namely a probability density function (pdf). Assuming that we can sample
random points following the pdf distribution $x_n \sim p(x)$, then the MC
integration becomes:
\begin{equation}
    \label{eqn:mc_est_is}
    E_N = \frac{1}{N} \sum_{n=1}^N \, \frac{f(x_n)}{p(x_n)}
\end{equation}

After this transformation, the new variance for the MC integral is given by \linebreak
$\sigma[f/p]/\sqrt{N}$, which is estimated by:
\begin{equation}
    \label{eqn:var_is}
    S_N^2 = \frac{1}{N} \sum_{n=1}^N \, \braces{\frac{f(x_n)}{p(x_n)}}^2 - E_N^2
\end{equation}
Zero variance can be achieved by choosing $p(x) = f(x)/I$, but, of course, this
cannot be done since we do not know the exact value of the integral. However,
the variance can be reduced by choosing $p(x)$ as close as possible to $f(x)$ or,
equivalently, requiring that the new integrand function $f(x)/p(x)$ is as flat as
possible. It should be noted, though, that choosing a $p(x)$ that is zero or
approximately zero in a region with non-zero $f(x)$ results in a divergence in
equation~\ref{eqn:var_is} and a consequent erroneous estimation of the MC
integral variance. Note that this behavior is not detected by the MC integration
algorithm if this region of divergence is small and the randomly sampled points
fall outside it. This argument signal that some attention should be paid when
choosing the function $p(x)$.

\paragraph{Multi-channel Monte Carlo}
\label{par:multi-channel-mc}

Importance sampling introduces a transformation that acts on the whole
integration domain flattening the integrand function and ensuring that random
sampling produces an efficient computation of the integral. However, when the
integrand function exhibits multiple sharp peaks in localized different regions
it is very unlikely to find a suitable change of variables that accommodates them
all. The key idea behind a multi-channel MC is that when the single peak
structures of $f(x)$ are known, it is possible to transform each contribution
with an independent transformation. Each transformation is called a channel
and defines a dedicated probability density function. An MC integration on a
single channel contributes to the final result only for the corresponding peak,
since the sampling is inevitably more focused around that region of the
integration domain and probably under-populating the others.

Multi-channel MC requires listing all the probability density functions $p_i(x)$
associated with each change of variables and to select each channel with a
probability $\alpha_i$ for $i = 1, \dots, m$, such that the $\alpha$s sum up
to unity. The desired integral is computed by substituting
$p(x) = \sum_{i=1}^m \alpha_i p_i(x)$ in equation~\ref{eqn:importance-sampling}:
\begin{equation}
    I = \sum_{i=1}^M \alpha_i \int_{\mathcal{M}} dP_i(x)\,\frac{f(x)}{p(x)}
\end{equation}
where $dP_i(x) = d\mu(x) \, p_i(x)$.
The algorithms are usually implemented by fixing the number of function evaluations
$N$ and selecting each channel with probability $\alpha_i$: approximately
yielding $N_i\approx \alpha_i N$ function evaluations for each channel. The MC
estimation of the integral $I$ is, then:
\begin{equation}
    E_N =
    \frac{1}{N} \sum_{i=1}^m \sum_{n_i=1}^{N_i}\frac{f(x_{n_i})}{p(x_{n_i})}
\end{equation}
and a corresponding variance, analogously to equation~\ref{eqn:var_is}:
\begin{equation}
    \sigma^2[E_N] = \frac{1}{N} \sum_{i=1}^m \alpha_i \int_{\mathcal{M}}
    dP_i(x)\,\braces{\frac{f(x)}{p(x)}}^2 - I^2
\end{equation}
Since the values of $\alpha_i$ do not affect the MC estimate of the integral, but
only the estimated variance of the integral, a careful choice of the $\alpha_i$
coefficients may lead to better results. An iterative method to adjust the
$\alpha_i$ in order to minimize $\sigma^2[E_N]$ is presented
in~\cite{Kleiss:1994}.

\paragraph{Control variates}

An alternative technique related to importance sampling is known as the method
of control variates: in this application, a function $g(x)$ is subtracted to the
original integrand $f(x)$, introducing a new integrand function $(f(x) - g(x))$
and leaving as the remainder term a possibly known integral of $g(x)$:
\begin{equation}
    I = \int_{\mathcal{M}} d\mu(x)\,f(x)
      = \int_{\mathcal{M}} d\mu(x)\,\big(f(x) - g(x)\big)
      + \int_{\mathcal{M}} d\mu(x) \, g(x)
\end{equation}
In principle, the $(f(x) - g(x))$ integration should be associated with a smaller
variance to benefit from the transformation. It is known that this method is
stabler than the importance sampling method, since the $g(x)$ function does not
introduce singularities in the integration function.

\subsection{Adaptive Monte Carlo: VEGAS algorithm}
\label{sub:vegas}

The variance-reducing techniques presented just above require some knowledge
of the integrand function to be exploited. It would be desirable, instead, to
design methods that act in an automated way: given a function on the integration
domain, they choose the best strategy to enhance algorithm efficiency. This is
exactly the purpose of adaptive techniques, namely those algorithms in which the
integrand function properties are learned at runtime. In particular,
VEGAS~\citep{Lepage:1978,Lepage:1980} is a tool that employs importance sampling
in high dimensions to iteratively shape an estimate of the optimal pdf:
\begin{equation}
    p_{\mathrm{best}} (x) =
    \frac{\abs{f(x)}}{\int_{\mathcal{M}}d\mu(x)\,\abs{f(x)}}
\end{equation}
A multi-dimensional histogram of the absolute value of the integrand function is
computed at each iteration. The number of bins along each dimension is kept
constant during the process, while gradually adjusting their width. The objective
is to produce a finer grid in regions providing large contributions to the
integral and a coarser grain in the integration domain portions with
"low-activity". As a consequence of the \emph{curse of dimensionality}, a
$d{-}$dimensional histograms require an exponential number of bins with respect
to the number of dimensions $d$ to cover the integration domain. Therefore,
scalability in the VEGAS algorithm is ensured considering only separable
functions for $p(x)$:
\begin{equation}
    p(x) = p_1(x_1)\cdot p_2(x_2) \cdot \ldots \cdot p_d(x_d)
\end{equation}
This way, it is possible to work independently along each dimension with the
marginals probability functions $p_i(x_i)$.

The algorithm acts in two phases: exploration and evaluation. VEGAS first
explores the integration domain to adapt the grid to the integrand function shape,
starting from a naive uniform distribution for $p(x)$ and adapting the widths of
the bins following the criteria described above: after some repetitions a stable
grid configuration is reached. At this point, the grids are frozen and the
evaluation phase starts: several MC integrations are performed and their results
are aggregated into a final output. Given $M$ integral evaluations, the
intermediate estimations for the integral $E_i$ and its variances $S^2_i$ are
averaged into the final quantities $\bar{E}$ and $S^2_{\bar{E}}$ with the
following rules:
\begin{equation}
    \bar{E} = S^2_{\bar{E}} \, \sum_{i=1}^M \frac{E_j}{S^2_i},
    \qquad
    \frac{1}{S^2_{\bar{E}}} = \sum_{i=1}^M \frac{1}{S^2_i},
\end{equation}
The intermediate $E_i$ and $S^2_i$ are computed with the usual formulae for
importance sampling, equations~\ref{eqn:mc_est_is} and~\ref{eqn:var_is}
respectively. The algorithm also performs checks on the intermediate evaluations
to ensure that the $S^2_i$ are reliable estimates of the MC error. For
this purpose, internally, a $\chi^2$ per degree of freedom is computed with the
following equation:
\begin{equation}
    \label{eqn:vegas-check}
    \chi^2/\mathrm{dof} =
    \frac{1}{M-1}\sum_{i=1}^M \frac{(E_i - \bar{E})^2}{S^2_i}
\end{equation}
Consistent estimates give $\chi/\mathrm{dof}$ values approximately equal to $1$.

\subsection{Sample generation}
\label{subsec:sample-gen}

MC integrators assume the ability to randomly sample points
in the integration domain and evaluate the corresponding integrand function
values: a naive approach draws points uniformly on the integration domain to
update the estimates for the integral and its variance (see
equation~\ref{eqn:mc_formula}). Modern scientific libraries already have
implemented algorithms to sample from uniform as well as many other notable
probability distributions~\citep{Lonnblad:1994,Brun:1997}. Unfortunately, this
is not sufficient, since a basic Monte Carlo integration is known to perform
poorly when complex integrands are involved. The importance sampling technique,
then, provides a method to gain efficiency requiring that sample points are drawn
from a chosen probability density function that is close as possible to the
integrand. In the next paragraphs, we introduce the most widely used methods to
produce samples following a given pdf.

In the following, the cumulative distribution function for a $1{-}$dimensional
random variable $X$ following a probability distribution function $p(x)$ is:
\begin{equation}
    P(x) = \mathrm{Prob}\big(X \leq x\big) = \int_{-\infty}^{x} dt\,p(t)
\end{equation}
The definition can be extended naturally to higher dimensional random variables.

\subsubsection{Inverse transform method}

The inverse transform method gives a recipe to draw random numbers following an
arbitrary $p(x)$. For each random variable $X$ following the pdf $p(x)$ it can
be defined a new random variable $U = P(X)$ that follows the uniform probability
in the unit interval. It is possible, then, to invert the formula finding the
inverse function of the cumulative distribution, which always exists since $F$
is monotonically increasing by construction. Provided that the $P^{-1}$ can be
written analytically, then, sampling a uniform random number in the unit
interval $u$ and computing:
\begin{equation}
    \label{eqn:inverse-method}
    x=P^{-1}(u)
\end{equation}
yields a number distributed according to the desired pdf function $p(x)$.
\begin{figure}
    \centering
    \subfigure[Continuous distributions case.]{
      \centering
      \label{subfig:inverse_cont}
      \myincludegraphics[width=0.47\textwidth]{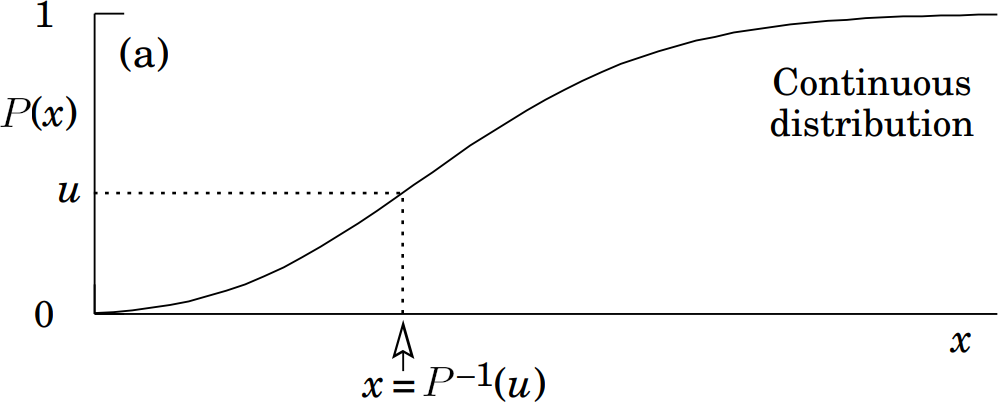}
    }
    \hfill
    \subfigure[Discrete distributions case.]{
      \label{subfig:inverse_disc}
      \myincludegraphics[width=0.47\textwidth]{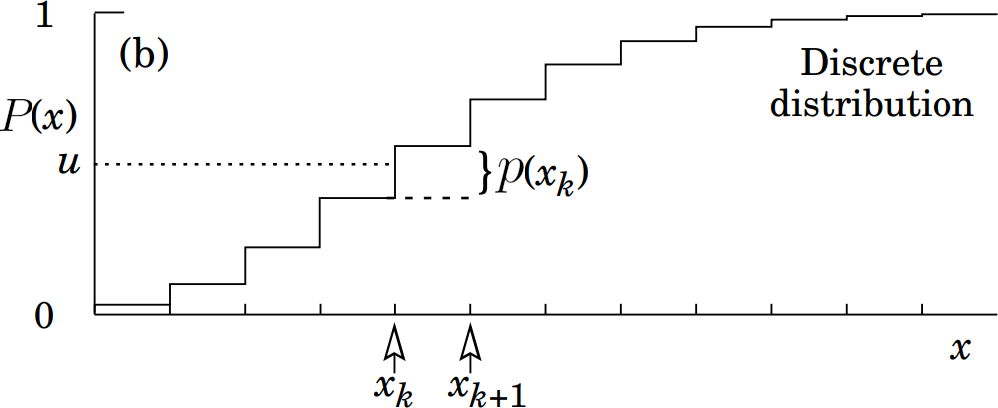}
    }
    \caption{The inverse transform method for sampling random numbers according
    to some probability distribution function $p(x)$.}
    \label{fig:inverse-method}
  \end{figure}
Figure~\ref{fig:inverse-method} depicts the idea behind the inverse transform method for 
both continuous and discrete random variables. In the latter case, the cumulative
function is step-like and the desired output $x_k$ is the one that satisfies:
\begin{equation}
    P(x_k) \leq u < P(x_{k+1})
\end{equation}

\subsubsection{Acceptance-rejection method}
\label{ssubsec:accept-reject}

The inverse transform method is strongly limited by the fact that the inverse
function must be known exactly. Therefore, Von Neumann developed an
algorithm~\citep{Forsythe:1972} to sample random numbers according to a pdf $p(x)$
when the inverse transform method hypothesis cannot be fulfilled. The idea relies
on the assumption that $p(x)$ is bounded from above by $C$ times a pdf $h(x)$:
$p(x) \leq C h(x)$. Both $p(x)$ and $h(x)$ are normalized, hence $C \geq 1$
follows. The choice of $h(x)$ should be done to be able to generate random
samples from it easily: often this is set to the uniform distribution.

The method requires knowledge of the pdf $p(x)$ only and works as follows. First,
generate a single $x {\sim} h(x)$. Second, uniformly draw a random number $u$ in
the unit interval. Then, check the following inequality:
\begin{equation}
    u C h(x) \leq p(x)
\end{equation}
If the equation is satisfied accept $x$, reject it otherwise. Repeating the
procedure many times produces a large set of random numbers following the $p(x)$
distribution.
\begin{figure}
    \centering
    \subfigure[The original method.]{
      \centering
      \label{subfig:acc-rej-original}
      \myincludegraphics[width=0.47\textwidth]{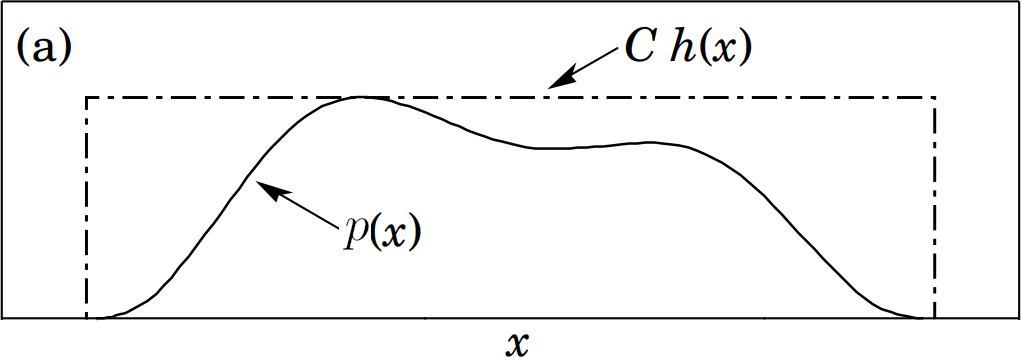}
    }
    \hfill
    \subfigure[The improved method.]{
      \label{subfig:acc-rej-imp}
      \myincludegraphics[width=0.47\textwidth]{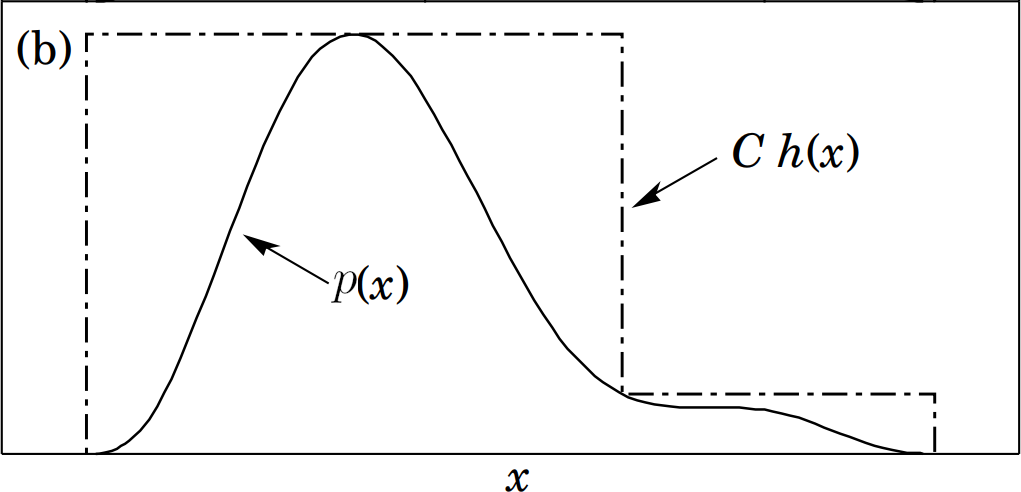}
    }
    \caption{The acceptance-rejection method  for sampling random numbers
    according to some probability distribution function $p(x)$.}
    \label{fig:acc-rej}
\end{figure}
The efficiency of this method is constrained by the probability at which the
acceptance happens: if a point $x$ is generated in a region where
$p(x) \ll C h(x)$, it is very likely that the point will be rejected. That's why
figure~\ref{subfig:acc-rej-imp} suggests a trick to improve the naive
acceptance-rejection method: the quantity $C$ can be chosen piecewise constant,
in order to keep the ratio of $p(x)$ and $C h(x)$ as close as possible to unity.

\subsubsection{The Metropolis algorithm}

The approaches presented in the last two paragraphs are rather simple and they
are not employed for realistic applications and complex use cases.
Quite often it is required to generate numbers following multi-dimensional
probability distributions. When the number of dimensions $d$ grows large, a
Markov Chain Monte Carlo (MCMC) technique, like the Metropolis
algorithm~\citep{Hastings:1970}, performs way better than the naive \linebreak
acceptance-rejection method.

The Metropolis algorithm is essentially a random walk in the space where the
multi-variate pdf $p(x)$ is defined: it produces a chain of states
$x_1,\dots,x_n$, that eventually follows the $p(x)$ function. This happens if
two assumptions of ergodicity and detailed balance are verified. Ergodicity
means that every possible state of the system can be reached within a finite
number of steps: this allows to pick a random initial state being sure that the
equilibrium state will be reached at some point. Detailed balance is a
statement of reversibility of the system: it claims that the probability flow
outgoing from state $x_{n-1}$ to the next state in the chain $x_n$ equals the
ingoing probability flow from $x_n$ to $x_{n-1}$. Formally, it is defined by the
following equation:
\begin{equation}
    \label{eqn:detailed-balance}
    p(x_{n-1}) \pi(x_{n-1}, x_n) = p(x_n) \pi(x_n, x_{n-1})
\end{equation}
where $\pi(x_{n-1}, x_n)$ is the transition probability to step from state
$x_{n-1}$ to state $x_n$ and, of course, $p(x_n)$ is the probability of being in
state $x_n$.

Each iteration of the Metropolis algorithm consists of the following list of
actions.
\begin{enumerate}
    \item At step $n$, randomly generate a new candidate state $x_{n+1}$.
    \item Compute the discriminative quantity
    $\Delta S = -\log \frac{p(x_{n+1})}{p(x_n)}$.
    \item If $\Delta S < 0$, accept the candidate state as the new state.
    \item Else, accept the candidate state with probability $p(x_{n+1})/p(x_n)$.
    \item Repeat from the first point.
\end{enumerate}
Point $3$ and $4$ can be absorbed into the definition of a single transition
probability \linebreak $\pi(x_n, x_{n+1}) = \min(1, e^{-\Delta S})$, that satisfies the
detailed balance condition. The intuitive explanation of the algorithm is that
the system spends more time in a higher probability state, rather than in points
associated with low probability regions. In particular, if the new state has a
higher probability than the old one, it gets certainly accepted, conversely less
probable states are accepted only with probability $p(x_{n+1})/p(x_n)$.

In practice, this algorithm initially starts in a random state and evolves
naturally after some iterations towards an equilibrium state where the detailed
balance condition holds. The ergodicity property ensures that convergence to the
equilibrium is reached independently of the starting point. However,
this process might take some time and, for particularly complex use cases, the
system might be trapped in metastable states for an unknown number of steps.
Moreover, successive states are not independent at equilibrium since the system
exhibits a typical correlation length $\xi$. Therefore, it is really difficult
to obtain an unbiased sample of points. A solution might be to save only those
states in the chain with a time distance proportional to $\xi^2$, which is known
as the decorrelation time in random walk theory. The main problem with this
technique is the fact that systems can acquire very large correlation lengths,
especially while experiencing phase transitions.

\subsection{Phase space generation from particle collisions}
The methods described above find many applications in particle physics. Both in
fixed order calculations and event generators the problem arises to generate
phase space \linebreak points, namely the $4{-}$momenta of $n$ particles with masses
$m_1,\dots,m_n$. The goal is to generate sets of particles $4{-}$momenta
$p_1,\dots, p_n$ for collision events following physical distributions.

The $n{-}$particle phase space is given by the following quantity:
\begin{align}
    \label{eqn:phase-space}
    d\Phi_n (P; p_1,\dots,p_n) &=
    (2\pi)^4 \delta^{(4)}\braces{P - {\textstyle\sum_i p_i}}
    \prod_{i=1}^n \frac{d^4p_i}{(2\pi)^4} \,
    (2\pi)\delta(p_i^2-m_i^2) \Theta(p_i^0) \\
    &= (2\pi)^4 \delta^{(4)}\braces{P - {\textstyle\sum_i p_i}}
    \prod_{i=1}^n \frac{d^3\mathbf{p}_i}{(2\pi)^3 2p_i^0}
\end{align}
where the $4{-}$dimensional Dirac delta expresses momentum conservation law.
In general, there are two basic approaches to the problem. The
first~\citep{Byckling:1971} is based on a recursive relationship producing a
sequential or hierarchical approach to phase space sampling. The second is a
democratic one dubbed RAndom Momenta Booster (RAMBO)~\citep{Kleiss:1986}, where
massless particles are produced in a flat phase space and subsequent boosting
transformations enforce the physical constraints to output the final
distribution.

\paragraph{Hierarchical approach}

A simple recursion rule can be derived from the $n{-}$body phase space equation.
The objective is to create an algorithm that allows the generation of events
containing $n$ final state particles with given masses $\{m_1,\dots,m_n\}$. The
output event will be obtained by a sequence of $1\to2$ particle branchings.

From the $p_1,\dots, p_n$ output momenta, consider the auxiliary momenta $q_i$
defined as the sum of all the output momenta with an index less or equal to $i$:
$q_i = \sum_{j=1}^i p_i$. Also, set the resonant mass parameters to
$M_i^2 = q_i^2$. Then, it is possible to insert the following two identities
in~\ref{eqn:phase-space}:
\begin{align}
    1 &= \int d^4q_{n-1} \delta^{(4)}(q_{n-1} - \sum_{i=1}^{n-1}p_i) \\
    1 &= \int dM^2_{n-1} \delta(q^2_{n-1}-M^2_{n-1})
\end{align}
which 
to obtain:
\begin{equation}
    \begin{split}
d\Phi_n (P; p_1,\dots,p_n) &= dM^2_{n-1}
    \textcolor{blue}{\delta(q^2_{n-1}-M^2_{n-1})\, d^4q_{n-1}}
    \textcolor{red}{\delta^{(4)}(q_{n-1}-\sum_{i=1}^{n-1}p_i)} \\
    & \textcolor{red}{(2\pi)^4} \textcolor{blue}{\delta^{(4)}
    \braces{P - {\textstyle\sum_i p_i}}
    \frac{d^3\mathbf{p}_n}{(2\pi)^3 2E_n}}
    \textcolor{red}{
        \prod_{i=1}^{n-1} \frac{d^4p_i}{(2\pi)^4}\, (2\pi)\delta(p_i^2-m_i^2)
    }\\
    &= dM^2_{n-1} \textcolor{blue}{\frac{d^3\mathbf{q}_{n-1}}{2M_{n-1}}
    \frac{d^3\mathbf{p}_{n}}{(2\pi)^3 2E_{p_n}} \delta^{(4)}
    \braces{P - p_n-q_{n-1}}} \\
    & \textcolor{red}{d\Phi_n (q_{n-1}; p_1,\dots,p_{n-1})} \\
    &= dM^2_{n-1} \,
    \textcolor{blue}{\frac{1}{2\pi}d\Phi_2(P;p_n,1_{n-1})} \,
    \textcolor{red}{d\Phi_n (q_{n-1}; p_1,\dots,p_{n-1})}
    \end{split}
\end{equation}
This equation underlines that the $n{-}$particle phase space can be reduced to
an $(n-1){-}$particle phase space through the radiation of a particle with
momentum $p_n$ by an intermediate state with mass $M_{n-1}$.

The recursive relation can be repeated introducing multiple sequential
branchings in the process to take care of all the external particles,
resulting in:
\begin{equation}
    \label{eqn:sequential-splitting}
    d\Phi_n = \frac{1}{(2\pi)^{n-2}} \,
    dM_{n-1}^2 \dots dM_2^2 \,
    d\Phi_2(n)\dots d\Phi_2(2)
\end{equation}
where $d\Phi_2(i) = d\Phi_2(q_i;p_i,q_{i-1})$ is a shorthand for the contribution
given by the decay $q_i \to p_i\,q_{i-1}$.

It is possible to work out the integration of the Dirac delta in the two body
phase space for a particle of momentum $q_i$ that decays in two others with
momenta $p_i$ and $q_{i-1}$ respectively, the result is:
\begin{equation}
    \begin{split}
        \label{eqn:2bodyps}
d\Phi_2(i) &= \frac{d^3\mathbf{p}_{i}}{(2\pi)^3 2E_{i}}\,
              \frac{d^3\mathbf{q}_{i-1}}{(2\pi)^3 2M_{i-1}}
              (2\pi)^4 \delta^{(4)}(q_i-p_i-q_{i-1}) \\
            &= \frac{\lambda^{1/2}(q_i^2,q_{i-1}^2,p_i^2)}{(2\pi)^2\, 8M_i^2}
            d\Omega_i
    \end{split}
\end{equation}
recalling that $q_i^2=M_i^2$ and where $d\Omega = d\cos\theta_i d\phi_i$ is the
solid angle differential element of the two outgoing particles $i$ and $i-1$ and
the characteristic $\lambda$ function is defined by:
\begin{equation}
    \lambda(x,y,z) = x^2 + y^2 + z^2 - 2xy - 2xz - 2yz
\end{equation}
which is symmetric under the $x \longleftrightarrow y$ exchange.

Inserting equation~\ref{eqn:2bodyps} into the recursive
relation~\ref{eqn:sequential-splitting} we obtain the result for the $n{-}$body
phase space with $n-1$ sequential splittings:
\begin{equation}
    \label{eqn:sequntial}
    d\Phi_n(P;p_1,\dots,p_n) = (2\pi)^{4-3n} 2^{3-2n}\,dM^2_{n-1} \dots dM^2_2 \,
    \prod_{i=2}^n \frac{\lambda^{1/2}(M^2_i,M^2_{i-1},p^2_i)}{2M_i} d\Omega_i
\end{equation}
where the masses $M_i$ in subsequent decays are constrained by the
kinematics to respect the inequalities:
\begin{equation}
    \mu_i \leq M_i \leq M_{i+1} - m_{i+1}
\end{equation}
where $\mu_i = \sum_{j=1}^i m_i$.

The ratio in the product of equation~\ref{eqn:sequntial} turns out to be equal to
the absolute value of the $3{-}$momentum $\abs{\mathbf{p}_i}$ in the rest
frame of the decaying intermediate particle $M^i_2$. Therefore, it is possible
to design the following algorithm to hierarchically generate phase space events
containing $n$ final state particles with $\{m_1,\dots, m_n\}$ masses.
\begingroup
\renewcommand{\arraystretch}{1.3}
\begin{table}
    \scriptsize
    \centering
    \begin{tabular}{l}
        \hline\noalign{\smallskip}
        {\bf Inputs:} Incoming momentum $P$, output masses ${m_1,\dots,m_n}$.\\
        {\bf Outputs:} Final state particle $4{-}$momenta $\{p_1^\mu,\dots, p_n^\mu\}$ weighted by $w$.\\
        \noalign{\smallskip}\hline\noalign{\smallskip}
Set $i \leftarrow n$, $q_i \leftarrow P$, $M_i \leftarrow \sqrt{q^2_i}$\\
{\bf While} $i>1$ {\bf:}\\
\qquad Lorentz transform to rest frame of $q_i$ with $\Lambda^\mu_{\;\nu}$\\
\qquad Sample uniformly $u_{i1}, u_{i2} \in [0,1]$\\
\qquad $\phi_i \leftarrow 2\pi u_{i1}$ and $ \cos \theta_i \leftarrow 2u_{12}-1$\\
\qquad {\bf If} $i \geq 3${\bf :}\\
\qquad\qquad Sample uniformly $u_{i3} \in [0,1]$\\
\qquad\qquad Set $M_{i-1} \leftarrow \sum_{j=1}^{i-1} \mu_{i-1} + u_{i3} \braces{M_i-\mu_i}$\\
\qquad {\bf Else:}\\
\qquad\qquad Set $M_{i-1} \leftarrow m_1$\\
\qquad Set $|\mathbf{p'}_i| \leftarrow \frac{\lambda^{1/2}\braces{M^2_i,M^2_{i-1},m^2_i}}{2M_i}$
and $\mathbf{p'}_i \leftarrow |\mathbf{p'}_i|\braces{\sin\theta_i \cos\phi_i, \sin\theta_i \sin\phi_i, \cos\theta_i}$\\
\qquad Define $4{-}$momenta ${p'}_i^{\mu} \leftarrow (\sqrt{\abs{\mathbf{p'}_i}^2 + m_i^2},\mathbf{p'}_i),
\,{q'}_i^{\mu} \leftarrow (\sqrt{\abs{\mathbf{p'}_i}^2 + M_{i-1}^2},-\mathbf{p'}_i)$\\
\qquad Lorentz transform back to the lab frame with $\Lambda_\nu^{\;\mu}$\\
\qquad {\bf If} $i = 2${\bf :}\\
\qquad\qquad $p_1 \leftarrow q_1$\\
\qquad $i \leftarrow i - 1$\\
Set the weight $w$ of the generated event to equation~\ref{eqn:sequential-weight}\\
        \hline\noalign{\smallskip}
    \end{tabular}
    \caption{The hierarchical approach to sample weighted $n{-}$body phase space
    configurations.}
    \label{tab:hierarchical}
\end{table}
\endgroup
The pseudocode describing the algorithm is contained in
table~\ref{tab:hierarchical} and produces $n$ particles momenta configurations
that should be weighted with the following quantity, which takes into account
the jacobians of the transformations induced by the uniform random sampling:
\begin{equation}
    \label{eqn:sequential-weight}
    w = \frac{(2\pi)^{3-2n} 2^{1-n}}{M_n \braces{M_2-\mu_2}}
    \prod_{i=2}^n \braces{M_i-\mu_i} \sqrt{\lambda\braces{M^2_i,M^2_{i-1},m^2_i}}
\end{equation}

\subsubsection{RAMBO}
\label{ssubsec:rambo}
In contrast to the hierarchical approach which generates weighted events, the
RAndom Momenta BOoster (RAMBO)~\cite{Kleiss:1986} produces a set of $n$ physical
$4{-}$momenta with total momentum $P$, drawing uniformly numbers from the unit
hyper-cube in $\mathbb{R}^{4n}$. The algorithm treats differently the production
of massless and massive particles.

We first discuss the massless case. The $n{-}$massless particle phase space element
is given by:
\begin{equation}
    d\Phi_n = (2\pi)^4 \delta^{(4)}\big(P - {\textstyle\sum_i p_i}\big)
    \prod_{i=1}^n \frac{d^4p_i}{(2\pi)^3} \, \delta(p_i^2) \Theta(p_i^0) 
\end{equation}
To derive the RAMBO algorithm, instead, we consider the alternative form:
\begin{equation}
    dR_n = \prod_{i=1}^n \frac{d^4p_i}{(2\pi)^3} \, \delta(q_i^2) \Theta(q_i^0) (2\pi)^4 f(q_i^0) 
\end{equation}
where we have replaced the $4{-}$dimensional Dirac delta expressing the momentum
conservation with a generic function depending on the temporal component of the
$q_i$ momenta, $f(q_i^0)$. This function is needed to keep the total phase space
volume finite. Integrating over the Dirac delta to implement the on-shell
condition of the particles, we obtain:
\begin{equation}
    dR_n = (2\pi)^{4-2n} \mathbf{x} f(\mathbf{x}) \Theta{\mathbf{x}} d^n\mathbf{x}
\end{equation}
which shows how the integrals over the different particle momenta decouple in
this case.

We define a Lorentz plus a scaling transformation connecting the RAMBO \linebreak
$4{-}$momenta and the physical $p_i$ ones. The transformation is characterized
by the vector $\mathbf{b}$ and the scaling factor $x$, as follows:
\begin{equation}
    \begin{split}
&\gamma = \frac{Q^0}{M}=\sqrt{1+\mathbf{b}^2}, \quad
a = \frac{1}{1+\gamma}, \quad x = \frac{\sqrt{P^2}}{M} \\
& Q^\mu = \sum_{i=1}^n q_i^\mu, \quad M = \sqrt{Q^2}, \quad
\mathbf{b} = -\frac{1}{M}\mathbf{Q}
    \end{split}
\end{equation}

Therefore, the transformations linking the two sets are
$xH_{\mathbf{b}^\mu(q_i)}$ and its inverse \linebreak
$\frac{1}{x}H_{\mathbf{-b}^\mu(p_i)}$, such that:
\begin{equation}
    \label{eqn:rambo-transform}
    p^0_i = x(\gamma q_i^0 + \mathbf{b}\cdot \mathbf{q}_i), \qquad
    \mathbf{p}_i = x \sbraces{\mathbf{q}_i + \mathbf{b}q_i^0 +
    a (\mathbf{b}\cdot \mathbf{q}_i)\mathbf{b}}
\end{equation}

It is possible to show that expressing the RAMBO phase space element in terms of
the momenta $p_i$, the following equation holds:
\begin{equation}
    \begin{split}
        \label{eqn:rambo-split}
        dR_n &= \frac{d^4p_i}{(2\pi)^3} \delta(p_i^2)\Theta(p_i^0)
        (2\pi)^4 \delta^{(4)}\braces{P-\sum_{i=1}^n p_i} \\
        &\cdot \braces{\prod_{i=1}^n f\braces{\frac{1}{x}H^0_{-\mathbf{b}}(p_i)}}
        \frac{(P^2)^2}{x^{2n+1}\gamma}d^3\mathbf{b}dx
    \end{split}
\end{equation}
This equation factorizes the RAMBO phase space in the usual physical $n{-}$body
phase space for massless particles plus a contribution $S_n$ given by the second
line of equation~\ref{eqn:rambo-split}, which depends on the choice of the
auxiliary function $f(x)$. Fixing $f(x) = e^{-x}$ and performing the integrals
over the $\mathbf{b}$ and $x$ parameters the $S_n$ equals to:
\begin{equation}
    S_n = 2\pi (P^2)^{2-n}
    \frac{
        \Gamma\braces{\frac{3}{2}}\, \Gamma(n-1)\, \Gamma(2n)
    }{
        \Gamma\braces{n + \frac{1}{2}}
    }
\end{equation}
which depends on the total invariant mass $P^2$ only.

Therefore, it can be possible to generate the $n$ massless particles with the
following receipt:
\begin{enumerate}
    \item Generate $n$ $4{-}$momenta $q_i^\mu$ with isotropic angular
    distribution and energy sampled from the $q^0_ie^{-q_i}dq_i^0$ probability
    density function. This can be done by sampling uniformly $4n$ random numbers in
    the unit interval and combining them with:
    \begin{equation}
        \begin{split}
            c_i = 2u_{i1} - 1, \quad
            \phi_i = 2\pi u_{i2}, \quad
            q_i^0 = -\log(u_{i3} u_{i4})\\
            q_i^1 = q_i^0 \sqrt{1-c^2_i}\cos\phi_i, \quad
            q_i^2 = q_i^0 \sqrt{1-c^2_i}\sin\phi_i, \quad
            q_i^3 = q_i^0 c_i
        \end{split}
    \end{equation}
    \item  Transform the set of RAMBO momenta $q_i$ into the physical ones $p_i$
    with the help of the transformation appearing in
    equation~\ref{eqn:rambo-transform}.
    \item Attach to the produced configuration the flat weight:
    \begin{equation}
        \label{eqn:rambo-w0}
        w_0 = (2\pi)^{4-3n} \braces{\frac{\pi}{2}}^{n-1}
        \frac{(P^2)^{n-2}}{\Gamma(n)\Gamma(n-1)}
    \end{equation}
\end{enumerate}

In the massive case, there exists a similar approach that exploits the massless
case generation with a further transformation that allows providing $4{-}$vectors
associated with massive particles. As a result, the event weight is no longer
constant over the phase space, but depends on the generated momenta. The
algorithm is modified in the following way.
\begin{enumerate}
    \item Generate a set of momenta for the massless particles $p_i$.
    \item Compute the massive particle momenta $k_i$ from the rescaled massless
    momenta $\xi p_i^\mu$:
    \begin{equation}
        k_i^\mu = \braces{\sqrt{m_i^2 + (p_i^0)^2}, \xi \mathbf{p}_i}
    \end{equation}
    where the parameter $\xi$ is a constant quantity, constrained by the equation
    $\sqrt{P^2} = \sum_{i=1}^n \sqrt{m_i^2 + (\xi p_i^0)^2}$, which should be
    usually solved numerically.
    \item Attach the momentum configuration dependent weight $w= w_0 w_m$ to the
    event, where $w_0$ is given by equation~\ref{eqn:rambo-w0} and:
    \begin{equation}
        w_m = (P^2)^{2-n} \braces{\sum_{i=1}^k \abs{\mathbf{k}_i}}^{2n-3}\,
        \braces{\prod_{i=1}^n\frac{\abs{\mathbf{k}_i}}{k_i^0}}
        \braces{\sum_{i=1}^n \frac{\abs{\mathbf{k}_i}^2}{k_i^0}}^{-1}
    \end{equation}
\end{enumerate}

%% file: chapters/partI/chap2/chap2.tex
\chapter{Introduction to deep learning and its physics applications}
\label{chap:introcs}
\thispagestyle{plain}

\minitoc

In this chapter we review the basic concepts of machine learning (ML),
introducing learning algorithm and neural networks from the basics to the
definitions of the state-of-the-art techniques in this field of research. We
include also an overview of the most widely used optimization methods, exploited
to fit these algorithms on the input data.

We then collect the main artificial intelligence (AI) results in particle
physics by classifying the plethora of different models with respect to the
particular physics problem they aim to solve.

\input{chapters/partI/chap2/ai_base}

\input{chapters/partI/chap2/ai_review.tex}

%% file: chapters/partI/chap2/ai_base.tex
\section{Fundamental definitions and techniques of machine learning}
\label{sec:ml-intro}

\renewcommand{\myincludegraphics}[2][width=\textwidth]{
    \includegraphics[#1]{chapters/partI/chap2/ai_base_plots/#2}
}

In this section, we present the principles of machine learning. We first
define what a learning algorithm is and which kind of tasks ML tries to solve.
We inspire by the exhaustive exposure of~\cite[ch.~5]{Goodfellow:2016}. Then, we
introduce the fundamental building blocks of neural networks (NN), namely
artificial neurons, and how they are organized in feed-forward (FF) layers to
form a first simple NN example. We review the most widely used kinds of layers
in the literature, focusing, in particular, on the convolutional and attention
layers, which are the hot topics in the computer vision field. In the last part
of the section, we describe the main algorithms used to optimize neural networks.

\subsection{Learning algorithms}

\newtheorem*{defin}{Definition}

An ML algorithm is a model that can learn from data. A definition of what
such a learning algorithm dates back to 1997~\citep{Mitchell:1997}:
\begin{quoting}
\begin{defin}
    A computer program is said to learn from experience E with respect to some
    class of tasks T and performance measure P, if its performance at tasks in T,
    as measured by P, improves with experience E.
\end{defin}
\end{quoting}
In the following paragraphs, we briefly give intuitive descriptions and examples
of the abstract entities introduced in the quote: the task, the performance
measure and the experience.

\paragraph{The task}

The task refers to the goal of the learning process. It is important to underline
that learning is not identified with the task, but it is the way of achieving the
ability to perform the task. Tasks in ML are usually described by how the model
processes examples. An example is a collection of $n$ quantitative features,
measured from some object, that we want the model to inspect. This way we can
encode an example as a vector $\mathbf{x} \in \mathbb{R}^n$. For instance, we
treat a squared grayscale image as a vector in  $\mathbb{R}^{n\times n}$,
where each component is a $\SI{8}{\bit}$ integer, taking values in the $[0,255]$
range, representing the corresponding pixel's intensity.

The following list reports a summary of the most common machine learning tasks.
\begin{itemize}
    \item \textbf{Classification}: in this type of task, the program is asked to
    specify which of $m$ categories some input belongs to. Hence, the model will
    attempt to learn a function $f: \mathbb{R}^n \to \{1 \dots m\}$,
    which maps an example $\mathbf{x}$ to its category $y=f(\mathbf{x})$.
    \item \textbf{Machine Translation}: in this type of task, the network takes
    as input a sequence of symbols in some language and aims to convert them from
    that native language to another. ML algorithms are employed also in natural
    language translations, like English-Italian translations.
    \item \textbf{Density estimation}: in this type of estimation problem, the
    model is trained to output a function
    $p_{model}:\,\mathbb{R}^n \to \mathbb{R} $, where
    $p_{model}(\mathbf{x})$ can be interpreted as a probability density function
    on the space where the examples were drawn from.
\end{itemize}

\paragraph{The performance metric}

To assess the ability of the algorithm to accomplish some task, we have to define
a measure that quantifies the performance. The performance metric
happens to be designed specifically for each task the model should be optimized
for. For classification and transcription, the accuracy $acc.$ is often a good
choice, namely the ratio between the number of examples where the model makes a
correct prediction over the total. Alternatively, we can measure the error rate,
also called $0-1$ loss, which is equal to $1-acc.$. Of course, the metrics are
task-specific and sometimes it is desirable to design a custom loss function
that reflects which features we would like the model to learn, rather than which
system's behaviors we want to penalize.

A central concept in ML is generalization: we do not want the algorithm to fit
the distribution of the input data, we would like it to learn features underlying
data, instead. This is necessary to achieve good performance of examples that
are completely new to the model, that is when performing inference on new data.
It is common, then, to have two distinct collections of examples (datasets): the
training dataset and the test one. The former is exploited by the model to tweak
its parameters, while the latter is used to assess its performance.

\paragraph{The experience}

ML algorithms, depending on what kind of experience they are allowed to receive
during the learning process, can be grouped into three main categories:
supervised, unsupervised and reinforcement learning models.
\begin{itemize}
    \item Unsupervised learning algorithms experience a dataset containing many
    features. The model has to capture this information, learning the true
    probability distribution function $p(\mathbf{x})$ underlying the examples.
    The key idea is that the optimization of the cost function and the data
    themselves are enough to accomplish the given task.
    \item Supervised learning algorithms are trained on examples with known target
    labels. That is, each point in the dataset comes with information about the
    truth: the model is adapted through a trial and error process to output the
    desired value. For instance, the MNIST~\citep{Li:2012} database is a
    collection of $60,000$ grayscale $28\times28$ pixels images of handwritten
    digits, plus a vector of labels that identifies the correct category of each
    image. In this training mode, a model tries to predict the correct label $y$
    from an example $\mathbf{x}$, or, in other words, it tries to reproduce the
    conditional probability density function $p(y|\mathbf{x})$.
    \item In reinforcement learning algorithms, an agent has to learn how to
    interact with an environment to maximize a reward function. Models of this
    kind have to learn to take decisions to be successful. Therefore, robotics
    represents a natural field of application of these techniques.
\end{itemize}
The term supervised arises from the fact that the model is taught by the labels
what to do, while in unsupervised learning the database completely lacks this
information. It is worth noting that these categories are not formally defined and
always well separated as there are models that can be used to accomplish
different tasks. We would also like to mention that other variants of the
learning paradigm exist, such as semi-supervised learning, meta-learning and
multi-instance learning to name a few.

\paragraph{Generalization}

As we stated above, the central challenge in ML is to build models which perform
well on new unseen data. This idea remarks the difference between an optimization
problem and an ML algorithm: the former is a process in which we seek the model's
best configuration in parameter space to reduce the training error,
while in the latter we want the generalization error (also called test error) to
be small. The generalization error is the expected value of the error on a new
unseen input. Practically, it can be evaluated by averaging over the performance
achieved on the collection of examples called test dataset.

In general, it is not possible to gain insights on the test error knowing just
the training error, since train and test datasets represent two distinct
collections. To be able to affect the generalization error with training,
we make two claims on how the datasets are collected. First, the datasets should
contain independent examples. Second, these examples should be sampled from the
same probability distribution $p_\mathrm{data}$, dubbed data generating
distribution. The two hypotheses of independent and identically distributed
examples ensure that the training error sets an upper limit for the
generalization error. It is possible, then, to influence the final performance
of an ML algorithm by monitoring and reducing two fundamental quantities: the
training error itself and its gap from the test error.

\begin{figure}
    \centering
    \subfigure[Underfitting]{
        \myincludegraphics[width=0.3\textwidth]{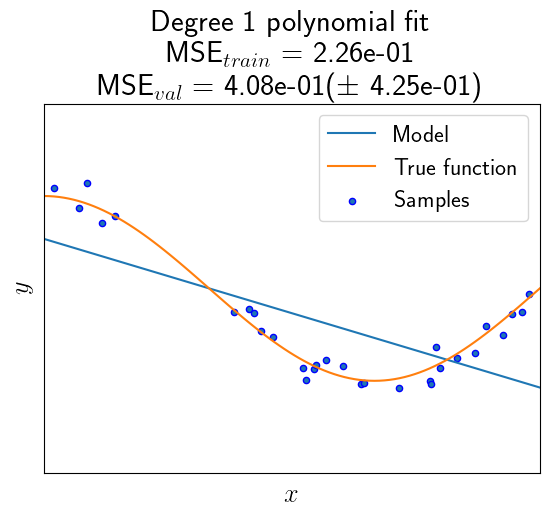}
    }
    \hfill
    \subfigure[Good generalization]{
        \myincludegraphics[width=0.3\textwidth]{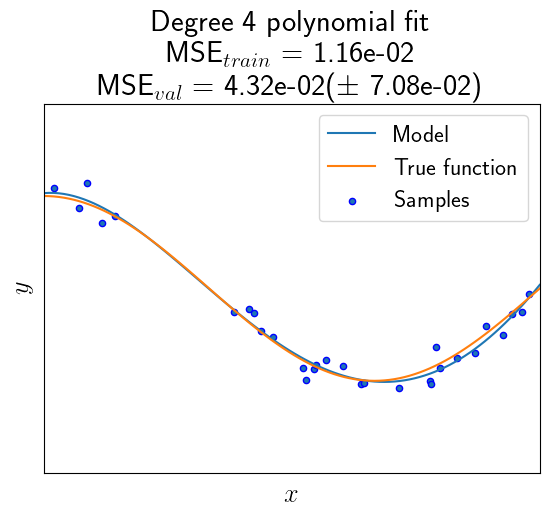}
    }
    \hfill
    \subfigure[Overfitting]{
        \myincludegraphics[width=0.3\textwidth]{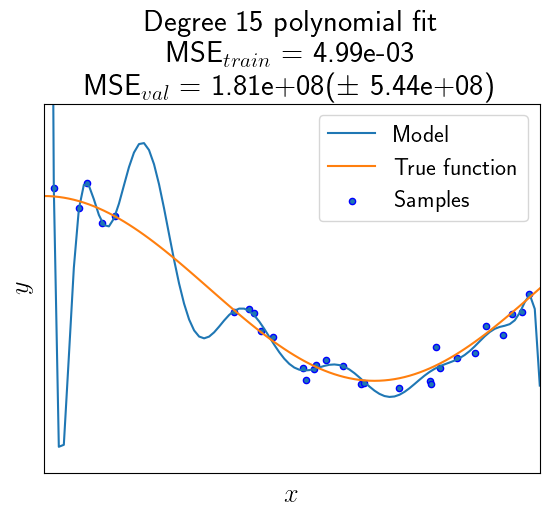}
    }
    \caption{Three limit cases. \emph{Left panel}, the linear output model has
    not captured the curvature present in the data. Training errors are high:
    underfitting. \emph{Central panel}, the model fits well all the data:
    good generalization. \emph{Right panel}, the model interpolates the samples
    with a complicated function with low training error, but high validation
    error: overfitting. The figure source code is inspired from the
    Scikit-learn~\cite{Pedregosa:2011} package documentation.}
    \label{fig:good-fit}
\end{figure}

When we do not manage to reduce one of such measures, we face two classical
undesired behaviors of an ML algorithm, called respectively underfitting and
overfitting. The former means that the model's parameters are not optimized
enough to perform the task and more training steps are required to lower the
fundamental quantities. The latter, instead, means that the algorithm is not
learning the true data-generating distribution, but it is fitting the training
points with a complex function. Borderline cases are displayed in
figure~\ref{fig:good-fit}.

To better explain the concept, we can intuitively think about a student.
Underfitting corresponds, of course, to the situation in which the student
learned the topic just superficially. The student, when is able to generalize,
has mastered the subject and can ingeniously apply what he has learned, is the
best; whereas when he just parrots back the lesson, he is simply overfitting the
issue. 

\begin{figure}
    \centering
    \myincludegraphics[width=0.5\textwidth]{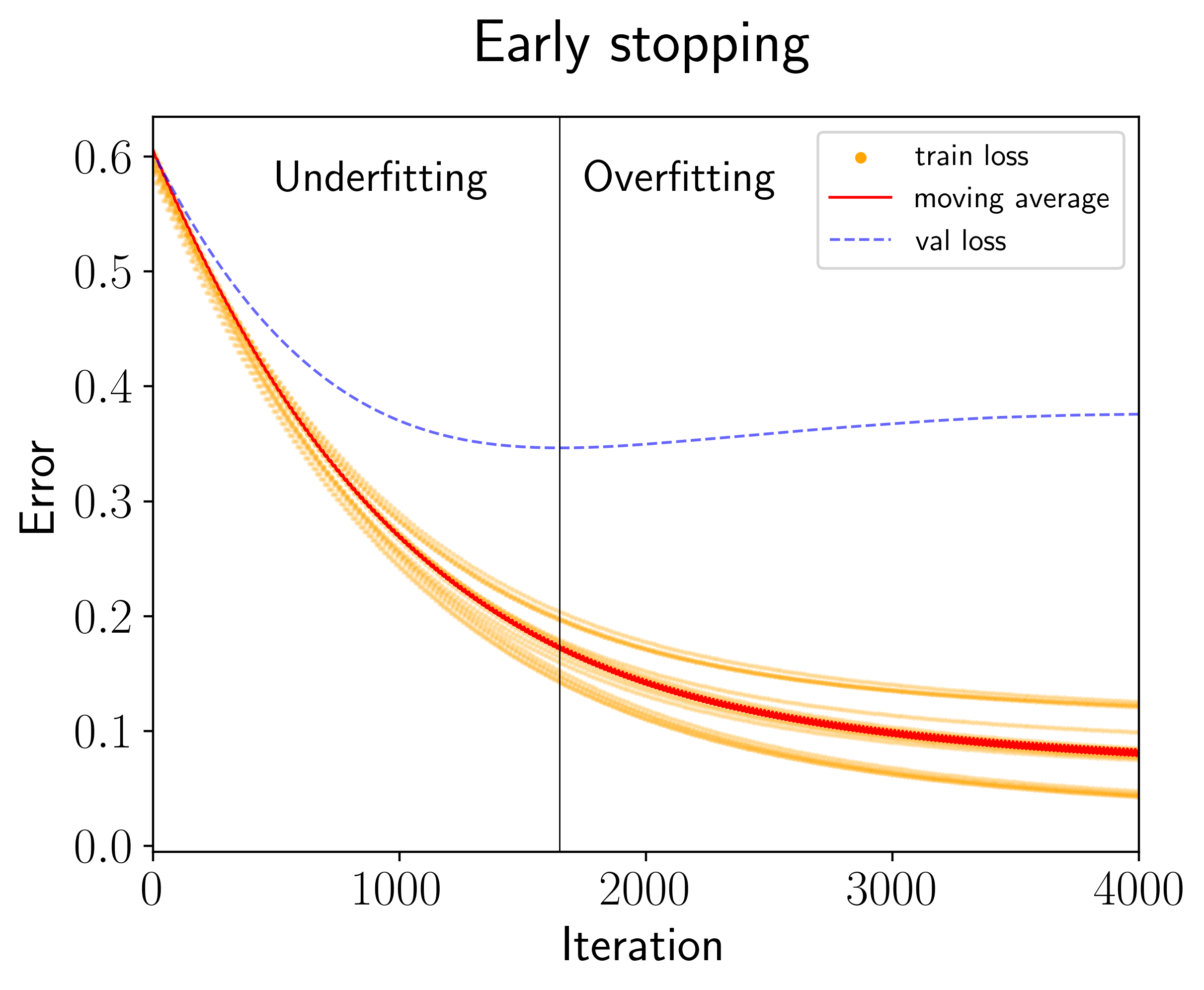}
    \caption{Typical trends of training and generalization (over the
    validation set) errors during training. Eventually, the training curve
    flattens (orange points and red curve), while the validation loss starts
    increasing. This behavior signals overfitting and the black line marks the
    best moment to stop the optimization algorithm, namely, as soon as the
    generalization gap starts increasing.}
    \label{fig:early-stopping}
\end{figure}

Within the ML field, we protect ourselves from these two bad behaviors, by
stopping the optimization algorithm at the right time. The question is, of course,
to understand which is the right time. A common technique used to address this
problem is early stopping. At each optimization iteration, both the training
error and the generalization error are evaluated. The training error is given by
the loss function value, while an estimate for the generalization error is
computed on a separate set of examples, usually called the validation dataset. The
validation dataset is a collection of examples that are not used for
optimization but are generally employed to tweak some un-trainable parameters
of the model. Figure~\ref{fig:early-stopping} shows the typical trend of the
error functions during training: according to the early stopping prescription,
the training should be stopped when the gap between training and validation error
starts increasing. This roughly provides a trade-off between underfitting and
overfitting regimes.

We have described what an ML algorithm is following Mitchell's
definition~\citep{Mitchell:1997}. We linked it to a statistical modeling
procedure, where the algorithm optimizes its output by processing vectorial
examples and tries to grasp the true underlying data generating distribution
$p_\mathrm{data}$. Since this distribution is usually quite complicated,
we have to be sure that our model will be capable of correctly reproducing it.
This motivations, supported by the universal approximation
theorem~\citep{Hornik:1989,Zhou:2020} lead us to introduce neural networks (NNs)
as function approximants that can achieve the goals of an ML algorithm.

\subsection{Neural networks}

We introduce the concept of the NN explaining the characteristics of artificial
neurons, named like this because their behavior resembles that of a biological
neuron. Furthermore, neurons can be arranged
together to form layers, which are the building blocks of neural networks, that
in turn can accomplish an incredible variety of tasks, if properly designed and
tuned. We overview the different types of layers widely employed in the
literature and explain how it is possible to fix a very large number of
parameters to allow the process of learning. With this in mind, we hint
at the main ideas behind the most common optimization algorithms.

An artificial neuron is defined as a function
$f_{\mathbf{w},b}:\, \mathbb{R}^n \to \mathbb{R}$ that maps a
collection of $k$ input signals $\mathbf{x} = \{x_0,\dots,x_{k-1}\}$ to an
output:
\begin{equation}
    y = f_{\mathbf{w},b} (\mathbf{x}) = \varphi_{\mathbf{w},b}(w_j x_j+b)
\end{equation}
where $\mathbf{w}$ is a vector of $k$ weights and $b$ is a real coefficient
called bias; in general, we refer to these quantities as the model's parameters
$\mathbf{\theta}$. It is interesting to note the role of the so called activation
function $\varphi$, because it introduces the opportunity to have non-linearities
in an otherwise affine transformation. Table~\ref{tab:activations} lists the most
common activations used by ML models.

\newcommand{\lrelu}{
    $\varphi_\alpha(x)=
    \begin{cases}
        \alpha x & \text{is $x < 0$} \\
        x & \text{if $x \geq 0$}
    \end{cases}$
}

\newcommand{\elu}{
    $\varphi_\alpha(x)=
    \begin{cases}
        \alpha (e^x - 1) & \text{is $x \leq 0$} \\
        x & \text{if $x \geq 0$}
    \end{cases}$
}

\begingroup
\renewcommand{\arraystretch}{1.7}    
\begin{table}
    \centering
    \scriptsize
    \begin{tabular}{llc}
\noalign{\smallskip}\hline\noalign{\smallskip}
Name & Support & Equation \\
\noalign{\smallskip}\hline\noalign{\smallskip}
Linear & $\mathbb{R}$ & $\varphi(x) = x$\\
Rectified Linear Unit (ReLU) & $\mathbb{R}^+$ & $\varphi(x) = \max(0, x)$\\
LeakyReLU & $\mathbb{R}$ & \lrelu\\
Exponential Linear Unit (ELU) & $(-\alpha, \infty)$ & \elu \\
Hyperbolic tangent & $(-1,1)$ & $\varphi(x) = \tanh(x)$ \\
Logistic (a.k.a. sigmoid)& $(0,1)$ & $\varphi(x) = \sigma(x) = 1/\braces{1+e^{-x}}$ \\
Softmax & $(0,1)$ & $\varphi_m(x_k) = e^{x_k} / \braces{\sum_{i=1}^{m} e^{x_i}})$ \\
        \noalign{\smallskip}\hline
    \end{tabular}
    \caption{List of the most common activation functions. The parameter $\alpha$
    in ReLU and ELU is an input positive constant. The SoftMax activation is typical
    of multi-class problems: $\varphi_m(x_k)$ is the probability to obtain the
    $k{-}$th category out of a total of $m$ classes.}
    \label{tab:activations}
\end{table}
\endgroup

\begin{figure}
    \begin{minipage}{0.5\textwidth}
        \centering
        \myincludegraphics[height=4.25cm]{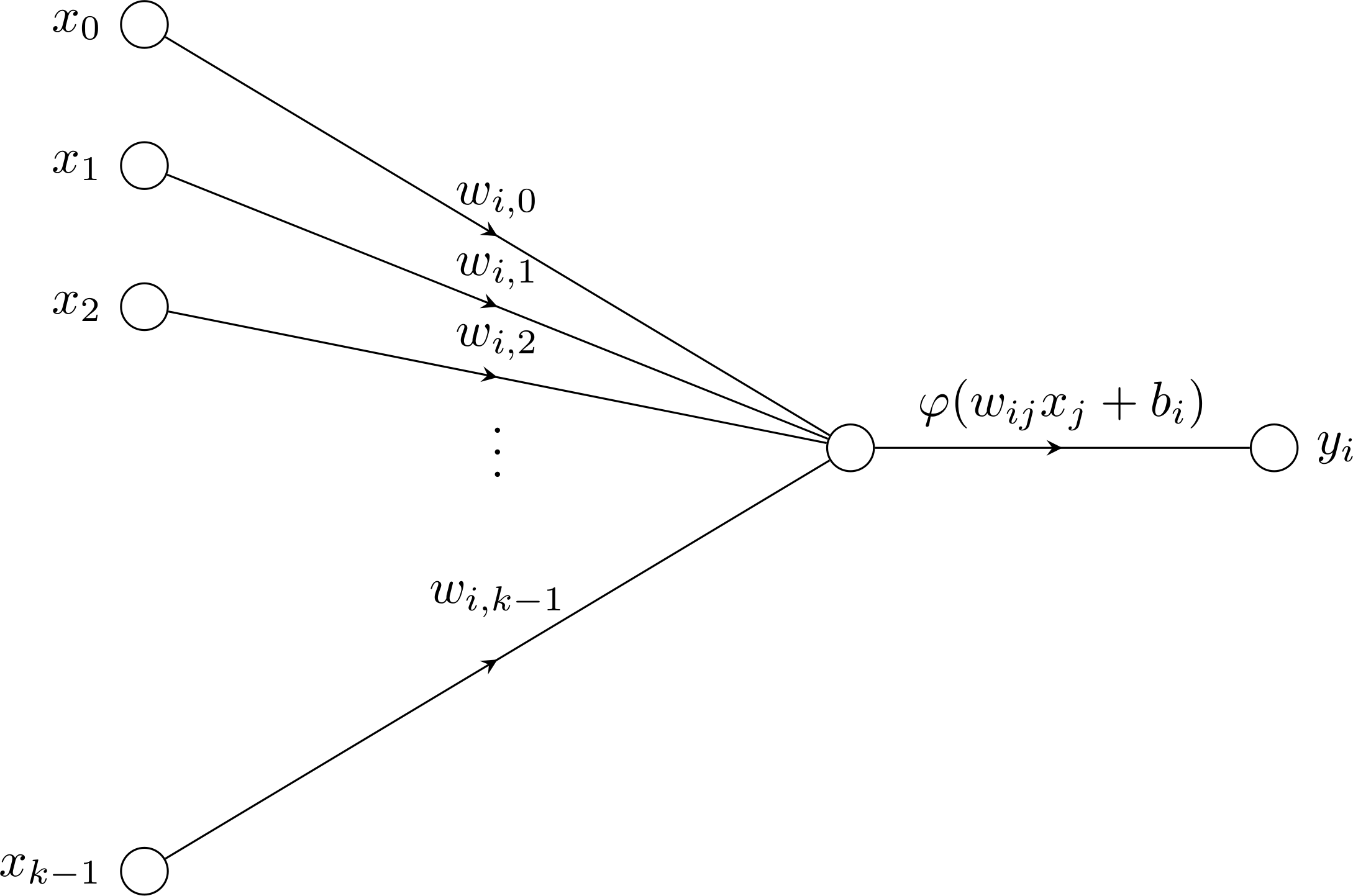}
        \caption{An artificial neuron.}
        \label{fig:neuron}
    \end{minipage}
    \hfill
    \begin{minipage}{0.45\textwidth}
        \centering
        \myincludegraphics[height=4.25cm]{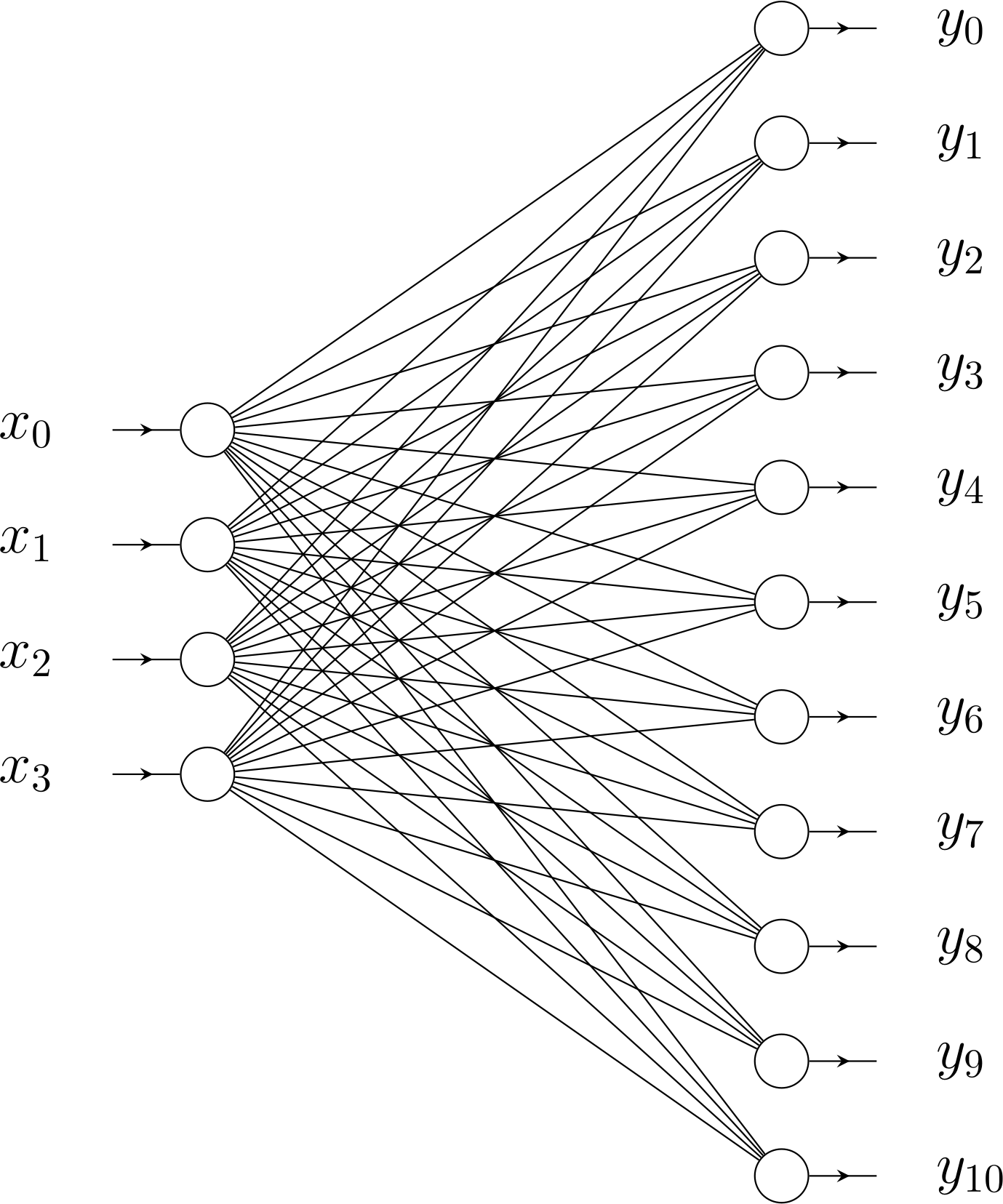}
        \caption{A Feed-Forward or dense layer.}
        \label{fig:fflayer}
    \end{minipage}
\end{figure}

It is possible to group a set of $n$ neurons to form a dense layer.
Hence, the vector of $w_j$ weights and the bias $b$ become an $n \times k$ matrix
$w_{ij}$ and an $n$-dimensional vector $b_i$, respectively. The layer has now $n$
outputs:
\begin{equation}
y_i = f (o_i) = \varphi(w_{ij} x_j+b_i) \quad \mathrm{for} \, i = 0, \dots ,n-1
\end{equation}
where we employed Einstein's convention of implicitly summing over repeated
indices. Figure~\ref{fig: neuron} pictorially displays the $i{-}$th neuron inside a
layer and figure~\ref{fig:fflayer} plots a collection of neurons to form a feed
forward or dense layer.

We now look to several classes of layers that employ alternative operations to
process inputs. Such layers are more specialized than the general feed-forward
one: each of them has a peculiar structure and therefore is best suited to
solve particular tasks.

\paragraph{Convolutional layers}
\label{par:cnns}

Convolution is a mathematical operation on two functions of a real argument. Let,
then, $f ,g:\mathbb{R} \to \mathbb{R}$ be two real valued functions.
Their convolution is the function $(f\ast g):\mathbb{R} \to \mathbb{R}$,
defined by the following mapping:
\begin{equation}
    t\longmapsto (f\ast g)(t) =  \int_{-\infty}^{+\infty} f(\tau) g(t-\tau) d\tau
\end{equation}
In general, we can look at convolution, as an operator that applies a filtering
function $g$ on an input function $f$: $g$ is called kernel in ML applications
and the output is sometimes named the feature map. Furthermore, if $g$ is also a
probability density function (pdf), the output $f \ast g$ is an average of $f$
weighted with the pdf $g$ ,or, equivalently, the expected value of $f$ by means
of $g$.

\begin{figure}
    \centering
    \subfigure[An RGB image represented as a grid.\label{subfig:image-grid}]{
        \myincludegraphics[width=0.45\textwidth]{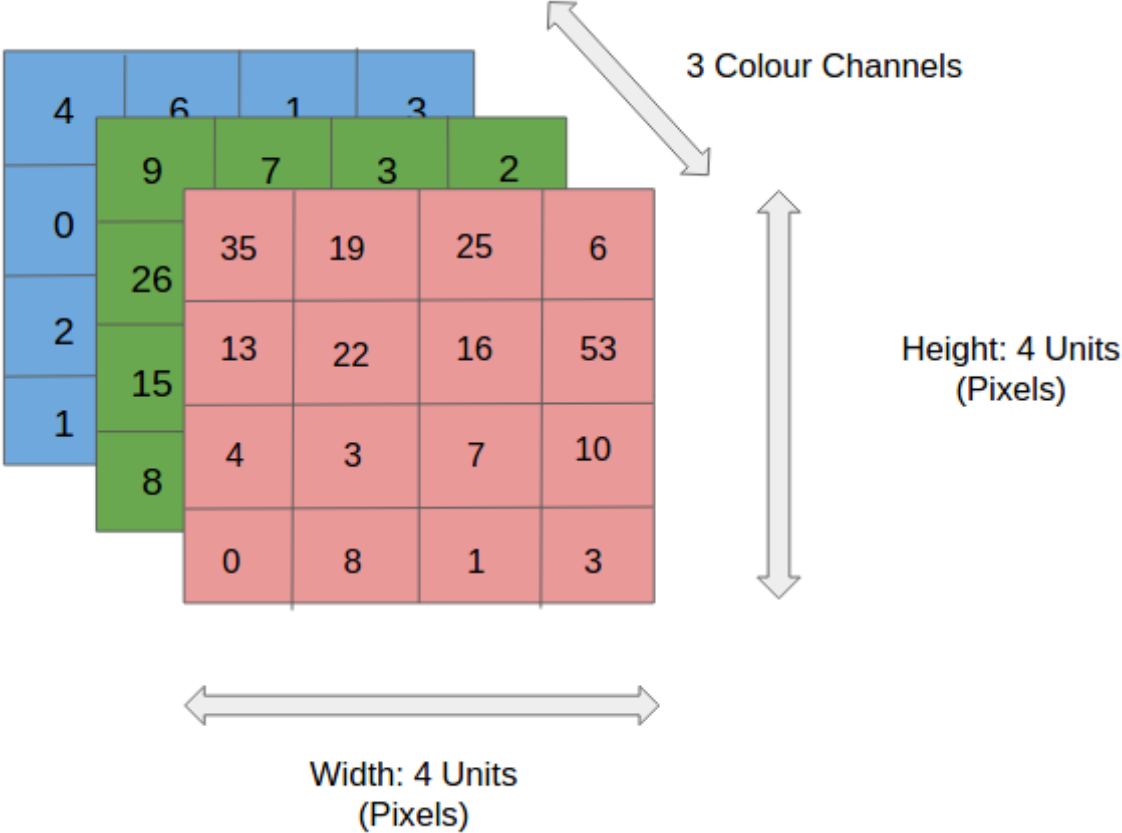}
    }
    \hfill
    \subfigure[Convolution on a pixel.\label{subfig:convolution}]{
        \myincludegraphics[width=0.45\textwidth]{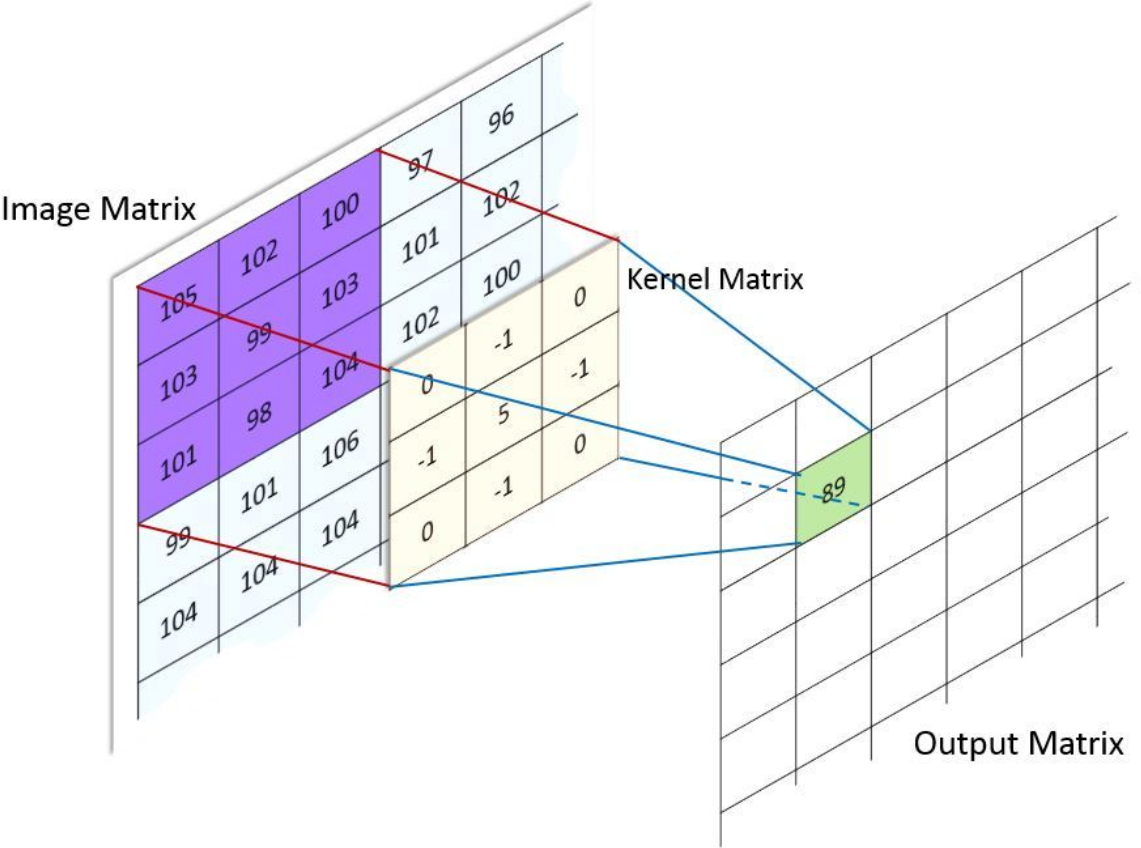}
    }
    \caption{Images, both grayscale and RGB, are represented by rank 3 tensors.
    Convolution is the state-of-the-art operation in ML for image processing.}
    \label{fig:conv-layer}
\end{figure}

To apply convolution within an ML framework, we introduce a discretized
form of this operation: the arguments of real functions become integer indices.
Convolutional layers are usually employed, with great success, to process images,
which can be seen as grids of pixels described by rank $3$ tensors, as in
figure~\ref{subfig:image-grid}: the first two indices identify the row and the
column in the grids, while the third one refers to the pixel's channels, which
mixes primary colors. One-channeled images are grayscale, otherwise
descriptions with three (RGB for red, green, and blue) or four channels (RGBA for
red, green, blue and alpha, which measures transparency) are suited for colored
ones.

The convolutional kernel $K$ contains all the layer's information and it
is represented by a tensor with the following index structure: the first two
labels refer to the size of the filtering window (row and columns), the third one
runs from $1$ to the number of channels in the input image $c_{in}$, while the
last corresponds to the number of channels in the output image $c_{out}$.
Convolution is the operation in which we apply these multi-dimensional filters
to different subsets, with shapes equal to the filtering window, of consecutive
pixels in the input image. Mathematically we write:
\begin{equation}
    O_{i,j,k} = \sum_{l=0}^{n_r-1} \sum_{m=0}^{n_c-1} \sum_{n=1}^{c_{in}}
    I_{i\times r+l,j\times s+m,n} \, K_{l,m,n,k}
    \label{eqn:convolution}
\end{equation}
where we introduced the possibility to have strides $r$ and $s$.
Figure~\ref{subfig:convolution} visually shows the convolution operation with
single-channeled input, kernel and outputs.

The stride parameters tell the model not to inspect each consecutive subset of
pixels: the convolution in this case skips respectively $r$ and $s$ image's cells
in each direction before taking again the convolution operation. The stride option
affects the information overlap between near pixels in the output image: having
minimum strides in each direction, namely equal to one, ensures that the maximum
amount of information is retained within the output image. Nonetheless, this is
computationally expensive and sometimes, in fact, does not improve the
performance of the model. The choice of these parameters is in any case
dataset-specific.

Looking at equation~\ref{eqn:convolution}, it is clear that there is an issue
when the kernel deals with cells next to the boundaries: there, the sum's indices
would go out of range for the input image. A solution consists in carrying on the
convolution operation only until the kernel lies entirely inside the image. This
option is called \texttt{valid} convolution by ML libraries. Of course, the
output image will be shrunk in comparison to the input one. The opposite
behavior is the \texttt{same} convolution, in which the layer implicitly
zero-pads the image to have input and output images of the same shape. The
\texttt{same} option has the drawback that pixels near the borders of the input
image influence a smaller amount of cells in the output than the ones in the
middle. Optimality between these alternatives, of course, is not an absolute
fact, but strongly depends on the input data: the best choice lies, in general,
somewhere between the \texttt{valid} and the \texttt{same} modes.

Three key ideas support the introduction of convolutional layers in neural
networks: sparse interactions, parameter sharing and equivariant representations.
In the following, we give a brief introduction to these concepts, to motivate
the intense usage of convolutional layers in the literature.

Figure~\ref{fig:fflayer} showed the connections established by a dense layer
with the input; in particular, we highlight that each neuron inside the layer is
linked with each component of the input vector. Due to this aspect, dense layers
are also called fully connected. The number of weights used by a single layer
scales linearly with the dimensionality of the inputs.  Moreover, images span a
two-dimensional space and their number of pixels grows quadratically with the
size of the image edge, even containing millions of pixels in the case of
high-resolution pictures. It is clear, then, that inspecting images with fully
connected layers can be very computationally expensive, both in terms of storing
the whole amount of weights and working out the matrix multiplication operations.

\begin{figure}
    \centering
    \myincludegraphics[width=0.7\textwidth]{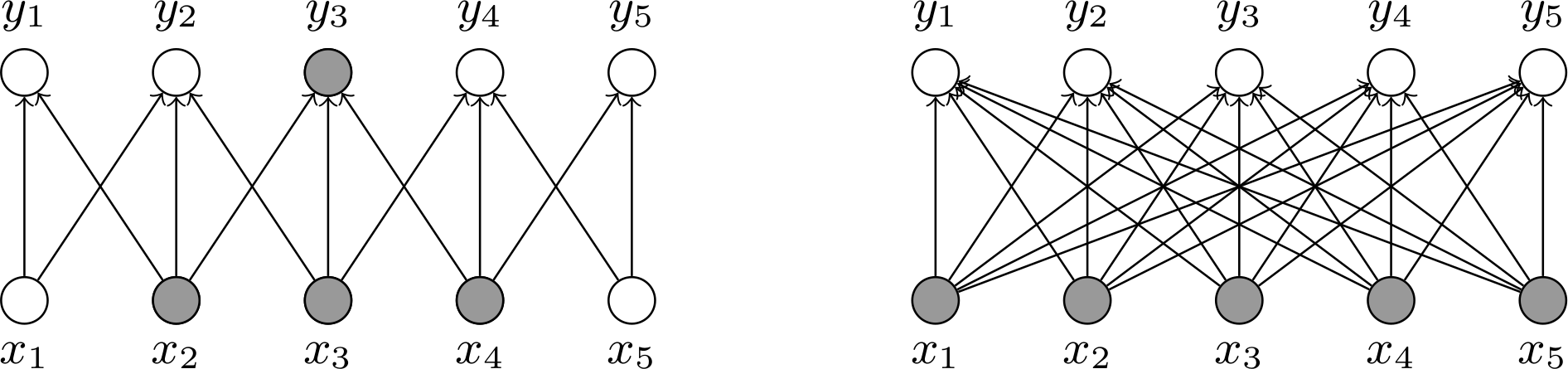}
    \caption{Sparse interactions vs full connectivity. Grey shaded circles the
    central node receptive field. \emph{Left}: convolutional layer with kernel
    size equal to three, stride one and same padding. \emph{Right}: fully
    connected layer. Each neuron in the output is linked to every neuron in the
    input. Even with this simple model, we have a large number of edges in the
    graph.}
    \label{fig:sparse-connectivity}
\end{figure}

The introduction of convolutional layers allows having a smaller number of
weights, depending only on the size of the kernel K, which is often way smaller
than the input image. We refer to this property of a convolutional layer
saying that it presents sparse interactions or sparse connectivity. The dense
layer indeed can be seen as a fully connected graph with nodes corresponding to
the input and output vector components. The graph associated with the
convolutional layer, instead, can be obtained by the previous one retaining only
the edges that connect pixels closer than the kernel size. This is optimal for
image analysis since in this field we are often interested in looking for
patterns arising within a small portion of the image: linking together very
distant pixels might be just an overshooting. Sparse connectivity is the natural
way to address these issues. Figure~\ref{fig:sparse-connectivity} graphically
reviews the concept.

We previously mentioned that storing the enormous number of weights of dense
layers may become expensive in terms of memory usage. Convolutional layers
provide a simple answer to this problem, called parameter sharing. When training
a fully connected layer, the model has to learn the correct weight for each link
in the graph. In convolutional layers, instead, the model has to learn a small
set of kernel weights and then re-apply them (this consists in sharing parameters)
to inspect each portion of the image, resulting in a dramatic reduction of the
total amount of memory needed to store the model.

Due to this form of parameter sharing, the model inherits the well-desired
property called translation equivariance. In particular, we say that a
function $f$ is equivariant to a function $g$ if $f(g(x)) = g(f(x))$. In the
present case, convolution is equivariant to translations means that the
application order of the two processes on the input image does not matter: if we
slightly move the input image and then compute the convolution, the result will
be the same as if we made the convolution and then shifted the output. We can
also say that a convolutional layer looks for certain features in the input, no
matter where they are.

Convolutional layers represent very useful and efficient tools to analyze images,
but they usually come with another operation that modifies further their output
values. This operation is called pooling. Different types of pooling layers exist,
but they all exploit the same idea: they replace the value of each output unit of
a convolutional layer with a statistical metric that aggregates information from
the nearby pixels. Of course, different metrics are possible: the most used are
max pooling~\cite{Zhou:1988}, average pooling, weighted average pooling and $L^2$
norm pooling. Pooling is useful because it makes the output invariant under small
translations of the input. Indeed, if we apply max pooling over a small area,
results would be the same if we shifted by a small amount the input image before
because the maximum would be obtained inside the same small area.

\paragraph{Attention mechanism}

The introduction of the attention mechanism~\citep{Vaswani:2017} marked a \linebreak
milestone in the field of deep learning, especially among the Natural Language
Processing (NLP) models. This particular kind of layer was, indeed, firstly
designed and exploited in Transformers architectures to handle sequences of data
representing sentences of words. However, given its great success, a plethora of
variants have been proposed to either provide improvements on the standard
technique or adapt the model to other input structures, such as images. In
particular, the Hugging Face\footnote{
    \url{https://huggingface.co/docs/transformers/index}\\
    \url{https://github.com/huggingface/transformers}}
community published a large collection of such models~\citep{Wolf:2020}, which
includes the majority of the most widely used architectures. In this paragraph we
give an overview of the original work on attention, leaving its alternatives to
the interested reader.

\begin{figure}
    \begin{minipage}{0.5\textwidth}
        \centering
        \vspace{2cm}
        \myincludegraphics[height=4cm]{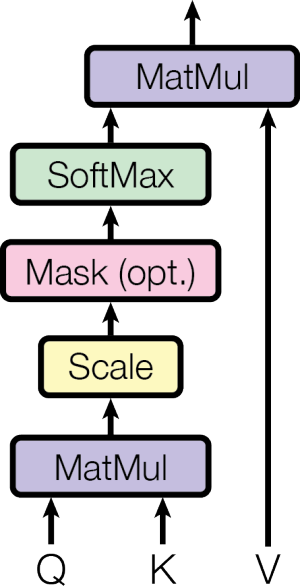}
        \caption{The attention mechanism.}
        \label{fig:attention}        
    \end{minipage}
    \hfill
    \begin{minipage}{0.5\textwidth}
        \centering
        \myincludegraphics[height=6cm]{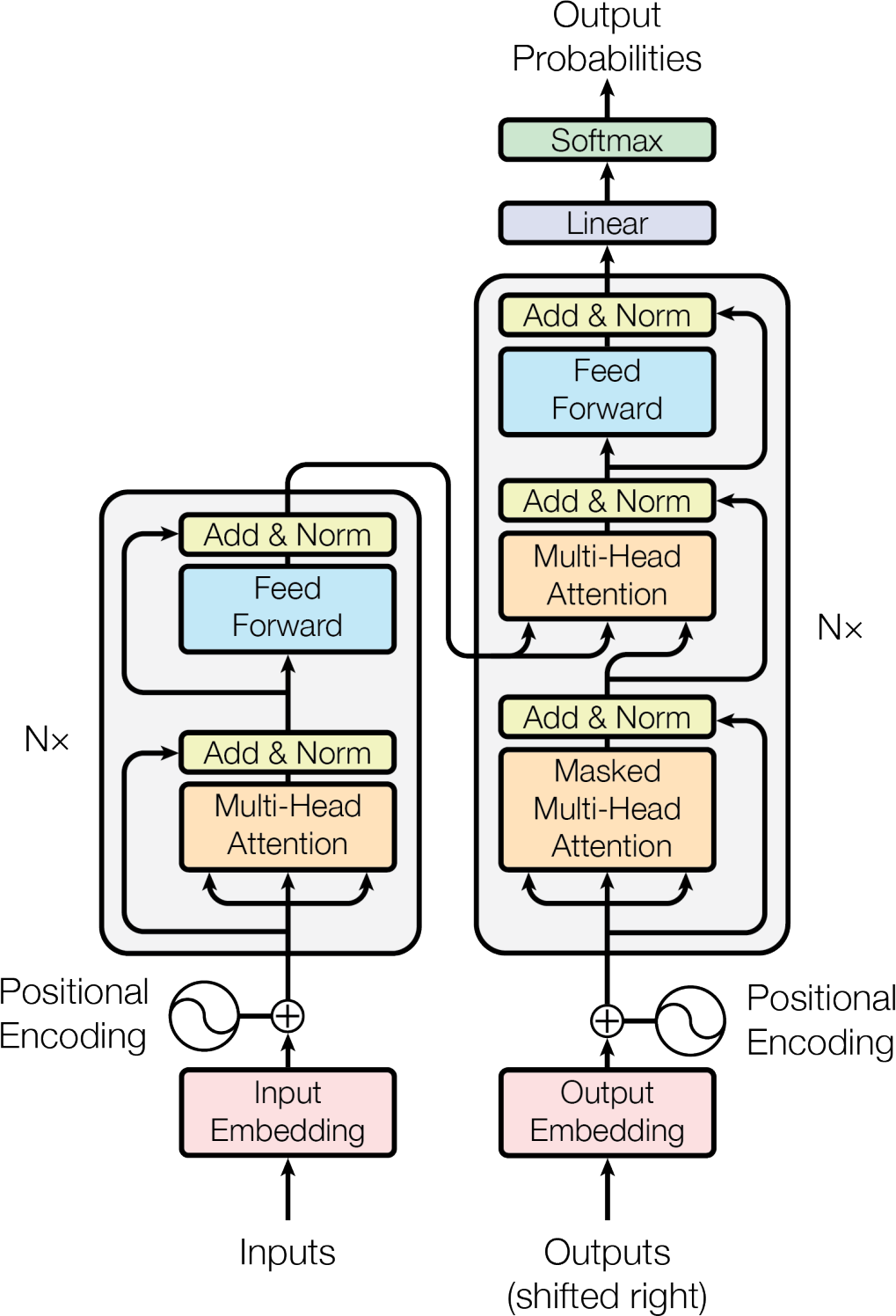}
        \caption{The Transformer model.}
        \label{fig:transformer}
    \end{minipage}
\end{figure}
The attention mechanism works as in figure~\ref{fig:attention}: it accepts three
inputs named key $\mathbf{k} \in \mathbb{R}^{L\times d_k}$, query
$\mathbf{q} \in \mathbb{R}^{L\times d_q}$ and value
$\mathbf{v} \in \mathbb{R}^{L\times d_v}$, respectively. The inputs are encoded
in matrices, representing sequences of length $L$ where each element is a vector
with an input-dependent dimension $d_{k/q/v}$.
The input vectors are initially transformed into a projection space by three
independent matrices of trainable parameters: $W_k \in \mathbb{R}^{d_h\times d_k}$,
$W_q \in \mathbb{R}^{d_h\times d_k}$ and $W_v \in \mathbb{R}^{d_{o}\times d_v}$.
Notice that both queries and keys are projected into a space with the same
dimensionality.

After this first transformation, the query
$Q = W_q \, \mathbf{q} \in \mathbb{R}^{L\times d_h}$
and key $K = W_k \, \mathbf{k} \in \mathbb{R}^{L\times d_h}$ quantities are
multiplied together to form a matrix of attention weights according to the
following equation:
\begin{equation}
    A_{ik} = \mathrm{SoftMax}\braces{\frac{QK^T}{\sqrt{d_h}}}_{ik} =
    \frac{
        \exp\braces{Q_{ij}K_{kj}/\sqrt{d_h}}
    }{
        \sum_{m=1}^L \exp\braces{Q_{ij}K_{mj}/\sqrt{d_h}}
    } 
    \quad \in \mathbb{R}^{L\times L}
\end{equation}
where the quantity $\sqrt{d_h}$ acts as a normalization factor in the $Q\,K^T$
matrix product.

The $\mathrm{SoftMax}$ operation is included to ensure that each row vector of
the attention matrix is correctly normalized and represents a probability density
function on the sequence space $\mathbb{R}^L$. The final output of the attention
layer is given by the matrix multiplication of the attention matrix and the
projected value vector:
\begin{equation}
    \mathrm{Attention}(Q, K, V)_{ij} = A_{ik} V_{kj} 
    \quad \in \mathbb{R}^{L\times d_o}
    \label{eqn:attention}
\end{equation}
As a consequence, each row in the attention matrix is a vector of weights
that mixes the elements of the value input. In other terms, the element
$A_{ik}$ can be interpreted as the amount of attention that the model pays to the
$k{-}$th word of the value sentence to produce the $i{-}$th token of the output.
The most widely used form of attention is, in fact, the self-attention, which is
obtained from equation~\ref{eqn:attention} setting the inputs as
$\mathbf{q} = \mathbf{k} = \mathbf{v}$.

The attention operation showed incredible flexibility and tremendously improved
the performance of language models, allowing them to scale towards larger and larger
architectures containing billions of parameters. The success of the attention
operation lies in the fact that it is the learning process itself that gives
importance to some features of the inputs rather than others. Moreover, the entire
input sequence is allowed to influence the output, providing improved model capacity
and expressiveness. The attention operation can be easily parallelized since it
is mainly comprised of matrix multiplications: therefore, a multi-head attention
version has been defined by the original authors, where multiple attention layers
work in parallel to obtain output sequences which are concatenated and eventually
passed through a last matrix of weights for further mixing.

Besides the main advantages of introducing the attention mechanism, which also
include the ability to handle inputs of varying sizes, the construction operation
of the weight matrix does not scale well with the length of the input sequence.
Indeed, the complexity grows as $\orderof{L^2}$ and large amount of memory is
needed to hold the computation if the length of the sequence reaches hundreds or
thousands of elements. Efforts to achieve a more efficient attention computation
have been made in the literature: among them, we note the
Performer~\citep{Choromanski:2020}, where the authors try to approximate the
softmax function with kernel methods to achieve linear complexity of
the overall attention transformation.

The attention mechanism has been also exploited in the treatment of kind of data
different than $1{-}$dimensional sequences of words: the Visual Transformer
(ViT)~\citep{Dosovitskiy:2020} and its variants successfully applied these models
to computer vision tasks. The main difficulty for attention-based models in
processing images is the fact that, as opposed to CNNs, they completely lack
geometrical inductive biases. This implies that training is more difficult and
unstable, especially in the early stages of optimization.

\subsection{Optimization methods}

In the previous sections we discussed several kinds of network layers, which are
the building blocks of neural network architectures. Layers can be stacked
one on top of the other to design a custom pipeline that computes the desired
outputs. Neural networks are comprised of input and output layers, plus a certain
number of hidden layers that sequentially connect the first two. In general, we
can represent a neural network by a directed graph whose nodes are the different
layers and the edges encode the data processing flow.

Feed-Forward neural networks (FFNN) form, in particular, directed acyclic graphs
comprised of fully connected layers. For this simple form of network, it is
possible to define the width of each hidden layer as its number of neurons as
well as the depth of the network, being the number of hidden layers. The width
and the depth influence the number of trainable parameters $\mathbf{\theta}$ in
the model and, hence, its complexity. A NN can be viewed as a family of functions
$\{f(\mathbf{x})\}_{\mathbf{\theta}}$ as $\mathbf{\theta}$ varies in a
high-dimensional parameter space. Theoretically, as stated by the universal
approximation theorem~\citep{Hornik:1989,Zhou:2020}, a NN can approximate any
continuous function on a compact subset of $\mathbb{R}^n$ with a particular
choice of $\mathbf{\theta}$. Flexibility, then, makes NNs fundamental tools in ML
algorithms, where we try to guess the data-generating distribution. The challenge
consists, of course, in finding the best parameter configuration with an
efficient training algorithm. Such a process is called optimization.

In this section, we explain the main ideas behind optimization algorithms found in
ML literature: we take a look at gradient descent, momentum-driven and adaptive
optimizers. As stated above, the improvements of a model during training are
assessed by computing the value of a performance function associated with it, often
called cost or loss function $\mathcal{L}$. As a general prescription, the
problem of learning is cast in the form of an optimization problem: the goal is
to find the point in the multi-dimensional parameter space that corresponds to
the minimum of the cost function.

The optimization is usually accomplished in three different steps.
\begin{enumerate}
    \item Feed the model with a batch of examples from the dataset and obtain
    the corresponding outputs (feed-forwarding).
    \item Find the gradient of the loss function with respect to the model's
    parameters (back-propagation~\citep{Rumelhart:1986}). This is done with
    an automatic differentiation algorithm, always implemented by ML libraries.
    \item Update the model's weights according to a particular updating rule
    (depends on the optimization algorithm chosen).
\end{enumerate}
An epoch of training is the time taken to pass all the training dataset data
into this algorithm. The whole training consists of multiple epochs, generally
until the loss function converges around a minimum value. We refer to this
process as gradient descent.

An important quantity that influences the training is called batch size. The first
step of the algorithm does not say how many samples should be contained in
a batch for the forward pass. The usual approach consists in dividing the dataset
into mini-batches of fixed size and shuffling randomly the training points at the
end of each epoch. This introduces random fluctuations in the optimization process
helping the algorithm to explore better the weights space. The number of examples
in each mini-batch, namely the batch size, can drift the process towards two
alternative behaviors. If the batch size is too low, subsequent gradient updates
will tend to fluctuate and provide opposite contributions, yielding algebraic
cancellations that stop the overall optimization. On the other hand, a batch size
that is too high will average out the information carried by each example,
leading to sub-optimal results.

Step $2$ and $3$ are driven by back-propagation, which is at the core of the
learning process: derivatives of the loss function with respect to the
model's trainable parameters are computed by applying the chain rule from the output
backward to the inputs of the model. The overall gradient is a vector in the
space of the weights and it is exploited to jump from one configuration of the
model to another that provides the highest possible negative change in the loss
function. The following equation describes the update rule for the Stochastic
Gradient Descent (SGD) algorithm, which gives the simplest recipe to implement
the optimization process:
\begin{equation}
    \mathbf{\theta} \longleftarrow \mathbf{\theta}
    - \eta \nabla_{\mathbf{\theta}}\mathcal{L}
    \label{eqn:sgd}
\end{equation}
Of course, the gradient computed during a batch update is only an estimate of
the loss function associated with the task: the better the data, the better the
approximation of the true gradient and the easier the optimization.

\begin{figure}
    \centering
    \hspace{0.1\textwidth}
    \subfigure[Too large learning rate]{
        \myincludegraphics[width=0.3\textwidth]{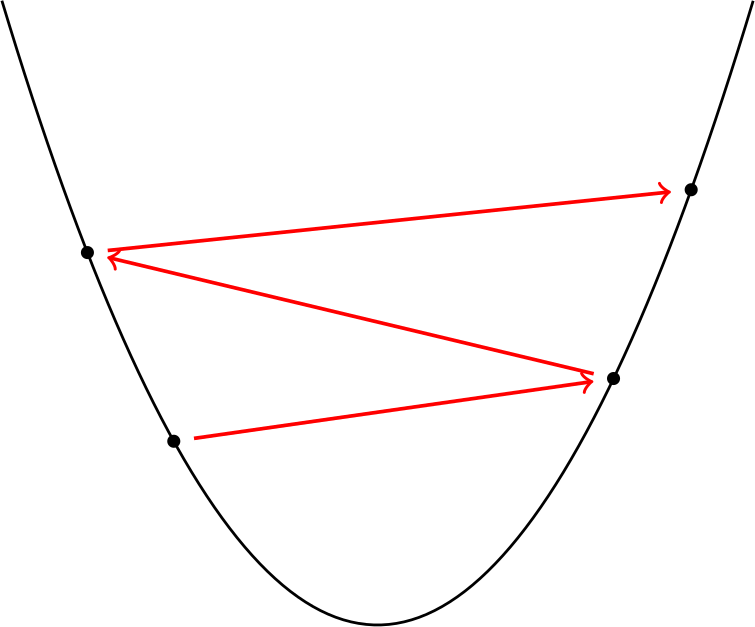}
    }
    \hfill
    \subfigure[Too small learning rate]{
        \myincludegraphics[width=0.3\textwidth]{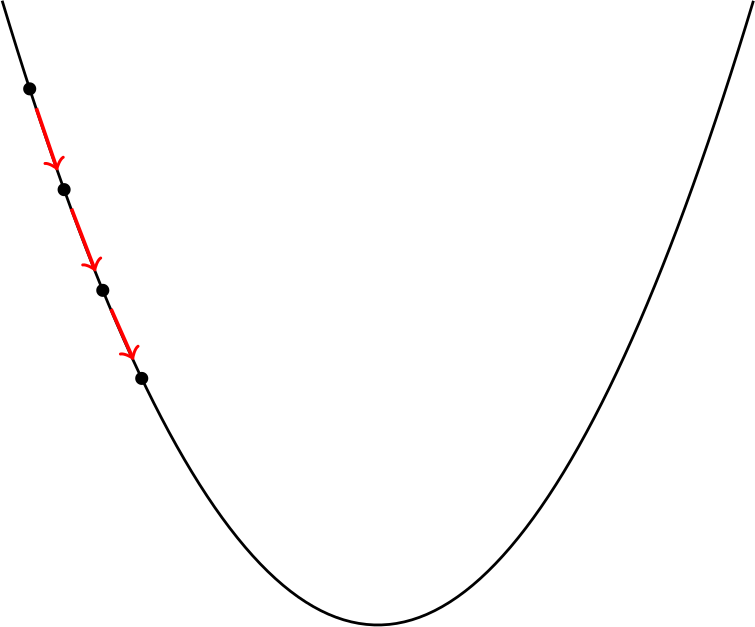}
    }
    \hspace{0.1\textwidth}
    \caption{Different learning rate behaviors: one-dimensional problem with
    a quadratic loss function. This example is meaningful since every function can
    be approximated by a quadratic polynomial if we sit sufficiently close to a
    minimum.}
    \label{fig:learning-rate}
\end{figure}
The parameter $\eta$ in equation~\ref{eqn:sgd} is called learning rate and it is
arguably the most important non-trainable parameter of a neural network. The
learning rate controls the process of descending the gradient and must be
fine-tuned for every architecture, dataset and task. Two undesired behaviors can
arise when the learning rate is not properly set. Figure~\ref{fig:learning-rate}
sketches what happens in a simple one-dimensional case with a quadratic loss
function, if the learning rate is too large or too low. In the former case, the
training progresses too slowly and hardly converges to the minimum. In the latter,
the network parameters receive big increments and the loss function oscillates
around the minimum value, actually never reaching it. These effects get
incredibly enhanced in the optimization of non-convex multi-dimensional problems,
causing training failures. Other issues, such as instability of the optimization,
vanishing and exploding gradients motivate the research of more effective
algorithms.

\begin{figure}
    \centering
    \myincludegraphics[width=0.5\textwidth]{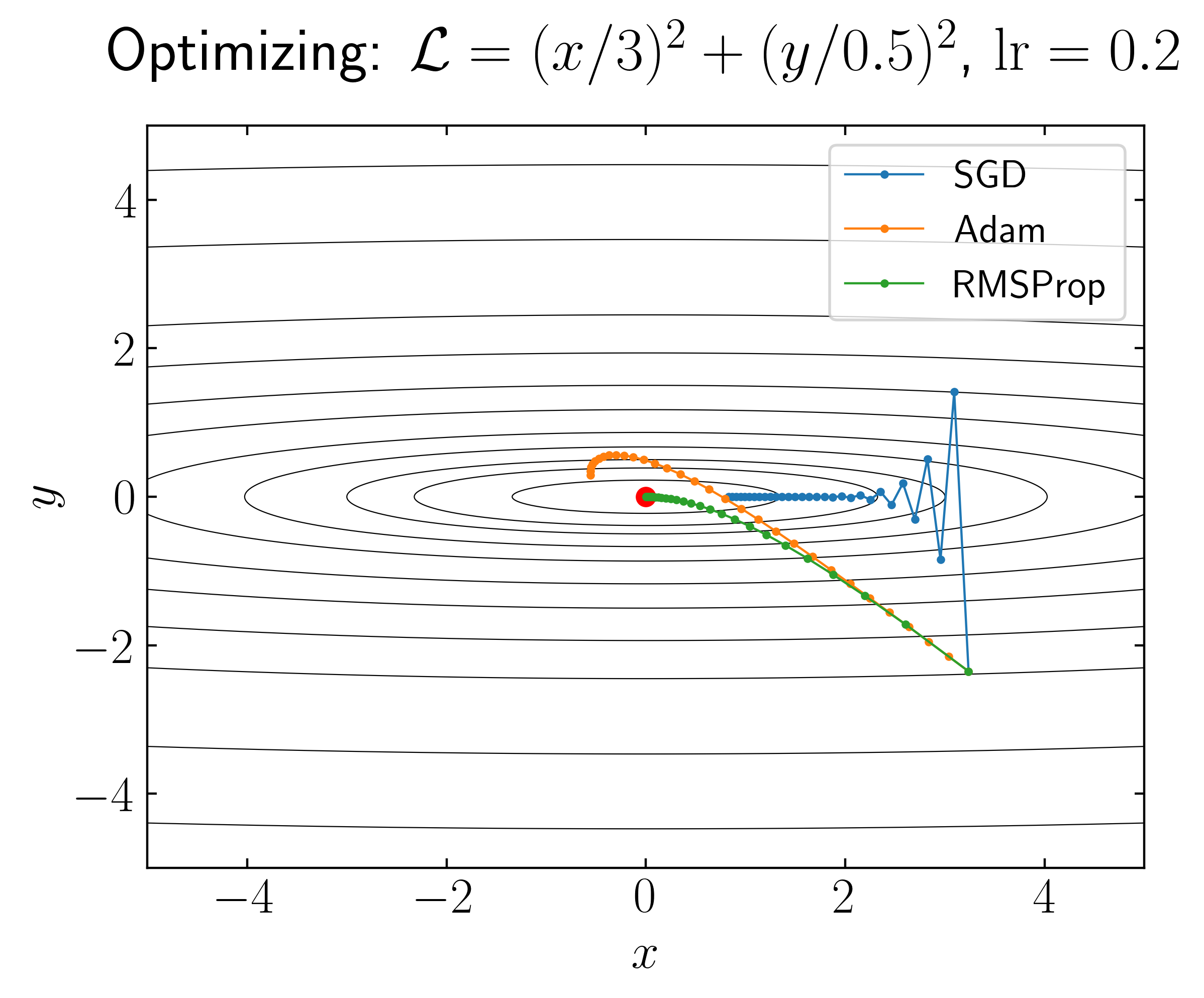}
    \caption{Optimization of an elliptical loss function with different
    optimization algorithms. The red point is the loss function global minimum.
    Gradients give great contributions in the vertical direction, presenting in
    an oscillating pattern and leading to slow convergence of SGD. RMSProp and
    Adam algorithms solve this issue through momentum.}
    \label{fig:optimizers}
\end{figure}

We introduce two of the most widely used optimization algorithms: RMSProp and
Adam. SGD performs badly when it has to come across a path in the loss function
landscape where gradients in one direction are greater than gradients in all
other directions (like ravines). In figure~\ref{fig:optimizers} we draw this
situation in a two-dimensional problem. Ravines often occur around local minima
and cause the algorithm to oscillate in the direction of the steepest change,
making only very little progress along the rift. A technique called momentum was
invented to reduce oscillations and provide a stabler convergence to the
minimum. As a result, the stability of this method allows setting learning rates
to larger values, speeding up the algorithm. The idea behind momentum is driven
by classical point dynamics: if a ball is thrown down a hill, it accelerates
increasing its momentum and going downhill faster and faster. When momentum is
high, it is more difficult for the ball to make sharp turns in the wrong
direction.

RMSProp optimizer, first introduced by Geoffrey Hinton in the lectures of his
Coursera class
\footnote{\url{http://www.cs.toronto.edu/~tijmen/csc321/slides/lecture_slides_lec6.pdf}},
implements these ideas. During the $t$-th step, it adds a fraction $\gamma$ of
the update vector relative to the last step $\mathbf{v}(t-1)$ to the current
gradient update vector:
\begin{subequations}
\begin{align}
    \mathbf{v}(t) &= \gamma \mathbf{v}(t-1) + \eta \nabla_{\theta} \mathcal{L}\\
    \mathbf{\theta} &\longleftarrow \theta - \mathbf{v}(t)
\end{align}
\end{subequations}
where $\eta$ is the learning rate as usual and $\gamma$ is called momentum.
The momentum term is usually set to a $0.9$ or a similar value. This way the update
vector $\mathbf{v}$ account for the sum of all the past collected gradients
weighted with the exponential dumping parameter $\gamma$:
\begin{equation}
    \mathbf{v}(t)= \gamma^t v(0)+\eta \sum_{k=0}^t \gamma^{t-k} g_t
\end{equation}
where $g_t$ is a shorthand for the gradient with respect to the network's weights
at step $t$, $\nabla_{\theta_t} \mathcal{L}$. The problem with this method is
that as more steps are taken in the same direction, $\mathbf{v}$ keeps increasing
and when the minimum is reached its value is too high for the ball to be slowed
down at the optimum point.

Nesterov accelerated gradient (NAG) method tries to face this problem, by
computing gradients of the loss function not in the current position $\theta$,
but approximately in the  position where the ball will be after the update:
\begin{subequations}
\begin{align}
\mathbf{v}(t) &= \gamma \mathbf{v}(t-1) + \eta \nabla_{\theta-\gamma\mathbf{v}(t-1)} \mathcal{L}\\
\theta &\longleftarrow \theta - \mathbf{v}(t)
\end{align}
\end{subequations}

Different algorithms were proposed in ML literature after the introduction of
RMSProp. The development of optimization methods is an active research field,
because new proposals always appear trying to gather all the benefits from
previous methods while introducing new advancements. The adaptive moment
(Adam)~\citep{Kingma:2017} estimation optimizer is one of the most used
algorithms in the ML literature. It is a method that computes adaptive learning
rates for each parameter, storing an exponentially decaying average of past
squared gradients $v_t$ along with a moving average of past gradients $m_t$
(similar to momentum). The algorithm is based on the following equations:
\begin{subequations}
\begin{align}
g_t & \longleftarrow \nabla_\theta \mathcal{L}\\
m_t & \longleftarrow \beta_1 m_{t-1} + (1-\beta_1)\cdot g_t\\
v_t & \longleftarrow \beta_2 v_{t-1} + (1-\beta_2) \cdot g_t^2\\
\hat{m}_t & \longleftarrow m_t/(1-\beta_1^t)\\
\hat{v}_t & \longleftarrow v_t/(1-\beta_2^t)\\
\theta_t &
\longleftarrow \theta_{t-1} - \eta \cdot \hat{m}_t/(\sqrt{\hat{v}_t}+\epsilon)
\end{align}
\end{subequations}
where parameters $\beta_1$ and $\beta_2$ control the exponential decay rate of
gradients and are usually set to $0.9$ and $0.999$, respectively. $\epsilon$
is a small regularizing factor to prevent division by zero, default
implementation suggest to set it to $10^{-8}$. $\eta$ is the learning rate.

$m_t$ and $v_t$ are estimates of the first and the second raw moments, their
unbiased forms $\hat{m}_t$ and $\hat{v}_t$ are used to update network's
parameters. Initial values for $m_0$ and $v_0$ parameters are fixed to $0$. In
the ball example going downhill, Adam method can be compared to a heavy ball
moving with friction that prefers flat minima in the error surface. Since this
is an adaptive method, learning rate schedules are not needed, because the
magnitude of the updates is automatically adjusted according to the form of the
loss function.

%% file: chapters/partI/chap2/ai_review.tex
\section{Review of AI applications in particle physics}
\label{sec:ai-review}

\renewcommand{\myincludegraphics}[2][width=\textwidth]{
    \includegraphics[#1]{chapters/partI/chap2/ai_review_plots/#2}
}

This section summarizes the main applications of deep learning in particle
physics. The presented literature is inspired by general and specialized reviews
collected in~\cite{hepmllivingreview}, in
particular~\cite{Guest:2018, Psihas:2020}.

\subsection{Motivation}

Particle physics produces huge datasets. For example, LHC collects data from
protons, organized in bunches colliding at ${\sim}\SI{40}{\mega\hertz}$
frequency, with $\orderof{10^8}$ sensors. Each collision produces a large number
of particles, whose properties must be measured and stored. Gathering this
enormous amount of data might give physicists enough statistics to study
interesting rare events. These facts highlight that not only the quantity of
collected data is immense, but also its dimensionality. Therefore, machine
learning is a set of techniques of paramount importance in this scenario,
providing automation in data processing and dimensionality reduction of
such information.

For years, physicists in the High Energy Physics (HEP) domain investigated
machine learning techniques like neural networks, support vector machines,
genetic algorithms and predominantly boosted decision trees (BDTs) implemented
in the TMVA framework~\citep{Voss:2009}. This approach was based on the idea of
engineering high-level low-dimensional quantities from raw detector data to be
fed as multiple inputs to multivariate analysis (MVA) and provided important
boosts in many data analysis tasks. However, it was clear that reducing the input
dimensionality consisted in discarding a large part of potentially interesting
information, leading to inherently limited algorithms. As a consequence, these
tools often struggled to provide competitive performance in applications where
the dimensionality gap between raw data and extracted features grew large.

Starting in 2012, the computer science community achieved important results in \linebreak
training big neural networks~\citep{Krizhevsky:2012,LeCun:2015,Schmidhuber:2015},
converging to models able to provide outperforming solutions against traditional
approaches. These publications set the stage for further investigation of deep
learning techniques in many other research fields, including particle physics.
Moreover, this explosion of research activity was helped by the recent
technical improvements in hardware accelerators and their spread as consumer-grade
products, granting high-quality computational power at affordable
prices. In HEP, this wave mostly translated into the idea that engineered features,
designed at cost of time and great intellectual effort, could have been replaced
by high-dimensional low-level raw information if processed by deep enough models.

Besides producing large datasets, the particle physics field is especially suited
for the proliferation of deep learning applications thanks to the availability of 
labeled datasets from Monte Carlo event generators. These programs aim to simulate
the physics world employing probabilistic laws, accurately describing particle
interactions hierarchically from the sub-atomic scales, all the way up to include
the macroscopic long-range effects of physics theories.~\cite{Agostinelli:2002,Sjostrand:2006,Bahr:2008,Gleisberg:2009,Alwall:2014}
represent modern examples of Monte Carlo event generators. The role of
the artificial intelligence tools in this picture is often to grasp the
probability laws of nature from sets of observations (like particle momenta and
charge) and estimate the corresponding Monte Carlo truths (such as the type of a
particle or even an interaction between particles in the event).

The rest of the section is dedicated to an overview of the main results in
particle physics obtained with deep learning models. We split the plethora of
models proposed in the literature by their sector of application. Among physics
at colliders, we identify four main areas: jet physics, tracking, fast simulation
and anomaly detection. In this work, we focus on AI applications to jet physics
and tracking being entangled with the reconstruction process at colliders. We
remark that fast simulation is mainly achieved with the implementation of
Generative Adversarial Networks (GAN)~\citep{Goodfellow:2014}: the model generates
the specific detector response with a fast inference pass of the GAN
generator, producing physical distributions from synthetic random numbers.
The anomaly detection applications, instead, are mainly devoted to beyond
Standard Model (BSM) searches through the identification of exotic signatures of
events or tensions between data and theory that signal the presence of some
new physics mechanism. In this area, model-dependent searches aim to identify
new kinds of particles or interactions through classification, such as
in~\cite{Baldi:2014,Cerri:2018,Bernreuther:2020,Cogollo:2020}, while model
independent approaches~\citep{DAgnolo:2018,DAgnolo:2021} design ad-hoc
strategies to look for new physics with a model agnostic approach.
Conversely, in dealing with non-collider physics, we restrict
our attention to the advancements in deep learning tools for neutrino physics
only, given its prominent role in the present thesis work.

\subsection{Jet physics}

Events at HEP colliders are interpreted in the Quantum Chromo Dynamics (QCD) \linebreak
framework, namely building a hierarchical picture, which approximates the
underlying physics into several non-interacting regimes that happen at different
time scales. Due to the Heisenberg uncertainty principle, also different
characteristic energy ranges are associated with these subprocesses.

Hence, each event is first associated with a hard scattering subprocess
involving partons found inside the protons and interacting at high energy. Then,
all the initial and final state particles are subject to a process called parton
showering, which takes place at lower energy and is computed within the Monte
Carlo software. The showering algorithm consists in recursive branchings, where
each initial or final state particle undergoes multiple subsequent splittings in
a $1\rightarrow2$ fashion, resulting in a tree structure. At this stage, a large
number of particles is created and eventually, their momenta directions are mostly
focused in a collimated region around the particle initiating the shower, called
a jet. Finally, the output of such process can be fed as input to a hadronization
model, which takes into account infra-red (or long-range) effects of
non-perturbative QCD and builds a final realistic picture of the event. The
described algorithms are key ingredients for the Monte Carlo event generators,
which manage to link the predictions of physics theories with the outcomes of
the measuring experiments at colliders.

Machine learning applications to jet physics mainly involve classification
algorithms and include flavor tagging, jet substructure tagging, quark-gluon
tagging and pileup removal. All the tagging tasks are related to the
identification of the shower initiating particle from the knowledge of the
properties of either the final particles representing the tree leaves of the jet
or the whole tree nodes itself. Flavor tagging classifies the jet among heavy
($c$, $b$, $t$) or light ($u$, $d$, $s$) quarks, gluons or $\mathrm{W{/}Z{/}H}$
bosons. Jet substructure tagging, instead, discriminates between
$\mathrm{W{/}Z{/}H}$ and $t$ jets. Finally, quark-gluon tagging is the ability
to distinguish between the two kinds of particles contributing to the main
source of background in HEP events, sometimes named QCD background.

Pileup is a concept that roots in the design of accelerators machines: in
order to increase the probability to produce interactions, at colliders, bunches
of protons tightly packed together are smashed against each other, rather than
individually. The luminosity $\mathcal{L}$ is a measure of such compactness: the
higher the luminosity, the more the protons are squeezed together and increase the
number of collisions. As a consequence, it is likely that for each beam crossing,
more than one couple of protons scatters, emitting soft radiation at wide angles
named pileup (PU), as opposed to the interesting hard interaction often referred
as leading vertex (LV).
Pileup is extensively
studied at colliders and depends on the machine operative setup. For example, at
LHC, each PU vertex roughly contributes for
$\SI{0.6}{\giga\electronvolt/\radian^2}$ of the detector deposited
energy~\citep{Cacciari:2008,Cacciari:2009,Rubin:2010}. Hence, considering an
average number of pileup collisions per beam crossing of
$n_{\mathrm{PU}}{\sim} 100$ and characteristic jet radius of $R=1.0$, each jet
might suffer for about $\SI{200}{\giga\electronvolt}$ of pileup contamination.

The data collected by Atlas and CMS experiments up to run II estimated an average
$n_{\mathrm{PU}}{\sim} 20$, while for run III and HL-LHC, this quantity is expected
to increase to $n_{\mathrm{PU}}{\sim} 80$ and $n_{\mathrm{PU}}{\sim} 200$,
respectively. Pileup interactions modify the shape of the observables
reconstructed in the events, affecting jet properties like its overall momentum
and mass, rather than jet multiplicity in the event. Being able to design
automatic tools to mitigate those effects is expected to be one of the biggest
data analysis challenges during the forthcoming LHC phases. These considerations
justify the importance given to pileup mitigation strategies at colliders.

The next paragraphs will review the proposed techniques in the literature
concerning the described two areas of jet tagging and pileup removal. Both of
them try to present the advancements in the research activity as it evolved
during the last decades. from the point of view of the different data
representations of jet objects used as inputs of the several neural network
architectures proposed.

\subsubsection{Jet Physics - tagging}
Jet tagging mainly concerns classification algorithms.
Since the early $'90$s, shallow artificial neural networks have been used to detect the
type of jet-initiating particles. In these initial years, the predominant approach
was to feed neural networks, comprised of just a few layers, with event features
tailor-made for the specific task or, sometimes, by packing jet information into small vectors
of fixed size, containing the most representative characteristics of the object.

Following this idea,~\cite{Lonnblad:1990} exploited a neural network with $3$ fully
connected hidden layers with $6$ neurons each, to process the $4{-}$momenta of
the four leading particles within a jet, to discriminate between quarks and
gluons.~\cite{Behnke:1995}, instead, implemented a neural network to distinguish between
$b$ and $c$ jets al LEP, with the help of the Fortran77 JETNET 3.0 library~\citep{Peterson:1994},
which represented the de facto standard for machine learning in HEP physics during
those early years.

The advent of deep learning and the improvements in hardware accelerator technologies
paved the way for new strategies to solve the jet tagging problem.~\cite{Cogan:2014}
processed for the first time entire events through neural networks: their insight
was to encode calorimeter information, namely the particle deposited energy (or
equivalently its transverse momentum $p_T$) as a regular grid in the pseudorapidity
$\eta$ and azimuthal angle $\phi$ plane, forming an image. The pixels of the image
contained raw event information that can be used to compute discriminative quantities.
The authors implemented a recipe to compute the Fisher linear discriminant~\citep{Fisher:1936}
after some physics-inspired preprocessing of the images. The algorithm was tested
for $W$ boson tagging against QCD background, providing performance improvements
against the traditional discrimination method based on N-subjettiness
($\tau_2/\tau_1$)~\citep{Thaler:2010,Thaler:2011}.

Raw inputs-based neural network tools started being investigated extensively from
that point onwards. Examples can be found in top quark tagging tasks~\citep{Almeida:2015}
and jet substructure classification (namely, understanding if the considered jet
is due to a showering of a low-mass single particle or a massive particle decaying
into multiple fast-moving lighter objects producing overlapping
jets in the calorimeter, like for the $W\rightarrow qq$ process)~\citep{Baldi:2016}.
Another application of this framework has been presented by~\cite{Barnard:2016},
who studied the dependency of trained models on the Monte Carlo truths labels in
the training datasets. The key observation pointed out that the supervised learning
algorithm might bias the model predictions following the QCD approximations
employed by the specific generator used to collect the dataset, rather than focusing
on learning the underlying true laws of nature. The work raised the problem of
the interpretability of neural networks in the jet physics research field for the
first time, finding large discrepancies when testing the models on datasets produced
by different generators. The authors' final assertion underlined the need to
deeply understand how the input information is exploited to extract the output
and what assumptions a trained architecture relies on.

The calorimeter tower representation of~\cite{Cogan:2014} has then proven to be
a powerful representation of jets events, mainly thanks to the success of
Convolutional Neural Networks (CNNs)~\citep{LeCun:2015}.
Indeed,~\cite{deOliveira:2015} exploited CNNs to inspect the $(\eta,\phi)$ plane
deposited energy encoding of jet events. They proposed a network to identify
highly boosted $W$ bosons against the quark-gluon QCD background. The inputs were
initially cast to grayscale images (one channel only), however further
developments considered also multi-channel input images. In
particular,~\cite{Komiske:2016} proposed to build a three-channel RGB image
tensor stacking information from charged and neutral particles' transverse
momenta, plus the number of charged particles measured within each pixel area.

The standard calorimeter tower images were not the only image-like encoding that
has been studied in the literature: an alternative strategy has been given by
the Lund Jet Plane~\citep{Dreyer:2018}. It considers kinematic variables arising
while rewinding backward the Cambridge Aachen clustering
algorithm~\citep{Dokshitzer:1997,Wobisch:1998}, attempting to reconstruct a
de-clustering history of a jet. The output of this procedure is an ordered set of
variables that characterizes a jet object and can be seen as an image tensor.
According to the authors, this description should provide greater output
interpretability as well as discrimination power when employed in classification
tasks.

Although the image based successfully tackled the jet classification problem
multiple times, CNNs rely on the assumption that pixels form a perfect grid,
while it is known that actual detectors' geometry is not perfectly regular.
Moreover, jet images often contain sparse features which lead to inefficient
processing by convolutional kernels. Hence, different data representation
strategies have been investigated. A jet object is the result of a clustering
algorithm\footnote{A modern \texttt{C++} implementation of jet definitions and
clustering algorithms is given by the FASTJET 3.0~\citep{Cacciari:2011} library.},
which generates a list of jet constituent particles. As a consequence, it can
be represented as a sequence of tracks and vertices, forming an acyclic-directed
graph or, equivalently, a tree. The complication arising from adopting this
encoding scheme is mostly given by the variable length size of the sequences,
which cannot be handled by standard Feed Forward Neural Networks.

\cite{Guest:2016} overcame this difficulty by proposing a Recursive Neural Network
architecture comprised of Long Short Term Memory (LSTMs)~\citep{Hochreiter:1997}
cells, able to deal with variable-size inputs. The work takes into account other
solutions involving Feed Forward Neural Networks supported by input truncation
and zero padding. The authors presented a comparison of the different strategies
applying them to the problem of light ($u$, $d$, $s$, $c$) versus heavy quark
($b$) jet flavor classification, achieving similar performance for the different
models. Since then, several algorithms based on RNNs have been proposed to become
part of the Atlas~\citep{Atlas:2017gpy,Atlas:2017bcq} and
CMS~\citep{CMS:2017-005,CMS:2017-012} software stack and many more have been
published to exploit variable size inputs~\citep{Egan:2017,Cheng:2017}.

The introduction of RNNs allowed for the treatment of the jet as lists and trees
of particles. However, even if some natural ordering is obtained by clustering
like in the $k_t{-}$algorithm, this is just an approximation. Imposing an ordering
often means establishing a spatio-temporal relationship between particles to be
identified as a history producing a specific final state. However, quantum
mechanics principles break down the causality concepts of space and time relying
on probabilistic laws. Therefore, the most natural way to represent a jet object
would be to decouple from this artificial ordering and process it like an
unordered set of particles described by their $4{-}$momenta and quantum numbers.
Designing an architecture with the ability to deal with unordered sets of
particles would then be desirable. Graph neural networks, deep sets and point
clouds networks achieve this objective.

\citep{Henrion:2017} implemented a RelNet~\citep{Santoro:2017} to accomplish $W$
jet tagging against QCD background: particles are regarded as graph nodes and
the adjacency matrix is learned to aggregate information between nodes
through a message-passing operation.~\cite{Komiske:2018} constrained a network architecture
acting on deep sets, to build infra-red and collinear (IRC) safe observables: the information in each
particle observable is then aggregated with a global permutation-invariant operation.
The authors implement two different networks called EnergyFlow and ParticleFlow,
which consider IRC-safe and non-IRC-safe quantities, respectively.~\cite{Qu:2019}
proposed to to use the EdgeConv operation~\citep{Wang:2019} on the k-nearest
neighbors points of each particle in a point cloud. The point cloud jet
representation encodes an event as a matrix where each row represents a vector
of properties associated with each particle.

\begin{figure}
    \centering
    \myincludegraphics{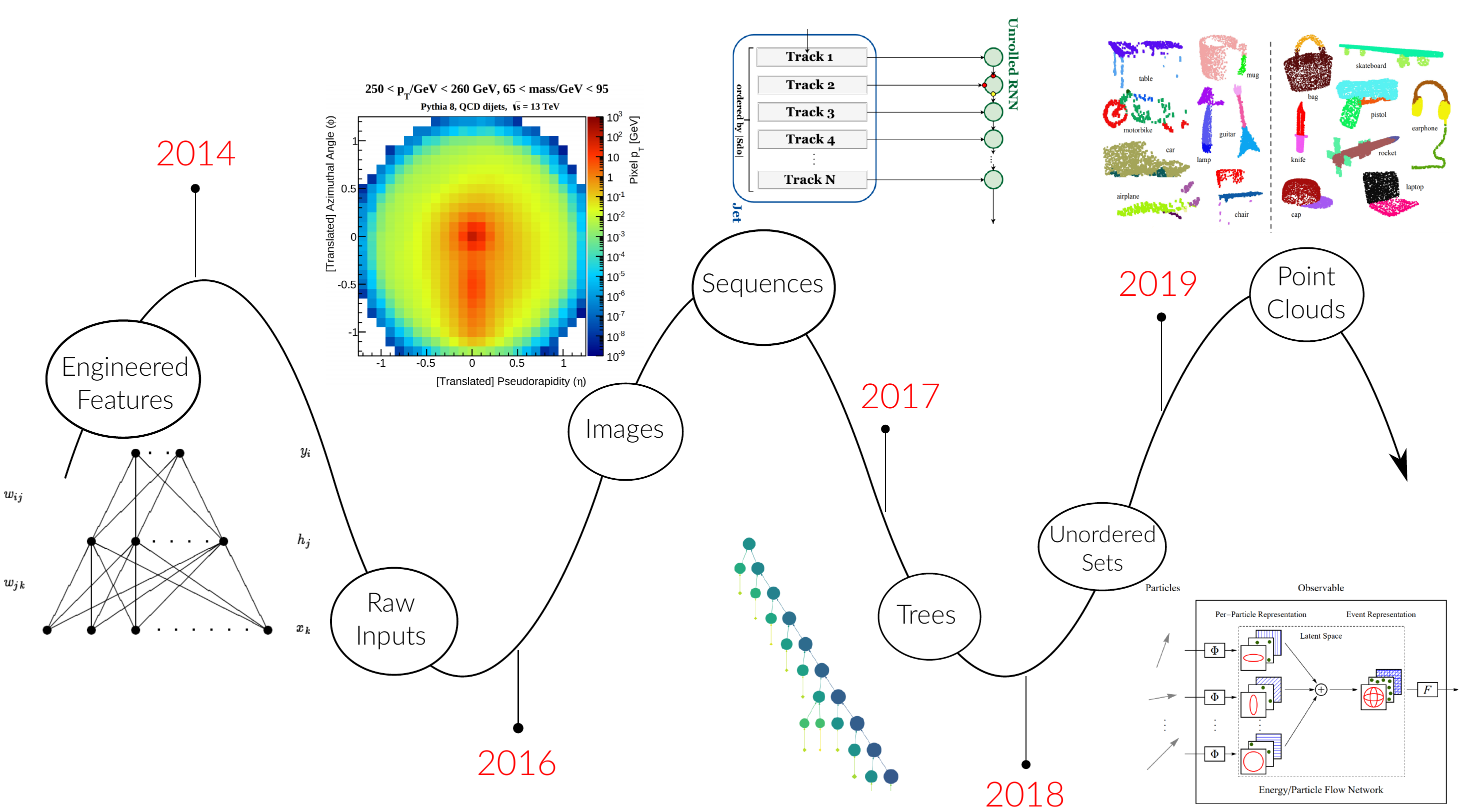}
    \caption{The data structure timeline of physics jets: only engineered
    features were used as input to neural networks before 2014; after~\cite{Cogan:2014},
    several encoding structures have been investigated to efficiently represent
    jets. The descriptive pictures in the chart are taken, in order of appearance,
    from:~\cite{Peterson:1994,deOliveira:2015,Atlas:2017gpy,Cheng:2017,Komiske:2018,Qi:2016}.
    }
    \label{fig:jet_inputs}
\end{figure}

\begin{table}
    \caption{Summary of the proposed architectures for jet classification. The table
    is inspired from~\cite{Larkoski:2020}.}
    \centering
    \label{tab:jet_classification}
    \begin{tabular}{c|c|c|c|c|c}
        \hline\noalign{\smallskip}
         & quark/gluon & $W{/}Z$ & $H$ & $b{/}c$ & $t$ \\
        \noalign{\smallskip}\hline\noalign{\smallskip}
        Image &\cite{Komiske:2016,Atlas:2017dfg,CMS:2017-027} & \cite{deOliveira:2015,Baldi:2016} & \cite{Lin:2018} & & \cite{Kasieczka:2017,Butter:2019} \\
        Sequences & \cite{CMS:2017-027} & & & \cite{Atlas:2017gpy} & \\
        Tree & \cite{Cheng:2017} & \cite{Louppe:2017,Andreassen:2018} & & & \cite{Butter:2019} \\
        Graph & & & \cite{Henrion:2017} & & \\
        Unordered set & \cite{Komiske:2018} & & & & \\
        Point Cloud & \cite{Qu:2019} & & & & \cite{Qu:2019,Butter:2019} \\
        \noalign{\smallskip}\hline
    \end{tabular}
  \end{table}

Figure~\ref{fig:jet_inputs} shows how the neural network input representations for
jet physics \linebreak evolved through time. Table~\ref{tab:jet_classification}, instead,
summarizes all the relevant applications of machine learning and deep learning
to jet physics.

\subsubsection{Jet Physics - pileup mitigation}

Pileup contribution from charged particles can be removed almost completely thanks
to the excellent vertex resolution at the ATLAS and CMS
detectors~\citep{Chatrchyan:2014,Atlas:2010gaa,Atlas:2010lca}. These particles are
identified and removed from the event with the charged-hadron subtraction (CHS)
procedure~\citep{CMS:2014}.
The challenge comes from pileup radiation due to neutral particles, which must
be taken into account with specialized algorithms. The rich literature on traditional
methods can be categorized on the level of detail these tools act on. A first
technique, known as jet areas subtraction~\citep{Cacciari:2007}, relied on
calibrating jet level information, scaling its $4{-}$momentum by a relevant factor.
However, this procedure did not manage to mitigate the pileup contribution effectively
for the computation of several subjet observables.

Therefore, other algorithms
have been proposed to act on the internal jet structure, namely at the subjet level.
Examples of such tools are usually classified as jet constituents pre-processing,
jet or event grooming, subjet corrections and constituent corrections. Grooming,
in particular, progressively removes jet constituents contaminated by pileup,
cutting the tree description of a jet arising from clustering algorithms through
filtering~\citep{Butterworth:2008}, pruning~\citep{Ellis:2009,Ellis:2010} and
trimming~\citep{Krohn:2009}. SoftKiller~\citep{Cacciari:2014}, instead, is a
popular event-level grooming algorithm that equally divides the $(\eta,\phi)$ plane in patches
of a certain area and imposes a cut-off $p_T^{cut}$ on the transverse momentum cumulated
on the patches, such that half of the patches are radiation free. This tool
has been used by several works as a benchmark to test the goodness of the proposed models.

Finally, the most advanced pileup mitigating algorithms act at the deepest level, working on a
particle-by-particle basis~\citep{Bertolini:2014,Berta:2014,Atlas:2017pfq}. Among
those, an excellent example is PUPPI, which evaluates a scaling factor for each particle
$4{-}$momentum in the event, by computing a local shape variable $\alpha$, which
collects information about each particle neighborhood. The $\alpha$ distribution
for charged particles, for which pileup information is known thanks to the CHS method,
can be exploited to extract the scaling weight for each neutral particle. The
net effect is to correct jet and subjet observables of interest
for physics analysis as if the pileup effects have been switched off.

Machine learning applications for pileup removal mainly act at the particle level,
since, as already discussed in this chapter, they can extract useful information
from the low-level description of events. The first application of this framework
was \linebreak PUMML~\citep{Komiske:2017}, a Convolutional Neural Network to inspect RGB
images in the $(\eta,\phi)$ plane.
The three RGB channels convey information about the transverse momenta of all
neutral particles, all charged pileup particles and all charged leading vertex particles,
respectively. The architecture is trained in a supervised way to output the missing
$p_T$ of the neutral leading vertex particles. Performance comparisons were presented
against SoftKiller and PUPPI algorithms for the reconstruction of mass and transverse
momentum distributions of the LV jets.

Other models have been proposed in subsequent years, trying to take advantage of
the different technologies developed in the computer vision research field:~\cite{ArjonaMartinez:2018}
introduced PUPPIML, a network working on a graph representation of the event.
After subtracting the charged pileup particles, the remaining ones are arranged
in a graph where all pairs of particles closer than a fixed radius $R_1$ in the
$(\eta,\phi)$ plane (default value is $R_1=0.3$) are connected by an edge.
This graph is processed by several Gated Recurrent Units (GRU)~\cite{Cho:2014}
and outputs a binary score for each particle to discriminate between leading vertex
and pileup. The authors claimed performance improvements up to ${\sim}30\%$ of PUPPIML
against PUPPI on the resolution of jet-related quantities and even higher ones with
respect to SoftKiller.

PUMA~\citep{Maier:2021} exploits the attention mechanism~\citep{Vaswani:2017}
to tackle the pileup mitigation task in realistic detector scenarios, corresponding to
extreme setups with $n_{PU}{\sim}200$. The performance was tested against classical benchmarks,
like CHS and PUPPI, showing large improvements in the key reconstructed jet variable
distributions. The authors judged this work as an important achievement in showing
the usefulness of statistically-learned algorithms during the HL-LHC phase.

Beyond the supervised algorithms presented in the paragraphs above, some alternative
approaches have been proposed.~\cite{Carrazza:2020} implemented a grooming procedure
within a reinforcement learning (RL) framework: a jet is represented as a binary tree graph
where each node $i$ is described by a Lund plane derived variable $\mathcal{T}^{(i)}$,
containing the state vector observed by the RL agent as well as a pointer to the
the parent node and the two child ones. The algorithm concerns applying recursively
a policy function $\pi_g$ to all the nodes in the graph. The policy function 
outputs the probability to groom or not a node in the tree, which determines
the action of the RL agent on the environment. The agent is trained through a
smooth reward function carefully designed to optimize the resolution of kinematic
variables both at the graph and node level, such as the mass of the resulting jet
or the fact that a node contributes to the wide-angle soft radiation (PU) rather than to the
hard-collinear emission (LV), respectively.

A semi-supervised learning approach for Graph Neural Networks, named Graph SSL, has
been investigated by~\cite{Li:2022}. The main advantage introduced by this technique
is the possibility to train directly on real detector data, without the need of
Monte Carlo truth labels. The algorithm is based on supervised training to learn
charged particles' properties, while inference is done on neutral particles, which
represents the main challenge in the identification of pileup. A careful masking
procedure is required to train effectively on charged particles as if they were
neutral ones. This method allows for avoiding the complex issues regarding the
dependence of the models on Monte Carlo datasets and the high costs in terms of
simulation time to reproduce physics processes with Monte Carlo generators. The
authors benchmarked Graph SSL against PUPPI and observed performance improvements
both for the accuracy in the LV-PU identification at the particle level and
regarding the resolution of the reconstructed jet quantities.

\subsection{Non-collider physics (neutrino)}
This section presents the applications of deep learning in neutrino physics. For
sake of brevity, we restrict our attention to experiments focusing on neutrino
oscillations only. Nevertheless, a large number of experiments are concentrating
their efforts in this field, as explained in section~\ref{sec:nu_phys}. In the
field of neutrino physics, deep learning is mainly investigated as a tool for
event classification and automated reconstruction algorithms. The former task is
well established in the physics community since the end of the $20^{th}$ century
as a robust strategy to select signal and reject background events, while the
latter is still an open issue and many techniques are currently being inspected.

The first neural network application to neutrino event classification is given
by~\cite{Brice:1996} in the context of the SNO experiment. The network contains
a modest $\orderof{700}$ number of trainable parameters employing a
shallow feed-forward neural network with $\orderof{30}$ inputs engineered on
detector hit patterns and count a single hidden layer with $20$ neurons, to
distinguish between four classes of neutrino interactions:
\begin{itemize}
  \item Charge current (CC): $\nu_e + \isotope[2]{H} \rightarrow p + p + e^-$;
  \item Electron scattering (ES): $\nu_x + e^- \rightarrow \nu_x + e^-$;
  \item Chlorine neutral current (NC): $n + \isotope[35]{Cl} \rightarrow \isotope[36]{Cl} + \gamma$;
  \item Deuteron neutral current (ND): $n + \isotope[2]{H} \rightarrow \isotope[3]{H} + \gamma$.
\end{itemize}

The investigation of artificial neural networks eventually spread among neutrino
physicists. The main cause of this success can be found in the detector data format:
most of the detectors built for detecting neutrinos produce image-like
data, which can be processed with the help of modern computer vision and convolutional
neural network techniques. As a consequence, two decades after the SNO paper, the NO${\nu}$A
collaboration~\citep{Ayres:2007} proposed to build a CNN to identify neutrino
background interactions~\citep{Aurisano:2016}. The network, named Convolutional Visual Network (CVN),
is comprised of two \linebreak GoogLeNet~\citep{Szegedy:2015} separate branches inspecting
$(x, y)$ and $(y,z)$ hit projections, respectively. The two resulting output tensors
are concatenated and fed into a classifier to extract the desired multi-class score.
It is interesting to notice that the two views are not concatenated along the
channel axis like in RGB images: the authors recognize that each coordinate pixel
in 2D projection would overlap unrelated features, as they do not refer to the
same $(x, y, z)$ spatial 3D point. The network is trained to compute the $\nu_e$
appearance and $\nu_\mu$ disappearance rates. This work marked a milestone in the
field since it became the first neural network-based analysis whose results were
included in a physics publication~\citep{Adamson:2017}.

\begin{figure}
  \centering
  \myincludegraphics{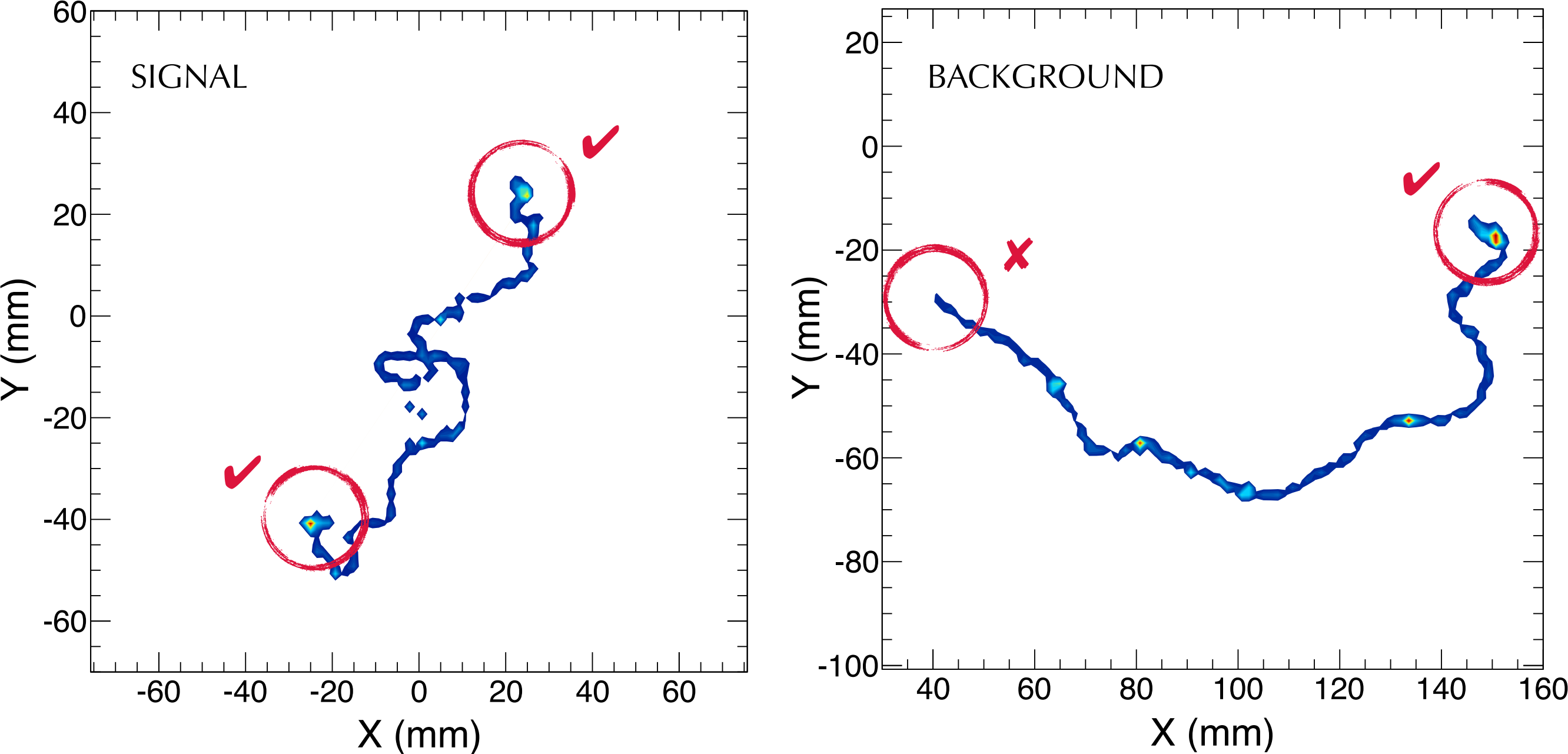}
  \caption{Monte Carlo simulation of signal ($0\nu\beta\beta$ decay of \isotope[136]{Xe})
  and background (single electron of energy equal to the $Q$ value of \isotope[136]{Xe})
  events in gaseous xenon at $\SI{15}{\bar}$. The picture shows the energy deposition
  heatmap colored from blue (low deposition) to red (large deposition).
  The signal consists of two electrons emitted from a common vertex, resulting in
  a region (blob) of intense deposition at both track ends. Conversely, the background
  shows one blob only, leading to cut-off-based discrimination, if the two 
  blobs are properly reconstructed.
  }
  \label{fig:0vbb_blob}
\end{figure}

The GoogLeNet architecture has also been exploited by~\cite{Renner:2016} to search
for neutrino-less double beta decay $0\nu\beta\beta$~\citep{Schechter:1981} process
at the NEXT experiment. In this application, as opposed to the NO${\nu}$A one,
three 2D projections of event images are concatenated like RGB images. The authors
highlight an improvement against the traditional "blob" discrimination method,
see fig.~\ref{fig:0vbb_blob}.
The techniques just reviewed try to process images with CNNs, achieving better
performance than baseline methods. In recent years, several other articles and
many experimental collaborations showed interest in developing CNN-based classifiers,
showing the performance superiority of this approach compared to the traditional
methods of event
classification~\citep{Aurisano:2016,Racah:2016,Acciarri:2016,Aiello:2020,Abi:2020,Abratenko:2020}.

CNNs have also proved useful in several tasks of reconstruction. First, they
have been used to tackle regression problems, namely to predict the interacting neutrino
energy value~\citep{Hertel:2017} or its direction in the frame of
reference of the detector~\citep{Aiello:2020,Liu:2020}. Then, they helped in
identifying non-empty activity regions, drawing bounding boxes around interactions
to discard uninteresting parts of the input images: this technique is called
Faster-Region Convolutional Neural Network (Faster-CNN). Alternatively,~\cite{Domine:2020}
exploit a model inspired by the U-Net architecture~\citep{Ronneberger:2015} to precisely locate track
end-points and shower vertices. Finally,~\cite{Adams:2018,Yu:2020,Acciarri:2021}
implemented CNNs aiming at segmenting the input images to assign each pixel to a
type of particle drifting in the detector, detecting Region Of Interest
(ROI) coordinates in raw data in ${2}$-dimensional planes and ${1}$-dimensional
channels, respectively.

Although Convolutional Neural Network models are the de-facto standard in image
processing, neutrino detectors often collect data with special features that
cause these techniques to be inefficient. The majority of the events recorded by
these experiments contain sparse long ${1}$-dimensional tracks with locally
dense features. The result is that large portions of such images are empty,
leading to a waste of computational resources when inspected with convolutional
filters: those filters, indeed, transform equally both the empty spaces and
signal regions. Additionally, the huge quantity of sensors in these detectors
gathers information into high-resolution images with $\orderof{10^6}$ of pixels,
that barely fit the memory constraints of modern hardware accelerating devices.

\begin{figure}
  \centering
  \subfigure[Normal ${2}$-dimensional convolution: single kernel transformation.]{
    \centering
    \label{subfig:conv}
    \myincludegraphics[width=0.35\textwidth]{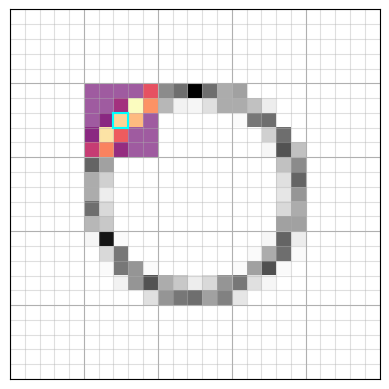}
  }
  \qquad\qquad
  \subfigure[Sparse ${2}$-dimensional convolution: single kernel transformation.]{
    \label{subfig:sparse_conv}
    \myincludegraphics[width=0.35\textwidth]{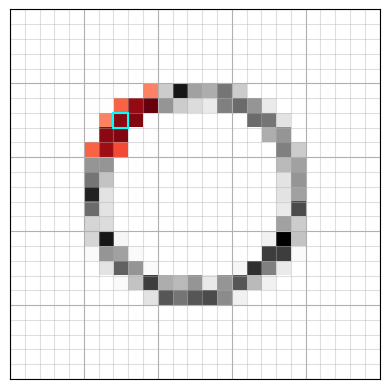}
  }
  \caption{The cyan pixel highlights the current convolution pixel. The normal
  convolution kernel operates on all pixels within the kernel window, while the
  sparse one acts on non-zero neighboring pixels only.}
  \label{fig:sparseconv}
\end{figure}

During the last few years, then, the neutrino community has dedicated a great
effort to design better encodings and experiment novel techniques to analyze such data.
In this picture, Sparse Convolutional Neural Networks (Sparse CNNs)~\cite{Graham:2018proc}
and Graph Neural Networks (GNNs)~\cite{Scarselli:2009} have been investigated.
Two architectures based on Sparse CNNs, acting with convolutional filters on non-zero
pixels only, have been implemented by~\cite{Domine:2019,Koh:2020}. The operation,
depicted by figure~\ref{subfig:sparse_conv}, allows to store the event data
in an efficient sparse format and dramatically decrease the number of operations
required by each convolutional layer forward pass.

On the other hand, GNNs provided performance improvements in processing data from detectors
with irregular geometry like IceCube~\cite{Choma:2018} and JUNO~\cite{Qian:2021}.
Besides this success, even if the data graph encoding is not always a natural
choice when dealing with either image or point cloud data, several
works~\citep{Drielsma:2020,Alonso-Monsalve:2020,Hewes:2021} showed good results
implementing these architectures for event classification and other reconstruction
tasks.

The use of GNNs is subject to two major issues. First, there is no standard 
choice of encoding neutrino data into a graph. Luckily, the sparsity of neutrino
images allows identifying detector hits as graph nodes, resulting in graphs of
manageable sizes (usually up to a few thousand nodes). Node connectivity, instead,
is use-case dependent. The majority of the authors reviewed in this paragraph use similar
approaches with small modifications: they rely on some distance metric, computed between
each pair of nodes $i$ and $j$ in the graph, and a pre-defined cut-off value $d_{cut}$
above which no edge between the corresponding nodes is drawn. Alternatively,
they propose to weight each edge with a normalized version of the distance metric value
itself, such that distant nodes have a suppressed information flow in the network.
The second problem related to this approach is the additional overhead represented
by the graph construction operation: this is often done through dedicated algorithms,
like in~\cite{Drielsma:2020,Qian:2021}, and must be repeated for all events inevitably
increasing the pre-processing wall-time.


In table~\ref{tab:neutrino_ai} we collect the main deep learning applications to
neutrino oscillation experiments found in the literature. The table groups the
works published by several collaborations into five task categories:
\begin{itemize}
  \item event classification, which encompasses event topology, interaction
  classification and background rejection;
  \item regression, grouping neutrino energy reconstruction and neutrino
  direction reconstruction;
  \item object detection, that collects interaction localization (vertex
  reconstruction, \linebreak bounding box drawing around pixel activity), track end-point
  localization and shower starting-point localization;
  \item graph operations, that include background rejection (graph
  classification), clustering (node classification), 3D reconstruction (graph
  cleaning through node classification) and primary particle classification
  (edge classification);
  \item segmentation, receiving contribution from pixel-level particle
  identification, instance segmentation and region of interest (ROI) finding.
\end{itemize}

\begin{table}
  \caption{Review of the deep learning for neutrino physics publications. The
  first column identifies which detector the publication focuses on.
  PilarNet~\citep{Adams:2020} is a general-purpose open dataset for LArTPCs data.
  Note:~\cite{Domine:2019} was published before the PilarNet~\citep{Adams:2020}
  dataset but deals with similar data and objectives. The citations are
  color-coded based on the neural network type implemented in the relative work:
  \color{olive}FFNNs, \color{red} CNNs, \color{green} GNNs, \color{orange}
  Hexagonal CNNs, \color{cyan} Sparse CNNs, \color{teal} Quantum CNNs
  \color{black}.
  }
  \centering
  \label{tab:neutrino_ai}
  \footnotesize
  \begingroup
  \hypersetup{hidelinks}
  \begin{tabular}{l|c|c|c|c|c}
    \hline\noalign{\smallskip}
    & Event classification & Regression & Object detection & Graph & Segmentation \\
    \noalign{\smallskip}\hline\noalign{\smallskip}
    SNO & \color{olive}\citep{Brice:1996} & - & - & - & - \\
    NEXT & \color{olive}\citep{Martin:2016} & - & - & - & - \\
    Daya Bay & \color{red}\citep{Racah:2016} & - & - & - & - \\
    NO${\nu}$A & \color{red}\citep{Aurisano:2016} & \color{red}\citep{Hertel:2017} & - & - & - \\
    MicroBooNE & \color{red}\citep{Acciarri:2016,Abratenko:2020} & - & \color{red}\citep{Acciarri:2016} & - & \color{red}\citep{Adams:2018}\color{blue}~\citep{Abratenko:2020yze} \\
    KM3NeT/ORCA & \color{red}\cite{Aiello:2020} & \color{red}\cite{Aiello:2020} & - & - & - \\
    DUNE / pDUNE & \color{red}\cite{Abi:2020}~\color{teal}\cite{Chen:2020} & \color{red}\cite{Liu:2020} & - & \color{green}\cite{Hewes:2021} & \color{red}\cite{Yu:2020} \\
    JUNO & \color{olive}\cite{Clerbaux:2020} & \color{green}\cite{Qian:2021} & \color{red}\cite{Qian:2021} & \color{green}\cite{Qian:2021} & - \\
    SuperFGD (T2K) & - & - & - & \color{green}\cite{Alonso-Monsalve:2020} & - \\
    IceCube & \color{green}\cite{Choma:2018} & \color{orange}\cite{Abbasi:2021} & - & \color{green}\cite{Choma:2018} & - \\
    ArgoNeuT & - & - & - & - & \color{red}\cite{Acciarri:2021} \\
    PilarNet & - & - & \color{red}\cite{Domine:2020} & \color{green}\cite{Drielsma:2020} & \color{cyan}\cite{Domine:2019,Koh:2020}\\
    \noalign{\smallskip}\hline
  \end{tabular}
\endgroup
\end{table}

\subsection{Tracking}
Tracking is a central process of reconstruction at colliders, it consists in
grouping detector hits within an event produced by a charged particle interacting
in the inner detector region and moving inside a static magnetic field. The
traditional approach is based on four different phases: hit clustering, track
seed finding, track building and track fitting. The present discussion gives a
brief overview of the traditional method employed to solve the tracking problem
and it is inspired by specialized reviews on tracking strategies at
LHC,~\cite{Fruhwirth:2000,Ragusa:2007,Fruhwirth:2011}.

The tracking process consists in sequentially reducing with clustering algorithms the
number of data from $\orderof{10^8}$ detector readout channels, to $\orderof{10^4}$
hits containing energy depositions and finally to $\orderof{10^3}$ tracks
per event. The hierarchical approach starts with hit clustering, which
consists in finding the $3{-}$dimensional locations of hits and the corresponding
deposited energies from the pixel-level raw data readouts.

After this first stage,
the two most computationally expensive steps take place. First, the hits in the
inner detector are processed to identify triplets, which consist of the minimum
number of points to estimate two important track parameters, namely the curvature
and the perigee with respect to the center of the interaction region. The three hits
in each triplet form a seed for the final track. Therefore, this step fixes the
final track multiplicity.

Second, once the seeds are selected, the proper track construction process starts:
the trajectory is sequentially extrapolated from the triplet from the
inner to the outer layers of the detector. Many pattern recognition techniques
have been designed to tackle this problem, ranging from global methods, such as
conformal mapping and Hough transform~\citep{Kalviainen:1995}, to local ones,
like the track road methods. However, the most efficient algorithm in use is the
Kalman filter~\citep{Kalman:1960,Catlin:1989,Fruhwirth:1987}. A more refined
version of the original algorithm, the Combinatorial Kalman Filter~\citep{Mankel:1997},
is leveraged to build tracks from seeds, including the possibility to keep track
of branching when multiple candidate points are identified within the same layer
and eventually, discard the fake tracks with high efficiency.

The final stage of the tracking problem, namely track fitting, requires estimating
the track parameters for each reconstructed trajectory. These include the location
of the interaction vertex, the direction of the track along with its curvature and
the momentum associated with the interacting particle. Moreover, tests to remove outliers
that do not belong to the track are performed in this final phase to
further refine the output. This technique achieves almost perfect performance,
meaning that the investigation of new methods is devoted to optimizing the existing
software implementation and trying to reduce CPU usage time.

However, the next generation High Luminosity LHC (HL-LHC) phase~\citep{Schmidt:2016},
starting from 2026, will see an increase in the current luminosity setup of the
Large Hadron Collider by a factor of $10$, putting these low-level reconstruction
tools under enormous stress. It is expected, in this collider configuration,
a great improvement in the hit detector occupancy and the particle tracking
software should be able to manage charged particles at a rate of
$\orderof{\SI{50}{\MHz}}$. The traditional approach does not scale at such
regimes. A naive solution would be to limit the reconstruction to detector
regions around specific calorimetry depositions compatible with rare signatures
like leptons or jets with high $p_T$. However, this approach will completely
neglect other phenomena that might hide in discarded regions, like low $p_T$
ones. Hence, alternative methods are currently under investigation.

In this context, the HEP.TrkX project~\citep{heptrkx:2016} aims to study deep learning
solutions to the particle tracking issue. The main outcome has been a
model~\cite{Farrel:2017} combining CNNs and Recurrent Neural Networks (RNNs),
mainly employing Long Short Term Memories (LSTMs) cells, to
reconstruct tracks within a simplified detector simulation. The generated data
involve straight-line tracks and neglect all other kinds of physical complexities,
such as track curvature, material effects and detection inefficiencies.
The model is trained to solve two tasks in particular: a $2{-}$dimensional
single-track reconstruction starting from seeded hits and an end-to-end
estimation of the track parameters without any seeding.

Detector data are projected onto two axes representing the
detector layer and the channel within each specific layer. The tool
opens for the possibility of encoding irregular layer geometries of varying size
with two strategies: either zero padding the input to retrieve a regular rectangular
grid, or through an autoencoder-like architecture that embeds each layer input
into a fixed-size vector representation with the help of a dense network, followed
at the end of the pipeline by another fully connected layer that projects back the
output into the original layer dimensionality.
The data encoding based on bi-dimensional images has also been exploited
by~\cite{Tsaris:2018}, the key idea is again to model the recursive track-following
approach of the Kalman filter through an LSTM, showing promising results in a
semi-realistic detector simulation.

Other approaches based on Graph Neural Networks (GNNs) have been introduced
by~\cite{Farrell:2018} driven by the observation that the image-like representation
of the data would not be able to manage realistic use cases matching the HL-LHC conditions.
Indeed, the collider and detector updates will provide high-dimensional and sparse
data due to the increased number of detector layers built with irregular geometries,
which would probably cause inefficiencies in the standard approaches with CNNs.
The authors advocate the investigation of methods acting on the space-point
representation of data, instead, involving variable amounts of hits per event and
exploiting the full detector resolution.

\begin{figure}
  \myincludegraphics{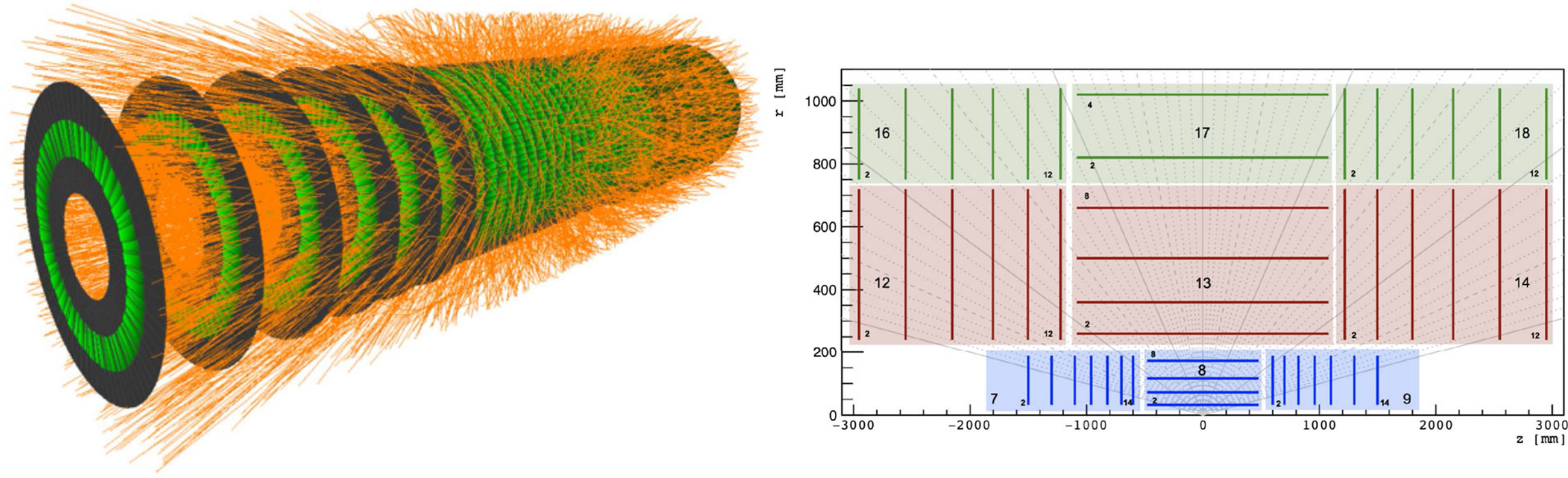}
  \caption{A sample event from the simulated TrackML dataset. The left panel
  shows a collision event with the TrackML detector picture and the right image
  is a schematic representation of the upper half of the detector projected on the
  $\mathrm{r{-}z}$ plane. The figure is taken from~\cite{Ju:2021}.}
  \label{fig:trackml}
\end{figure}
In $2018$, the TrackML competition~\citep{Amrouche:2020} took off within the HEP
community with the intent of finding the best candidate for the future particle
tracking algorithm. The desired feature of such a tool would be to achieve the
best performance score across several metrics reflecting the need to target high
reconstruction efficiencies with the fastest algorithm in terms of inference time.
Figure~\ref{fig:trackml} shows an event example from the TrackML dataset,
highlighting a large number of tracks to be reconstructed and the complex
detector design.

Following this competition, HEP.TrkX evolved into the Exa.TrkX
project~\citep{exatrkx:2019} which investigated a wide variety of models to
solve the task, mainly through GNNs~\citep{Ju:2020,Choma:2020}. The project
finally published an article~\citep{Ju:2021} summarizing the GNN pipeline on the
TrackML dataset, towards a first validation on ATLAS and CMS real detector data.
The potential of GNNs has also been exploited on implementations for specific
hardware acceleration, mainly provided by Field Programmable Gate Arrays
(FPGAs)~\citep{Elabd:2021}.

%% file: chapters/partII/PartII.tex
\part*{Part II\vspace{0.5cm}\\Monte Carlo event generators on hardware accelerators}
\addstarredpart{\large Part II : Monte Carlo event generators on hardware accelerators}

\adjustmtc[1]

\thispagestyle{empty}

%% file: chapters/partII/chap3/chap3.tex
\chapter{Interpolating parton distribution functions}
\label{chap:pdfflow}
\thispagestyle{plain}

\minitoc

\renewcommand{\myincludegraphics}[2][width=\textwidth]{
    \includegraphics[#1]{chapters/partII/chap3/plots/#2}
}

In this chapter, we discuss the implementation of
PDFFLow~\citep{Carrazza:2020pdfflow,Rossi:2021}, a Parton Distribution
Function (PDFs) access tool. We first briefly introduce PDFs in HEP and LHAPDF,
the state-of-the-art framework for PDF values interpolation, LHAPDF. The last
section of the present chapter contains a detailed description of our novel
implementation of such kind of technology. We show that PDFFlow software can
leverage hardware acceleration to speed up the computation while obtaining the
same outputs of the previous solutions.

\section{Parton Distribution Functions}

Parton Distribution Functions are of paramount importance in HEP since they
universally describe the inner content of hadrons and their partonic structure.
At leading order in perturbation theory, PDFs $f_i(x;Q^2)$ represent the probability
to extract a parton of flavor $i$ with a given momentum fraction $x$ from a hadron,
when probed at an energy scale $Q^2$. At higher orders, this naive probability
interpretation is not true anymore, since PDF positivity cannot be ensured.
Nevertheless, these functions respect at all orders particular relationships, known
as sum rules, which constrain their normalization, since integrating over the
momentum fraction of the parton and summing all the possible parton flavors yield
the momentum of the parent hadron:
\begin{equation}
    \label{eqn:sumrule-1}
    \sum_i \int_0^1 dx \, f_i(x;Q^2) = 1
\end{equation}
Further, the baryon number conservation leads to the following equation:
\begin{equation}
    \label{eqn:sumrule-2}
    \int_0^1 (f_i(x,Q^2) - \bar{f}_i(x,Q^2)) = n_i
\end{equation}
where $i$ takes values on the parton flavors and counts the valence partons of the
specific baryon involved in the formula: for example, for the proton the
non-zero $n_i$s are $n_u = 2$ and $n_d = 1$. For completeness, $\bar{f}_i$ denotes
the anti-quark PDFs for flavor $i$.
Equations~\ref{eqn:sumrule-1} and~\ref{eqn:sumrule-2} are intended to be satisfied
at all orders for every fixed factorization scale $Q^2$. For fixed $x$ the evolution
of the PDFs with respect to the factorization scale $Q^2$ is given by a set of
partial differential equations known as the DGLAP equations.

PDFs are a crucial object for both theoretical and experimental aspects of collider
physics: they must be convoluted with the partonic cross section $\hat{\sigma}_{i,j}$
to compute the total event cross-section:
\begin{equation}
    \sigma(Q^2) = \sum_{i,j} \int_0^1 dx_1 dx_2 \,
    f_i(x_1, Q^2) \, f_j(x_2, Q^2) \, \sigma_{i,j}(x_1,x_2,Q^2)
\end{equation}
where the sum runs on all the partonic flavors of the two partons $i$ and $j$ picked
from the two colliding hadrons, respectively.
The PDFs cannot be derived from first principles only, since they encode the
long-distance effects of non-perturbative QCD. Therefore, they are fitted on
experimental data by PDF fitting collaboration following different assumptions and
methodologies~\citep{Harland-Lang:2014,Ball:2017,Yuan:2019}. The measurement of
PDFs is influenced by several parameters like renormalization and factorization scales
as well as the running of the QCD strong coupling $\alpha_s$: hence, PDFs are grouped
into sets containing replicas, taking into account the variations of these quantities.

\section{LHAPDF: the Les Houches Accord PDF}
\label{sec:lhapdf}

During the $90$s, PDFLIB~\cite{Plothow-Besch:1992} has been the main software
dedicated to accessing PDF values. Written in the Fortran language, it originally
gathered about $100$ different PDF sets. However, the consistent production of
measurements by PDF fitting collaborations like CTEQ and MRST highlighted that
storing the PDF values within the library itself was not an option due to the
unbearable memory requirements. Therefore, the first versions of the LHAPDF
software~\cite{Whalley:2005,Bourilkov:2006} tackled the issue proposing to store
only PDF starting values at low $Q^2$ scale and evolving them through the DGLAP
equation to reach high energies.

Again, this solution was eventually surpassed around the mid${-}2000$s mainly due to
the large amount of custom code required by each different PDF parametrization
to be included in the framework and the need to modernize the algorithms for the $Q^2{-}$evolution.
As a consequence, the PDF collaborations decided to provide external files collecting
entire PDF grid values of fitted points in the $(x, Q^2)$ plane together with routines
to access them and interpolate values between grid knots. This choice decoupled
the PDF values to the LHAPDF library, but at the same time filled it with a plethora
of methods to read the multitude of available formats. Given the Fortran static memory allocation,
this resulted in large memory portions of code never accessed by the user focusing
on some specific methods only.

A decade later, LHAPDF 6~\citep{Buckley:2015} finally ported the software
to C++ language, also providing a Python interface. The program is built around
the object-oriented programming philosophy, in a way that allows the user to
easily extend the library to incorporate custom code. The PDF function values
are accessed through a powerful cascading metadata system following a universally
agreed format, allowing the PDF fitting collaboration to provide new measurements
independently from the LHAPDF software release development. PDF grid values can
be accessed by instantiating a \texttt{PDF} class object representing parton density
functions for several parton flavors and calling its \texttt{PDF::xfxQ(\dots)}
and \texttt{PDF::xfxQ2(\dots)} methods: the only difference being in the energy
scale argument, which should be squared in the latter case. Note that the output
value is always in the $xf(x,Q^2)$ format, which is the standard in PDF sets.

The agreed format for presenting the PDF grids contains values at discrete knots
in the $(x,Q^2)$ plane. Points position in the grid is PDF set dependent and usually
ranges from very small momentum fractions $x$ of the order of $\mathcal{O}(10^{-9})$
to $1$ and from few $\SI{}{\GeV}$ to $\mathcal{O}(\SI{100}{\TeV})$ in the $Q$ energy scale.
For example, the \texttt{NNPDF31\_nlo\_as\_0118} PDF set shows $x \in [10^{-9}, 1]$
and $Q \in [1.65, 10^5] \SI{}{\GeV}$. The knots are not sampled uniformly, to allow
PDF discontinuities across quark mass thresholds, like in the case of the $b$ quark
whose mass $m_b = \SI{4.92}{\GeV}$. The grid limits are accessed via the \texttt{PDF::xMin(),
PDF::xMax()} and \texttt{PDF::QMin(), PDF::QMax()} methods. Sometimes a PDF grid
is organized into multiple sub-grids.

\begin{figure}
    \centering
    \subfigure[
        The LHAPDF default log-bicubic interpolation method. Only the points addressed
        by solid circles enter the calculation of $xf(x,Q^2)$.
    ]{
      \centering
      \label{subfig:logbicubic}
      \myincludegraphics[width=0.4\textwidth]{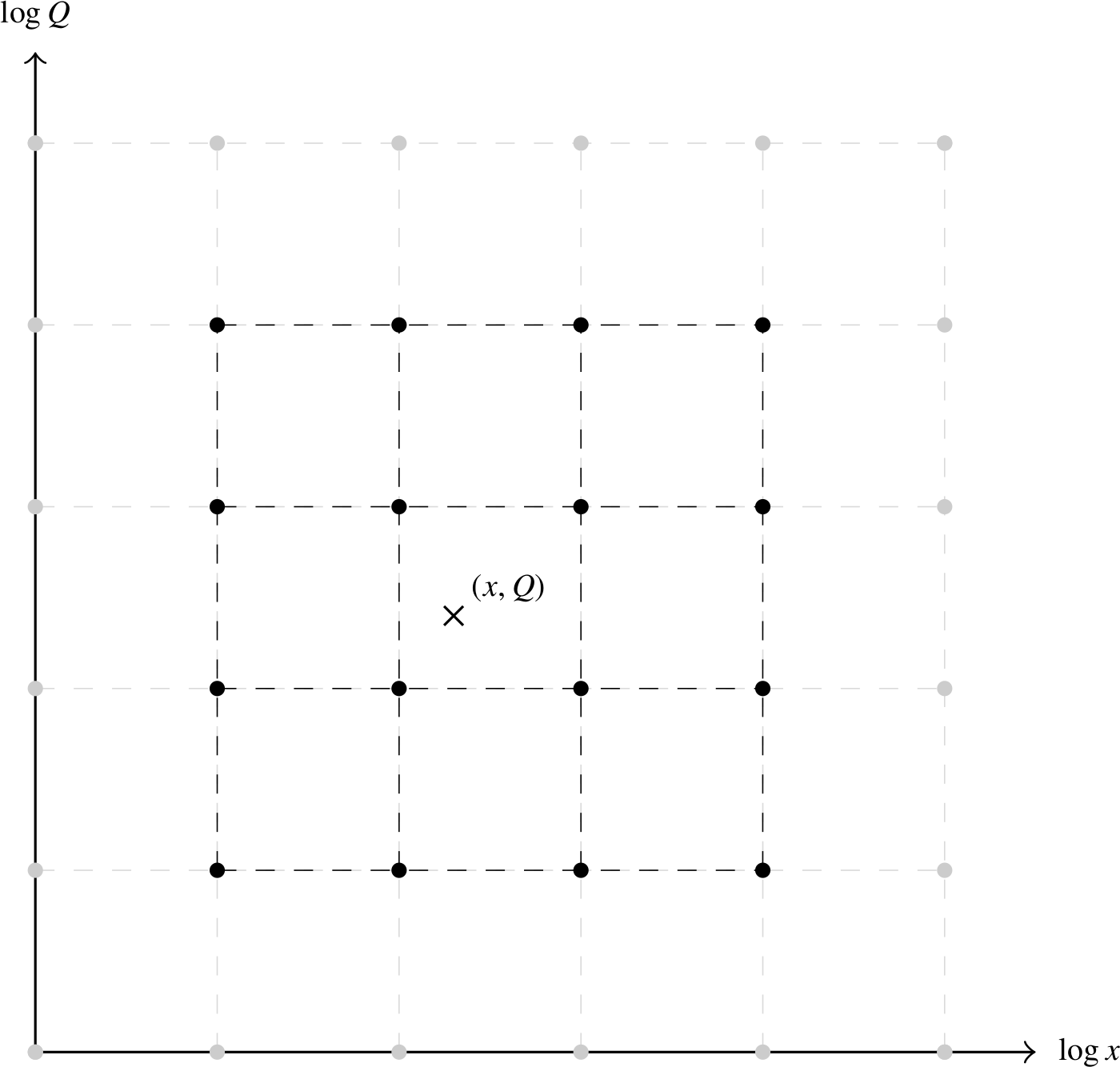}
    }
    \qquad\qquad
    \subfigure[
        The interpolation and extrapolation regions of \texttt{NNPDF31\_nlo\_as\_0118}.
        The PDF grid is organized into $2$ sub-grids separated at the $b$ quark mass
        threshold.
    ]{
      \label{subfig:subgrids}
      \myincludegraphics[width=0.4\textwidth]{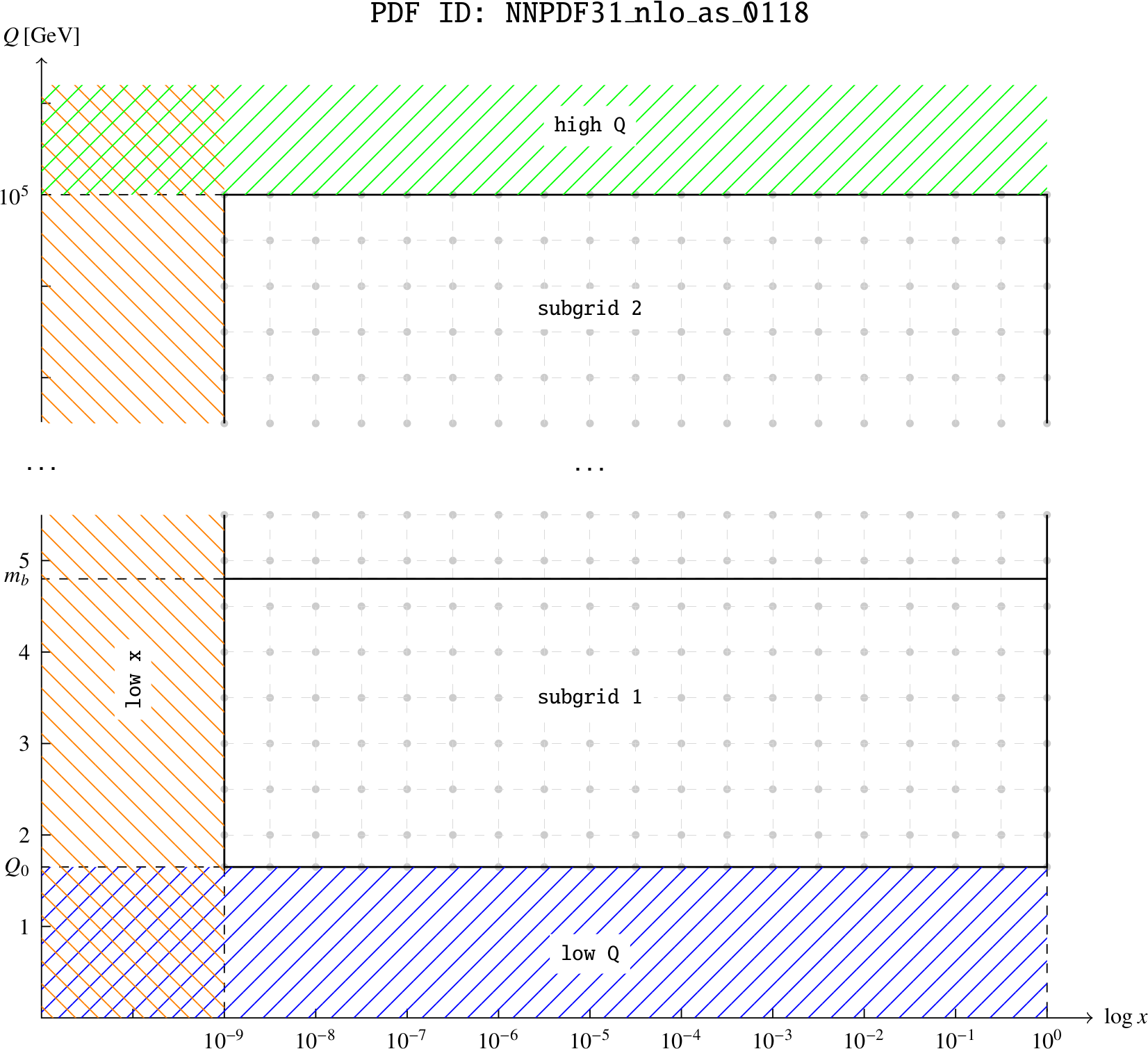}
    }
    \caption{Scheme of the LHAPDF algorithm for PDF values access. The grid knots
    spacing is uniform for convenience only, real PDF grids are not uniform in
    the $(\log x, \log Q)$ plane.}
    \label{fig:lhapdf}
\end{figure}

A value $xf(x,Q^2)$ within the grid boundaries can be retrieved by interpolating
neighbor node values with a specific method. The default behavior implemented
by LHAPDF is to interpolate plane values through cubic Hermite splines in the
$(\log x, \log Q)$ space. This algorithm requires sampling $4$ neighbor knot
values as in figure~\ref{subfig:logbicubic} and performing a log-bicubic
interpolation.
In regions close to the grid boundaries, where the number of neighbor points is not
enough to perform such interpolation, the software automatically switches to bilinear
interpolation, which requires $2$ neighbor knots only.

Outside the grid boundaries, some extrapolation method is required. Multiple \linebreak choices
are available within the LHAPDF 6 library, the simplest one being to raise an error
if a point outside the grid range is queried. Another option is to freeze the grid
edges, returning the PDF value at the closest grid knot. However, the default extrapolation
method is the legacy extrapolation system from LHAPDF 5, proposed by the MSTW collaboration:
figure~\ref{subfig:subgrids} shows the regions where the three behaviors for the low${-}x$, low${-}Q$
and high${-}Q$ extrapolation take place.

When a low${-}x$ value is queried for a specific
energy scale $Q$, a continuation to $x$ is guaranteed by linear extrapolation
from the two lowest $x_0$ and $x_1$ knots in the PDF grid.
If the PDF values are sufficiently positive, namely they both exceed a threshold 
set to $10^{-3}$, their logarithm is used in the extrapolation, otherwise, $xf$ is
exploited. The same method is implemented for the high${-}Q$ region extrapolation.

For the low${-}Q$ value extrapolation a different method is employed. The LHAPDF
framework interpolates the anomalous dimension $\gamma(Q^2)$, namely \linebreak
$\gamma(Q^2) = \partial \log xf(x,Q^2)/\partial Q^2$,
between the value at $Q_{min}$ and $1$ for $Q \ll Q_{min}$ with the following rule:
\begin{equation}
    xf(x,Q^2) = xf(x,Q^2_{min}) \braces{Q^2 / Q^2_{min}}^{\frac{\gamma(Q^2_{min})- 1}{Q^2_{min}} Q^2 + 1}
\end{equation}
This ensures that the following limiting behaviors are verified:
\begin{equation}
    xf(x,Q^2) \to
    \begin{cases}
        xf(x,Q^2_{min}) \braces{\frac{Q^2}{ Q^2_{min}}}^{\gamma(Q^2_{min})} & \text{for} \ Q^2 \to Q^2_{min} \\[10pt]
        xf(x,Q^2_{min}) \frac{Q^2}{Q^2_{min}} & \text{for} \ Q^2\to 0
    \end{cases}
\end{equation}
In regions where both the $x$ and $Q$ values need extrapolations, first the corresponding
$x{-}$algorithm is performed, then the results are passed through the continuation
function for $Q{-}$values.

Integral to the parton distribution function description is the correct evolution
of the strong running coupling $\alpha_s(Q)$, which depends on the energy scale $Q$
at which it is evaluated. Therefore, the PDF sets are accompanied by a sample
of measurements of $\alpha_s$ at different energies and LHAPDF implements interpolating
and extrapolating methods similar to the PDF ones. The provided grid exhibits knots
on the $\log Q$ line. The interpolation algorithm is again a cubic Hermite spline,
while for $Q > Q_{max}$ a constant value $\alpha_s(Q_{max})$ is returned and
for $Q > Q_{min}$ a logarithmic gradient continuation similar to the low${-}x$
behavior of the PDF is implemented.

\section{\pdfflow: interpolating PDFs on GPUs}
\label{sec:pdfflow}

\subsection{Methodology}
\label{subsec:pdfflow-methodology}

The \pdfflow package~\citep{Carrazza:2020pdfflow,Rossi:2021} follows the
\vegasflow~\citep{Carrazza:2020vegasflow,vegasflow_package} concept to vectorize
operations on the program inputs.
The neat result is that transformations that should have been done sequentially
on independent points are now done altogether on the whole sample. As a consequence,
scalar input arguments of the functions become vectors with the size of the number of
desired events to be computed. Hardware accelerators, like GPUs, provide massive
speed-up factors in highly parallel environments like the present one. However,
the programming skills required to port the code and optimize it into a suitable
format for such devices represent a strong limit for the current applications.

We present a prescription to port the code into a format compatible with hardware
accelerators leveraging the Python language and Google's Machine Learning TensorFlow
library~\citep{Abadi:2015}. Indeed, the \pdfflow package is implemented by exploiting
TensorFlow primitives, which are natively designed to be GPU-friendly. Therefore,
the tool automatically inherits the TensorFlow portability on hardware accelerators.
We wrap each custom function code with the \texttt{tf.function}
decorator triggering the computation of a \texttt{tf.Graph}: this induces a transformation
on Python static data types and operations into \texttt{tf.Tensor} and \texttt{tf.Operation}
primitives, respectively. The process automatically builds an implicit
operational graph, which optimizes the code and runs seamlessly
on multiple devices like multi-thread CPUs, GPUs and, if the operations are
supported by the TensorFlow library version, even on TPUs. A little overhead in
the running time is introduced by this function conversion to load the corresponding
\texttt{tf.Graph} in memory. Luckily this step is done only once, namely when
the piece of code wrapped by\texttt{tf.function} is executed for the first time.

\begin{figure}
\begin{minipage}{0.49\textwidth}
    \begin{tiny}
    \begin{minted}[autogobble]{python}
import numpy as np
import tensorflow as tf

def py_function(x, y):
  """The Python function with list comprehension."""
  return [
    sum(xx * yy[0] for xx, yy in zip(x_row, y))
    for x_row in x
    ]

def np_function(x, y):
  """The NumPy function."""
  return np.matmul(x, y)

input_signature=[
  tf.TensorSpec(shape=[10, None], dtype=tf.float32),
  tf.TensorSpec(shape=[None, 1], dtype=tf.float32),
]

@tf.function(input_signature=input_signature)
def tf_function(x, y):
  """The TensorFlow function."""
  return tf.matmul(x, y)

    \end{minted}
    \end{tiny}
    \caption{Matrix multiplication functions code snippet. The \texttt{@tf.function}
    decorator requires to define the shape of the input tensors, where \texttt{None}
    signal a variable length dimension.}
    \label{fig:matmul-code}
\end{minipage}
\hfill
\begin{minipage}{0.49\textwidth}
    \centering
    \myincludegraphics[width=\textwidth]{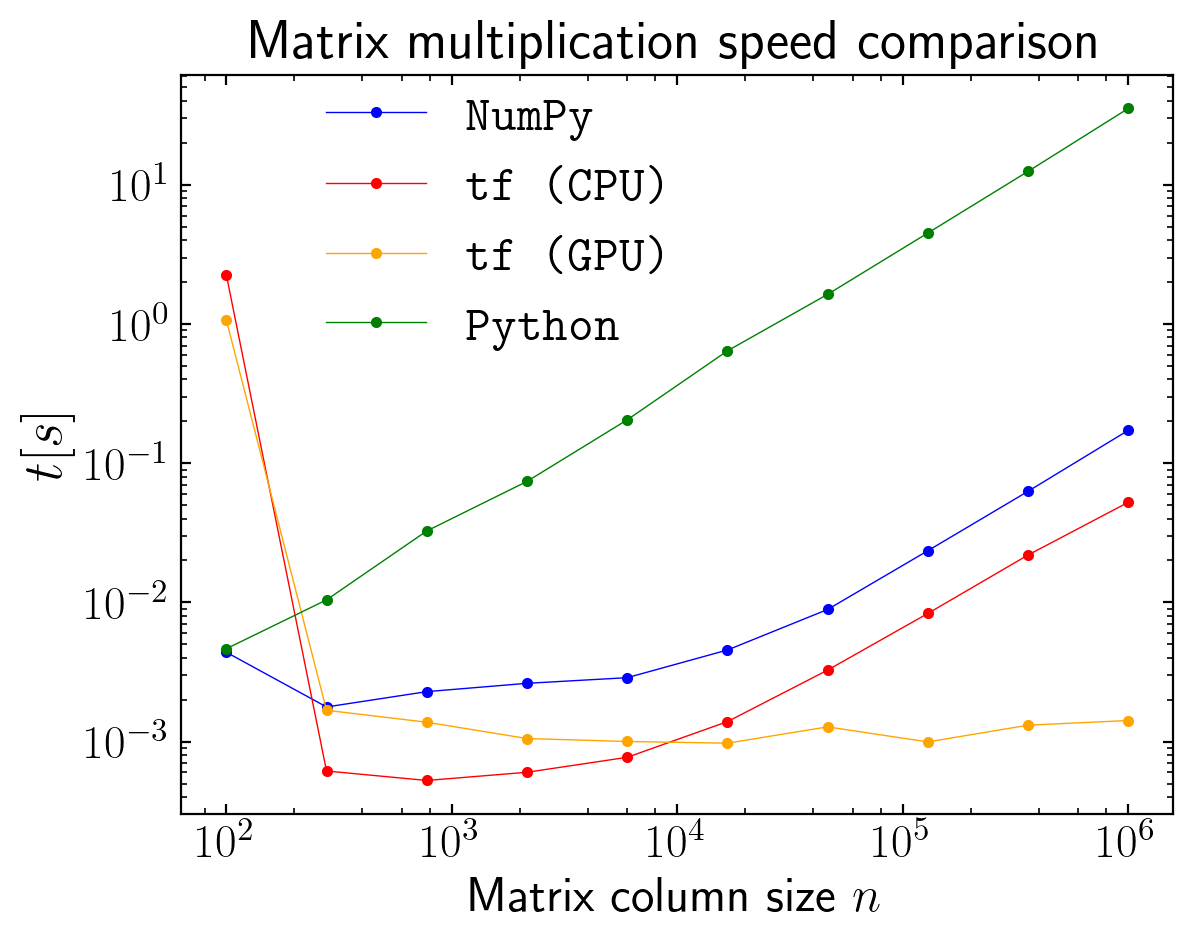}
    \caption{Time comparison of different matrix multiplication implementations.}
    \label{fig:tf-bench}
\end{minipage}
\end{figure}

To show the capabilities of this method, we design a toy example to assess
the speed of the matrix multiplication operation for different implementations.
We multiply a matrix of size $100\times n$ times a vector of length $n$ of random
numbers sampled from the normal distribution. We let the size $n$ span a wide range
of values from $10^2$ to a maximum of $10^6$. Table~\ref{tab:tf-bench} collects the
hardware and software specifics employed in this study. Note that this represents
the latest stable version of the software available on the Python Package Index
(PyPI)~\citep{pypi} at the time of the \pdfflow article~\citep{Carrazza:2020pdfflow}
publication (July 2021). However, the present discussion is valid for all the
\texttt{v2.x} TensorFlow versions.
\begin{table}
    \caption{Description of the system used for the matrix multiplication benchmark:
    hardware and software information.
    }
    \centering
    \label{tab:tf-bench}
    \footnotesize
    \begin{tabular}{ll|ll}
        \hline\noalign{\smallskip}
        Software & Version & Hardware specific & Value \\
        \noalign{\smallskip}\hline\noalign{\smallskip}
      Python & \texttt{v3.10} & CPU model & Intel Xeon Gold 6130 \\
      NumPy & \texttt{v1.23} & CPU cores & 64 @ $\SI{2.10}{\GHz}$\\
      TensorFlow & \texttt{v2.10} & CPU RAM & 192 GB \\
      && GPU model & Nvidia Tesla V100 \\
      && GPU memory & 32 GB \\
      \noalign{\smallskip}\hline
    \end{tabular}
\end{table}
We test for different implementations of the matrix multiplication function in Python
explicitly defined in figure~\ref{fig:matmul-code}:
\begin{itemize}
    \item a matrix multiplication implemented in pure Python code with list comprehensions;
    \item a NumPy~\cite{Harris:2020} implementation exploiting the \texttt{np.matmul} function;
    \item a TensorFlow implementation wrapping the code with the \texttt{@tf.function} decorator.
\end{itemize}
Figure~\ref{fig:tf-bench} shows the behavior of the different implementations of
the matrix multiplication as a function of the matrix column size $n$. The TensorFlow
implementation runs either on CPU or GPU, while the others are CPU only. We note
that the maximum size for the parameter $n$ is strictly dependent on the device
memory hosting the computation: the GPU will raise an Out Of Memory (OOM) error
if the size of the tensors exceeds the available memory, thus interrupting the job.
On the other hand, the CPU is more flexible and tries, nevertheless, to handle
the computation at the price of a great slowdown: this is due to the CPU
exhausting the RAM memory and trying to exploit the hard disk memory instead.
The buses connecting the RAM and the CPU ensure way faster communications than
the ones between the CPU and the hard disk: this motivates the performance
breakdown.

As a rule of thumb, the expected memory used by an algorithm can roughly be
estimated by summing the memory allocated for each tensor created in the routine.
However, TensorFlow always reserves some small amount of memory (about
$\orderof{\SI{100}{\mega\byte}}$) for internal operations depending on the
algorithm itself. Even neglecting this little memory overhead, an a priori
estimation of the memory used by a script might be difficult to compute,
especially for complex operations. Therefore, the user should be approximately
aware of the amount of memory he needs and possibly run memory tests before
deploying his tool.

The plot in figure~\ref{fig:tf-bench} displays that the first calls to the
\texttt{tf.function} decorated code present an overhead time due to the
\texttt{tf.Graph} construction; the subsequent function executions, instead,
manifest the improvements of these operations against the more common Python and
NumPy versions. Another feature that is evident in the graph is the almost
constant performance of the TensorFlow program running on the GPU device:
indeed, this behavior is expected until the whole GPU memory gets exhausted.

As a final remark, the TensorFlow CPU curve has the best
performance for low sizes of the input matrix, even exceeding the GPU one in
this region. This effect is probably caused by the fact that the CPU has to place
the tensors on the GPU before launching the computation and the elapsed time
can be comparable to the effective computing time for small-size operations.

In the rest of this chapter, we discuss the performance improvements brought by
the introduction of this prescription into state-of-the-art software for
parton distribution function value access.

\subsection{Software design and benchmarks}

\begin{figure}
    \centering
    \myincludegraphics[width=0.4\textwidth]{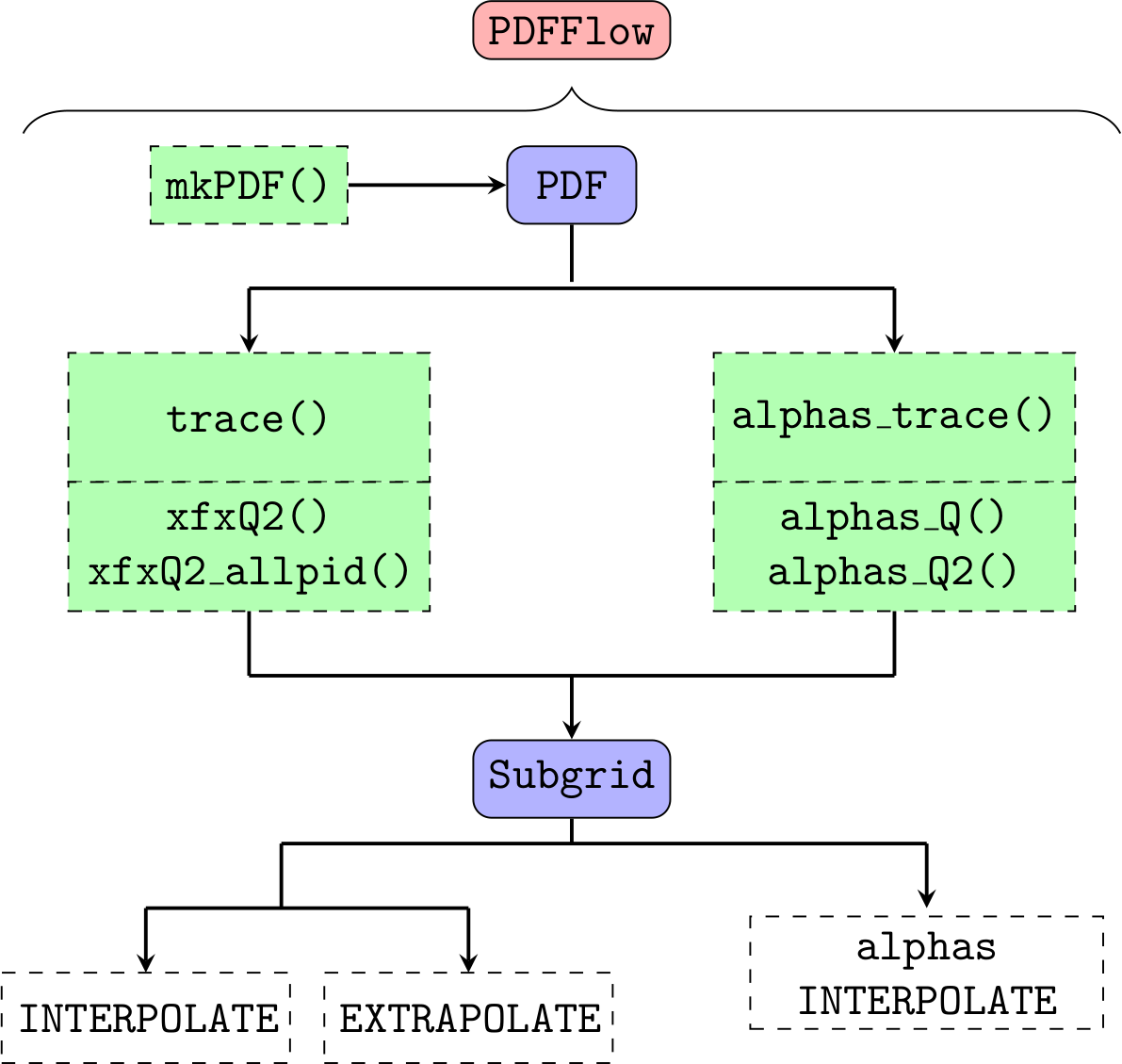}
    \caption{\pdfflow flowchart. Blocks are color-coded as follows: red for
	the tool, violet for classes, green for functions and class methods,
	and white for interpolation algorithms.}
    \label{fig:pdfflow_design}
\end{figure}

Figure~\ref{fig:pdfflow_design} depicts the \pdfflow design, which follows the
LAHPDF6 concept.
The \texttt{mkPDF()} function instantiates the desired PDF representation, given
by the \texttt{PDF} class. A \texttt{PDF} object stores all the quantities and
algorithms needed for the interpolation of both the PDF itself and the strong
running coupling $\alpha_s$. Notable member methods in the class are the trace
methods (\texttt{trace} and \texttt{alphas\_trace}) and interpolating methods
(two for the PDF and two for $\alpha_s$), contained in the green dashed boxes in
the figure. The tracing methods trigger the building of the \texttt{tf.Graph}
relative to the \pdfflow operations. Calling these methods before querying interpolation
points allows ahead-of-time compilation of all the functions declared within the tool.

The interpolating methods include a call to a \texttt{Subgrid} class object.
\texttt{Subgrid} stores PDF grid data and includes a switch to enable
interpolation on $\alpha_s$ grid knots. This class gives access to algorithms
that implement the actual computation of \pdfflow outputs, represented by white
boxes in the flowchart and briefly described below.

The interpolation procedure implemented in \pdfflow follows the prescription
originally implemented in LHAPDF6, namely the log-bicubic interpolation in
terms of $x$ and $Q^2$ and the extrapolation methods presented in section~\ref{sec:lhapdf}.
The PDF data files stored in the LHAPDF directories are directly loaded into
\texttt{tf.Tensor} objects so it is not necessary to install new sets or formats.
The interpolation algorithms compute independently query points belonging to
different sub-grids of the PDF set. Special care is taken about regions in the
$(x,Q^2)$ plane close to quark mass thresholds and grids $x$ edges, where the
minimum number of knots required for bicubic interpolation is not available.

Similar to the PDF interpolation procedure, the evaluation of the running
of the strong coupling, $\alpha_s(Q)$ is performed using a log-cubic interpolation
with constant extrapolation from the $(\alpha_s(Q),Q)$ nodes stored in the PDF
metadata file. The implementation includes the improved treatment of the
sub-grids mechanism and takes into account the impact of flavor thresholds on
$\alpha_s(Q)$ evolution.

We introduce now the interpolation accuracy and performance benchmarks results between \pdfflow
\texttt{v1.0} and LHAPDF \texttt{v6.3.0} libraries. All the studies presented here and
in the next section, where we discuss actual physics examples, are done exploiting
the hardware pointed out in table~\ref{tab:hardware-pdfflow}.
\begin{table}
    \begin{scriptsize}
    \centering
    \begin{tabular}{c|l|c|c|c|c}
        Device & CPU model & CPU cores & CPU RAM & GPU(s) model & GPU memory \\ \hline
        C & Intel i7-6700K & 4 @ 4-4.2GHz & 16GB @ 3000MHz & Nvidia RTX2080 & 8GB\\
        \hline
        P0 & AMD 2990WX & 32 @ 3-4.2GHz & 128GB @ 3000MHz & - & - \\
        \hline
        P1 & Intel i9-9980XE & 18 @ 3-4.4GHz & 128GB @ 2666MHz& Nvidia TITAN V & 12GB\\
         &  &  &  & Nvidia RTX2080TI & 12GB\\
         \hline
        P2 & Intel Xeon Gold 6126 & 6 @ 2.6-3.7GHz & 20GB @ 2133MHz & Nvidia V100 (2x) & 32GB\\
        \hline
    \end{tabular}
    \end{scriptsize}
    \caption{Description of the systems in which the different codes have been run.}
    \label{tab:hardware-pdfflow}
\end{table}
The consumer-grade hardware (C) consists of a standard desktop computer with
gaming level specifics. Different research groups have access to professional
grade hardware which is better suited for the kind of computation described in
this part of the thesis. In particular, this corresponds to many-threaded CPUs
and GPUs with enough memory to hold the necessary kernels for very complicated
computations. For the CPU-based calculation, we use the P0 system with a
medium-level processor in terms of clock speed, while for the GPU-based
calculations we use two different machines: P1 with a very powerful processor,
which greatly reduces the latency of the calculation for CPU-based operation
such as the accumulation of the final results, and P2, a less powerful CPU and a
more limited RAM size which can add an important overhead to the communications
between the CPU and the GPU. In exchange, the V100 GPUs have greater memory size
which reduces the frequency of communications between the main memory and the device.

To measure and compare the PDF interpolation accuracy between \pdfflow
and LHAPDF, we define a relative difference:
\begin{equation}
    r_{i}(x,Q) = \frac{| x f_i^{\rm \pdfflow}(x,Q) - x f_i^{\tt LHAPDF}(x,Q)|}{| x f_i^{\tt LHAPDF}(x,Q)| + \epsilon},
    \label{eqn:pdfflow-diff}
\end{equation}
where $x f_i (x, Q)$ is the numeric value of a PDF flavour $i$ evaluated at a
given momentum fraction $x$ and energy $Q$, and $\epsilon=10^{-16}$ is a ratio
stabilizer.
\begin{figure}
    \centering
    \subfigure[Gluon NNPDF3.1 NLO PDF.]{
      \centering
      \label{subfig:pdfacc_nnpdf_21}
      \myincludegraphics[width=0.47\textwidth]{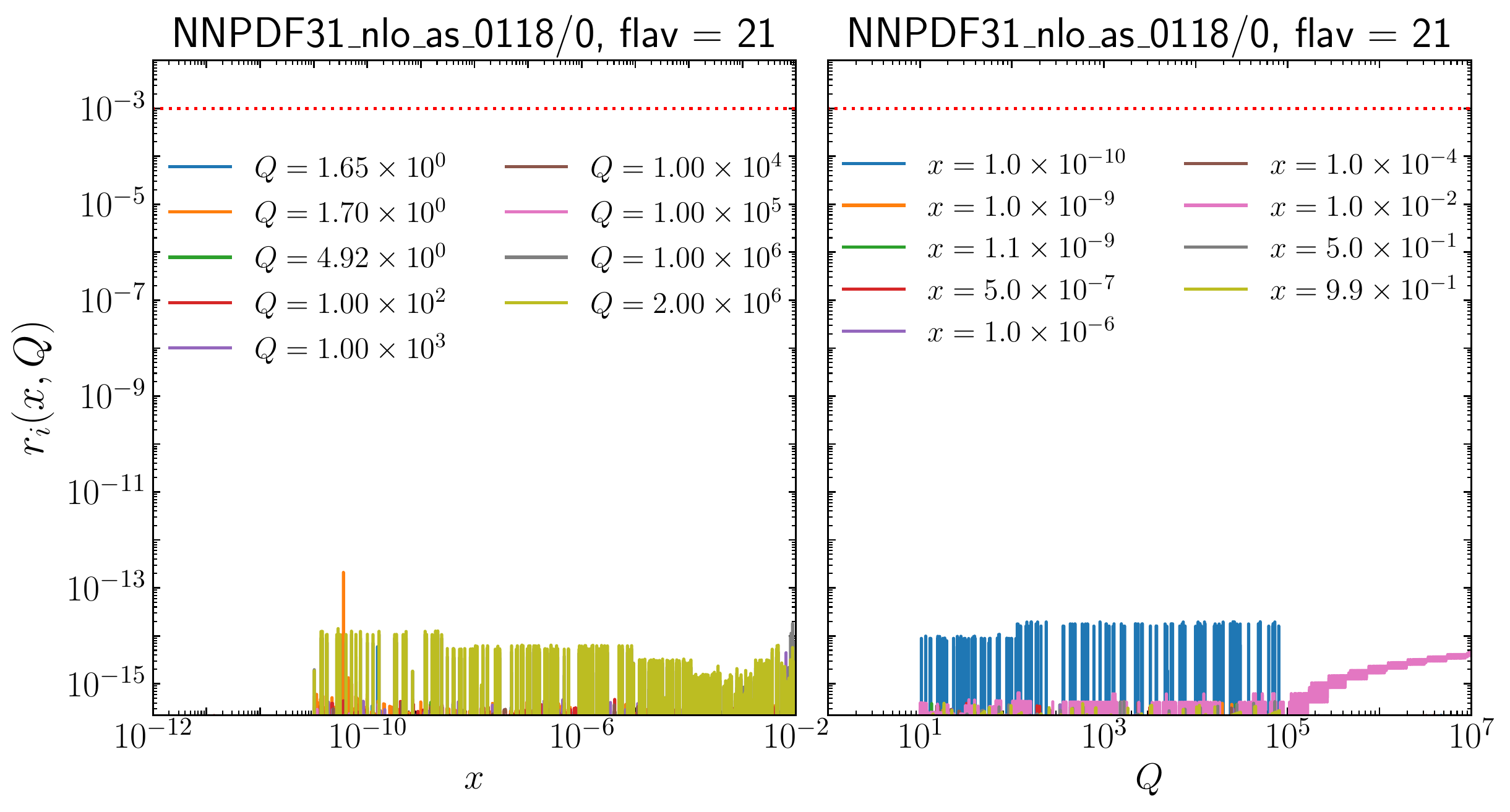}
    }
    \hfill
    \subfigure[Down quark NNPDF3.1 NLO PDF.]{
      \centering
      \label{subfig:pdfacc_nnpdf_1}
      \myincludegraphics[width=0.47\textwidth]{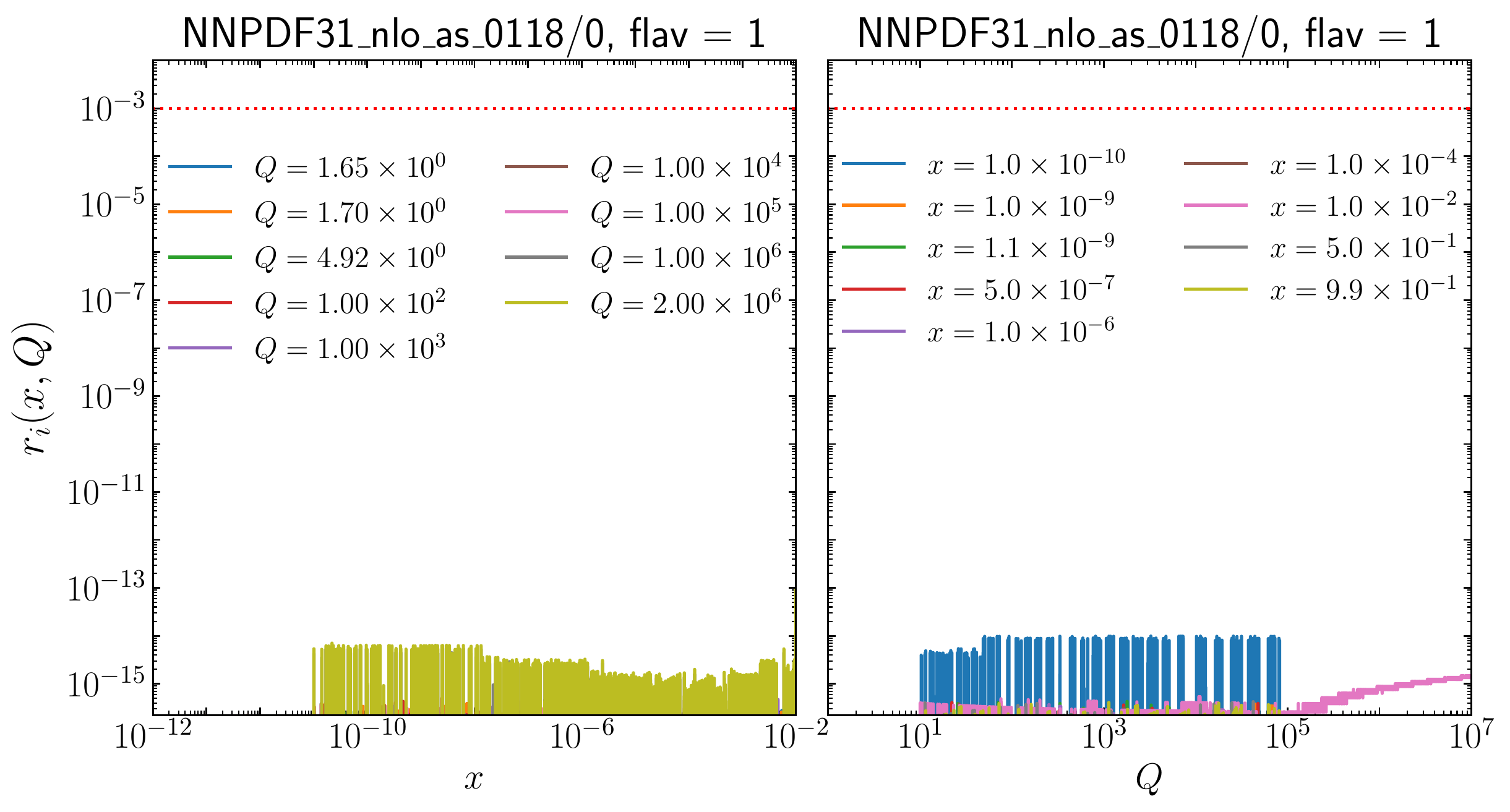}
    }    
    \caption{\pdfflow vs LHAPDF relative difference for the NNPDF3.1 NLO central
    PDF. In both the sub-figures, the first column refers to differences in a
    grid of $x$ points for fixed $Q$ values, while the second column shows
    differences in a grid of $Q$ values for fixed $x$.}
    \label{fig:pdfaccuracy_nnpdf}
\end{figure}
In figure~\ref{fig:pdfaccuracy_nnpdf} we plot this relative difference for the
NNPDF3.1 central PDF at Next-to-Leading Order (NLO) for the gluon and down quark
flavor values. In both cases, we plot first the error as a function of $x$ for
fixed values of the energy scale $Q$ and the opposite for the second image.
We prove that \pdfflow interpolated values are compatible with LHADPF ones, since
the error is several orders of magnitude lower than the fixed threshold of $10^{-3}$,
which was set by LHAPDF itself during the upgrade from version \texttt{v5} to
\texttt{v6}. Similar results can be achieved by comparing different PDF sets, we
present a complete set of numerical comparisons for the MMHT2014 NLO
PDF~\cite{Harland-Lang:2014} in figure~\ref{fig:pdfaccuracy_mmht} at the end of
the present chapter.

We remark that the pattern of the errors showed in all the presented accuracy
plots is determined by the rounding errors due to the representation of numbers
with finite precision in the machine. Another source of discrepancy might be given
by the different implementations of the low-level primitives between the specific
libraries exploited by the \pdfflow and LHAPDF tools. These small fluctuations
can be neglected as long as they prove to be much lower than the $10^{-3}$
threshold set above.

\begin{figure}
    \begin{minipage}[b]{0.49\textwidth}
        \centering
        \myincludegraphics[width=\textwidth]{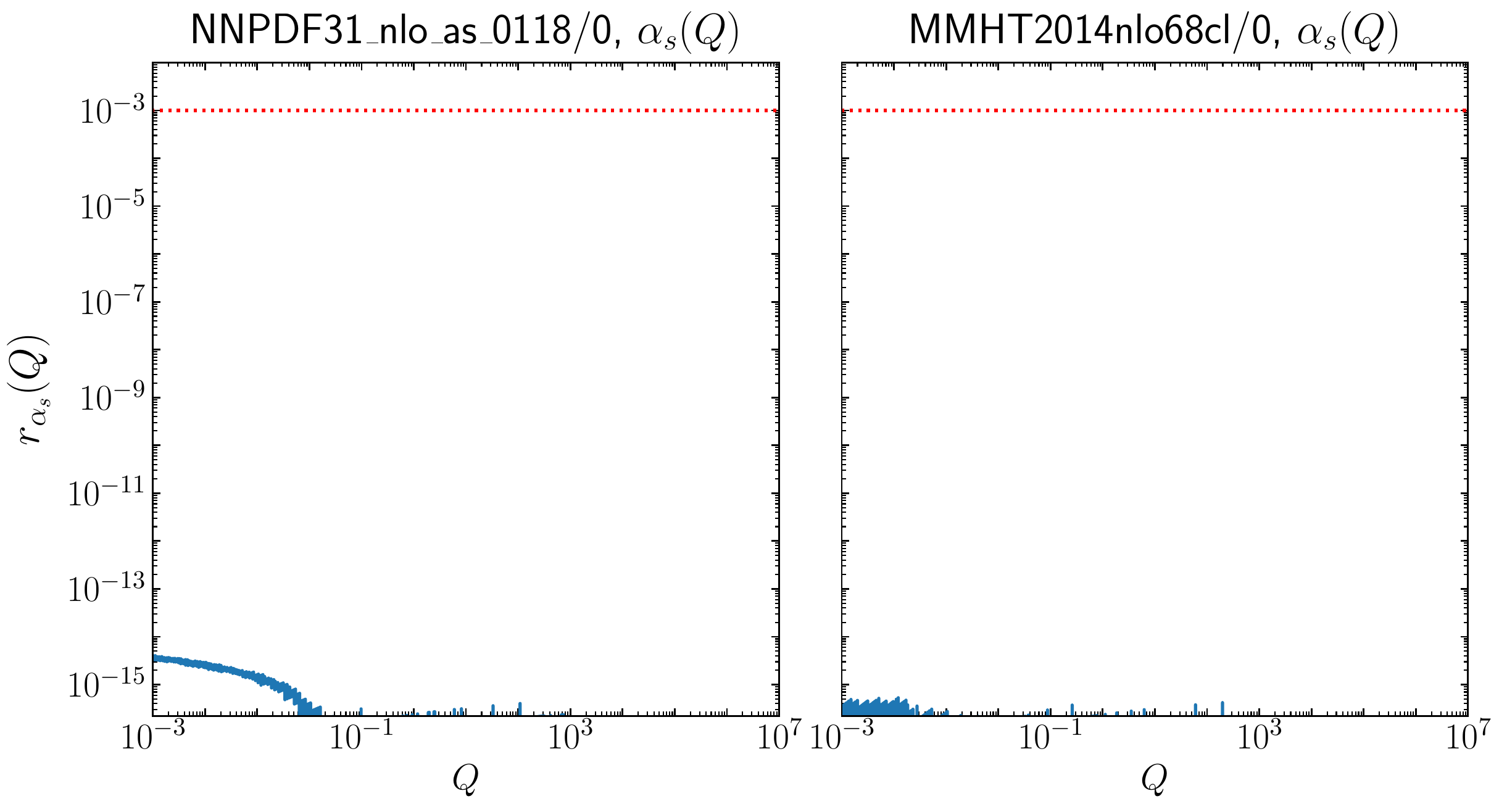}
        \caption{\pdfflow vs LHAPDF relative difference for $\alpha_s(Q)$
        interpolation. The left and right panels refer to NNPDF3.1 NLO and
        MMHT14 NLO sets, respectively.
        }
        \label{fig:alphas-accuracy}
    \end{minipage}
    \hfill
    \begin{minipage}[b]{0.49\textwidth}
        \centering
        \myincludegraphics[height=5cm]{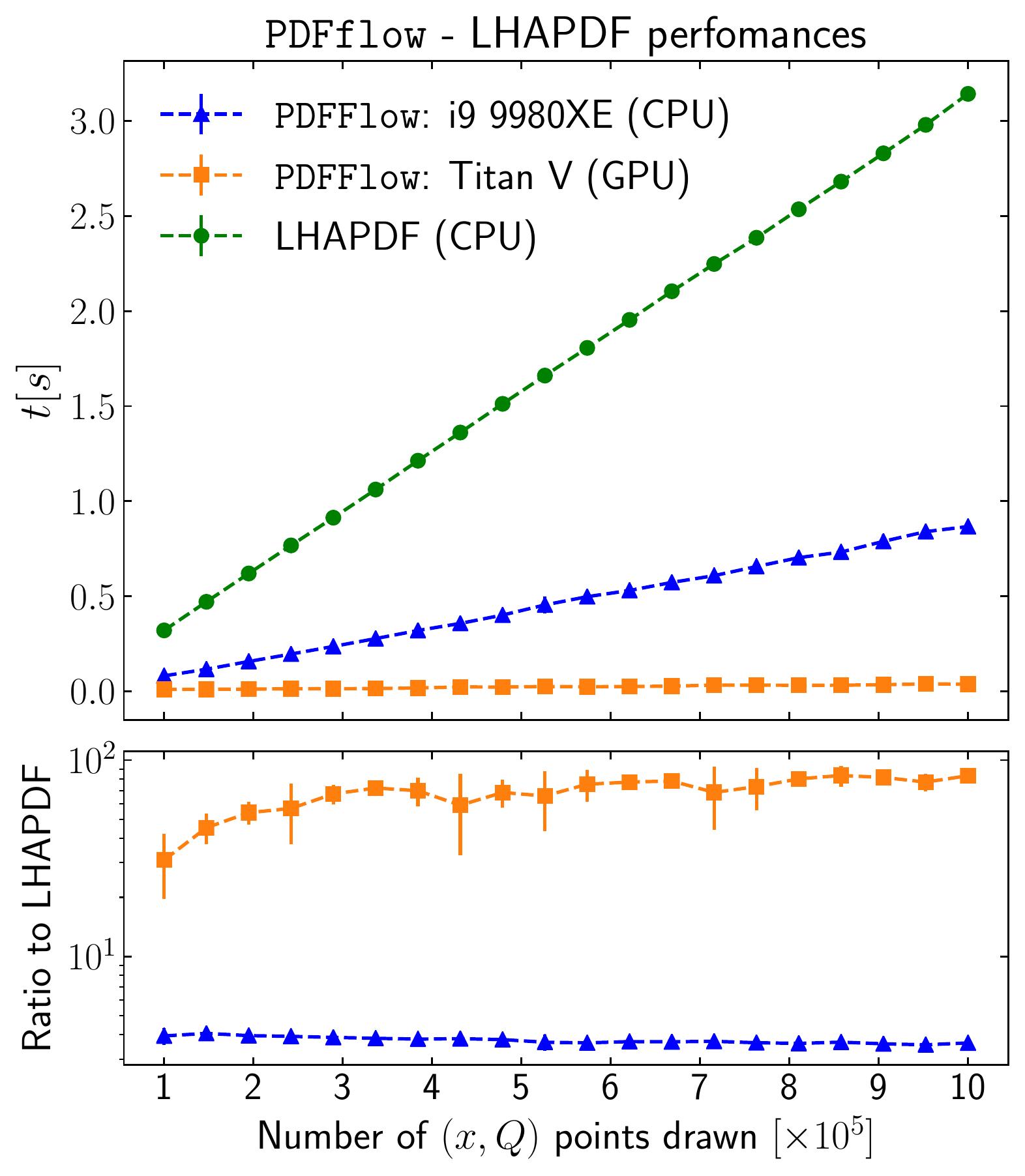}
        \caption{\pdfflow vs LHAPDF running time comparison. Top and bottom rows
        plot respectively the absolute and relative execution time as functions
        of the input size.}
        \label{fig:pdfflow-time}
    \end{minipage}
\end{figure}

A similar benchmark is made on the interpolation of the strong running coupling
$\alpha_s(Q)$. We compute the same relative error $r_{\alpha_s}(Q)$
with the help of equation~\ref{eqn:pdfflow-diff}, after replacing the $xf_i(x,Q^2)$
PDF value with $\alpha_s(Q)$. Figure~\ref{fig:alphas-accuracy} demonstrate that,
again, the values interpolated by \pdfflow are equal up to rounding errors to
the LHAPDF ones.
In terms of performance speed, we conduct a benchmark analog to the one presented
in section~\ref{subsec:pdfflow-methodology}: we measure the elapsed time to
interpolate sets of $(x, Q^2)$ points by \pdfflow and LHAPDF. The query points
are distributed within the PDF grid boundaries, whereas each set has a
logarithmically increasing size across different evaluations. The resulting data
are collected in figure~\ref{fig:pdfflow-time}. We observe a great performance
improvement when running \pdfflow default configuration on CPU, thanks to the
built-in multi-threading CPU support. Concerning GPU results, the performance
improvement is huge and opens the possibility to construct new models and
applications with parallel evaluations. Note that, as opposed to
figure~\ref{fig:tf-bench}, the time excess in the first call of the \pdfflow
interpolation function is not visible in this plot, since the \texttt{tf.Graph}
is built ahead of time invoking the \texttt{PDF.trace()} method.

\subsection{Physics examples}

This paragraph introduces several physics examples, developed in the HEP context,
that exploit the \pdfflow software operating in synergy with \vegasflow to evaluate
interesting quantities at different orders of perturbation theory.

\subsubsection{Single t-quark production at LO}

\begin{figure}
    \begin{minipage}{0.45\textwidth}
        \centering
        \myincludegraphics[width=\textwidth]{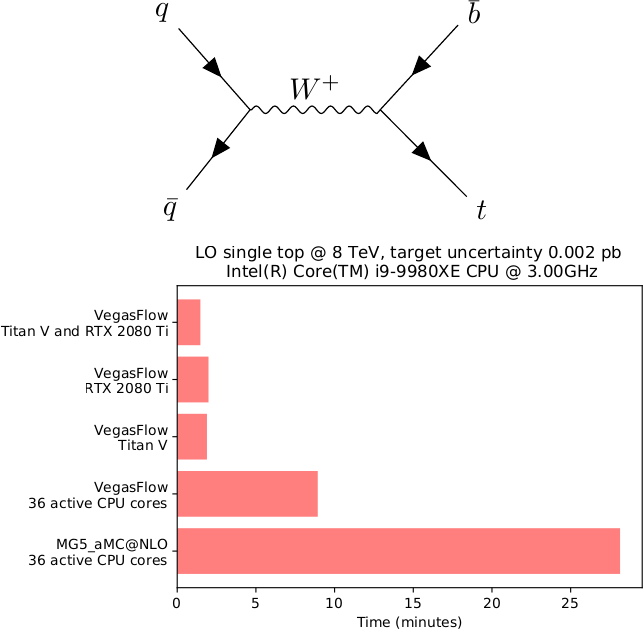}
        \caption{Single top quark production. Top row: Feynman diagram. Bottom row:
        comparison of the execution time between \pdfflow and \vegasflow
        code against \mgamc.}
        \label{fig:singletop}
    \end{minipage}
    \hfill
    \begin{minipage}{0.45\textwidth}
        \centering
        \myincludegraphics[width=\textwidth]{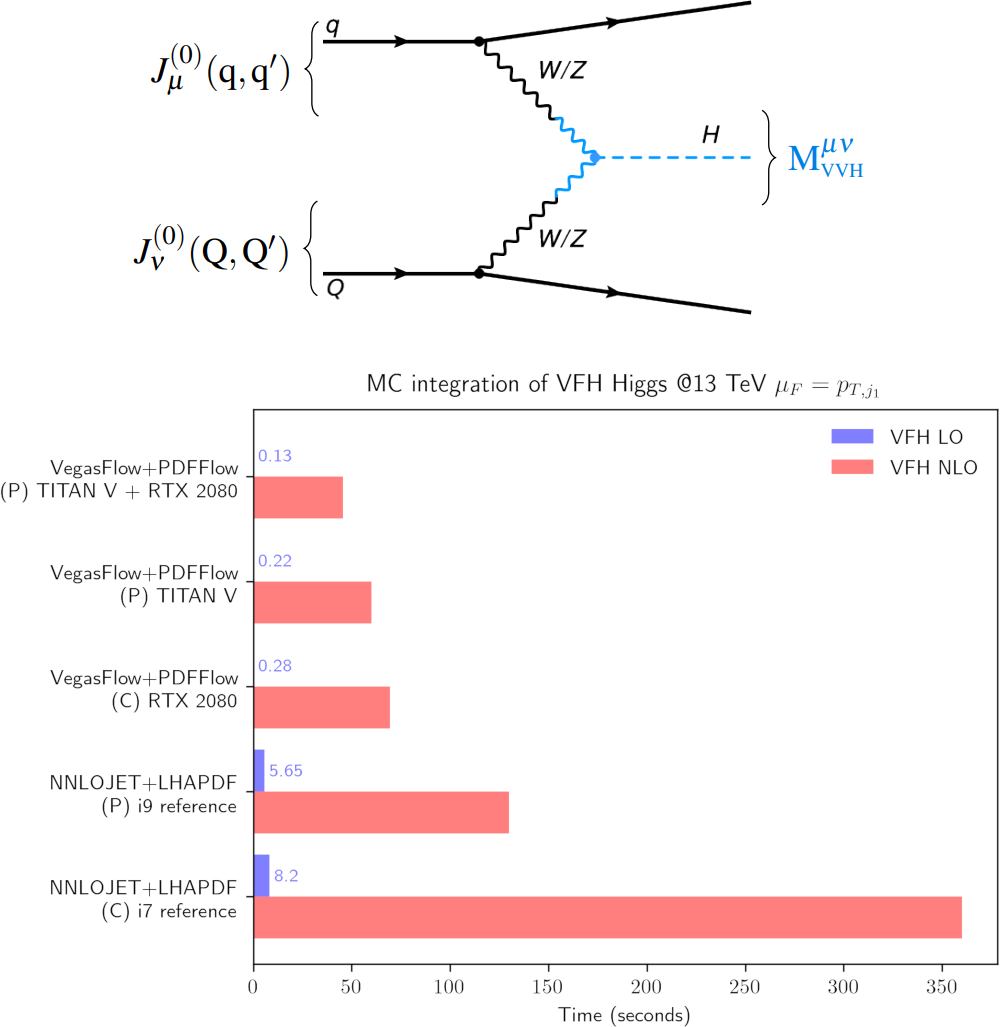}
        \caption{Quark initiated VBF Higgs production. Top row: Born Feynman diagram.
        Bottom row: comparison of the execution time between \pdfflow and \vegasflow
        code against NNLOJET plus LHAPDF.}
        \label{fig:vbf}
    \end{minipage}

    \caption{The examples figures}
\end{figure}
We present a calculation of the cross-section for single top quark $t$ production
process at Leading Order (LO) at LHC, namely $pp \to t + 1\mathrm{-jet}$. The only
contributing diagram is the $q \bar{q} \to t\bar{b}$ one depicted in figure~\ref{fig:singletop}.
The cross-section is returned by an integrator algorithm, computing the
integral over the phase space of the matrix element squared times the
luminosity factor, which contains the PDF evaluation. We assess the
performance of two alternative setups involving on one hand \vegasflow and \pdfflow
and on the other \mgamc~\cite{Alwall:2014} \texttt{v3.0.2} plus LHAPDF.
The two tools share the same physical parameters configuration, such as the top quark
mass $m_t = \SI{173.2}{\GeV}$ and the center of mass energy fixed at $\sqrt{s} = \SI{8}{\GeV}$,
matching the LHC end of Run 1 conditions. In both cases, the central member of the
NNPDF3.1 NLO set PDF is used. Figure~\ref{fig:singletop} collects the integration
time needed to reach an absolute precision of at least $2 \cdot 10^{-3} \SI{}{\pico\barn}$
and a relative error of $4\cdot 10^{-5}$ on the target cross section value.
We observe a great speed-up of the \vegasflow and \pdfflow runs on all the different
devices, especially for GPUs.

\subsubsection{Higgs production on VBF at NLO}

In the second example, we consider the Vector Boson Fusion (VBF) Higgs production
process at NLO. This higher-order computation provides an interesting benchmark
of our software, given the added complexity of the phase space integration.
Full implementation of a parton-level Monte Carlo simulator such as NNLOJET~\cite{Gehrmann:2018}
or MCFM~\cite{Campbell:2019} is beyond the scope of this work, which is to provide
a proof of concept for an NLO computation. We exploit, instead, a simplified
version~\cite{Cruz-Martinez:2018} of the NNLOJET process implemented in Fortran
95 programming language which uses LHAPDF as the PDF access tool.

We limit ourselves to the quark-initiated W-boson-mediated process: the Born
Feynman diagram is visible in the top panel of figure~\ref{fig:vbf}. At NLO, instead, a gluon is
radiated from any of the quark lines. This process is characterized by a non-trivial
phase space, containing divergences appearing when the intermediate quark
propagator goes on the mass shell. The singularities must cancel with the poles
in the corresponding virtual diagrams to have a finite result for the
fixed-order computation.
Phase space cuts as well as a suitable subtraction scheme have to be implemented
to regularize the divergent behavior of the real radiation integrand.
To be comparable with the original Fortran implementation
in~\cite{Cruz-Martinez:2018} we also choose the antenna subtraction method at
NLO as described in~\cite{GehrmannDeRidder:2005, Currie:2013}.

The NNLOJET code is heavily optimized for CPU and CPU-cluster usage so it provides a good benchmarking
ground for our python-TensorFlow implementation which is to be run on a GPU.
Although our strategies for phase space point sampling and subtraction methods are
not specifically dedicated to GPU computing and could represent suboptimal choices,
we achieve competitive performance against NNLOJET, leading us to believe that a fully
optimized implementation of NLO (and NNLO) computations on GPUs can lead to drastic
performance gains. The results are presented in the bottom plot of figure~\ref{fig:vbf}
for both consumer and professional-grade hardware. It is interesting to discover
that our implementation reaches a competitive time performance even on cheaper
devices, opening a notable discussion regarding economic and environmental (i.e.
power usage and carbon footprint) perspectives on the usage of our tool.

As a final remark, we note that our software stack is currently able to store and
run in parallel an MC integration involving an order of a million events. We believe
that this result can be further optimized for GPU computing, paving the way for
more complex integrands and virtual diagram structures at higher orders (NNLO) in
perturbation theory.

\subsubsection{Multi-PDF members evaluation}

The \pdfflow package includes the possibility to load multiple members of a PDF set
at once. This capability allows to access values in the $(x,Q^2)$ plane for all
the specified PDFs. This functionality targets the field of PDF determination:
theoretical predictions for experimental data points are computed through 
the convolution of \texttt{FastKernel} tensors~\cite{Bertone:2016} with PDFs
evaluated in a grid of $x$ points. As discussed in~\cite{Carrazza:2019}, we
expect performance improvements of \texttt{FastKernel}-like operations when running
parallel multi-PDF member evaluation on GPU. Such a result is particularly relevant
for fitting methodologies based on the NNPDF methodology, where PDF replicas could be
obtained simultaneously in a single GPU card.
\begin{table}
    \centering
    \begin{tabular}{c|c|c|c}
     $N_{\rm rep}$ &  {\tt LHAPDF} CPU & \pdfflow GPU & \pdfflow CPU\\
     \hline
     10 & 0.08s & 0.07s & 0.05s \\
     50 & 0.41s & 0.35s & 0.28s \\
     100 & 0.83s & 0.69s & 0.56s \\
     200 & 1.87s & 1.46s & 1.12s \\
     300 & 2.85s & 1.29s & 1.79s \\
     400 & 3.63s & 1.69s & 2.12s \\
     \hline
    \end{tabular}
    \caption{Time required to evaluate all $11$ flavours from $N_{\rm rep}$
    members of NNPDF3.1 NLO in a grid of $2415$ points in $x$, using the P1
    system.}
    \label{tab:fktable}
\end{table}

We test the multi-PDF option of \pdfflow against the usual LHAPDF implementation
which sequentially loops on single points. In table~\ref{tab:fktable} we show
the total evaluation time required to compute the $11$ flavors of NNPDF3.1 NLO,
namely the flavors for $\mathrm{d,u,s,c,b}$ quarks plus their anti-particles and
the gluon, for a different number of members $N_{\rm rep}$. In this benchmark, we
exploit the P1 system
to compute \texttt{FastKernel} tensors composed of a total of $2415$ points in $x$.
We highlight that \pdfflow on CPU and GPU times are always smaller when compared to
\texttt{LHAPDF} thanks to the parallel graph evaluation. On the other hand, GPU 
results are better in the large $N_{\rm rep}$ regime. We conclude that the
multi-PDF member evaluation implemented in \pdfflow may accelerate computations
where a large number of PDF members and $x$ points are required, thus opening the
possibility to perform a full PDF fit in a single GPU device.

\clearpage
\begin{figure*}
    \centering
    \myincludegraphics[width=0.475\textwidth]{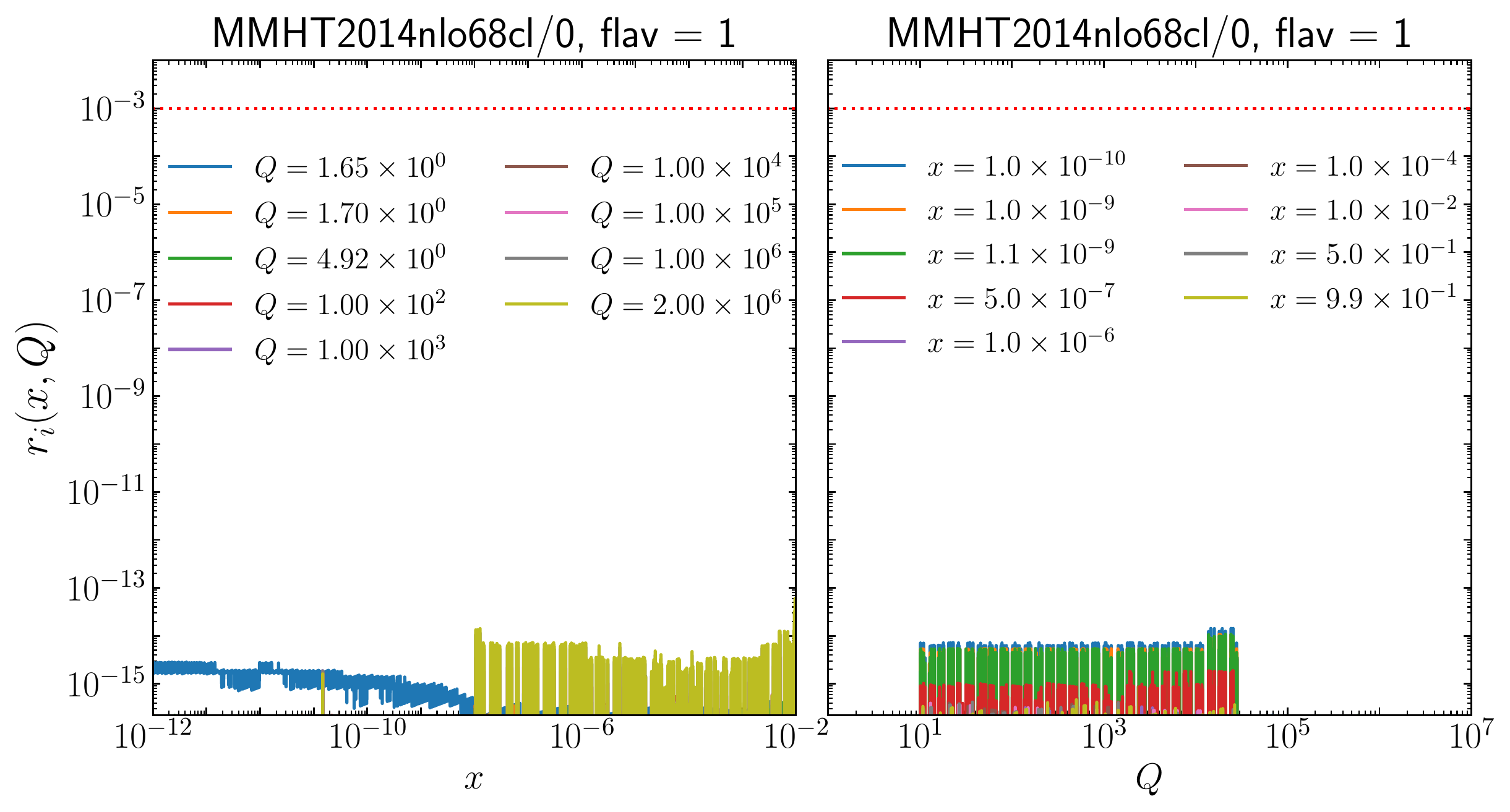}
    \myincludegraphics[width=0.475\textwidth]{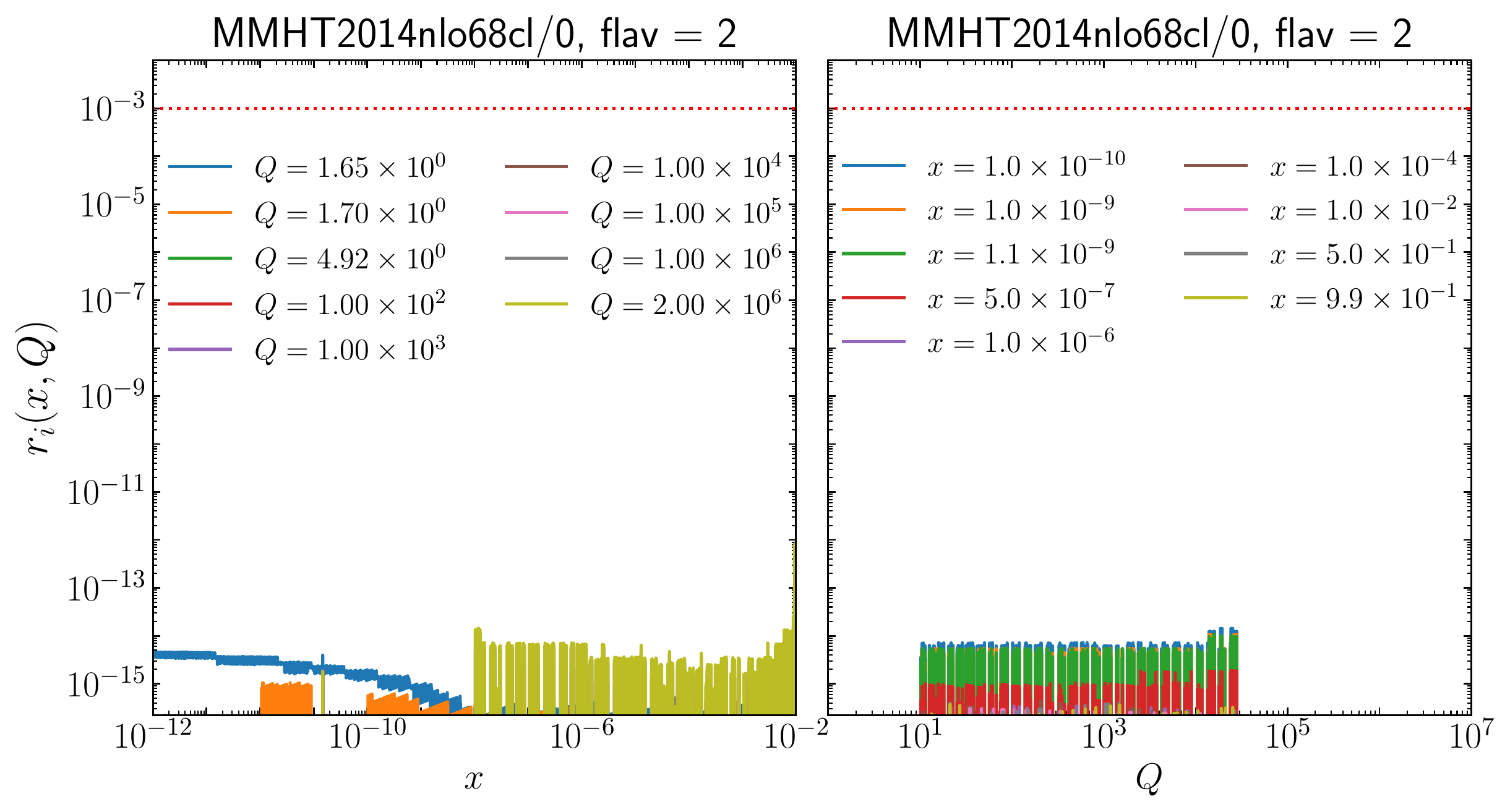}
    \myincludegraphics[width=0.475\textwidth]{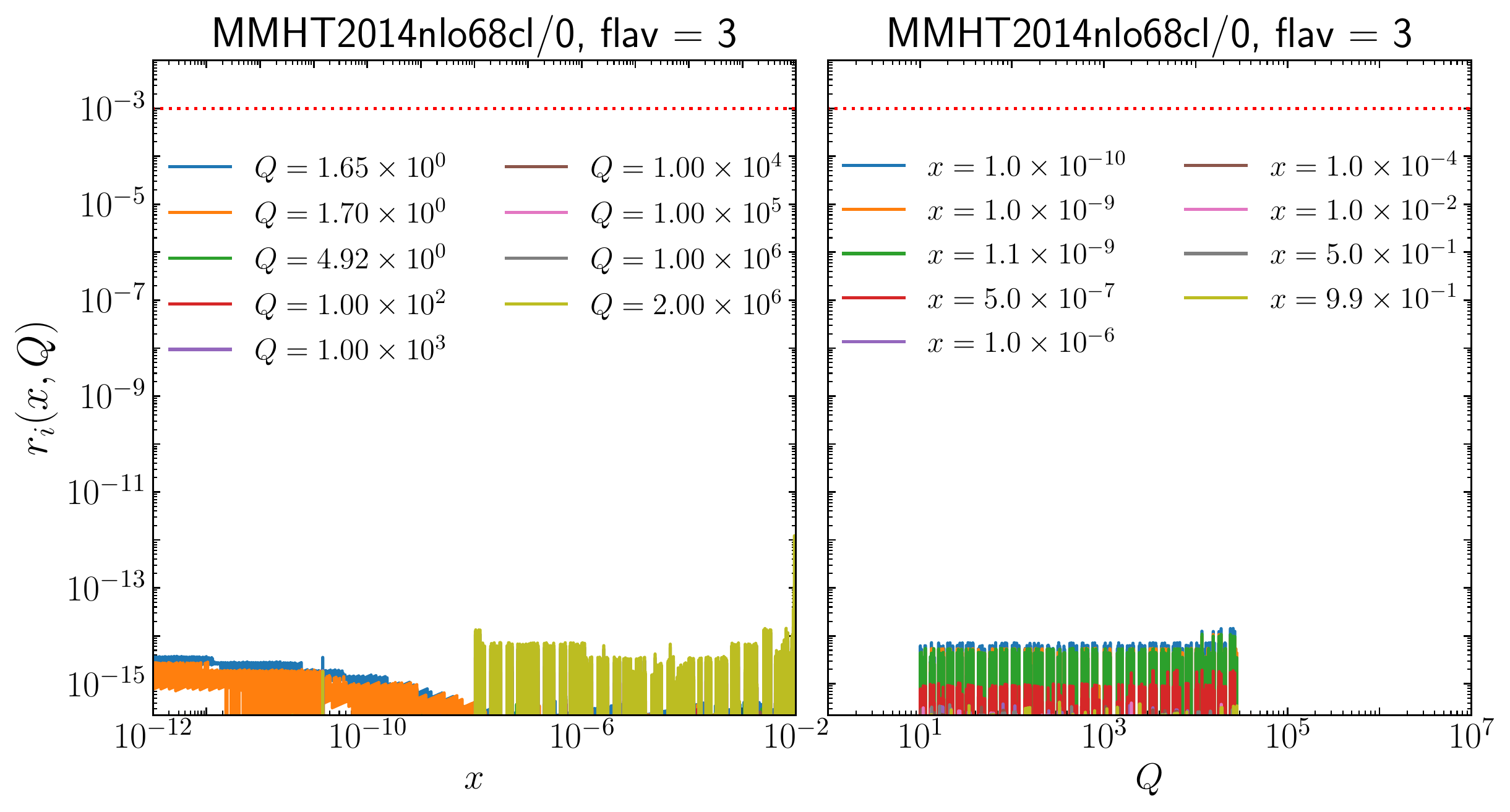}
    \myincludegraphics[width=0.475\textwidth]{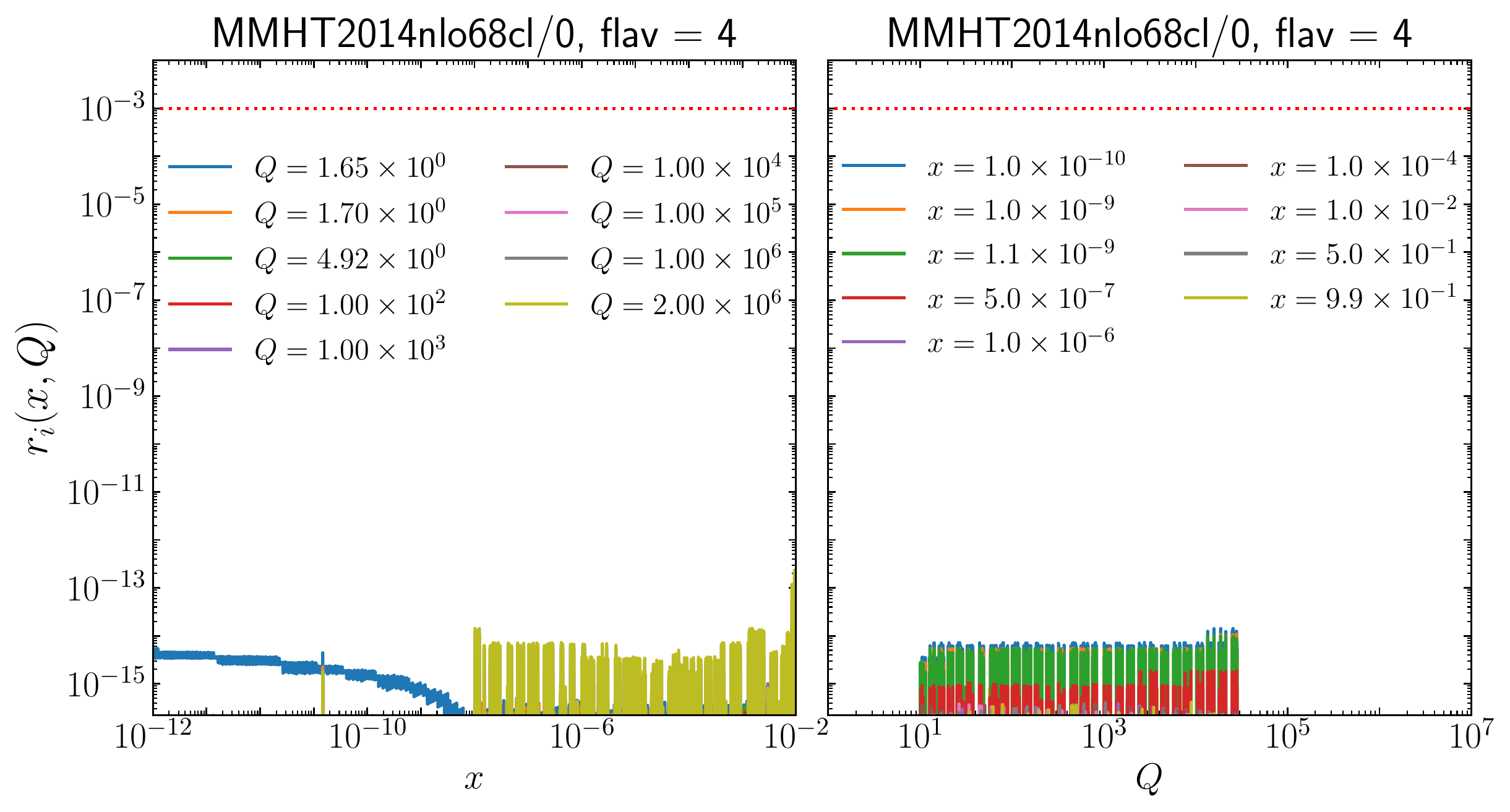}
    \myincludegraphics[width=0.475\textwidth]{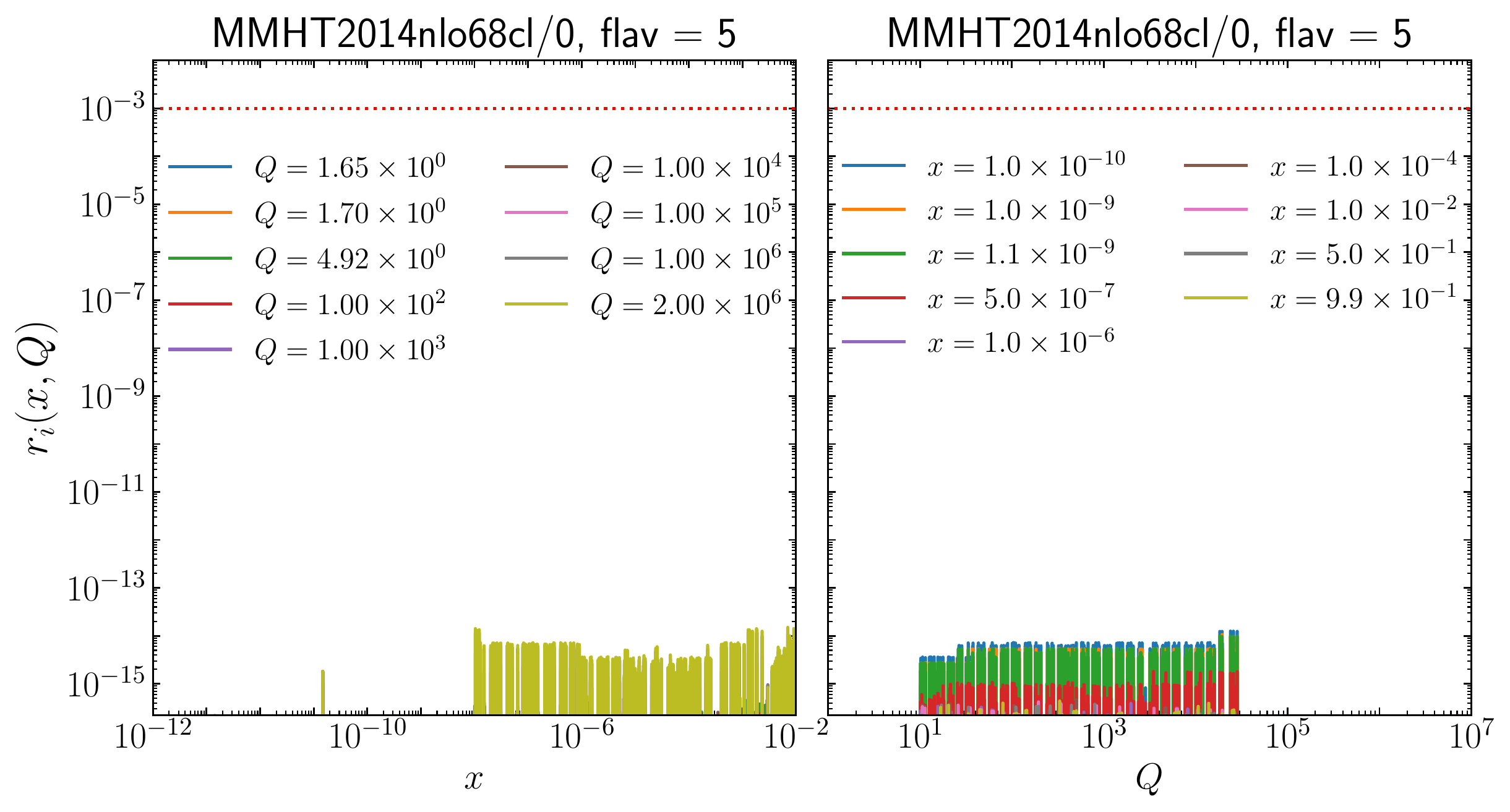}
    \myincludegraphics[width=0.475\textwidth]{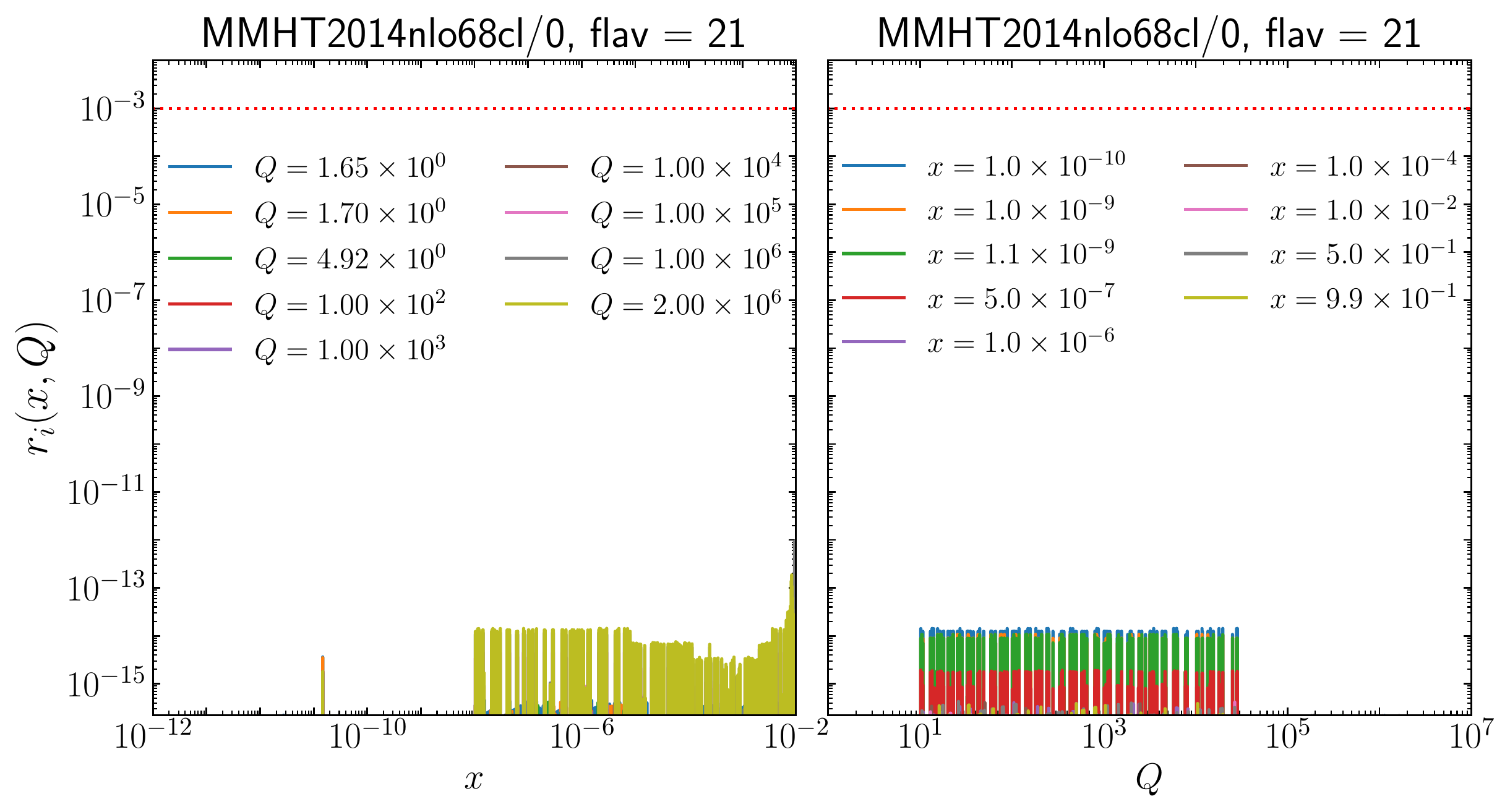}
    \myincludegraphics[width=0.475\textwidth]{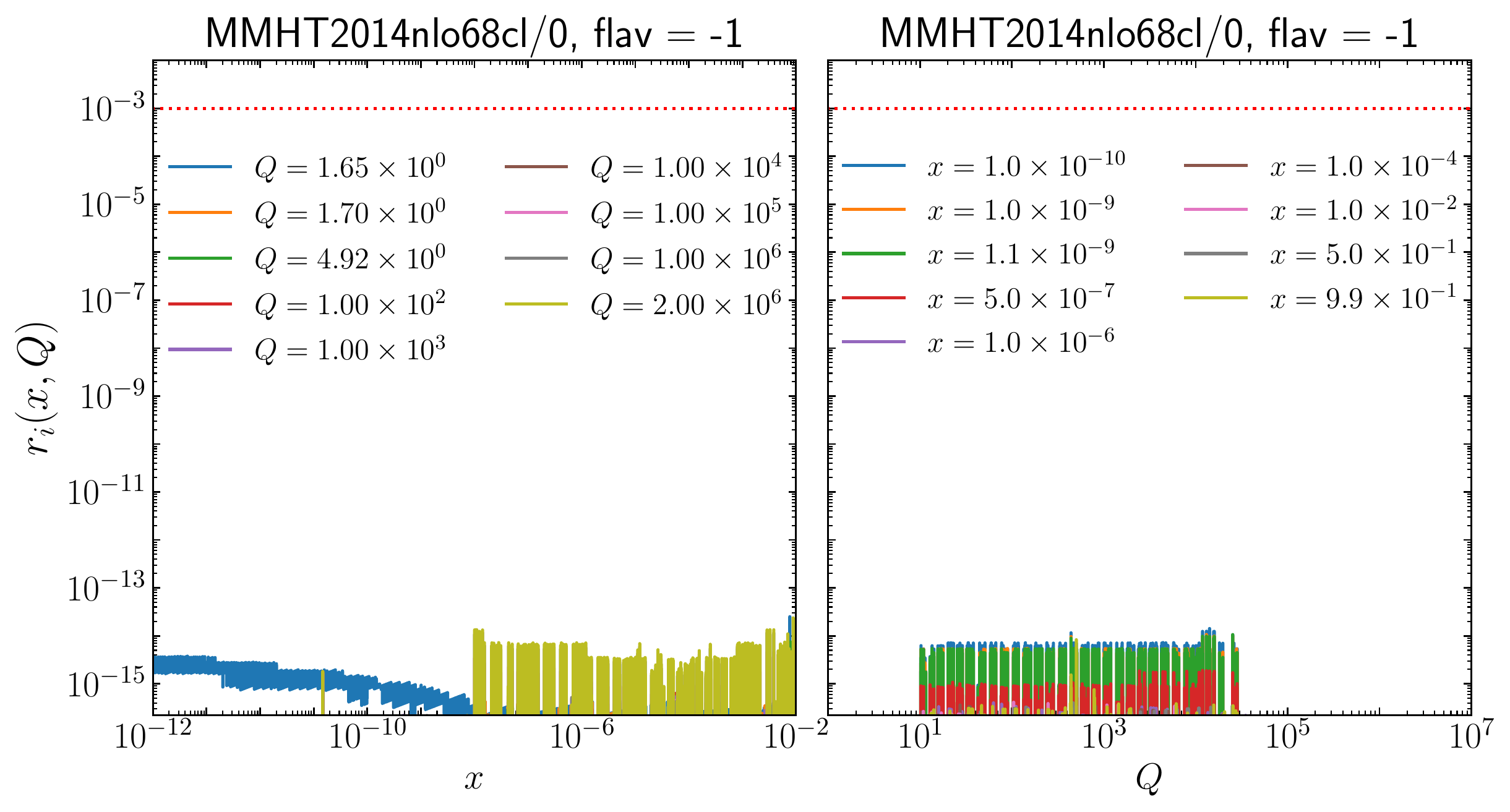}
    \myincludegraphics[width=0.475\textwidth]{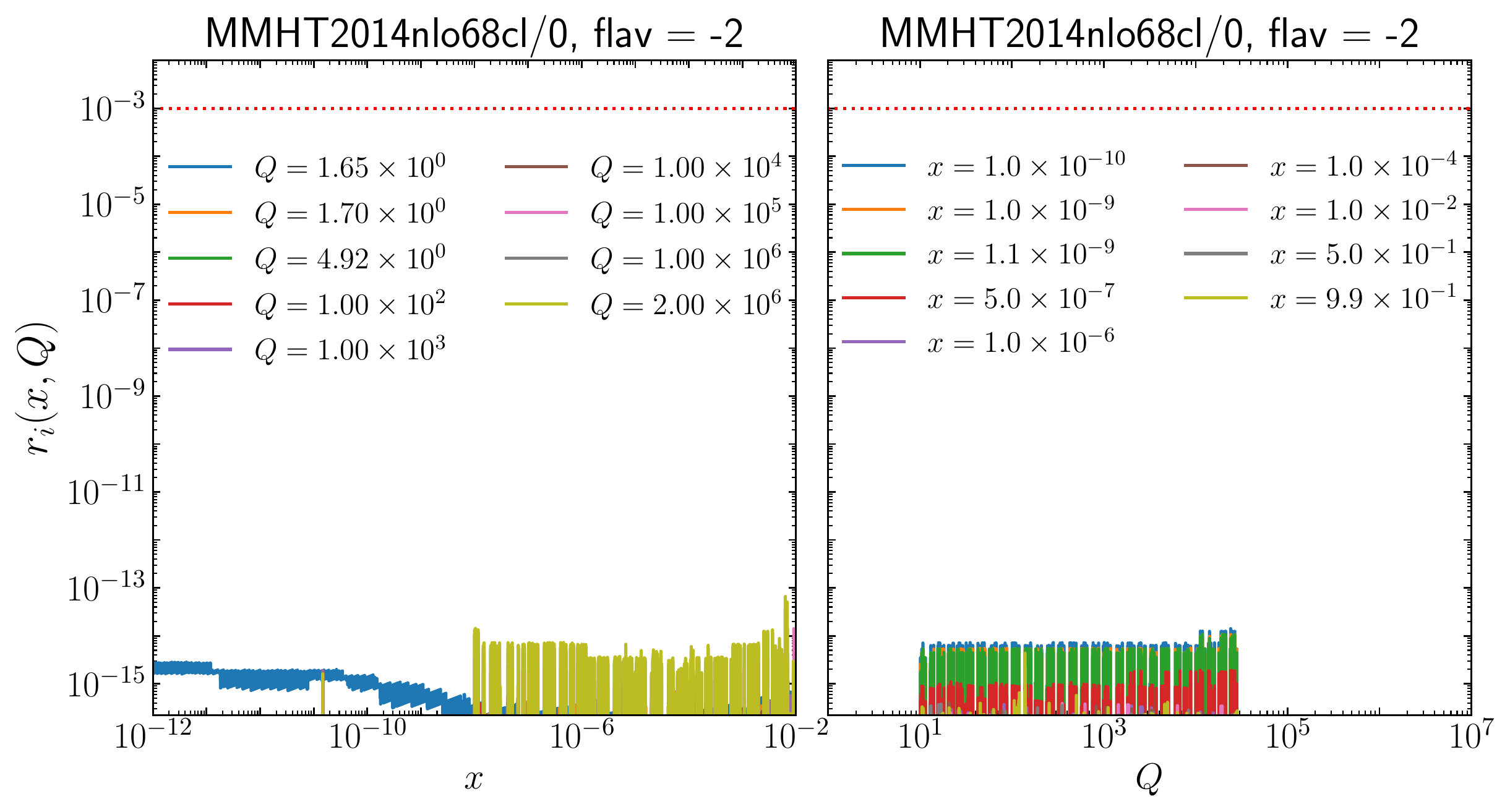}
    \myincludegraphics[width=0.475\textwidth]{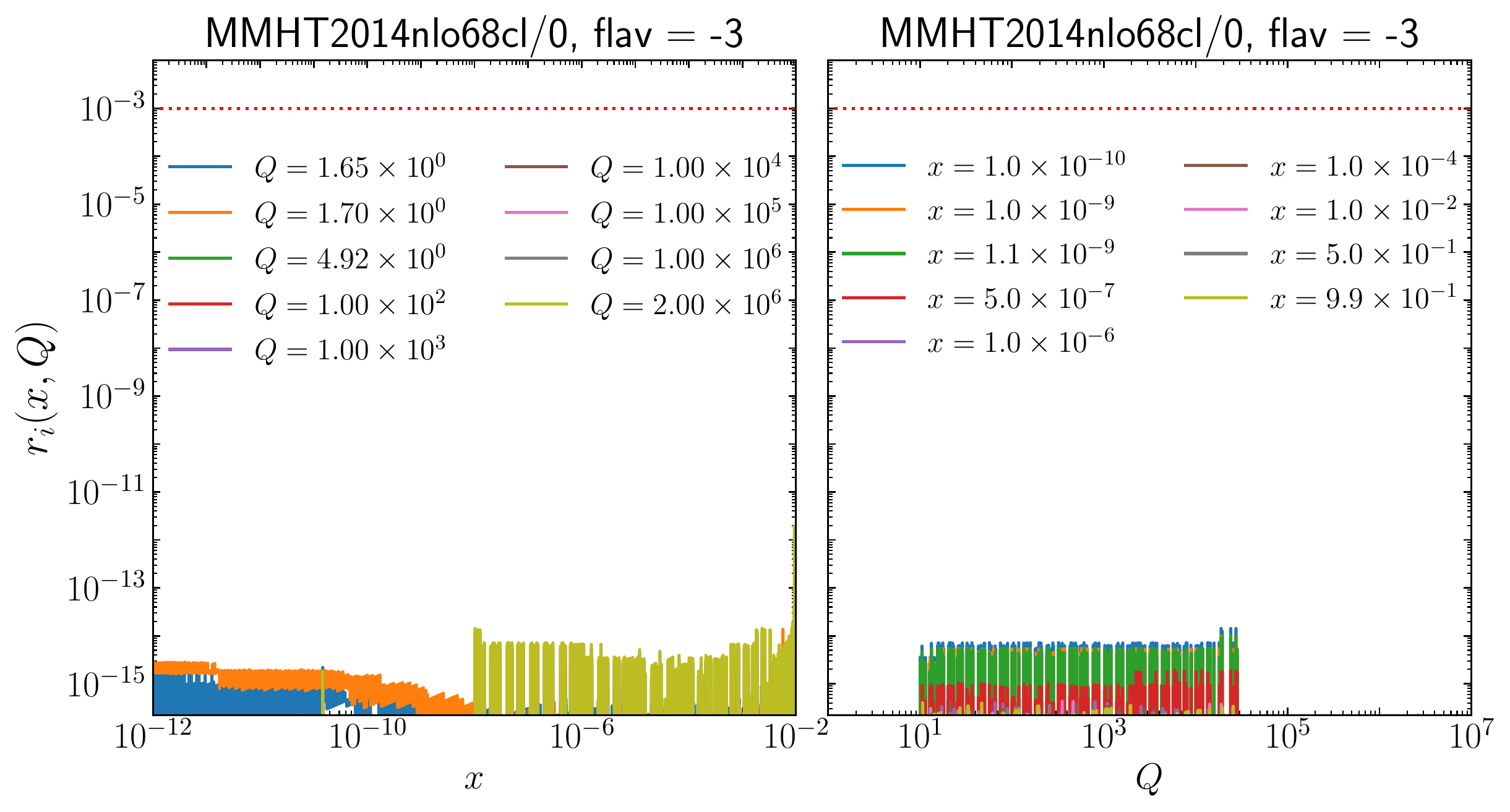}
    \myincludegraphics[width=0.475\textwidth]{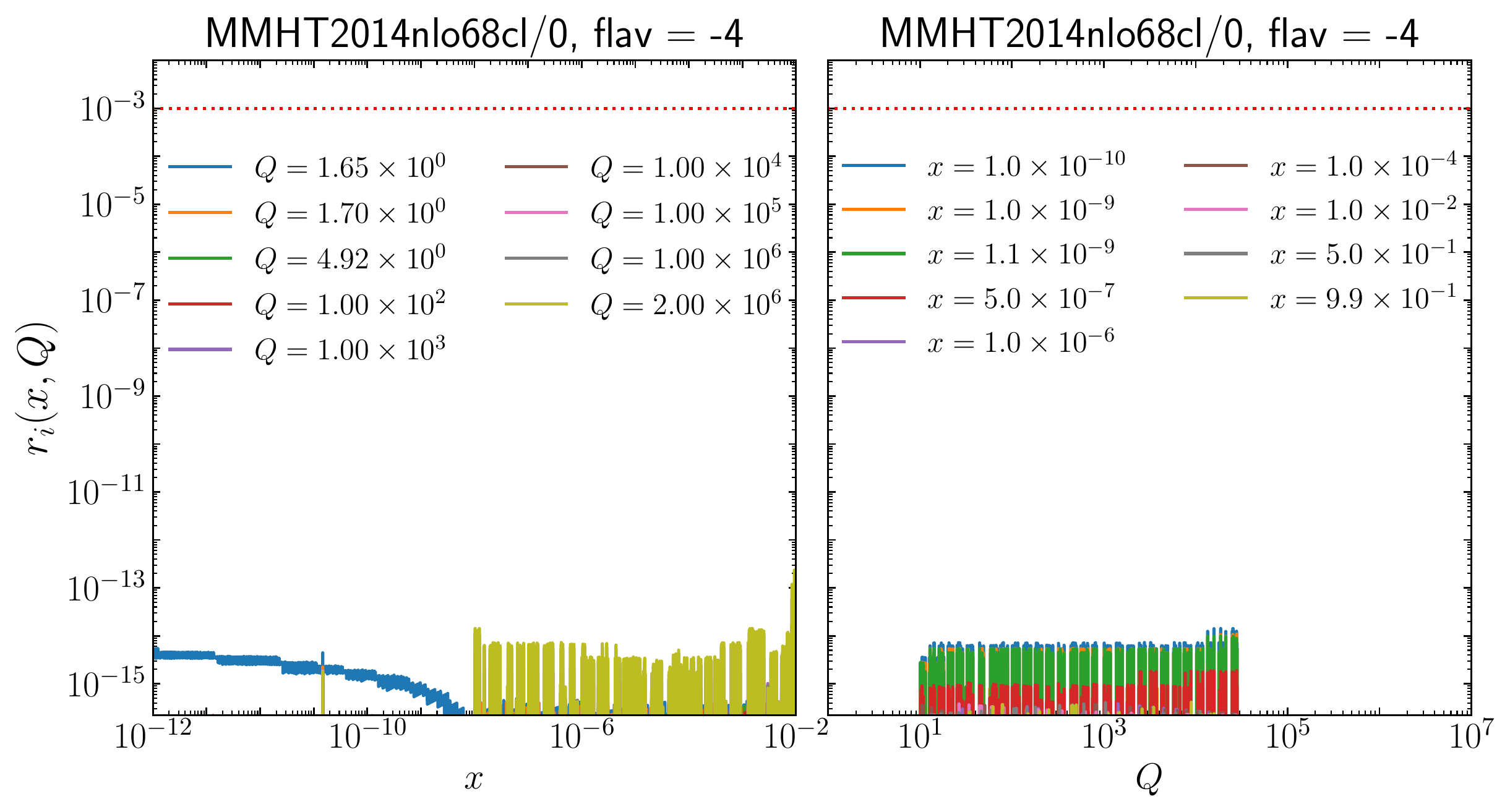}
    \caption{Relative difference between \pdfflow and LHAPDF
    (same as~\ref{fig:pdfaccuracy_nnpdf}) for the MMHT2014 NLO set for all
    flavors.}
    \label{fig:pdfaccuracy_mmht}
\end{figure*}

%% file: chapters/partII/chap4/chap4.tex
\chapter{\madflow: a Monte Carlo event generator on GPUs}
\label{chap:madflow}
\thispagestyle{plain}

\minitoc

\renewcommand{\myincludegraphics}[2][width=\textwidth]{
    \includegraphics[#1]{chapters/partII/chap4/plots/#2}
}

The \vegasflow and \pdfflow packages represent a first step towards a complete
general-purpose Monte Carlo framework for event simulation of particle physics
processes running across multiple devices and architectures, in particular, hardware
accelerators. In this section, we extend the ideas presented in chapter~\ref{chap:pdfflow}
to complete a full suite of tools able to automate the computation of leading order
cross sections and the generation of the associated unweighted events. We show how
to build a library with all the ingredients needed to implement a full Monte Carlo simulation
in a modern, extensible and maintainable way.

The most important result introduced in this chapter is a feasibility
study testing the ability of our software to successfully handle a high number
of Feynman diagrams, usually derived from processes with final state high jet multiplicities,
employing a full computation on GPU devices. We note that an
independent effort is currently being dedicated to port the \mgamc~\citep{Alwall:2014} software
on GPU, namely a project dubbed "Madgraph 4 GPU"~\citep{Valassi:2021,madgraph4gpu_package}.
The idea is to implement a library coded in CUDA programming language~\citep{cuda_toolkit}
to allow \mgamc to run on Graphics cards. The
\madflow~\citep{Carrazza:2021madflwpos,Carrazza:2021madflow} concept tackles the
issue from a different perspective, though: it aims to provide an MC simulator based on
modern software able to run automatically on different hardware setups without
requiring any specific architecture knowledge for either the developer and the user.
Furthermore, we note that CUDA-based libraries are compatible with the TensorFlow
framework: this, in principle, allows "Madgraph 4 GPU" matrix element calls to
be integrated into \madflow to gain the best performance out of the two worlds.
To date, up to the authors' knowledge, \madflow is the first attempt to implement
a full parton-level automated platform-agnostic MC event generator.

The chapter is organized as follows.
In section~\ref{sec:madflow-design} we initially describe the high-level design of the \madflow
package~\citep{madflow_package} and provide more details about its specific
modules in the subsequent paragraphs. Section~\ref{sec:madflow-examples} presents
all the physics examples implemented to test the capabilities of the new software.
We show radical improvements in all the benchmarks with respect to the
\mgamc state-of-the-art approach relying on CPU-only computations.

\section{The \madflow design}
\label{sec:madflow-design}

\begin{figure}
    \centering
    \myincludegraphics[width=\textwidth]{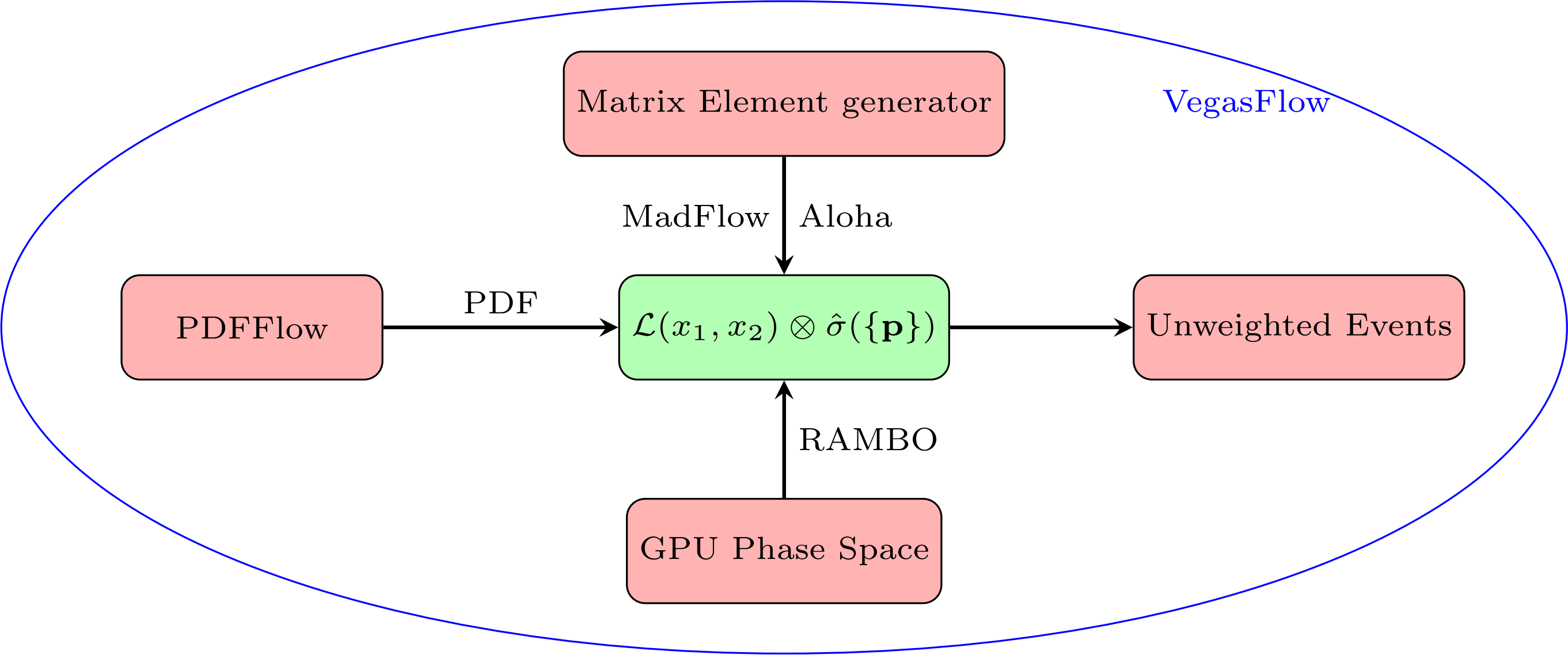}
    \caption{The \madflow tool-suite includes all the needed ingredients for MC
    event generation on hardware accelerators: on top of the \vegasflow MC integrator,
    the GPU phase space generator constructs physical points to be fed in the
    computation of the luminosity factor done with \pdfflow and the evaluation
    of the partonic cross section $\hat{\sigma}$ through the ME computation. The
    final results are presented as unweighted events.
    }
    \label{fig:madflow-suite}
\end{figure}
The generation of simulated events at hadron colliders, such as LHC, is driven by
the following master formula expressing the calculation of the cross-section $\sigma$
for a given process $p p \to X_n$ with $n$ partons in the final state:
\begin{equation}
    \label{eqn:mc_master_formula}
    \sigma (p p \to X_n) = \sum_{a,b} \int d\Phi_n (x_1, x_2, \mathbf{p})
    \, \mathcal{L}(x_1, x_2)  \abs{\mathcal{M} (a, b \to X)}^2
\end{equation}
where the sum runs over all the initial parton flavors and the differential phase
space depends on the initial parton momenta fractions $x_1$, $x_2$ plus the final
state $n$ partonic $4{-}$momenta $\mathbf{p} = p_1, \dots, p_n$. In this discussion
we neglect the parton shower and all the other complexities (such as
hadronization models, underlying event treatment etc.) related to a complete
description of a physical final state measurable experimentally.
Equation~\ref{eqn:mc_master_formula} instructs about the ingredients needed to
implement a full parton-level MC event generator:
\begin{enumerate}
    \item a phase space generator able to apply any fiducial cut and
    output the relevant $d\Phi_n$ contribution;
    \item a PDF and strong running coupling value access tool to compute the
    luminosity function $\mathcal{L}$;
    \item a general method to evaluate the Matrix Element (ME) modulus squared $\abs{\mathcal{M}}^2$;
    \item an MC integrator, defining a strategy to compute the final cross-section
    value;
    \item an algorithm to store the generated events in a suitable format for
    computing observables and histogramming differential distributions.
\end{enumerate}

To design a general-purpose MC generator that can exploit hardware
accelerators, all the items listed above must be carefully implemented to
take advantage of the parallel nature of the problem respecting the independence
of the transformations that each sampled point undergoes. In particular, \pdfflow
and \vegasflow already represent valid solutions to issues number $2$ and $4$,
respectively. The missing ingredients in the list require the implementation of
an efficient phase space generator able to load random points on GPU devices, a
vectorized implementation of the ME computation and, finally, a prescription to
save asynchronously such phase space points following physical distributions.
The fact that the storage
procedure acts asynchronously is crucial to avoid spoiling the performance improvements
gained from the parallel GPU computation through expensive I/O operations.
Figure~\ref{fig:madflow-suite} explicitly depicts all the modules that encompass the \madflow
framework.

\begin{figure}
    \centering
    \myincludegraphics[width=0.9\textwidth]{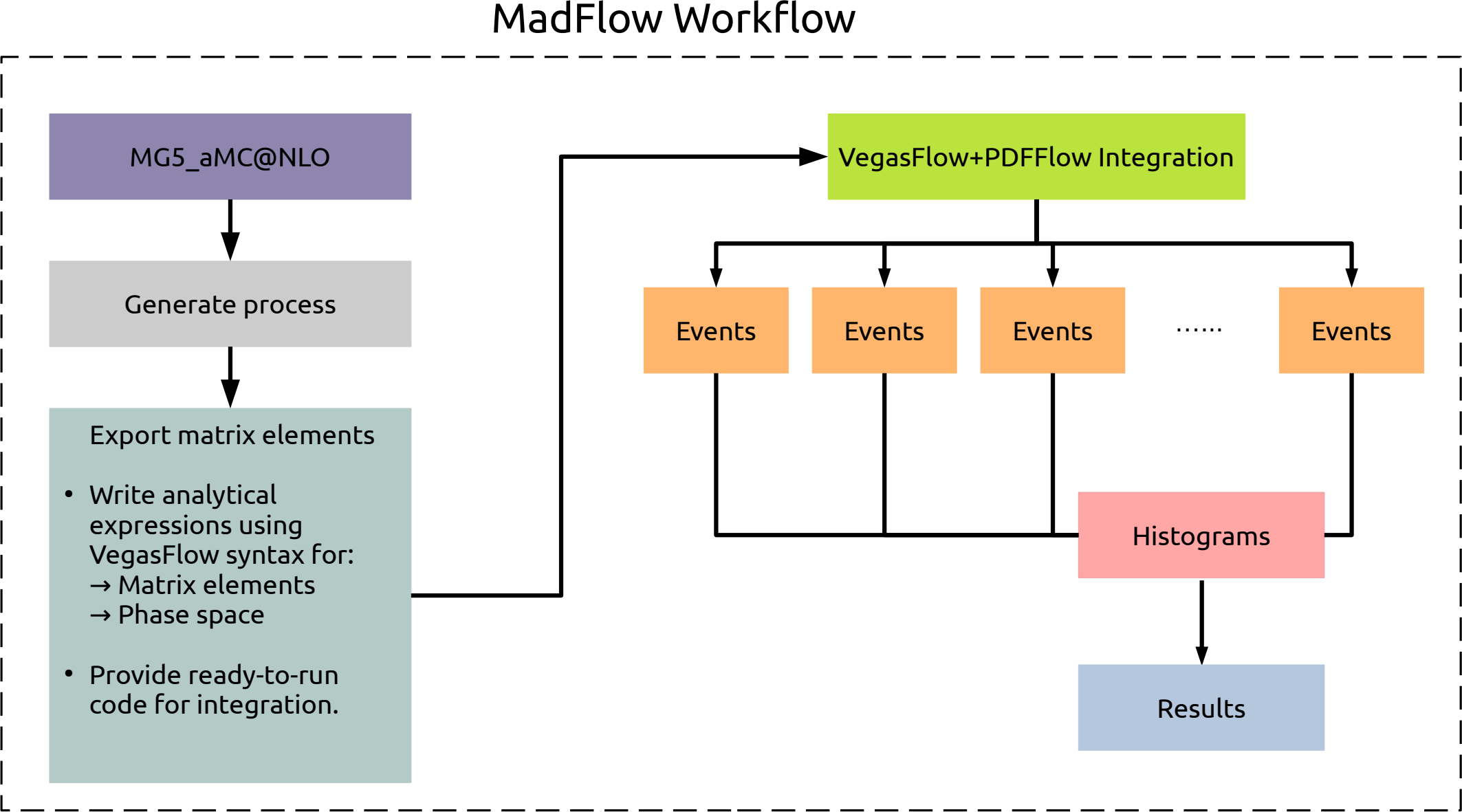}
    \caption{The \madflow design. The \mgamc plugin exports the code for the ME
    evaluation and connects the various \madflow components to perform the
    parallel MC integration. Eventually, unweighted events are produced along
    with the MC estimation for the cross-section.
    }
    \label{fig:madflow-design}
\end{figure}
Figure~\ref{fig:madflow-design} represents schematically how the \madflow package
works, which essentially involves two main steps: the process generation and the
actual MC integration. As an entry point, we rely on the \mgamc package with its
ability to write process-specific code for the generation of MC events: the
functions for the ME evaluation, see paragraph~\ref{subsec:me}, and the
integration routines are generated at runtime. Since the produced code must be
interfaced with \madflow, we implement an \mgamc plugin which ensures that the
syntax and the prescriptions exploited in the various tools connected to our
framework are respected. The plugin creates a script that runs the parallel MC
integration at LO within \madflow and outputs the unweighted events needed for
further analysis.

The goal of \madflow is to provide the foundation for future high-precision MC
simulations so they can efficiently take advantage of hardware development. In
its current version, \madflow provides the necessary tools for the automated
computation of LO calculations for any number of particles. Higher-order
computations can be implemented by building upon the provided LO template.
Parameters and model definitions are provided by {\sc \small MG5\_aMC}-compatible
parameter and run cards to allow a comfortable transition between the two frameworks
for both the user and the developer point of view.

\subsection{Evaluation of matrix elements routines}
\label{subsec:me}

\newcommand{\UFO}{{\sc\small UFO}~}
\newcommand{\feynrules}{{\sc \small FeynRules}~}
\newcommand{\MadGraph}{{\sc\small MadGraph5}~}
\newcommand{\MadGraphf}{{\sc\small MadGraph5}~}
\newcommand\madevent{{\sc\small MadEvent}~}

The \mgamc package is a Python meta-code, namely a code that produces other code.
The output script can be phrased into three different languages to easily accommodate
user needs: Python, C++ and Fortran. The idea is that the user first specifies the
model lagrangian and inputs the process of interest, which leads to an algorithm
to compute the corresponding amplitude.
Within the \mgamc interface, the \texttt{import model}, \texttt{generate} and \texttt{output}
commands produce the desired code to evaluate a query process.

As an example, the $p p \to t \bar{t}$ scattering is computed within the Standard Model
and written to the \texttt{my\_directory} folder with the following instructions:
\begin{minted}[autogobble]{bash}
import model sm
generate p p > t t~
output my_directory    
\end{minted}
The import statement loads the relevant \feynrules~\citep{Christensen:2008} package complying
with the Universal \feynrules Output ({\sc \small UFO}) format~\citep{Degrande:2011} describing
the query theory: a Python module listing the particles, the parameters and a set
of interactions fully specifying the underlying physics. The \feynrules were originally
thought as a package for {\sc\small Mathematica} calculations, however the introduction
of the \UFO format allowed the user to incorporate
them into \MadGraph to easily test new theories producing events and comparing
the results against experimental data in a quick and modular way.

The \texttt{generate} syntax triggers the search for the Feynman diagrams that
contribute to the query process. The program first lists all the possible topologies
compatible with the requested interaction. The invalid graphs should then be groomed:
an efficient recursive method to determine whether a graph is relevant or should
be discarded has been included in the framework since the release of the
\MadGraphf version~\citep{Alwall:2011}.

After all the diagrams are generated, the algorithm to compute the S-matrix element
is built with dedicated optimization techniques to enhance the performance of ME
evaluation. These methods rely on the computation of helicity amplitudes and color
decomposition to ensure that the complexity of the algorithm grows linearly
with the number of Feynman diagrams identified before, which, in turn, depend
factorially on the number of external particles in the process.

The techniques usually used for analytical ME results are based on Lorentz indices contraction 
and Dirac matrices trace technology. However, this approach is quadratically dependent
on the number of Feynman diagrams since all the cross-diagram interference terms
must be calculated individually. The overall number of such operations
forbids computing with this method the matrix elements involving high parton multiplicities,
typically beyond $2 \to 4$ processes. As a consequence, different algorithms are considered:
the HELicity Amplitude Subroutines (HELAS)~\citep{Murayama:1992} tool sums amplitudes
contributions for given helicities of the external particles, while
the QCD color-flow decomposition~\citep{Maltoni:2003} gather different terms into
gauge invariant groups named dual amplitudes. The ALOHA application~\cite{deAquino:2011}
automates this approach and is indeed incorporated into \mgamc.

The \texttt{output} request, finally, provides the user with the scripts produced
by ALOHA combining three different kinds of functions: external particles,
off-shell (connecting inner particle propagators with external legs through vertices)
and on-shell (joining together various pieces of the diagrams) routines.

To produce a ME evaluation program compatible with a vectorized MC event generator,
we design an \mgamc plugin, consisting of an exporter module that outputs the
components just described in a format fulfilling the requirements imposed by Tensorflow
primitives. Once the plugin is integrated into \mgamc, it is accessed with the following commands:
\begin{minted}[autogobble]{bash}
import model sm
generate p p > t t~
output pyout my_directory    
\end{minted}
which produces the desired \madflow code, instead of the usual \mgamc one, inside the
\texttt{my\_directory} folder.

Concerning the plugin implementation side, particular attention has been dedicated
to casting the ALOHA routines into a device-agnostic TensorFlow code. The
major complexity in the realization of the algorithm regards replacing the
wavefunctions routines abiding by the TensorFlow ControlFlow rules. The difficulty
comes from the fact that it is not easy to include conditional statements in
vectorized code. Moreover, GPU devices suffer in general from branching, since
the hardware is kept busy until both sides of a conditional are completed. In
the worst-case scenario, the majority of the GPU threads remain idle waiting for the
completion of an expensive operation acting on just a few points. The ALOHA vertices
rules, instead, being mostly comprised of algebraic operations are straightforward
to implement in this framework.

\begin{figure}
    \subfigure[$s{-}$channel $gg\to t \bar{t}$]{
        \centering
        \myincludegraphics[width=0.27\textwidth]{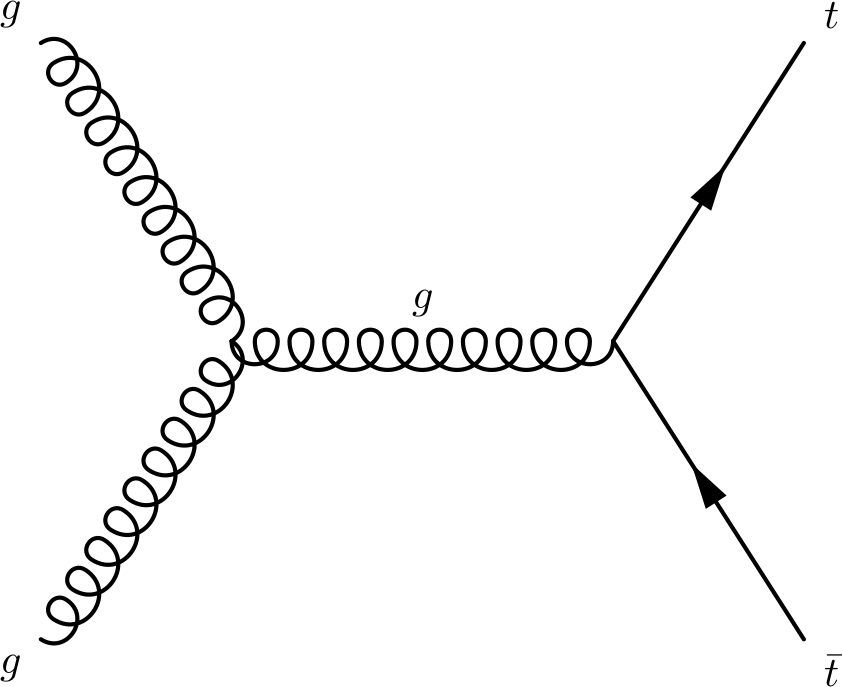}
        \label{subfig:gg_ttx1}
    }
    \hfill
    \subfigure[$t{-}$channel $gg\to t \bar{t}$]{
        \centering
        \myincludegraphics[width=0.27\textwidth]{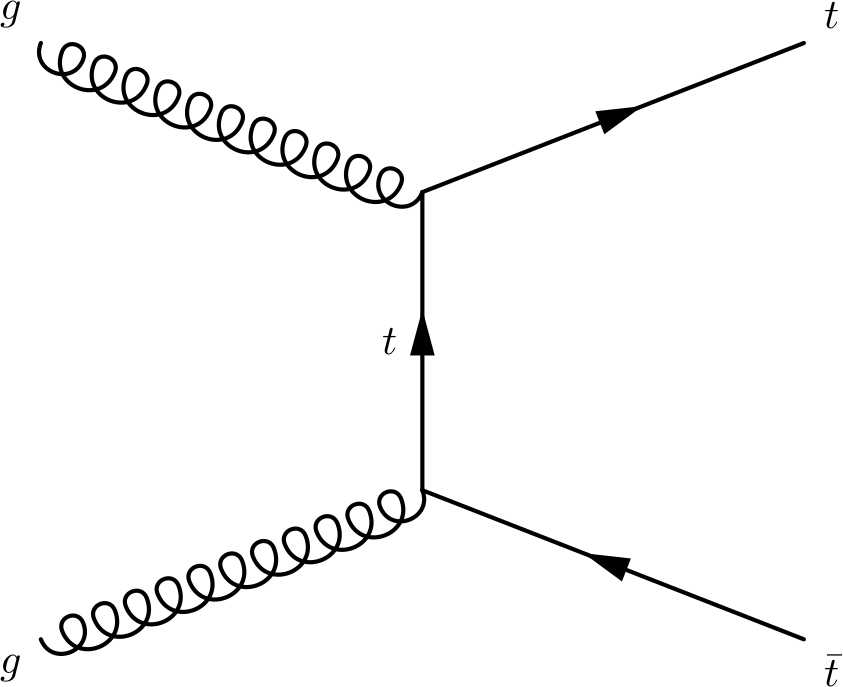}
        \label{subfig:gg_ttx2}
    }
    \hfill
    \subfigure[$u\bar{u}\to t \bar{t}$]{
        \centering
        \myincludegraphics[width=0.27\textwidth]{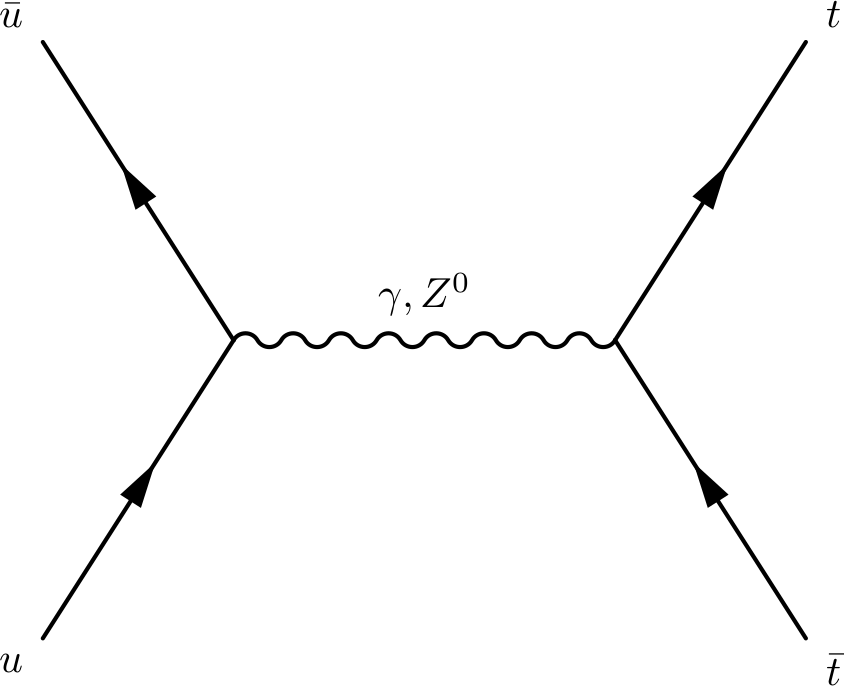}
        \label{subfig:uux_ttx}
    }
    \caption{The diagrams contributing to the $pp\to t\bar{t}$ process.}
    \label{fig:pp_ttx}
\end{figure}
We give here an example considering the $pp\to t\bar{t}$ process. There are two sub-processes
contributing to the final amplitude: $u \bar{u} \to t \bar{t}$ and $g g \to t \bar{t}$.
In turn, the amplitude with the gluon has both the $s$ and $t$ channels open.
Figure~\ref{fig:pp_ttx} shows the three leading order possible diagrams. The \madflow plugin
exporter, therefore, generates two Python files according to the two available
sub-processes: \texttt{matrix\_1\_uux\_ttx.py} and \texttt{matrix\_1\_gg\_ttx.py}.
Further, each matrix element file is accompanied by the relative \texttt{aloha\_<***>.py},
a script containing the ALOHA wavefunctions and vertices routines written with
TensorFlow primitives. Each matrix element file contains a \texttt{Matrix} class
implementing a \texttt{smatrix} method linking together the Feynman rules
to evaluate the corresponding ME with the techniques described above: namely, looping
over helicity amplitudes and aggregating them into gauge invariant groups according
to color decomposition. Finally, the exporter creates a \texttt{leading\_order.py}
file which collects the instructions to launch the \madflow computation, joining all
the pieces present in figure~\ref{fig:madflow-suite}. This includes calls to \vegasflow,
\pdfflow and the GPU phase space sampler that we describe in the following paragraph.

\subsection{Phase space generation}

The \madflow phase space generator implements a vectorized version of the RAMBO
algorithm (see section~\ref{ssubsec:rambo} for a description of this technique).
This method can produce a vector of physical $4{-}$momenta to be exploited
as input of the \texttt{smatrix} method of the ME. The approach produces a flat
phase space, democratically giving the same importance to all regions. As a result,
the ME topology is not taken into account, avoiding introducing excessive complexity
in the implementations at the expense of the MC integration efficiency.
Future releases of the \madflow package will focus on improving the sampling
technique to enhance the behavior of the MC generator especially when a large number
of final-state particles is requested.

\subsection{Unweighted event exporter}

The \madflow software focuses on producing samples of unweighted events following
physical distributions. This is, by definition, a key ingredient for an MC generator.
The standard output format for such events in HEP is the Les Houches Event (LHE) $3.0$
one~\citep{Andersen:2014}. Such a format is based on XML markup language to represent
generated events as a list of \texttt{<event>} tags containing the produced particles
along with their PDG code, $4{-}$momenta, mass and color flow information. The
workflow first generates weighted events, which means that each one contributes 
to the overall physical distributions differently according to its weight.
An unweighting procedure allows the production of a sample where each event has
the same importance. This implies that histogramming physical distributions can be done without
assigning a weight to each entry, hence the name unweighted events.

The collection of weighted events is done asynchronously to the matrix element
computation in \madflow. The reason is that communication between the hardware
accelerating device and the CPU host may slow down the hard computation. Therefore,
we implement a custom Python context manager named \texttt{LheWriter}, whose
\texttt{lhe\_parser} method constantly takes care of gathering the phase space
points sampled during the \vegasflow integration in a separate CPU thread to
eventually dump the corresponding LHE tags to file. We ensure that this operation
is done independently of the GPU calculations wrapping the call to the
\texttt{LheWriter.lhe\_parser()} method thanks to the TensorFlow \texttt{tf.py\_function()}
primitive.

The unweighting prescription is taken from \mgamc and works as an \linebreak
acceptance-rejection algorithm (see paragraph~\ref{ssubsec:accept-reject}):
after looking for the highest weight $w_{\mathrm{max}}$ in the list, the
algorithm produces a random number in the unit interval with uniform probability,
then a query event is accepted if its weight $w_{\mathrm{event}}$ is higher than
the product of the random number times the maximum weight in the sample. This
ensures that each event is accepted with a probability proportional to the ratio
of the weights $w_{\mathrm{event}} / w_{\mathrm{max}}$.

A consequence of this method is that if there are many events with low weight
compared to the maximum one, the algorithm acceptance efficiency
$\epsilon = N_{\mathrm{accepted}} / N_{\mathrm{tot}}$ will be low. This highlights
a sub-optimal phase space sampling behavior, yielding many ME evaluations at points
that give low contributions to the integral. We observe that the naive RAMBO approach
implemented in \madflow has, in fact, a rather poor efficiency $\epsilon \sim 5\%$,
meaning that future package releases will aim to improve this quantity by introducing
dedicated sampling techniques. Examples of improved sampling methods include weight
optimization of multi-channel Monte Carlo, previously described in
section~\ref{par:multi-channel-mc}, and single-diagram enhancement method already
implemented by \madevent~\cite{Maltoni:2002}.

\section{Physics examples}
\label{sec:madflow-examples}

In this section we present some usage examples of the \madflow package, demonstrating
the precision of the software in terms of the accuracy of the final observable
distributions. The \mgamc software represents the reference benchmark software in
all the experiments.

We consider the simulation of events at LO for hadronic processes at colliders with
center of mass energy of $\sqrt{s} = \SI{13}{\TeV}$. This configuration matches the
Run II LHC setup, which was running from $2015$ to $2018$. In the following different
processes are considered. All results presented in this study are obtained with the
software stack described in table~\ref{tab:software-madflow}.
\begingroup
\def\arraystretch{1.5}
\begin{table}
    \centering
    \begin{scriptsize}
    \begin{tabular}{l|c}
        Software & Version \\
        \hline
        \madflow & $0.1$ \\
        \vegasflow & $1.2.1$ \\
        \pdfflow & $1.2.1$ \\
        {\sc MG5\_aMC@NLO} & $3.1.0$ \\
        TensorFlow & $2.5.0$ \\
        Nvidia CUDA drivers & $11.3$ \\
        ROCm drivers & $4.2.0$ \\
        TensorFlow-rocm & $2.4.1$ \\
        \hline
    \end{tabular}
    \end{scriptsize}
    \caption{Description of the software used for the different \madflow experiments.
    The last two lines regard the software compiled for Radeon/AMD architectures.}
    \label{tab:software-madflow}
\end{table}
\endgroup

\subsection{\madflow accuracy}
\begin{figure}
    \subfigure[Top transverse momentum $p_T$ distribution.]{
        \centering
        \myincludegraphics[width=0.47\textwidth]{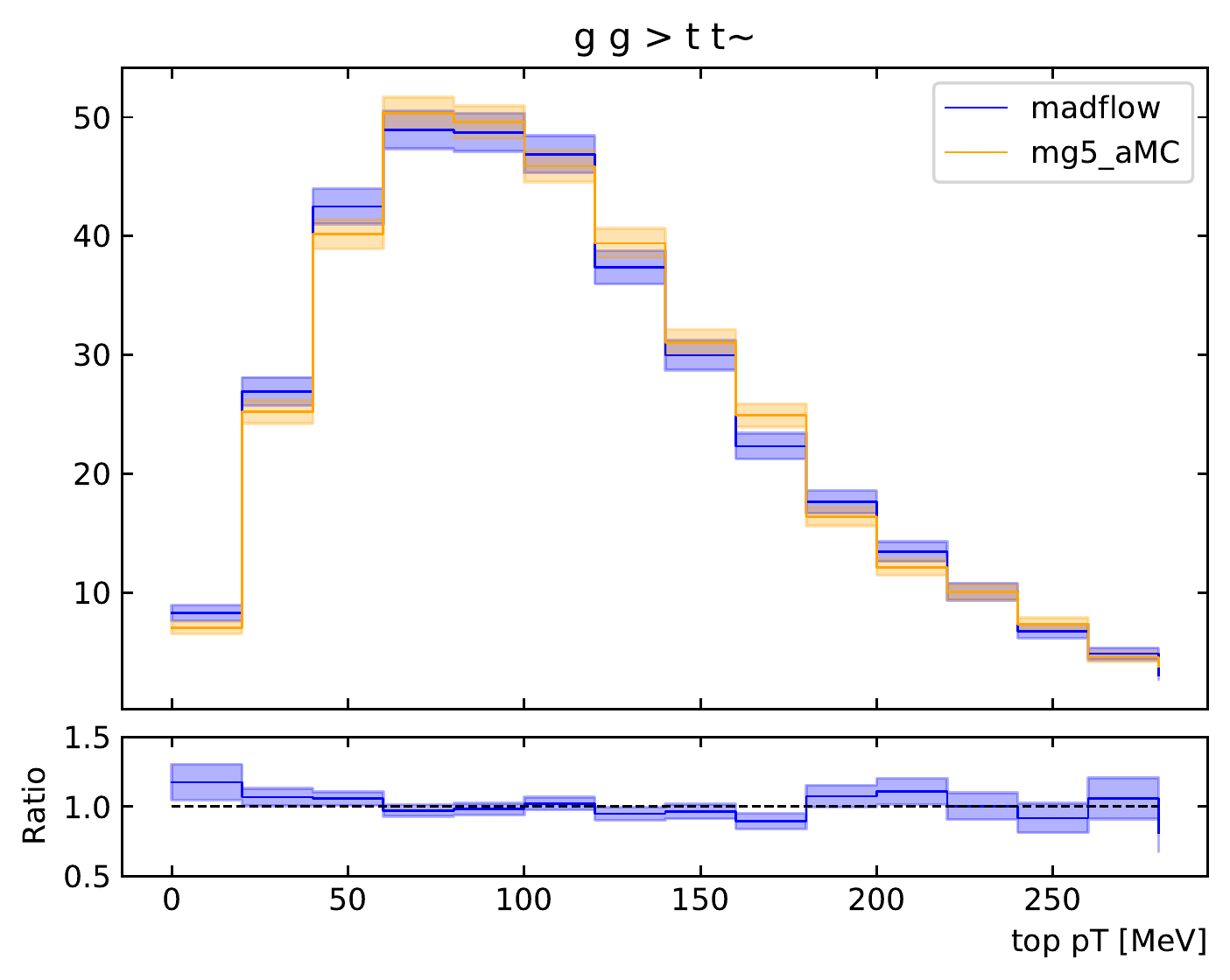}
        \label{subfig:gg_ttx_pt}
    }
    \hfill
    \subfigure[Top transverse pseudorapidity $\eta$ distribution.]{
        \centering
        \myincludegraphics[width=0.47\textwidth]{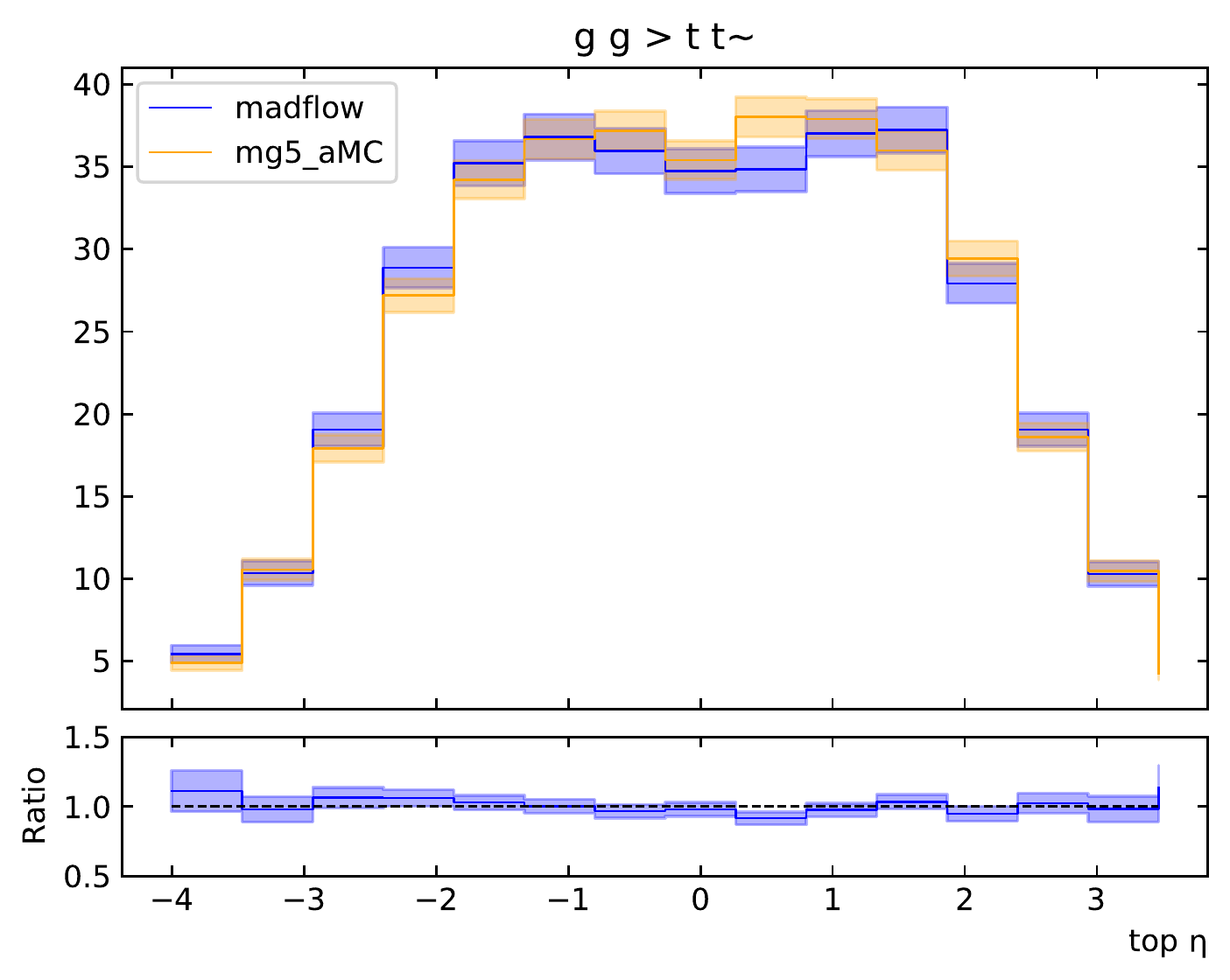}
        \label{subfig:gg_ttx_rap}
    }
    \caption{Leading order differential cross sections for $gg\to t \bar{t}$ process
    at $\sqrt{s} = \SI{13}{TeV}$. In both panels, the top row shows the histogram
    distributions output by \madflow and \mgamc, while the bottom one represents
    the ratio of the two to highlight the statistical agreement.}
    \label{fig:gg_ttx}
\end{figure}
We consider observable distributions for the $g g \to t \bar{t}$ process obtained
with both the original \mgamc implementation and the novel \madflow approach.
Figure~\ref{fig:gg_ttx} represents the differential distributions concerning
the final state top transverse momentum $p_{T,top}$ and pseudorapidity $\eta_{top}$.
The unity of measure of the top row plots is set to $\SI{}{\femto\barn}/\SI{}{\GeV}$
and $\SI{}{\femto\barn}$, respectively. The bottom row panels, instead, show the
ratio between the \madflow and \mgamc results. The pictures confirm a sufficient
level of agreement between the two implementations for the same level of target
accuracy between $2-5\%$ for each bin.

\subsection{\madflow performance}

\begin{figure}
    \myincludegraphics[width=0.47\textwidth]{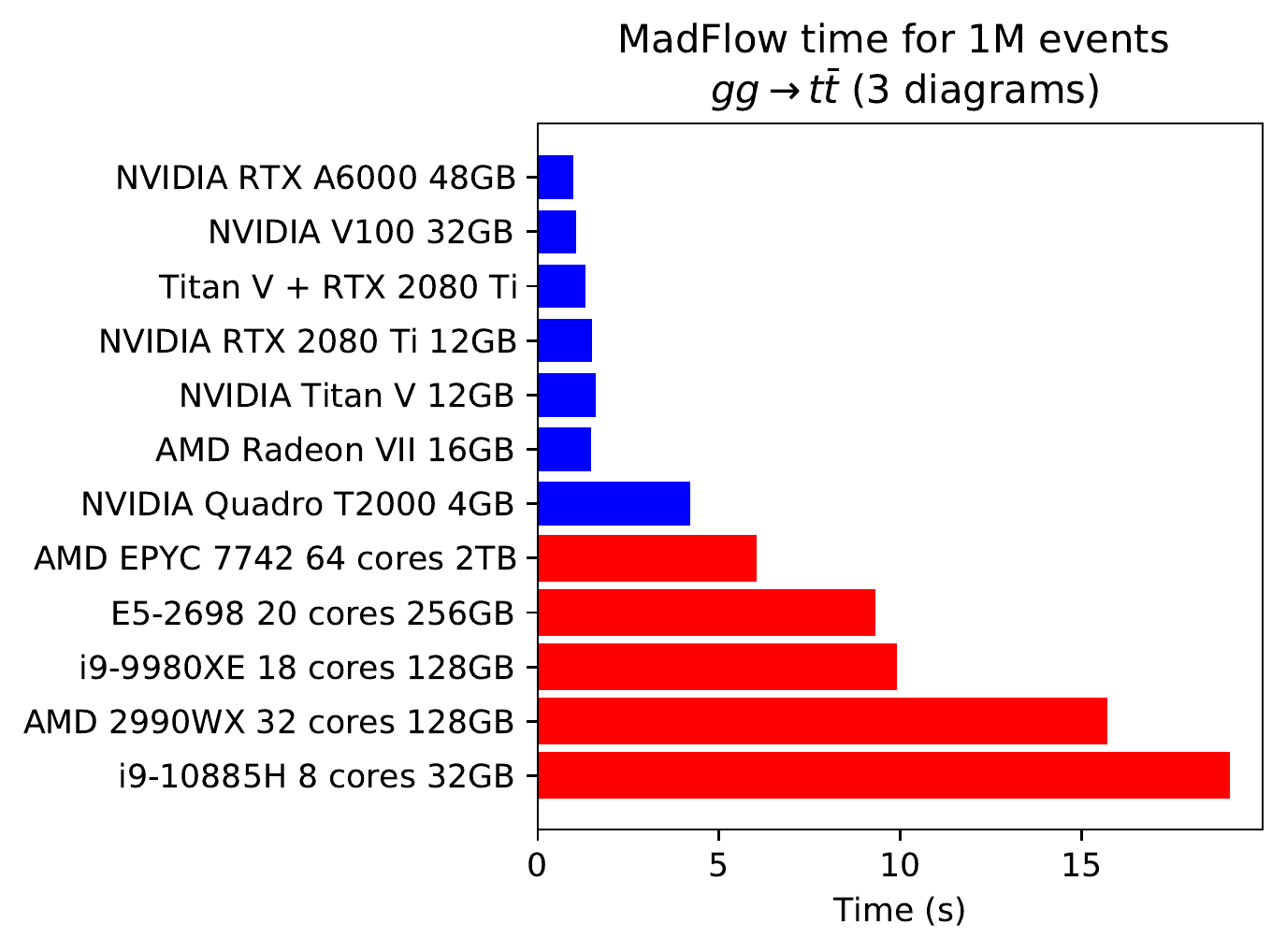}
    \hfill
    \myincludegraphics[width=0.47\textwidth]{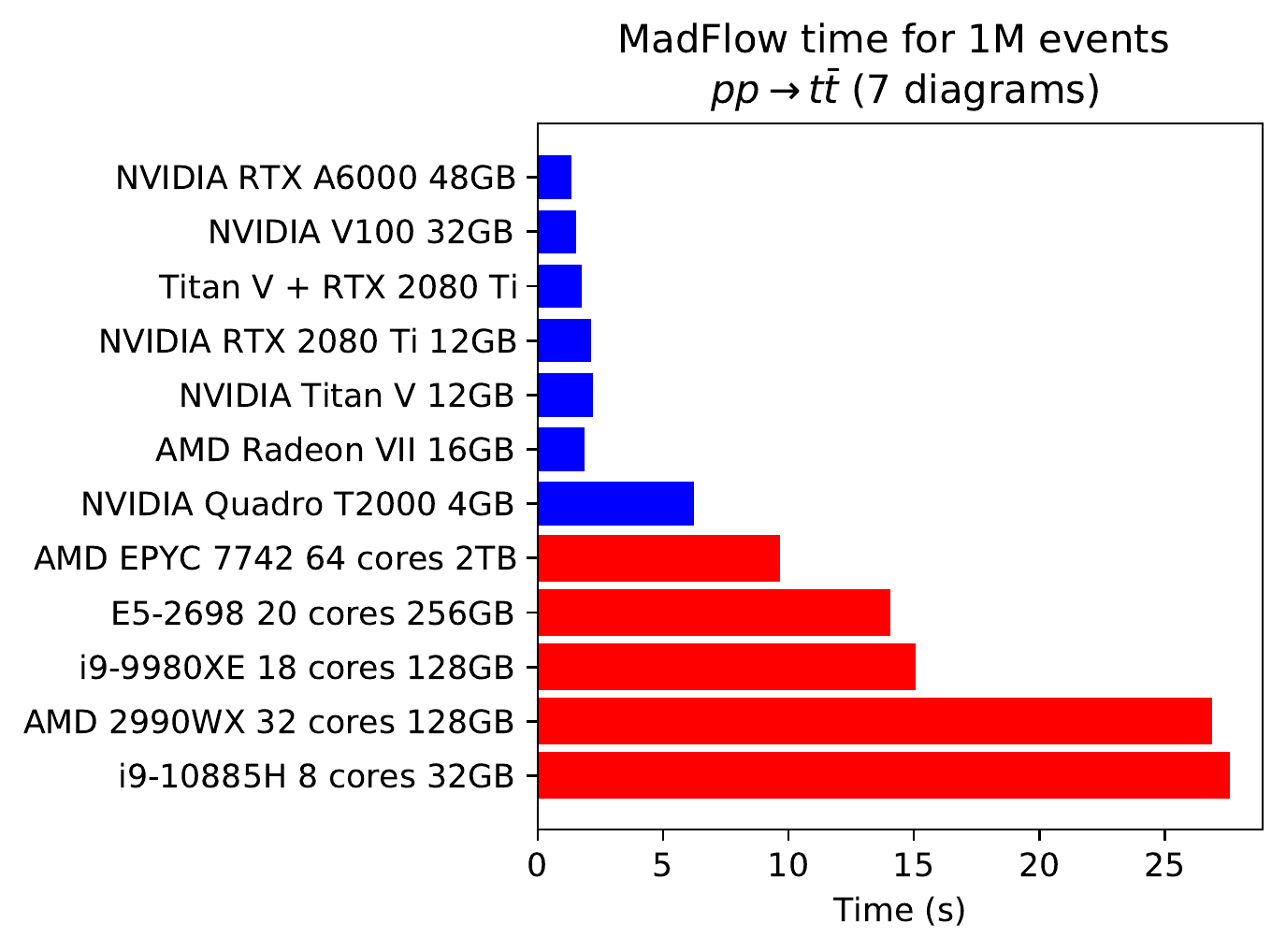}\\
    \myincludegraphics[width=0.47\textwidth]{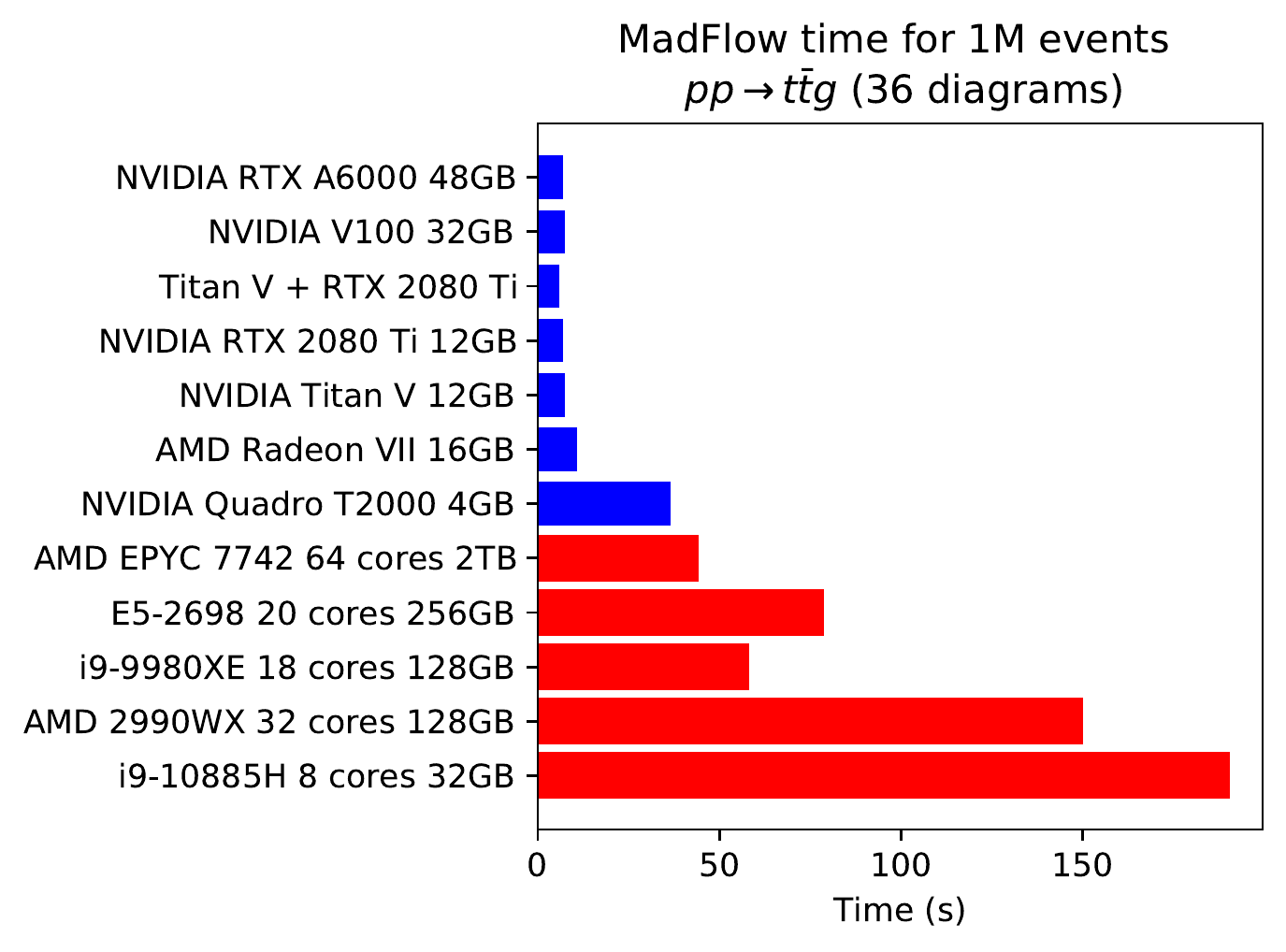}
    \hfill
    \myincludegraphics[width=0.47\textwidth]{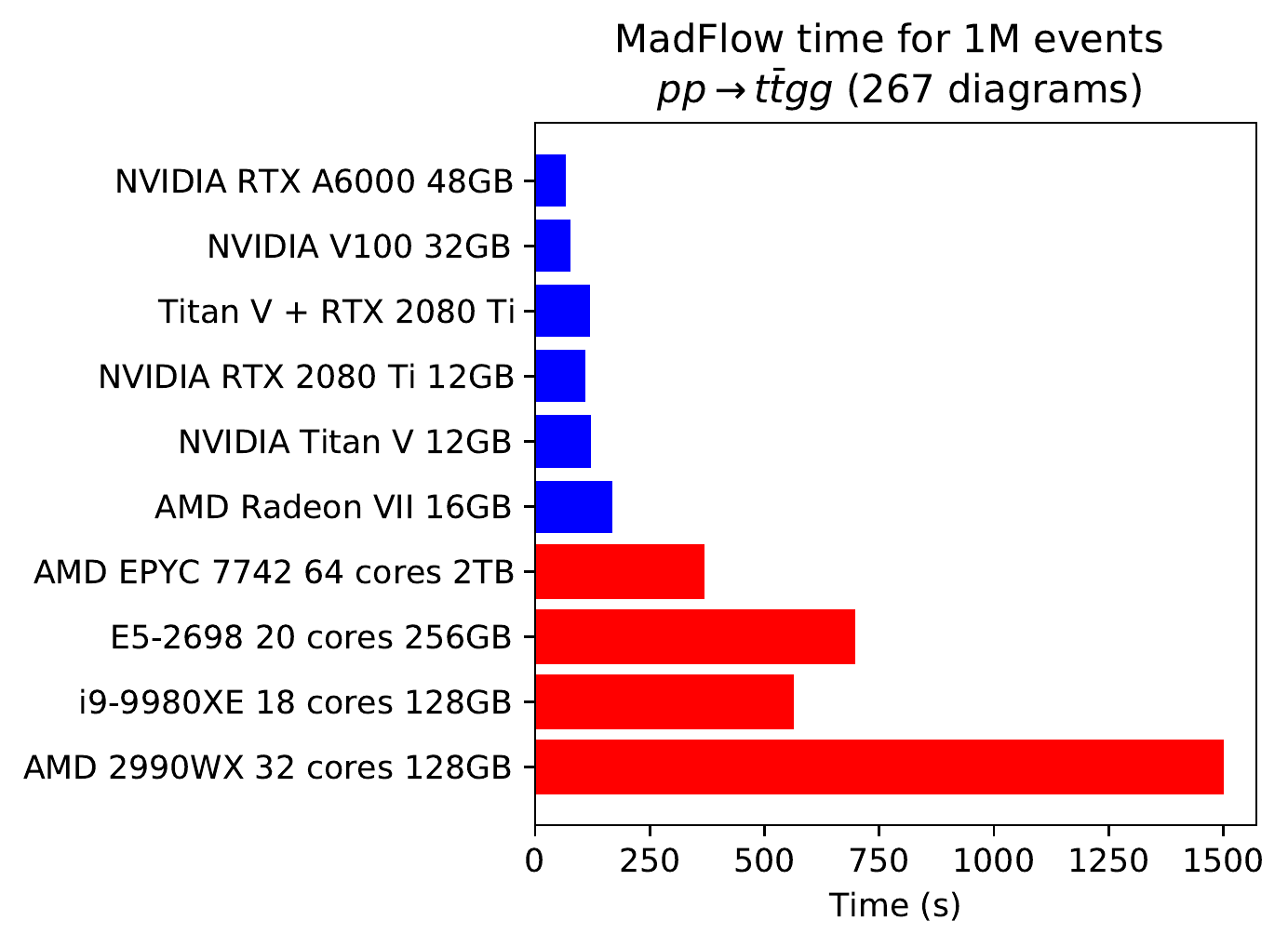}
    \caption{\madflow integration timings for different processes: $gg \to t\bar{t}$
    (top left), $pp \to t\bar{t}$ (top right), $pp \to t\bar{t}g$ (bottom left),
    $pp \to t\bar{t}gg$ (bottom right). Comparison of the results for consumer
    and professional-grade CPUs (red bars) and GPUs (blue bars) hardware. The plots
    report the available memory alongside each related device name. We note a
    systematic performance improvements for GPU cards.}
    \label{fig:madflow-performance1}
\end{figure}
We are interested in testing the speed of \madflow on different platforms.
Therefore we measure the total amount of time required to integrate $10^6$ events
for different processes with an increasing number of involved partons and, consequently,
factorially increasing the number of diagrams: we consider $gg \to t\bar{t}$ ($3$ diagrams),
$pp \to t\bar{t}$ ($7$ diagrams), $pp \to t\bar{t}g$ ($36$ diagrams),
$pp \to t\bar{t}gg$ ($267$ diagrams). In all the experiments we apply a cut on
the transverse momentum of the final state particles: $p_T > \SI{30}{\GeV}$.

In figure~\ref{fig:madflow-performance1} we summarize the results of the performed
benchmarks on various Intel and AMD CPU platforms (red bars) optionally hosting
Nvidia or AMD GPUs (blue bars) spanning from consumer to professional-grade hardware.
The blue bars underline the superiority of \madflow running on GPU cards. The Nvidia
Ampere architecture-equipped devices, such as the RTX A6000, represent the latest
generation of GPU technology and outperform the previous versions based on the
Tesla setup. The AMD brand with its AMD Radeon VII model, instead, is a good
alternative to the Nvidia solutions providing performance competitive with respect
to the most professional-grade GPUs at more affordable prices.

The red bars show the \madflow performance when running on CPU devices configured
to exploit all the available cores. Although we highlight the supremacy of GPU hardware,
we see that some top-level chips, such as the AMD Epyc 7742, can achieve
performance similar to consumer-level GPUs, like the Quadro T2000. Nonetheless,
we note that the code contains by no means any specific GPU optimization, which
would be a desirable feature when developing the software toward production mode.
As a final remark, we claim that the GPU speed-up highly depends on the number of
events requested per device, meaning that the maximum performance is achieved
when the computation exactly fills the entire available memory.

According to section~\ref{sec:madflow-design}, the goal of the \madflow
software is to provide hardware acceleration for future higher-order computations.
Of course, when including NLO or even NNLO contributions, the number of Feynman
diagrams substantially increases. It is then naturally interesting to test if
the program can successfully handle complex computations. To this extent, we
evaluate the performance of our code to simulate $10^5$ events for the
$pp\to t\bar{t}ggg$ process, which counts $2604$ Feynman diagrams. The results
shown in figure~\ref{fig:madflow-performance2} confirm that GPU performance are
competitive even accounting for such a huge number of contributions, remarking 
the strong potential of hardware accelerators for higher-order HEP simulations.
\begingroup
\def\arraystretch{1.5}
\begin{table}
    \begin{minipage}{0.4\textwidth}
        \myincludegraphics[width=\textwidth]{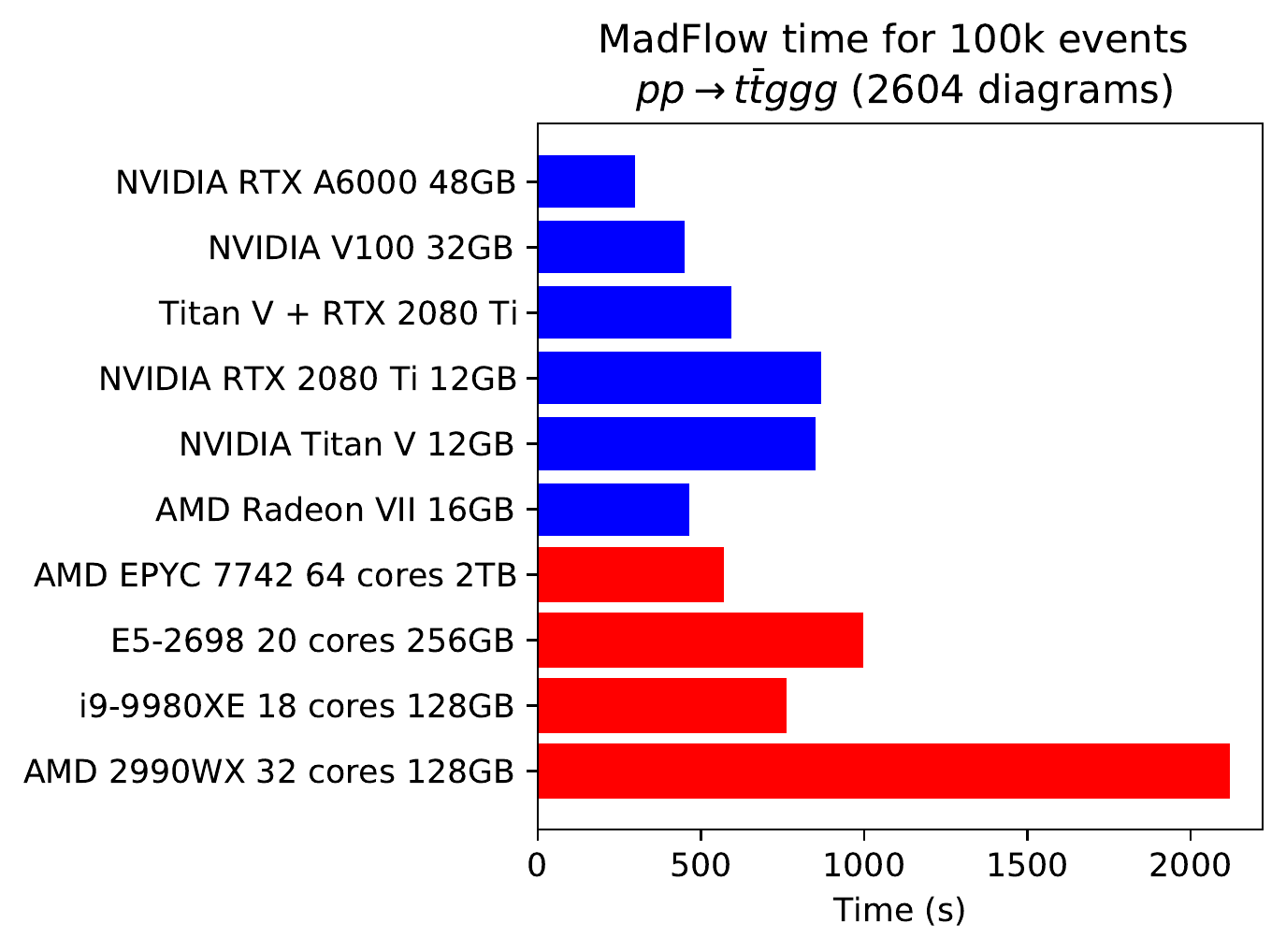}
        \captionof{figure}{$pp \to t\bar{t}ggg$ process \madflow timings.}
        \label{fig:madflow-performance2}
    \end{minipage}
    \hfill
    \begin{minipage}{0.5\textwidth}
        \centering
        \begin{scriptsize}
        \begin{tabular}{llll}
            \hline
            Process & \madflow & \madflow & {\sc MG5\_aMC} ($\SI{}{\micro\second}$) \\ 
            & CPU ($\SI{}{\micro\second}$) & GPU ($\SI{}{\micro\second}$) & \\\hline
            $gg \rightarrow t\bar{t}$ & 9.86 & 1.56 & 20.21 \\
            $pp \rightarrow t\bar{t}$ & 14.99 & 2.20 & 45.74 \\
            $pp \rightarrow t\bar{t}g$ & 57.84 & 7.54 & 93.23 \\
            $pp\rightarrow t\bar{t}gg$ & 559.67 & 121.05 & 793.92 \\
            \hline
        \end{tabular}
        \end{scriptsize}
        \caption{Comparison of event computation time for \madflow and \mgamc
        on Intel i9-9980XE system with $18$ cores and $\SI{128}{\giga\byte}$ of RAM for
        CPU simulation and Nvidia Titan V $\SI{12}{\giga\byte}$ for GPU simulation.}
        \label{tab:madflow-mg5-performance}
    \end{minipage}
\end{table}
\endgroup

The final test presented in this section designs a comparison between the \madflow \linebreak
and the \mgamc frameworks' performance. In table~\ref{tab:madflow-mg5-performance},
we report the total amount of time per event for the processes listed above
for an Intel i9-9980XE CPU system with $18$ cores and $\SI{128}{\giga\byte}$ of RAM
hosting an Nvidia Titan V $\SI{12}{\giga\byte}$. The results follow the trend
of the other benchmarks, presenting a dramatic increase in the number of evaluated
events per second by the GPU device as opposed to the CPU. At the same time,
going towards more complex processes and increasing the number of diagrams requires
also a larger amount of memory to hold the computation, decreasing the performance
gains granted by the GPU.

\section{Conclusion and outlook}

We have presented and tested a new approach for the generalization of Monte Carlo
event generation on hardware accelerators. The \madflow package encompasses a series
of tools implemented with fast and maintainable code that can express complex
analytical expressions into hardware-specific languages without the need to introduce
complicated operations. We tested the software's effectiveness to simulate multiple
scenarios on hardware accelerators. We plan to develop further the
\madflow algorithm, including specific optimization methods for vectorized phase
space sampling to enhance event generation efficiency and eventually make a first
step towards the automation of the Next-to-Leading Order computations taking
advantage of hardware accelerators.

%% file: chapters/partIII/PartIII.tex
\part*{Part III\vspace{0.5cm}\\Deep learning models for neutrino physics}
\label{part:dl4nu}
\addstarredpart{\large Part III : Deep learning models for neutrino physics}
\adjustmtc[1]

\thispagestyle{empty}

%% file: chapters/partIII/chap5/chap5.tex
\chapter{Denoising ProtoDUNE-SP Raw Data with deep learning}
\label{chap:dunedn}
\thispagestyle{plain}

\minitoc

\renewcommand{\myincludegraphics}[2][width=\textwidth]{
    \includegraphics[#1]{chapters/partIII/chap5/plots/#2}
}
\newcommand{\dunetpc}{\texttt{dunetpc}~}
\newcommand{\veight}{\texttt{v08\_24\_00}~}
\newcommand{\vnine}{\texttt{v09\_10\_00}~}
\newcommand{\xb}{\ensuremath{\mathbf{x}}}
\newcommand{\yb}{\ensuremath{\mathbf{y}}}

This chapter deals with the implementation of deep learning solutions for the
raw data denoising algorithm at ProtoDUNE Single Phase (SP) detector, the first
step of the event reconstruction pipeline at Liquid  Argon Time Projecting
Chamber (LArTPC) detectors. We first give an overview of the ProtoDUNE-SP detector
design and its geometry and including a characterization
of signal processing in LArTPCs, namely, how signals first get formed and then are
revealed inside such kind of experimental apparatus.

After that, we discuss the current implementation of the deconvolution method,
the state-of-the-art technique implemented by several neutrino experimental
collaborations. Finally, the last section of this chapter describes our novel
solutions for denoising ProtoDUNE-SP simulation data with deep neural
networks~\citep{Rossi:2022} and, in particular, with graph neural networks.

\section{The ProtoDUNE Single Phase design}
\label{sec:pdune}

The Deep Underground Neutrino Experiment (DUNE)~\citep{Abi:2020tdrI} is a major
experiment in the neutrino oscillation research field. It will be based between
Fermilab (Illinois) and the Sanford Underground Research Facility (SURF) in South Dakota,
the latter hosting the DUNE Far Detector
(FD)~\citep{Abi:2020tdrII,Abi:2020tdrIII,Abi:2020tdrIV}. The FD, with its four
modules containing a fiducial mass of \SI{10}{\kilo\tonne} of liquid argon will
be the largest Liquid Argon Time Projecting Chamber ever built. The
CERN Neutrino Platform, instead, currently hosts the ProtoDUNE-SP
detector~\citep{Abi:2017}, a prototype of the DUNE FD that faithfully reproduces
most of the FD components, but extrapolating the total LAr mass in scale $1{:}20$.

The realization of the ProtoDUNE detector is an essential part of the DUNE FD
development program and has four main objectives. First, to refine the production
and installation procedures of the FD detector modules.
Second, to test the basic detector performance to validate its design,
which is a crucial point in the roadmap of the DUNE FD realization.
Third, to calibrate the detector efficiency in revealing particles of a different
kind. This has to be done thanks to the collection of large samples of test beam
data, that in turn will allow implementing dedicated software solutions for the 
Monte Carlo simulation. Fourth, to prove the long-term operational stability of
the detector and prevent risks of continued operations on the DUNE FD modules.

\begin{figure}
  \begin{minipage}{0.45\textwidth}
    \centering
    \myincludegraphics[height=5cm]{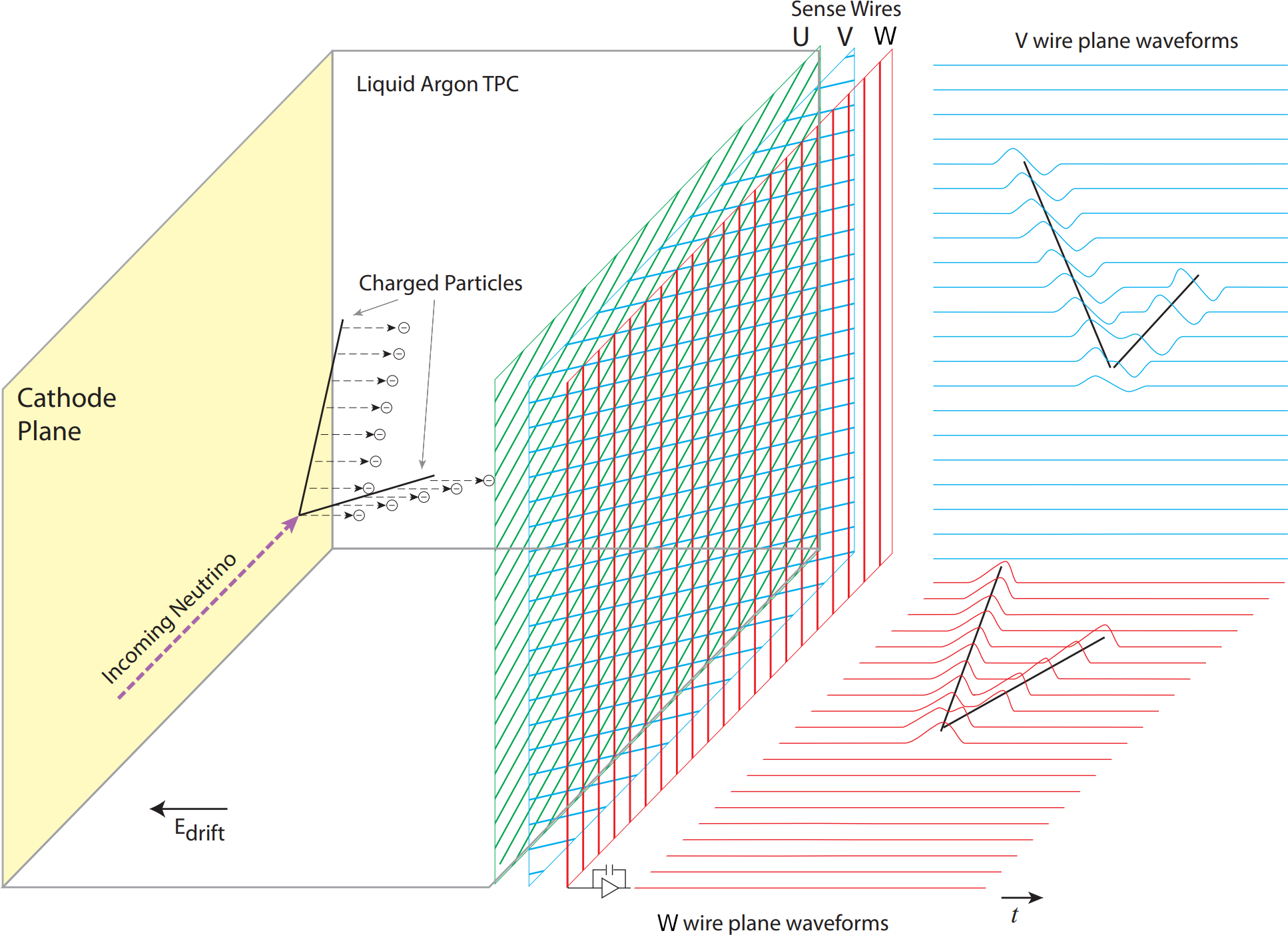}
    \caption{Sketch of the ProtoDUNE-SP detector. Only one drift chamber is
    depicted, the other extends to the left of the CPA.}
    \label{fig:pdune-detector}    
  \end{minipage}
  \hfill
  \begin{minipage}{0.45\textwidth}
    \centering
    \myincludegraphics[height=5cm]{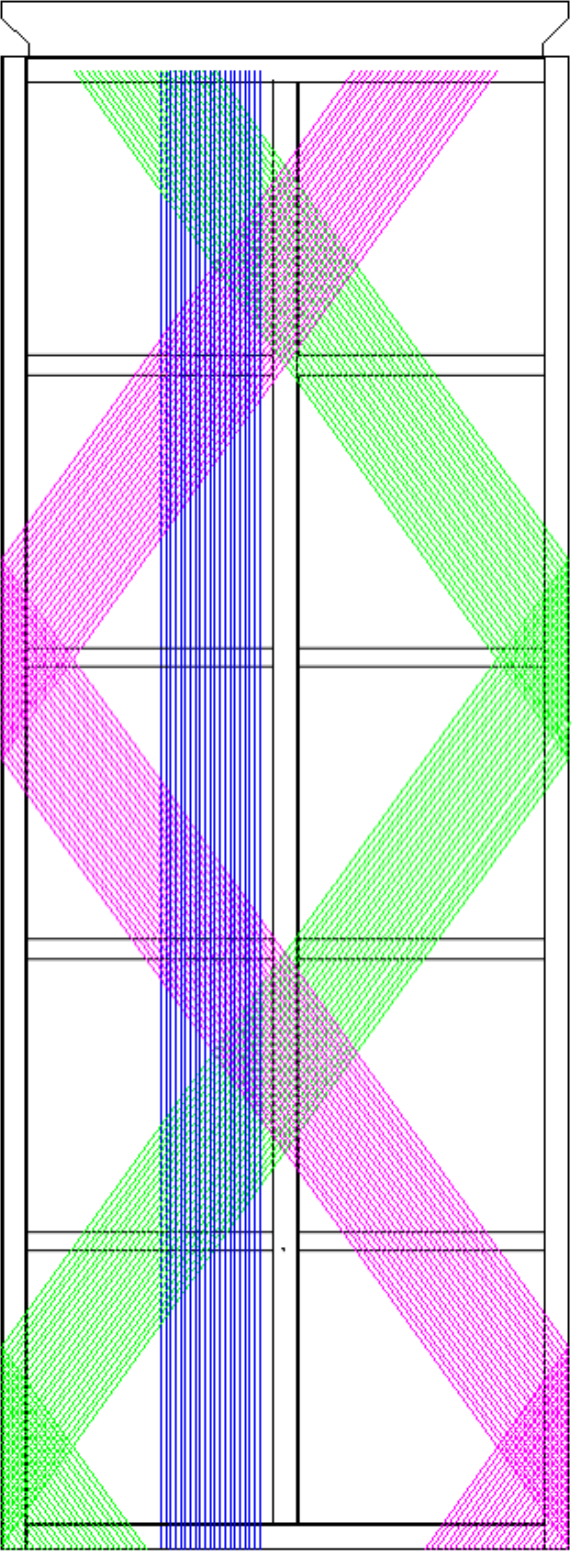}
    \caption{Wires wrapping around a ProtoDUNE-SP APA plane: $\mathrm{U}$,
    $\mathrm{V}$ and $\mathrm{W}$ planes in green, magenta and blue,
    respectively.}
    \label{fig:APAs}    
  \end{minipage}
\end{figure}
The ProtoDUNE-SP design is sketched in figure~\ref{fig:pdune-detector}: the active
TPC is a $\SI{6}{\m}$ high, $\SI{7}{\m}$ wide and $\SI{7.2}{\m}$ deep
box. A field cage surrounding the volume supports and isolates the detector.
The space is divided by a central cathode plane, named Cathode Plane Assembly
(CPA), that separates the two drift volumes. The sides of the detector, instead,
are paved by three adjacent $\SI{6}{\m}$ high and $\SI{2.3}{\m}$ wide Anode
Plane Assemblies (APAs). Each APA consists of a support where three parallel
planes of sense wires are wrapped around along different orientations to provide
three independent views of the events, allowing $3{-}$dimensional reconstruction.
Figure~\ref{fig:APAs} shows the orientation of the three $\mathrm{U}$,
$\mathrm{V}$ and $\mathrm{W}$ wire planes. The first two wire planes,
$\mathrm{U}$ and $\mathrm{V}$, are oriented forming angles of $35.7^\circ$ and
$-35.7^\circ$ with respect to the vertical, while the last plane $\mathrm{W}$
is directed along the vertical. A total of $2560$ readout channels are attached
to the APA wires, resulting in $15360$ channels included in the whole structure.

The CPA is kept at a tension of $\SI{-180}{\kilo\volt}$, while the APA wire layers
are held at $\SI{-370}{\volt}$ ($\mathrm{U}$ plane), $\SI{0}{\volt}$
($\mathrm{V}$ plane) and $\SI{820}{\volt}$ ($\mathrm{W}$ plane)
to establish a quasi uniform electric field of intensity $\SI{500}{\volt/\cm}$.
A report of the installation and the test beam runs taken before the LHC Long
Shutdown 2 a the end of 2018 is available in~\citep{Abud:2021}.

We present a qualitative overview of how an interaction
within the detector fiducial volume is revealed by the experimental apparatus
leading to the formation of the raw signal inside LArTPCs.
At the moment a particle enters the detector
fiducial volume, it may interact with Argon nuclei producing ionization electrons
and Argon ions pairs. The intense electric field provided by the CPA accelerates
the charged particles in opposite directions: the positive ions are attracted
towards the center of the cage against the CPA, while the electrons drift
in the opposite direction and get collected by the APA wires. We stress that
the ionization electrons induce current on the sense wires in the two first APA
wire planes $\mathrm{U}$ and $\mathrm{V}$, but are indeed collected only by the
last plane $\mathrm{W}$. Therefore, the $\mathrm{U}$ and $\mathrm{V}$ wire
bundles are often called induction planes, while we refer to the $\mathrm{W}$
one as the collection plane.

The motion of charges inside the detector volume generates a detector response,
given by the combined effects of different contributions. The overall simulation
within a LArTPC environment can be simplified into five different pieces that
should be convoluted together to provide the final detector measurement $M$:
\begin{equation}
  M = \braces{Depo \otimes Drift \otimes Duct + Noise} \otimes Digit
\end{equation}

The definition of the ingredients included in the equation above follows.
\begin{itemize}
  \item $Depo$: the initial distribution of ionization electrons created by
  energy depositions. It is modeled by point-like depositions $q_i$ at positions
  $\mathbf{r}_i$ and times $t_i$: \linebreak $Depo_i\braces{q_i, t_i, \mathbf{r}_i}$
  \item $Drift$: describes the drift of the initial ionization charge towards
  the APAs, due to the uniform electric field $\mathbf{E}$. The output is the
  number of electrons and their space-time distribution arriving at a distance of
  $x_{rp} = \SI{10}{\cm}$ from the APAs. The imaginary plane parallel to the APA
  at $x_{rp}$ distance is called response plane. The propagation of the initial
  charge until the response plane is characterized by diffusion along
  longitudinal and transverse directions producing a final charge distribution
  with longitudinal and transverse standard deviations:
  \begin{equation}
    \sigma_{L,T} \simeq \sqrt{2 D_{L,T} t_{drift}}
  \end{equation}
  where $D_{L,T}$ are diffusion coefficients in argon~\cite{Li:2016} and
  $t_{drift}$ is the time the electron takes to reach the response plane.
  The final space charge distribution is given by:
  \begin{equation}
    Depo_i \otimes Drift \to
    Depo_i\braces{
      q_i,t_i+t_{drift},\mathbf{r}_i\Big\rvert_{x=x_{rp}}, \sigma_L, \sigma_T
    }
  \end{equation}
  \item $Duct$: includes the field and electronics responses discussed in
  sections~\ref{subsec:field-response} \linebreak and~\ref{subsec:elec-response},
  respectively.
  \item $Noise$: the inherent electronics noise that must be added to the
  electronics readouts to produce a realistic measurement. It can be simulated
  as a random walk in the frequency domain to produce a time-dependent noise
  contribution for each readout channel.
  \item $Digit$: the digitized version of the electronics signal recorded by the
  detector.
\end{itemize}
In the following, we present a qualitative introduction to the $Duct$ components,
namely field response and electronics response, and the digitization procedure
done by LArTPCs. The discussion follows~\citep{Adams:2018ion1} which is tailored
on the MicroBooNE detector~\citep{Acciarri:2016mb}, but can be applied in general
to all LArTPCs including ProtoDUNE-SP.

\subsection{Field response}
\label{subsec:field-response}

The induced current $i$ on a wire by a charge $q$ moving in its proximity is
governed by the Ramo theorem~\citep{Ramo:1939} and its generalized
forms~\citep{Cavalleri:1971,Hamel:2008}:
\begin{equation}
  \label{eqn:ramo}
  i = -q \mathbf{E}_w \cdot \mathbf{v}_q
\end{equation}
where $\mathbf{v}_q$ is the drifting velocity of the charge and $\mathbf{E}_w$
is a weighting field, that is a function of the geometry of the electrodes. The
operative measure of the weighting field is done while keeping the target wire
at unit potential, setting all the others to ground, removing the drifting charge
and measuring the electric field in the space surrounding the electrode. The
generalized versions take into account non-linear effects and configurations in
which electrodes are placed within multiple dielectric mediums. The instantaneous
drift velocity of the charge in the case of the LArTPC depends mainly on the
geometry of the APAs and liquid argon temperature.

The qualitative behavior of the field response function is better understood
considering Green's reciprocity theorem: the moving charge $q_m$ induces
current on a sense wire $\mathrm{I}$, which, in turn, corresponds to a movement
of charges $Q_\mathrm{I}$ on the wire itself. The two quantities are related by
the formula:
\begin{equation}
  q_m \cdot V_m = Q_\mathrm{I} \cdot V_\mathrm{I}
\end{equation}
where $V_m$ is the potential induced at the position of the charge by the sense
wire kept at potential $V_\mathrm{I}$.

The induced current, or field response, can be obtained by taking the time
derivative of the charges moving on the electrode $\mathrm{I}$:
\begin{equation}
  i = \frac{dQ_\mathrm{I}}{dt} =
  q_m \cdot {\bm\nabla} V_w \cdot\frac{d\mathbf{r}}{dt}
\end{equation}
where we have introduced the dimensionless weighting voltage
$V_w = V_m / V_\mathrm{I}$, which assumes values in the unit range. The last
equation, in particular, is equivalent to equation~\ref{eqn:ramo} putting
$-\mathbf{E}_w = {\bm\nabla} V_w$. The integral through time of the current
induced on the electrode is given by the work done to bring the charge $q_m$
within the weighting potential from a starting position $A$ to an end position $B$:
\begin{equation}
  \label{eqn:ramo2}
  \int i dt = q_m \braces{V_{w}^{B} - V_{w}^{A}}
\end{equation}

\begin{figure}
  \centering
  \subfigure[
    Electron path drifts.\label{subfig:e-path}
  ]{
    \myincludegraphics[width=0.47\textwidth]{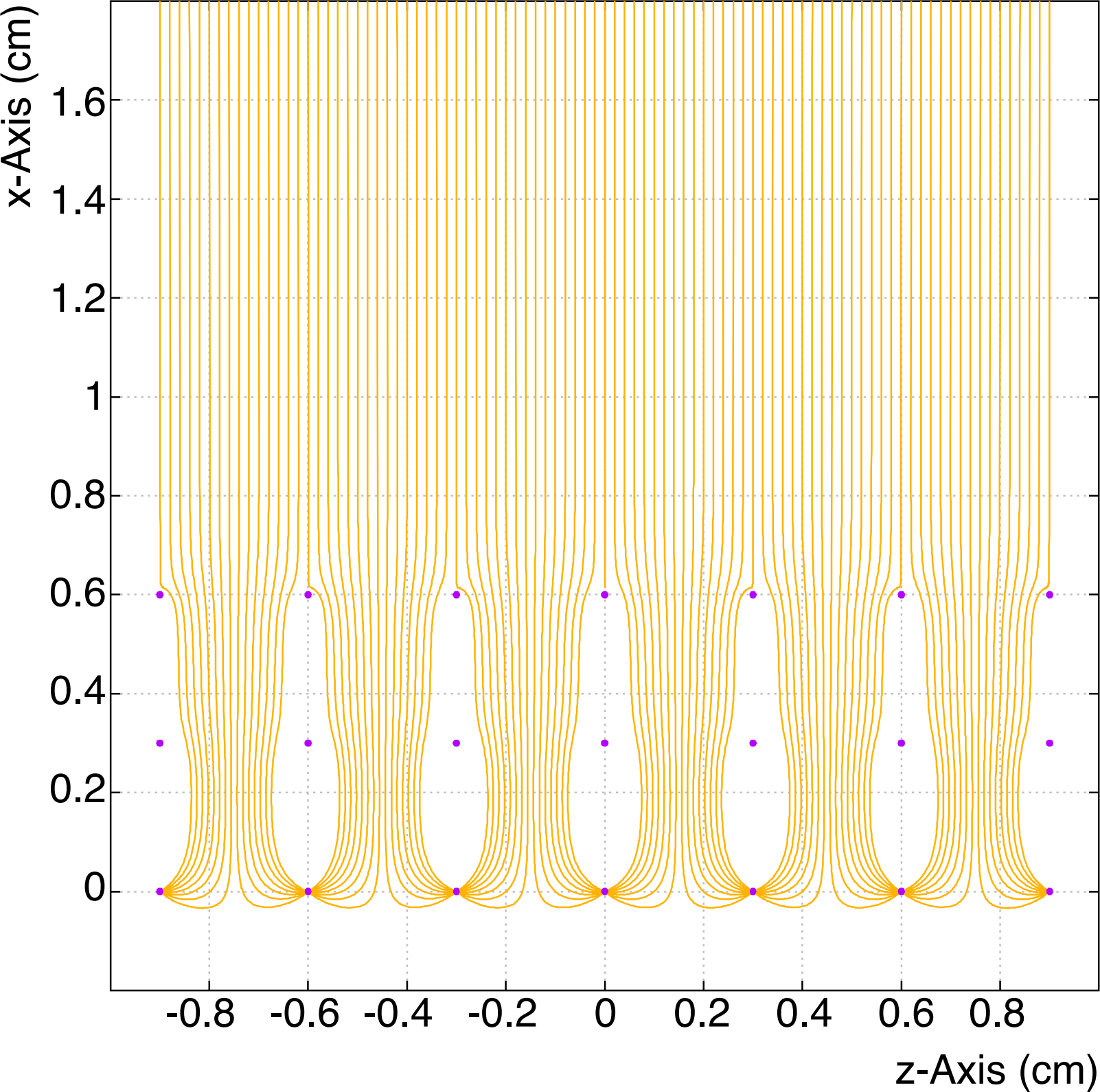}
  }
  \hfill
  \subfigure[
    Weighting potential $V_w$ on $\mathrm{U}$ plane wire.\label{subfig:vw-u}
  ]{
    \myincludegraphics[width=0.47\textwidth]{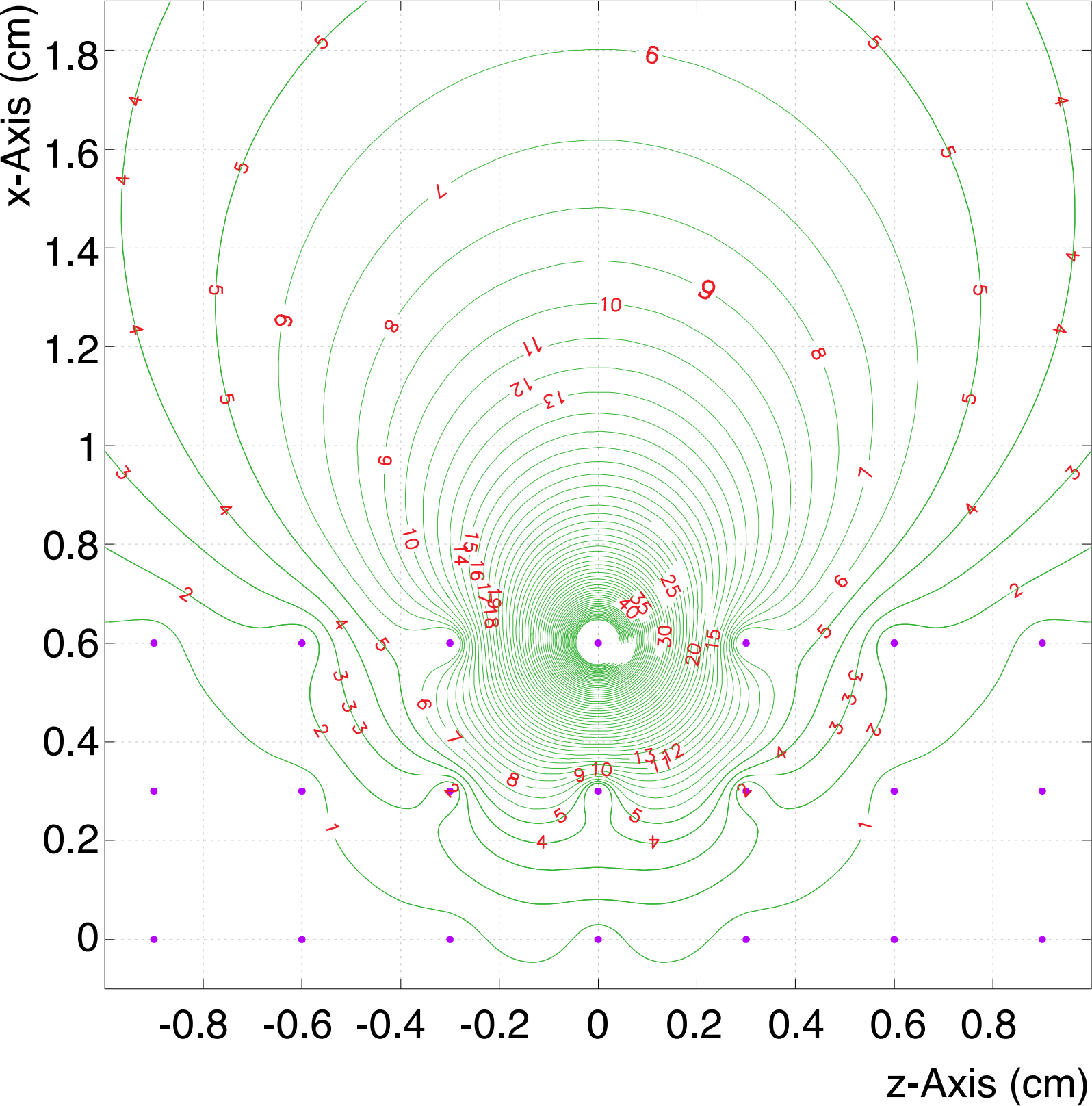}
  }\\
  \subfigure[
    Weighting potential $V_w$ on $\mathrm{V}$ plane wire.\label{subfig:vw-v}
  ]{
    \myincludegraphics[width=0.47\textwidth]{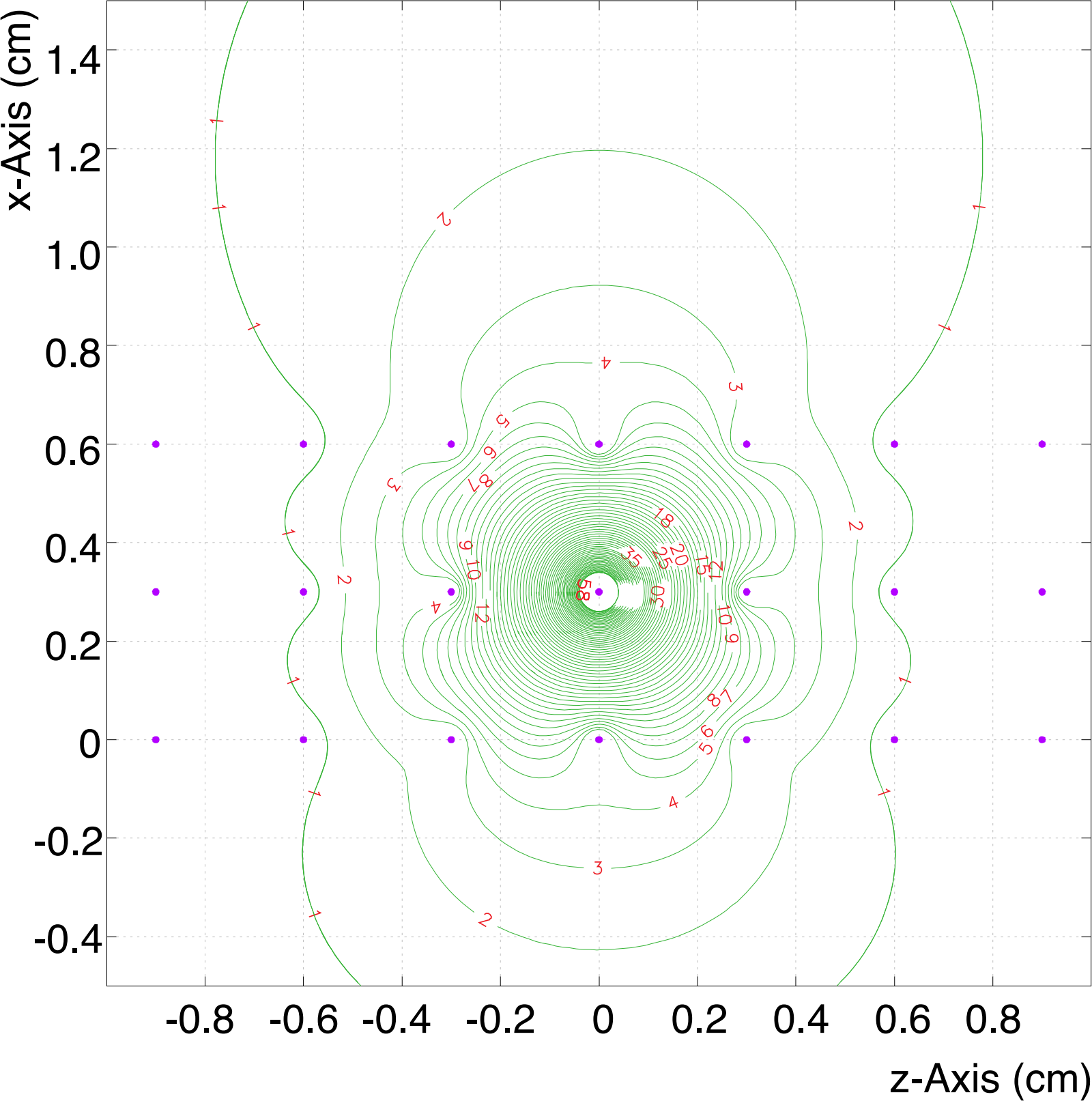}
  }
  \hfill
  \subfigure[
    Weighting potential $V_w$ on $\mathrm{W}$ plane wire.\label{subfig:vw-w}
  ]{
    \myincludegraphics[width=0.47\textwidth]{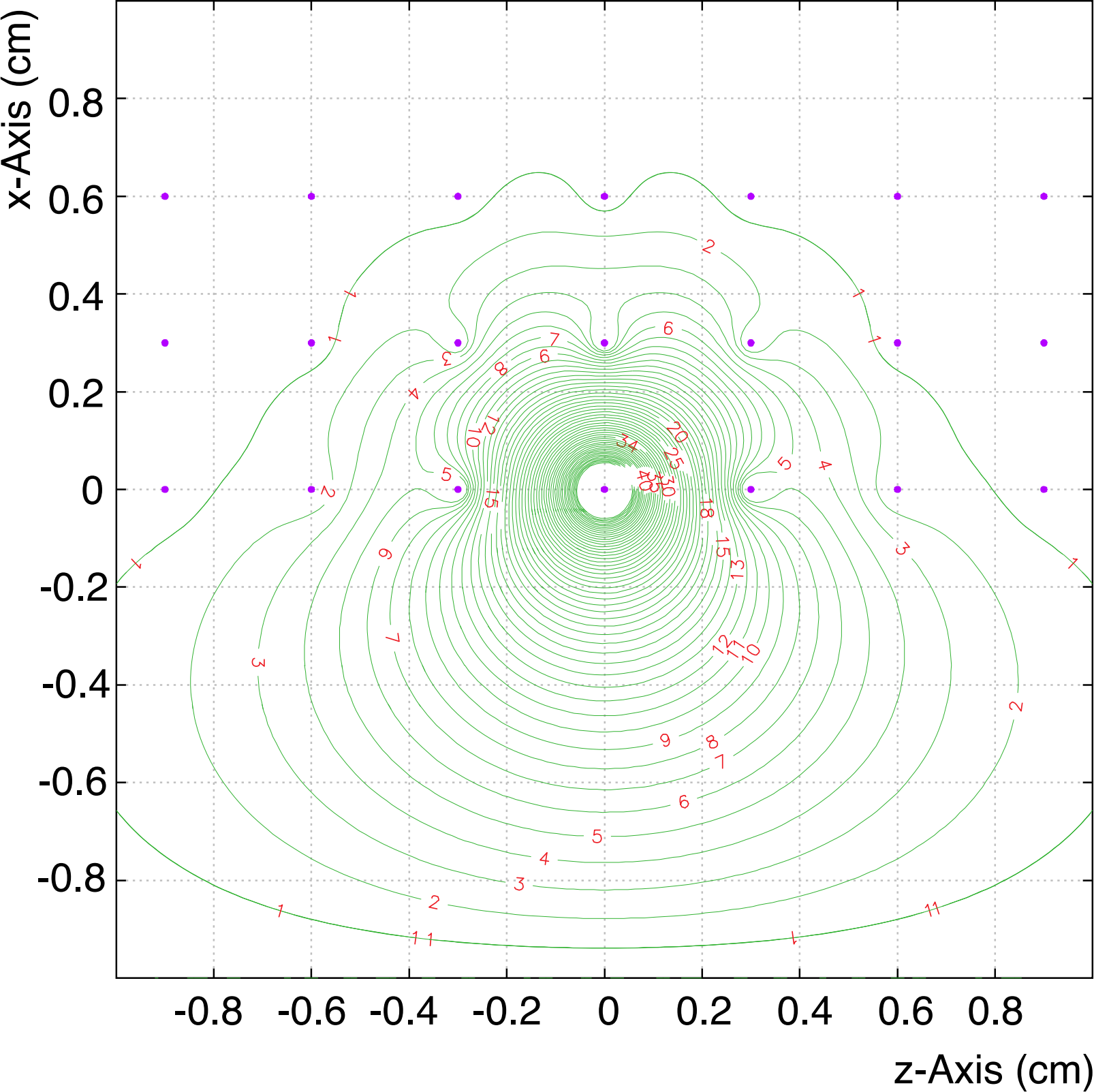}
  }
  \label{fig:garfield}
  \caption{The electron path drift and weighting potential in the Garfield
  simulation from~\cite{Adams:2018ion1}. The equipotential lines are labeled by
  numbers from $1$ to $60$ indicating the percentage with respect to
  $V_w^{\mathrm{max}}$, namely the electric potential on the considered wire.
  }
\end{figure}

In~\cite{Adams:2018ion1}, the field response is computed for a single ionization
electron through a simulation with the Garfield software~\citep{Veenhof:1998}.
The simulation assumes a $2{-}$dimensional wire geometry and nominal wire
potential compatible with the MicroBooNE detector setup, however, the results
presented here can be easily generalized to ProtoDUNE and other LArTPC detectors.
The authors, simulate an electron starting at a position $\SI{10}{\cm}$ away
from the first wire layer, namely $\mathrm{U}$ plane and approaching the APA.
Different electron paths are displayed in figure~\ref{subfig:e-path} depending
on the starting position, which is discretized along the direction transverse to
the drifting direction by a $\SI{0.3}{\mm}$ shift. The picture shows that, due to
the uniform electric field, the electron approaches the APA passing next to the
$\mathrm{U}$ and $\mathrm{V}$ planes (the first and second encountered layers)
and gets collected by the $\mathrm{W}$ collection plane.

Figures~\ref{subfig:vw-u},~\ref{subfig:vw-v},~\ref{subfig:vw-v} plot the simulated
weighting potential of a wire in the three different APA planes $\mathrm{U}$,
$\mathrm{V}$ and $\mathrm{W}$ respectively. The weighting potential is maximum
in the coincidence of the considered wire and decreases for increasing distance. The
plots highlight that each APA plane acts shielding the weighting potential of
the wires of the other APA layers. Indeed, figure~\ref{subfig:vw-v} has a $V_w$
that extends towards the inner region of the detector, but only small
${\sim}5\%$ fractions of the weighting field pass beyond the second line of wires.
The weight potential of the wire on the $\mathrm{V}$ plane is concentrated in the
inter-electrode space being shielded both by the first and the last APA layer.
The collection wires are shielded both by the two former layers, instead.

\begin{figure}
  \centering
  \myincludegraphics[width=0.5\textwidth]{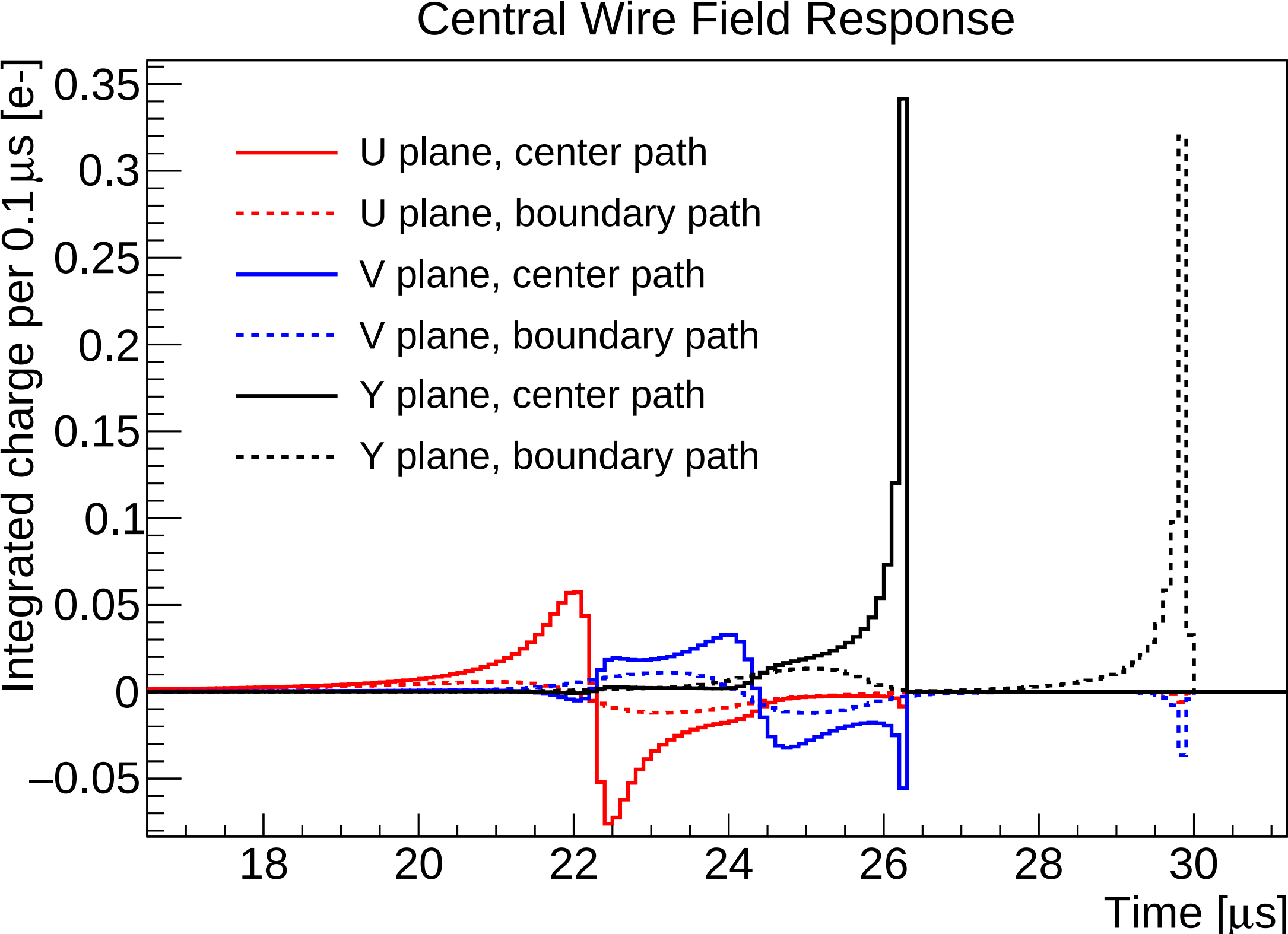}
  \caption{The waveforms due to single electron drift in the Garfield simulation
  by~\citep{Adams:2018ion1}. The solid waveforms refer to an electron starting
  position directly in front of the considered wires (center path). The dashed
  waveforms consider the shifted electron path starting from the boundary of the
  simulation volume. This shows that drifting particles influence the measured
  current also on distant wires.
  }
  \label{fig:waveforms}
\end{figure}

A consequence of equation~\ref{eqn:ramo2} is that each wire is sensible to the
induced current by an ionization electron only if it moves inside a region where
the weighting potential has a spatial increment or a decrement appreciably
different from zero. As an ionization electron approaches the first induction
plane, it moves towards higher regions of the associated weighting potential and
the induced current is positive. As soon as the particle moves away from the wire
and continues towards the far side of the detector, the induced current switches
to negative values and forms the typical bipolar sign for the induction waveforms
recorded by the LArTPC electronics, see figure~\ref{fig:waveforms}.

Once the electron passes the first induction wire, it starts inducing current
also on the collection wire. Since the paths of $q_m$ eventually end onto a
collection wire, the induced current on the $\mathrm{V}$ plane electrodes is
always positive, resulting in unipolar peaks in the waveforms in
figure~\ref{fig:waveforms}.

As a consequence, the integrated current on an induction plane wire for an
ionization electron originating on the cathode plane is zero because the drift
trajectory starts from a position where the weighting field is zero and ends
on a collection plane wire, where again $V_w = 0$. On the other hand, the
integrated current on a collection plane wire in the same situation equals to
a single electron charge, since the starting weighting potential is again zero,
but the final potential is equal to unity, referring to the collection wire.
In general, this is not true for ionization electrons originating in the inner
detector, since the starting weighting potential is not zero there. However,
the weighting potentials fall to zero rather quickly, reaching \orderof{10^{-2}}
values at distances of ${\sim}\SI{10}{\cm}$ away from the APAs.

When an ionization electron is created inside the LArTPC, also the corresponding
argon ion is generated. The ion drifts in the electron's opposite direction and,
in principle, influences the field response function. The contribution, however,
is negligible, since the drift velocity of the ions is far slower than the
ionization electrons due to their mass:
\begin{equation}
  v_{e^-}\sim\SI{1}{\mm/\micro\s}, \qquad
  v_{Ar^+}\sim\SI{5}{\mm/\s}
\end{equation}

\subsection{Electronics response and digitization}
\label{subsec:elec-response}

\begin{figure}
  \centering
  \subfigure[Pre-amplifier response function.]{
    \myincludegraphics[width=0.47\textwidth]{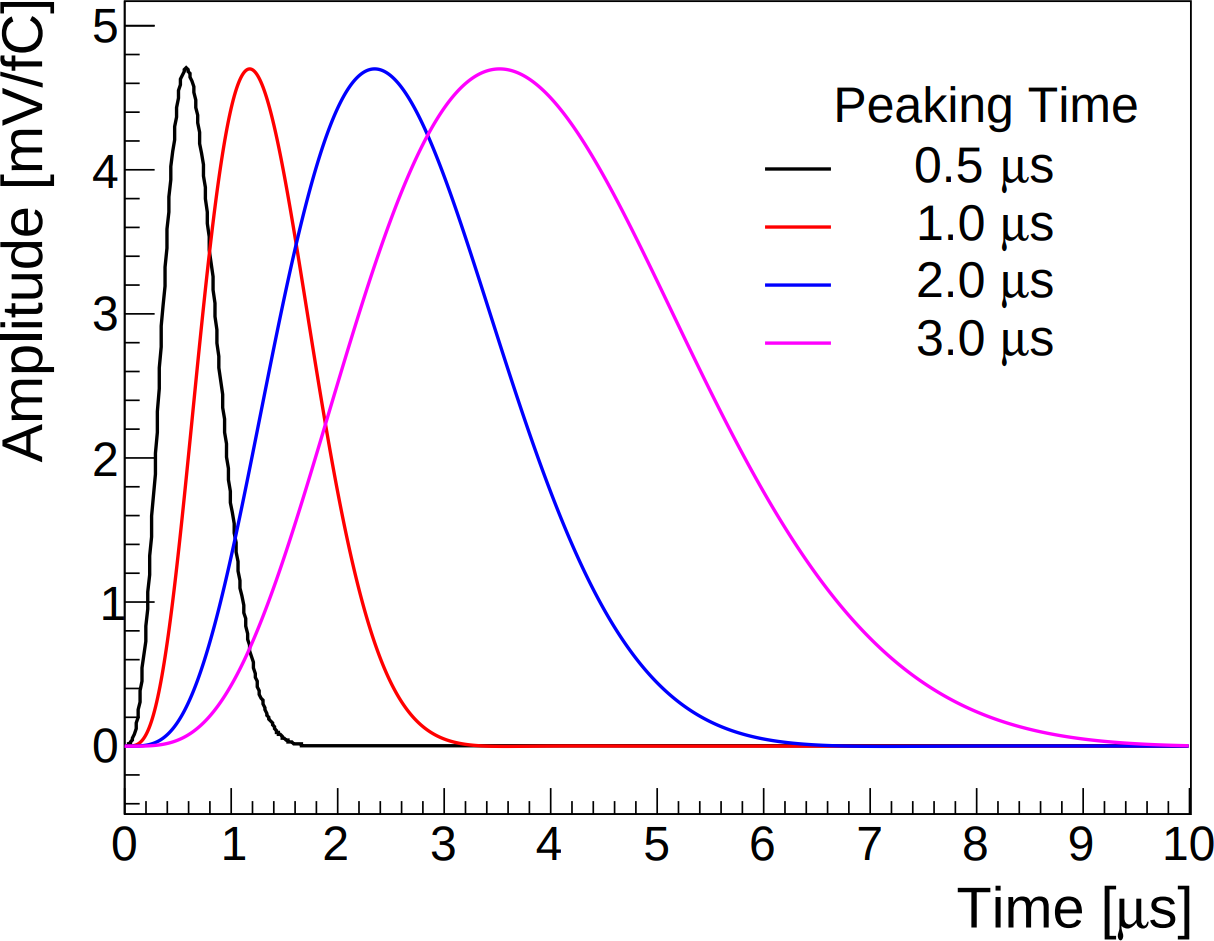}
    \label{subfig:pre-ampl}
  }
  \hfill
  \subfigure[RC filter response function.]{
    \myincludegraphics[width=0.47\textwidth]{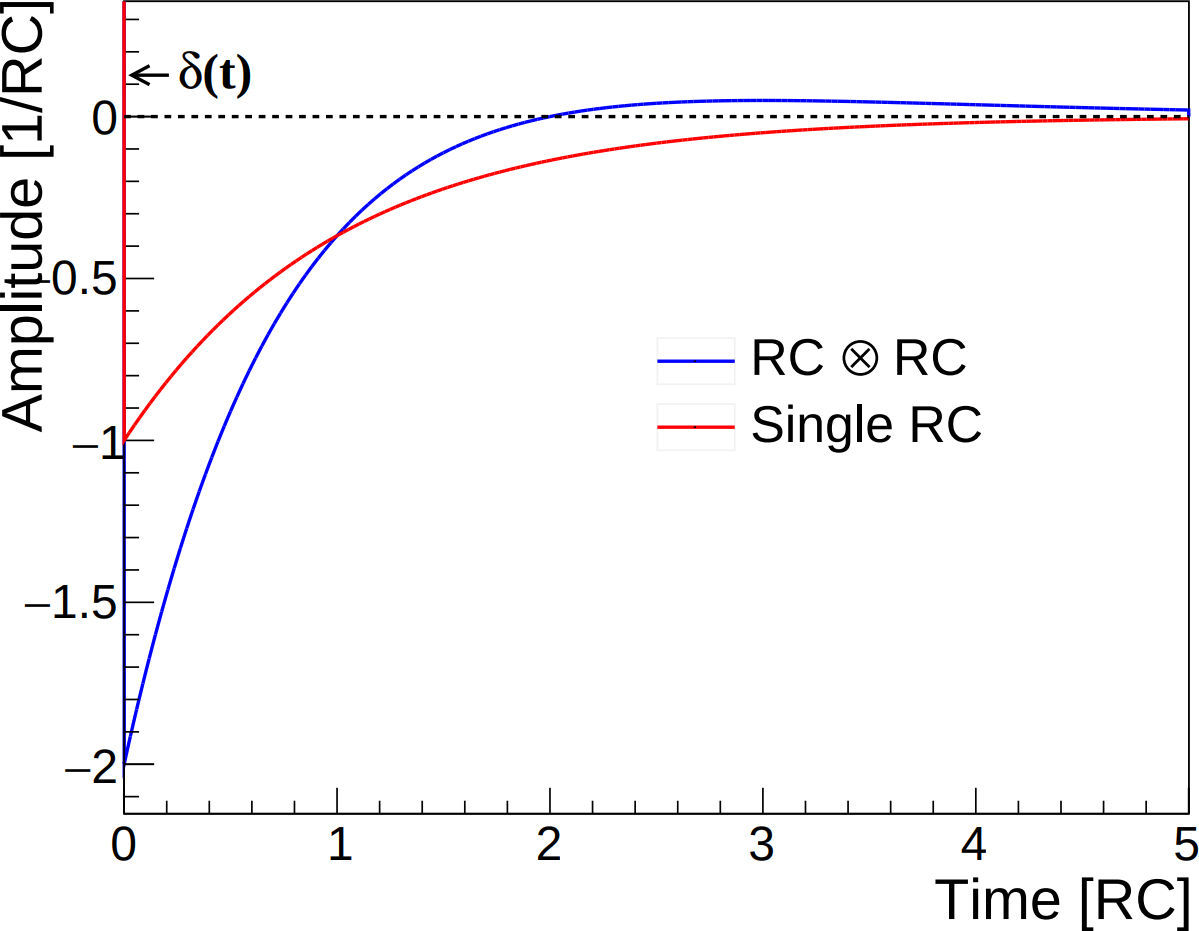}
    \label{subfig:rc}
  }
  \caption{The MicroBooNE electronics response to an impulse signal located at
  $t=0$: left panel plots the pre-amplifier response function for
  $\SI{4.7}{\milli\volt/\femto\coulomb}$. Both diagrams are taken from the
  study in~\citep{Adams:2018ion1}.}
\end{figure}

Once the induced current on a sense wire is received, a pre-amplifier amplifies
and shapes the signal. The pre-amplifier settings include the gain
and the peak time. These two parameters control the height, width and time
position of the peak measured from a unit input impulse at time $t=0$.
In particular, the gain is defined as the output waveform amplitude and, for
LArTPCs experiments such as MicroBooNE and ProtoDUNE-SP, there are four different
configurations of this quantity: $\SI{4.7}{\milli\volt/\femto\coulomb}$,
$\SI{7.8}{\milli\volt/\femto\coulomb}$, $\SI{14}{\milli\volt/\femto\coulomb}$,
$\SI{25}{\milli\volt/\femto\coulomb}$. The peaking time is defined as the time
difference between $5\%$ of the peak at the rising edge and the peak. Four
peaking times can be configured in the cold electronics setup in MicroBooNE and
ProtoDUNE-SP: $\SI{0.5}{\micro\s}$, $\SI{1.0}{\micro\s}$, $\SI{2.0}{\micro\s}$
and $\SI{3.0}{\micro\s}$. Figure~\ref{subfig:pre-ampl} shows the different
pre-amplifier responses for a fixed gain value of
$\SI{4.7}{\milli\volt/\femto\coulomb}$. Notice that the signal amplitude is
always the same for different peaking times since it is dependent on the gain
setting only. The four peaking time setups are important to satisfy the
Nyquist theorem~\citep{Nyquist:1928} for different sampling rates: an analogic
periodical signal can be correctly reconstructed only if the sampling frequency,
used to digitize the signal, is greater than double the highest frequency in the
signal spectrum.

The output of the pre-amplifier is further analyzed by two independent RC filters
to remove the baseline. The RC response function for a single impulse
located in the origin, $\delta(t)$, is displayed in figure~\ref{subfig:rc}. Both
the single RC behavior and the double independent $\mathrm{RC}\otimes\mathrm{RC}$
RC circuits responses are shown in the plot. The effect of the introduction
of such electronics is visible in the long or large waveforms, especially
for collection plane wire.

The output waveforms are digitized at a sampling rate of $\SI{2}{\MHz}$ for both
MicroBooNE and ProtoDUNE-SP and stored in a $\SI{12}{\bit}$ analog-to-digital
converter (ADC) values. This sampling rate corresponds to discretized readout
steps, usually called TDC ticks, of $\SI{0.5}{\micro\s}$ each. In particular,
the ProtoDUNE-SP event readout time window lasts $\SI{3}{\milli\s}$, yielding 
in each event an overall sequence of $6000$ ADC values per readout channel.
Each measured event by ProtoDUNE-SP can then be cast into an image-like matrix
of resolution $15360 \times 6000$, where the axes refer to the readout channel
and the TDC tick, respectively. In the LArTPC jargon, we often refer to each
pixel ADC value as a raw digit. The image representation of the ProtoDUNE-SP
data will be exploited in section~\ref{sec:dunedn} to introduce our novel deep
learning approach to raw data denoising.

\section{Signal processing and deconvolution method in LArTPCs}
\label{sec:deconvolution}

Noise has a great impact on the reconstruction efficiency in LArTPCs. The standard
noise removal strategy is deconvolution, a technique that we will explain in the
following. Sometimes, in real detector scenarios, excess noise may require
a dedicated noise filter design, like in the case of MicroBooNE~\citep{Acciarri:2017}.
Time or $1{-}$dimensional deconvolution has been introduced by~\cite{Baller:2017}
in the context of the ArgoNeuT experiment. The $2{-}$dimensional version of the
algorithm, which takes into account noise spatial dependencies between
neighboring wires, has proven more effective and has been exploited
by~\cite{Adams:2018ion1,Adams:2018ion2}.

A $2{-}$dimensional algorithm is needed because the weighting potential of a wire
in a LArTPC is appreciably non-zero in regions surrounding nearby electrodes.
As a consequence the charge moving in the volume influences with a sizeable
contribution not only the closest readout channels but also the neighboring
ones. For simplicity, we present the deconvolution method first in
$1{-}$dimension and then provide the $2{-}$dimensional extension.

Deconvolution is a mathematical technique used to reconstruct an original signal
$S(t)$, given a measured signal $M(t')$, which is the result of a convolution
operation between the original signal and a detector response function $R(t,t')$:
\begin{equation}
  M(t') = \int_{-\infty}^{+\infty} dt\, R(t,t') S(t)
\end{equation}
The detector response function equals the measured signal for an original
impulse signal $S(t) = \delta(t-t_0)$. The response function can be taken as a
time-invariant function depending on the time difference between reception and
measurement of signal: $R(t,t') = R(t-t')$. The convolution is turned into a
function multiplication in the frequency domain through the Fourier transform:
$M(\omega) = R(\omega)\cdot S(\omega)$. Therefore, in principle, it could be
possible to retrieve the original signal in the frequency domain by computing
the ratio:
\begin{equation}
  \label{eqn:fourier-sig}
  S(\omega) = \frac{M(\omega)}{R(\omega)}
\end{equation}

Taking the inverse Fourier transform allows to work out the original signal
waveform in the time domain $S(t)$. However, this is not always possible since
the measured signal includes the contribution of inherent electronic noise
that is unknown a priori and, further, in real detectors, the response function
$R(\omega)$ is rapidly decreasing for large frequencies $\omega$. As a
consequence of equation~\ref{eqn:fourier-sig}, the second effect yields a
vanishing denominator at high frequencies, resulting in a dominant noise
contribution in the reconstructed signal. The solution to these problems is to
include a filtering function $F(\omega)$ that mitigates the discussed effects:
\begin{equation}
  S_f(\omega) = \frac{M(\omega)}{R(\omega)} \cdot F(\omega)
\end{equation}
The reconstructed signal $S(t)$ retrieved with this equation is called
deconvolved signal.

Usually, a Wiener filter~\citep{Wiener:1964} is designed exploiting the expected
quadratic signal $\overline{S^2(\omega)}$ and noise $\overline{N^2(\omega)}$
functions in the frequency domain:
\begin{equation}
  \label{eqn:wiener-filter}
  F(\omega) = \frac{
    \overline{R^2(\omega)S^2(\omega)}
  }{
    \overline{R^2(\omega)S^2(\omega)} + \overline{N^2(\omega)}
  }
\end{equation}

Unfortunately, there are different problems in the implementation of this ideal
approach, mainly due to the impossibility of shaping exactly the expected signal
and noise frequency spectrum.
First, because the field response depends on the event topology, namely
different ionizing tracks inside the LArTPC produce different signals due to
$3{-}$dimensional effects.
Second, by definition, the filter spectrum at zero frequency is a quantity
lower than unity. The Fourier transform of a function equals the integral of the
original function in the time domain, then the quantity $S(\omega = 0)$ is the
integrated charge on a sense wire during a readout time window. The introduction
of the Wiener filter in equation~\ref{eqn:wiener-filter} leads to a
non-conservation of the number of ionization electrons on the wire:
\begin{equation}
  S(\omega = 0) = \int d\omega\,S(t) \equiv Q^\mathrm{tot}
  \hfill
  \to
  \hfill
  S_f(\omega = 0) = S(\omega) \cdot F(\omega) \Big\rvert_{w=0}
  \equiv Q_f^\mathrm{tot} < Q^\mathrm{tot}
\end{equation}
These issues drove the neutrino experimental collaboration to implement
Wiener-\linebreak inspired filters, rather than actual Wiener filters. Nonetheless, this
naive description allows to understand the core of the algorithms behind the
the main noise mitigation techniques at LArTPCs.

The $2{-}$dimensional extension of this approach takes into account the detector
response functions of wires nearby a central one. We consider the measured signal
$M_i(t_0)$ on the $i{-}$th wire at a time $t_0$:
\begin{equation}
  \small
  \label{eqn:2d-deconv}
  M_i(t_0) = \int_{-\infty}^{+\infty}dt\,
  \braces{
    \ldots
    + R_1(t_0-t)\cdot S_{i-1}(t)
    + R_0(t_0-t)\cdot S_i(t)
    + R_1(t_0-t)\cdot S_{i+1}(t)
    + \ldots
  }
\end{equation}
where $S_i$ represents the real signal waveform on the wire $i$. $R_0$ is the
average detector response function associated with the reference wire, while
$R_1$ is the average detector response linked to a wire one step next to the
reference wire. The average is taken on all the possible electron drift paths
that end on the considered wire. In principle, it is possible to include in the
sum the contributions provided by $n$ wires.

The Fourier transform of equation~\ref{eqn:2d-deconv} results in replacing the
convolutions with normal multiplications, allowing a description of the algorithm
through a matrix multiplication notation:
\begin{equation}
  \begin{pmatrix}
    M_0(\omega) \\ M_1(\omega) \\ \vdots \\ M_{n-2}(\omega) \\ M_{n-1}(\omega)
  \end{pmatrix} =
  \begin{pmatrix}
R_0(\omega)     & R_1(\omega)     & \dots  & R_{n-2}(\omega) & R_{n-1}(\omega) \\
R_1(\omega)     & R_0(\omega)     & \dots  & R_{n-3}(\omega) & R_{n-2}(\omega) \\
\vdots          & \vdots          & \ddots & \vdots          & \vdots          \\
R_{n-2}(\omega) & R_{n-3}(\omega) & \dots  & R_{0}(\omega)   & R_{1}(\omega)   \\
R_{n-1}(\omega) & R_{n-2}(\omega) & \dots  & R_{1}(\omega)   & R_{0}(\omega)   \\
  \end{pmatrix}
  \cdot
  \begin{pmatrix}
    S_0(\omega) \\ S_1(\omega) \\ \vdots \\ S_{n-2}(\omega) \\ S_{n-1}(\omega)
  \end{pmatrix}
\end{equation}
The problem of finding the real wire signals $S_i(t)$ can be solved through
inversion of the response function matrix and inverse Fourier transform.
Additionally, a filtering matrix can be applied to suppress high frequency
noises as in the $1{-}$dimensional case.

\begin{figure}
  \centering
  \myincludegraphics{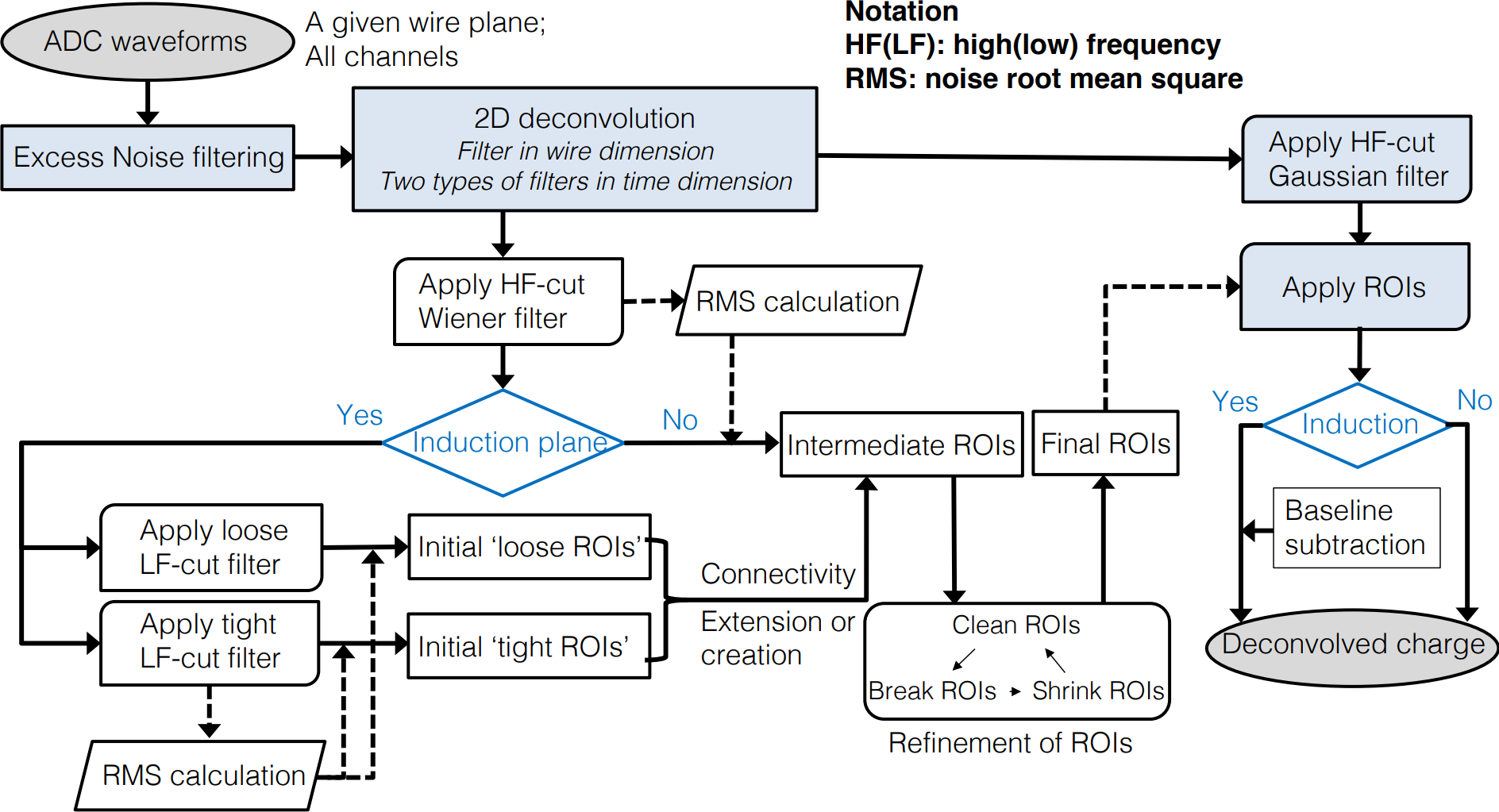}
  \caption{The signal processing flowchart is currently implemented by neutrino
  experiments such as MicroBooNE and DUNE. The figure is taken
  from~\cite{Adams:2018ion1}.
  }
  \label{fig:2d-deconv-workflow}
\end{figure}

In practice, in the real software for LArTPC signal processing the
$2{-}$dimensional deconvolution is followed by a region of interest (ROI)
finding and gaussian peak fitting in those regions. Further, low-frequency filter
cuts linear baseline subtraction are applied to induction plane waveforms to
compute the gaussian peak widths and the final deconvolved charge over each
electrode. Figure~\ref{fig:2d-deconv-workflow} depicts the signal processing
workflow implemented by MicroBooNE. The DUNE software employs an analogous
setup.

\section{Deep learning strategies for ProtoDUNE raw data denoising}
\label{sec:dunedn}

In this section, we introduce the novel deep learning approach to raw digit
denoising at ProtoDUNE-SP. We employ an approach based on the image
representation of ProtoDUNE-SP data. We first give an overview of the deep
learning models proposed to tackle the task, specifically with Graph Neural
Networks. Then, we proceed to describe the experiments we conduct on simulated
data at ProtoDUNE-SP, including a discussion on the training methods of our
Neural Networks. Finally, we present a comparison of the denoised outputs both
from our deep learning models and the baseline approach focused on deconvolution,
currently implemented within the DUNE software framework. This section in based
on the work we published in~\citep{Rossi:2022}.

\subsection{Proposed models}

Convolutional Neural Networks (CNNs), as described in section~\ref{par:cnns},
are the state-of-the-art technique in image processing. They are based on a sequence
of filtering kernels, whose parameters are trained with gradient descent optimization
to perform some tasks. The convolution operation outputs a feature vector for each
pixel of the input image. This vector is, indeed, a function of the pixel values
in a small neighborhood around the pixel, also named the receptive field. The receptive
field is constrained by the size of the kernel window. Stacking more convolutional
layers one on top of the other, namely increasing the depth of the CNN, is a method
to enlarge this quantity to make the output sensible to more information.
The expected result is an enhancement of the overall network performance. We present
an alternative approach to the problem: we enrich the model with different operations
hoping to increase the expressiveness of the internal representation. The key idea is
to exploit graph-like architectures to include non-local correlations between pixel
values in the transformation.

\subsubsection{Graph Convolutional Neural Network}

\begin{figure}
    \centering
    \myincludegraphics[width=0.4\textwidth]{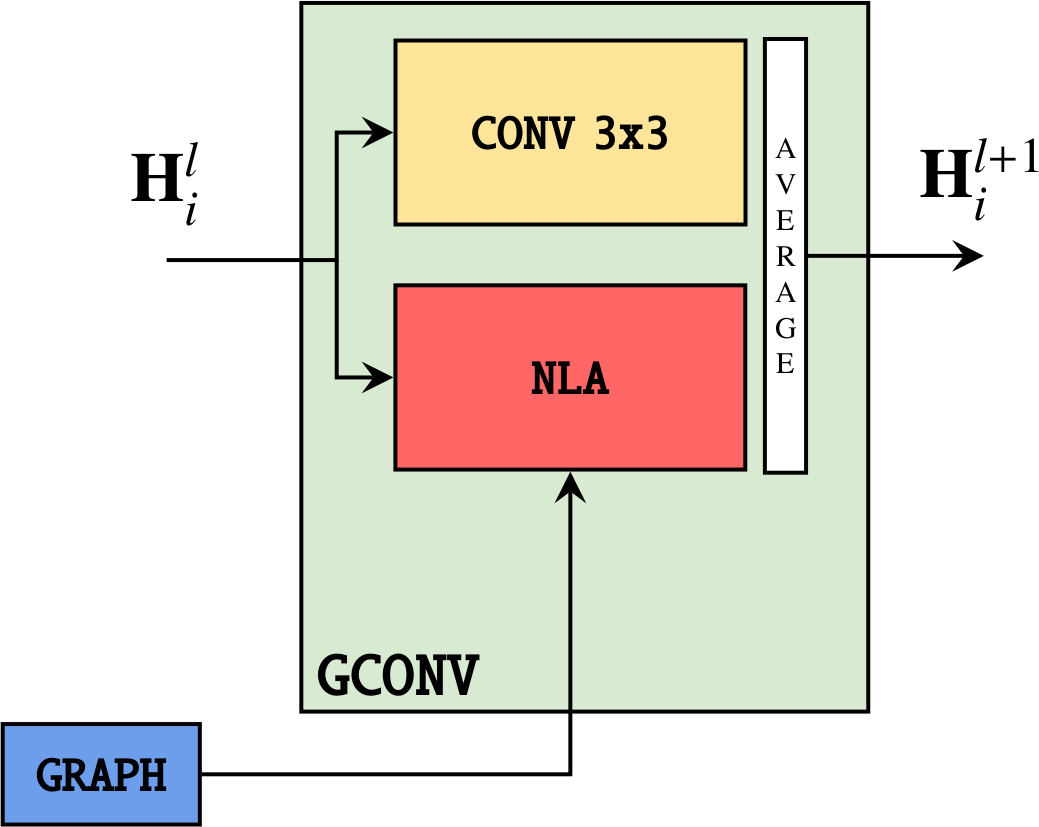}
    \caption{The GCONV layer. The input vector $\mathbf{H}^{l}_i$ is updated
    to $\mathbf{H}^{l+1}_i$ by means of NLA and 2D convolution operations. NLA
    relies on a previously computed KNN graph.}
    \label{fig:gconv}
\end{figure}

We implement a Graph Convolutional Neural Network (GCNN) adapted
from~\cite{Valsesia:2019image,Valsesia:2019deep}, which propose to introduce a new
operation in a denoising neural network called Edge Conditioned Convolution (ECC),
defined for the first time in~\cite{Simonovsky:2017}. We exploit a simplified
version of the ECC layer: the output representation is obtained pixel-wise averaging
a common convolution with a $3\times3$ filtering window and a Non-Local Aggregation
(NLA) operation. The insight is that the ordinary convolution inspects the values
of the neighboring pixels, while NLA exploits long-distance information. Figure~\ref{fig:gconv} shows the
described transformations wrapped by a new kind of layer, dubbed in the following 
Graph Convolution (GCONV).

The NLA builds a $k{-}$NN query graph connecting each pixel to its $k$ closest one
in feature space considering the Euclidean distance. Once the $k$ closest pixels
are identified, their feature vector is combined by a feed-forward layer
to output the updated representation. If at layer $l$, the $i$-th of an $n$-pixel
input image is described by the feature vector $\mathbf{H}^l_i \in \mathbb{R}^{d_l}$,
the result of an NLA operation is:
\begin{equation}
    \mathbf{H}^{l+1}_i = \sigma \Biggl( \frac{1}{|\mathcal{N}_i^l|}
                       \sum_{j\in \mathcal{N}_i^l} 
                        \mathrm{\Theta}^{l} \bigl(\mathbf{H}^l_i-\mathbf{H}^l_j \bigr)                           
                      + \mathrm{W}^l \mathbf{H}^l_i + \mathbf{b}^l
  \Biggr) \hfill \in \mathbb{R}^{d_{l+1}}
  \label{eq:nla}
\end{equation}
where $\mathcal{N}_i^l$ is the $k{-}$neighborhood of pixel $i$ in the space of all the
pixels feature vectors at layer $l$.
$\{\mathrm{\Theta}^{l} , W^l\} \in \mathbb{R}^{d_{l+1}\times d_l}$ and
$\mathbf{b}^l \in \mathbb{R}^{d_l+1}$ are trainable weights and biases shared
throughout pixels. $\sigma$ is the element-wise sigmoid function.

We observe that the operation above finds the $k{-}$nearest neighbors
for each pixel in the input graph, namely builds a $k{-}$NN graph. This requires
an amount of memory proportional to the area squared of the input image. The reason
is that the complexity of a $k{-}$NN query algorithm is quadratic in the number of
points since all the possible $N-1$ distances pairs must be checked and sorted for
each of the $N$ points in the graph, leading to an $\mathcal{O}(N^2)$ scaling behavior.
Our approach is to compute at once the $N\times N$ matrix of pixels pair distances and store it
in memory, after that only the top $k$ contributions for each pixel are considered, resulting
in a final $N\times k$ tensor. An alternative strategy would be to compute the matrix
looping on all the points saving in memory only the $N \times k$ matrix (with $k \ll N$).
Although the latter method optimizes the algorithm memory footprint avoiding
storing the intermediate large result, it sacrifices GPU execution time performance.

With this choice, we still have to take care of memory constraints. As an example,
considering a batch of $32$ input images of $128\times128$ pixels and employing single
precision $32-\SI{}{\bit}$ floating point numbers, the graph construction operation
burden is of order $\mathcal{O(\SI{30}{\mega\byte})}$. If the architecture involves
multiple graph building operations and we allow higher batch sizes
to stabilize the training process, a consumer-grade GPU with $\mathcal{O}(\SI{16}{\giga\byte})$
memory gets easily saturated. Given that the actual raw data from ProtoDUNE-SP are
way bigger than the $128\times128$ image in the example, we have to limit the model
input image size to just small crops of the original data, as explained in the
training paragraph of section~\ref{ssubsec:dunedn-train}.

\begin{figure}
  \myincludegraphics[width=\textwidth]{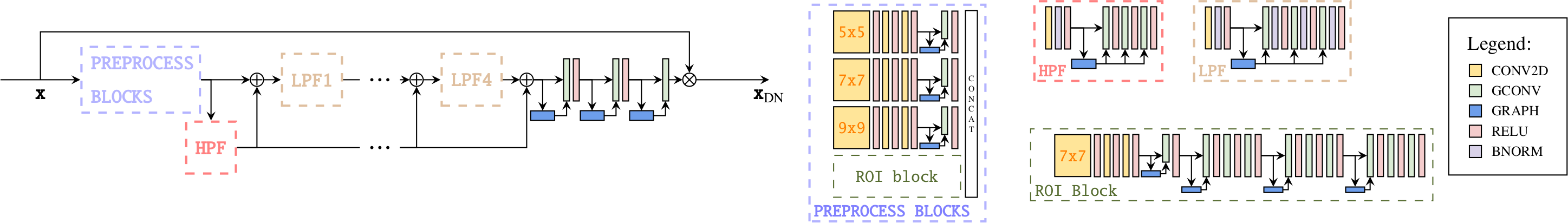}
  \caption{The GCNN neural network architecture. Input and output are batches of
  noisy $\xb$ and denoised \texttt{\textbf{x}$_{\mathrm{DN}}$} images,
  respectively. The network is organized as low-pass (LPF) and high-pass (HPF)
  filters as in~\cite{Valsesia:2019deep}. As explained in section~\ref{ssubsec:dunedn-train}
  we concatenate an ROI block, pre-trained on background vs signal binary segmentation,
  with multi-scale preprocessing layers.
  }
  \label{fig:gcnn}
\end{figure}
Figure~\ref{fig:gcnn} depicts the GCNN network. Note the final residual connection
in the top branch: the usual sum has been replaced by an element-wise multiplication.
The choice is motivated by the nature of the input images, which are mainly comprised
of long tracks separated by empty space. In regions without any signal, it could
be easier to learn how to clean the noise multiplicatively rather than additively.
The network, in principle, does not have to learn to perfectly profile the noise
and subtract it, instead, we can teach the model to predict a mask on the input
image to cut down the uninteresting regions by multiplying the pixels values with small
numbers.

\subsubsection{U-shaped Self-Constructing Graph Network}

As reported in the previous section, the main limitation of the GCNN network is
the memory burden due to the $k{-}$NN graph operation, allowing to feed only pieces
of the original image. The cropping method, unfortunately, prevents establishing
very long correlations between pixels and leads to poor time performance during
inference on large datasets, because the tiles must be sequentially fed into the
GPU and prevent full parallelism.

As an alternative approach to the GCNN, we build upon the work presented
in~\cite{Liu:2020scg}. The authors introduced the Self Constructing Graph-Network
(SCG-Net) model for image segmentation of satellite pictures. They first extract 
a low-dimensional representation of a high-resolution input image with the help of
a CNN. Then, two fully connected layers are trained to build a graph on the pixels,
returning the adjacency matrix as well as the vectors of features associated with
the nodes. At this stage, the graph is further transformed by a couple of graph
layers chosen between the spectral-based Graph Convolutional Network~\citep{Kipf:2016}
and the spatial-based Graph Isomorphism Network~\citep{Xu:2018}. Finally, the
original size image is retrieved through bilinear interpolation.

Upsampling through the interpolation method seems a natural approach when the images
contain dense features since there is a more gradual change in the pixel values
of a small neighborhood. In that context, a certain degree of blurred edges can be
considered acceptable and tolerated. Our use case, however, is rather peculiar in
this sense: we believe that acting on images containing sparse and localized features
potentially with a single upsampling operation washes out the fine-grained information
contained in the input. Therefore, we design an architecture that gradually
rebuilds the original image size in multiple steps.

\begin{figure}
  \myincludegraphics[width=\textwidth]{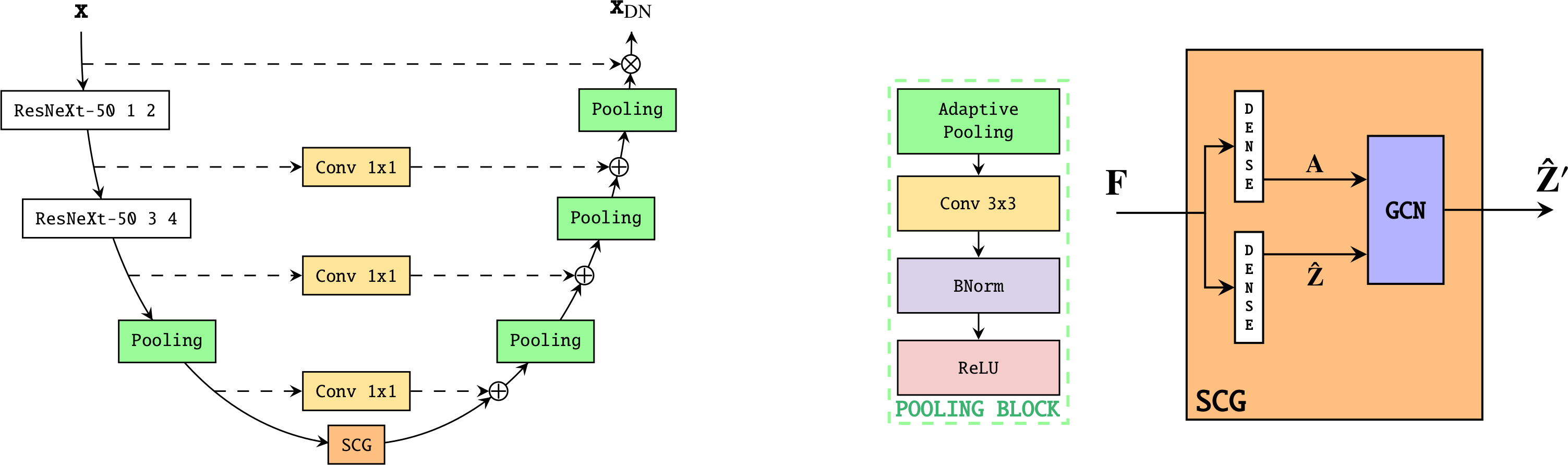}
  \caption{The USCG-Net architecture. The adaptive pooling layer has a two-fold
  utility: in the left branch it downscales the input, while in the right one,
  it provides upsampling. We employ the first $4$ blocks of a pre-trained
  \texttt{ResNeXt-50} in the shallowest layers of the network.
  The blocks respectively contain $9$, $12$, $18$ and $9$ convolutional layers.
  The horizontal dashed lines represent residual connections with a $1\times1$
  convolution for enhanced expressivity and are employed to match the number of
  image channels between the left and right branches. The SCG layer builds the
  adjacency matrix $\mathbf{A}$ and the vectors of node features $\mathbf{\hat{Z}}$
  to be passed into a GCN layer to output the final representation of the nodes
  $\mathbf{\hat{Z}'}$.
  }
  \label{fig:uscg}
\end{figure}

We introduce the U-shaped Self Constructing Graph Network (USCG-Net), where a
U-Net-like~\citep{Ronneberger:2015} network structure with residual connections
carries the information all the way through the data pipeline.
Figure~\ref{fig:uscg} shows the USCG-Net architecture. The pooling blocks contain
the Adaptive pooling layer, which is responsible for resizing its inputs. A
pre-trained \texttt{ResNeXt-50}~\citep{Xie:2017} with a $32\times4$ template is
employed to construct an initial feature map serving as input of the SCG layer.
The early-layer representation of the pre-trained network should be generic enough
to catch the spatial features of the inputs and drift the training process
during the initial phases of the optimization. We remark the residual skip
connections between the downsampling and upsampling branches of the network. The
usual sum has been replaced by a convolution with a $1\times1$ kernel to
increase the complexity of the network. Finally, the residual link displays the
same multiplication trick employed for the GCNN network.

The SCG layer is at the core of the network: it turns the image into a graph allowing
distant connections between pixels. The input is an image
$\mathbf{F} \in \mathbb{R}^{h\times w\times d}$ and maps it into the adjacency
matrix $\mathbf{A} \in \mathbb{R}^{n\times n}$ and node feature vectors
$\mathbf{\hat{Z}} \in \mathbb{R}^{n\times c}$ through two encoding dense layers.
In the previous formulae, $d$ and $c$ are respectively the input and output
channel dimensions and $n = h\times w$ is the number of extracted nodes. In our
experiment we further transform the graph encoding through the GCN layer, leaving
the investigation of other options to future works. $\mathbf{\hat{Z}'}$ can be
finally projected back into the full-size image
$\in \mathbb{R}^{h \times w \times c'}$ by the pooling blocks in the right
branch of the USCG-Net.

\subsection{Experimental results}

In this section, we describe the experiments conducted to validate our models. We
proceed to introduce the datasets generate to carry on the study, the training
methodology and the actual results.

\subsubsection{Datasets}

\begin{table}
  \centering
  \begin{tabular}{lll}
    \hline\noalign{\smallskip}
    \dunetpc & $n$ events & $p$ energy $(\SI{}{\GeV})$\\
    \noalign{\smallskip}\hline\noalign{\smallskip}
    \veight & 10 & $2$ \\
    \vnine & $70$ & $0.3$, $0.5$, $1$, $2$, $3$, $6$, $7$\\
    \noalign{\smallskip}\hline
  \end{tabular}
  \caption{Datasets for training and testing. The two samples differ in producer
  package version, size and event beam energies. The second dataset contains
  $10$ events for each $p$ energy specified.}
  \label{tab:dunedn-datasets}
\end{table}
We describe here the datasets employed to train and test the proposed
models. We remark that the present results have been tested on simulated events
only: the inclusion of detector data is beyond the scope of this first stage of
the investigation. We, therefore, simulate interactions within the ProtoDUNE-SP
detector through the LArSoft~\cite{Church:2014} framework, in particular with the
\dunetpc package. We consider events containing the interactions induced by a proton
beam of various energies and cosmic rays with the Argon targets. Our supervised
training approach is based on simulated raw digits. The simulation software includes
the possibility to switch off the electronics noise affecting the charge depositions:
we employ this information as target outputs for the denoising models.

\begin{figure}
  \centering
  \subfigure[Collection plane view]{
    \myincludegraphics[width=0.3\textwidth]{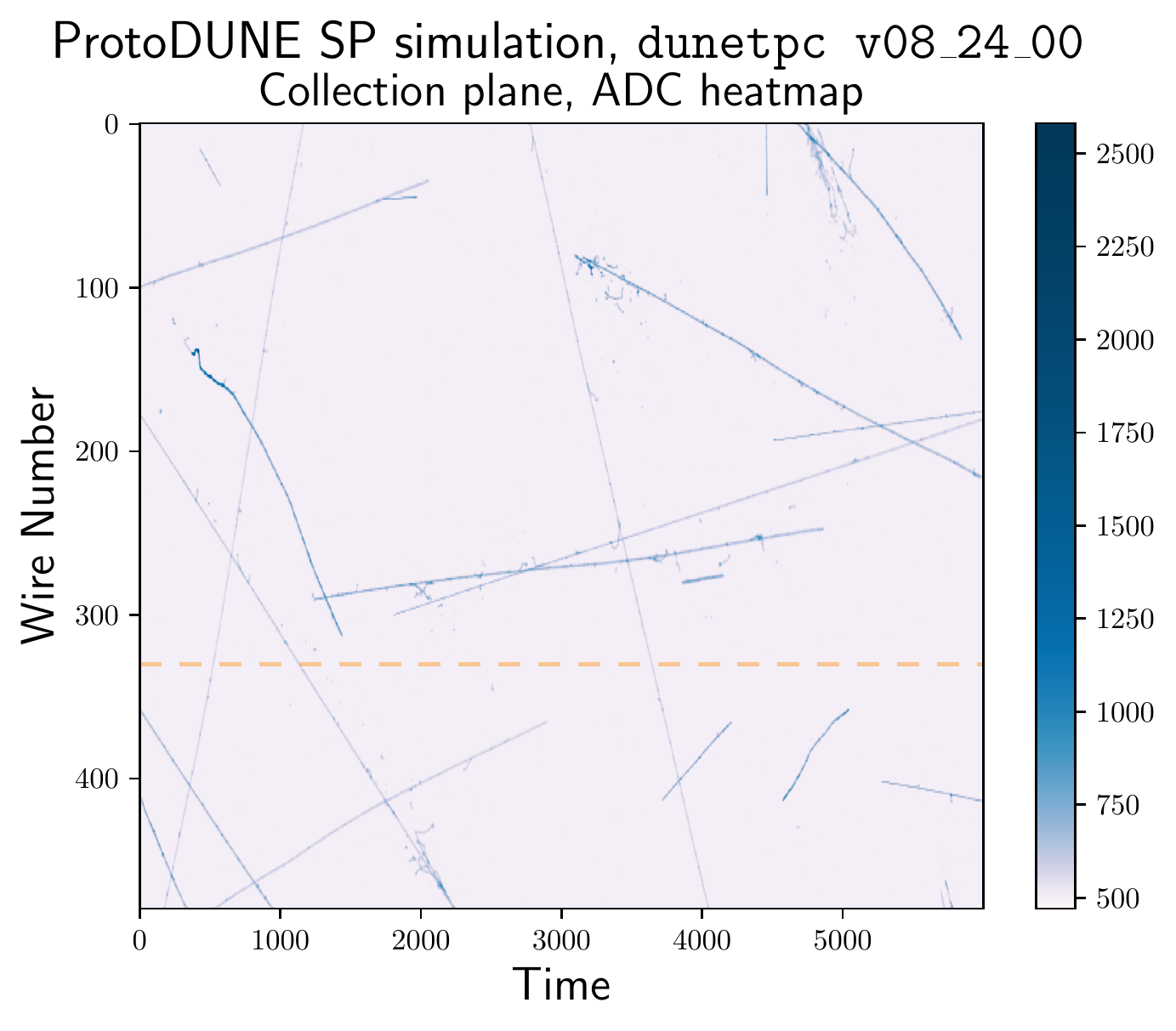}

  }
  \hfill
  \subfigure[Noisy waveform]{
    \myincludegraphics[width=0.3\textwidth]{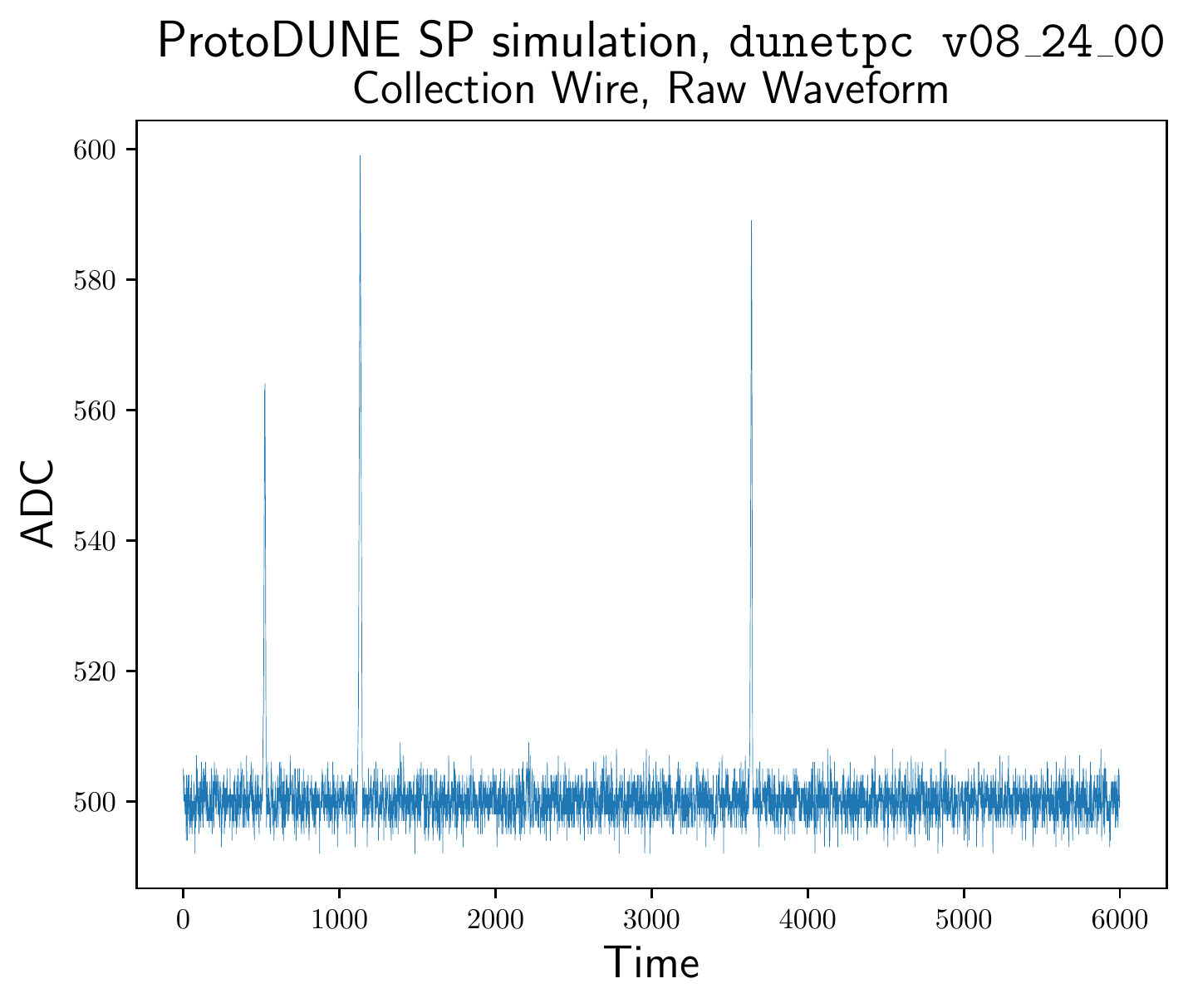}
  }
  \hfill
  \subfigure[Clear waveform]{
    \myincludegraphics[width=0.3\textwidth]{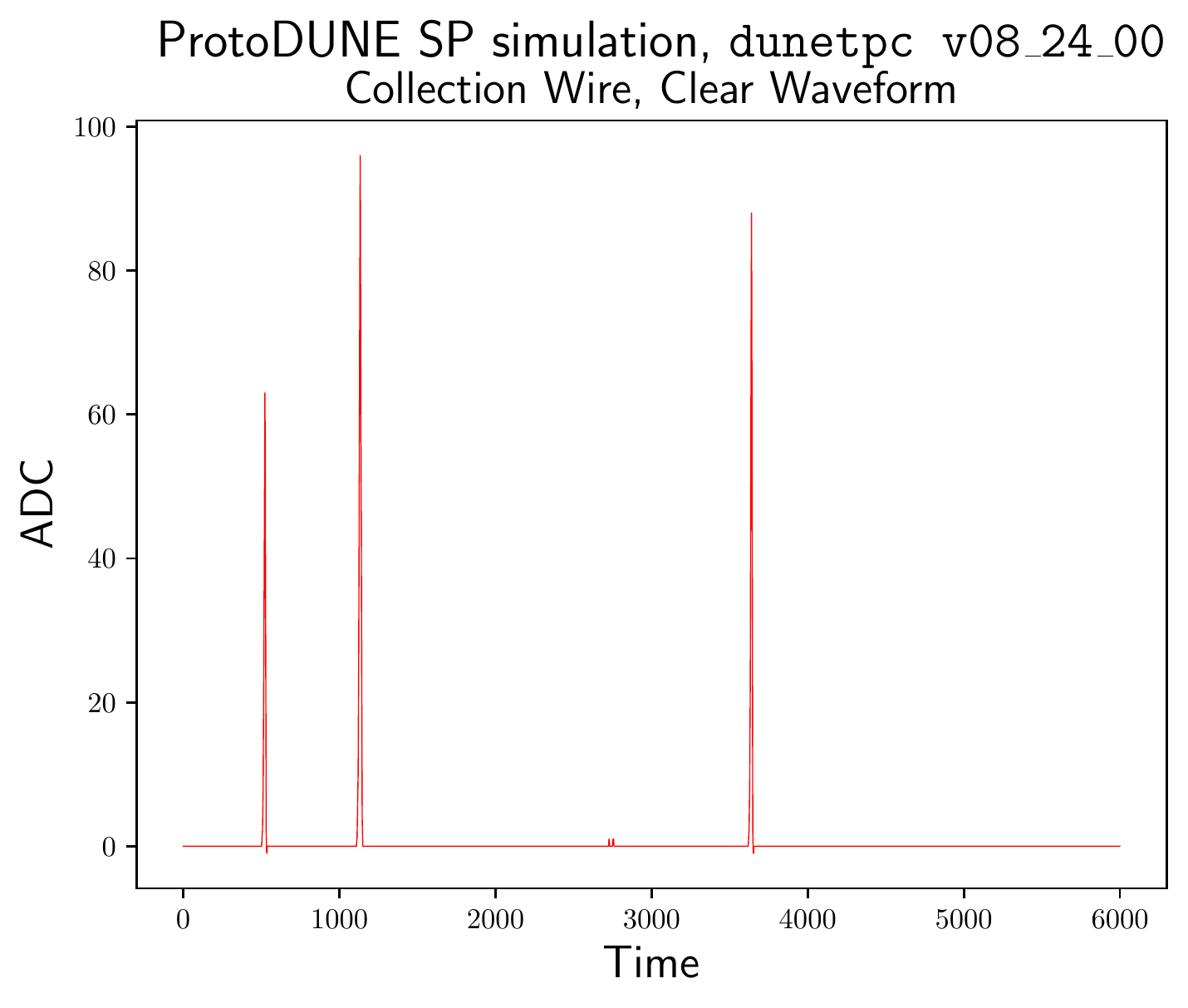}
  }
  \caption{Example taken from \dunetpc \veight dataset. The horizontal orange
  dashed line in the first panel marks the channel to extract the waveforms from.}
  \label{fig:data v8}
\end{figure}

\begin{figure}
  \centering
  \subfigure[Collection plane view]{
    \myincludegraphics[width=0.3\textwidth]{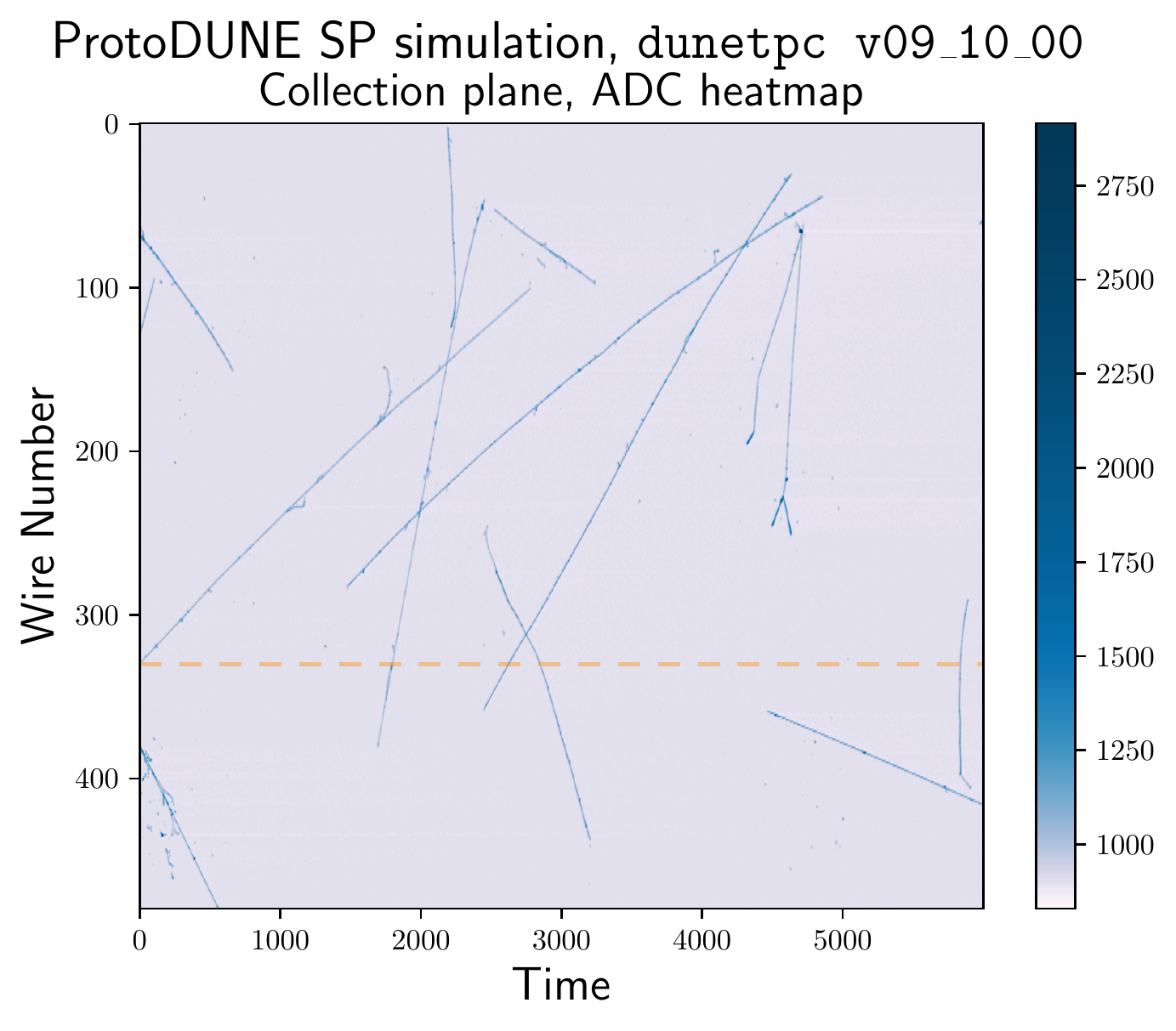}

  }
  \hfill
  \subfigure[Noisy waveform]{
    \myincludegraphics[width=0.3\textwidth]{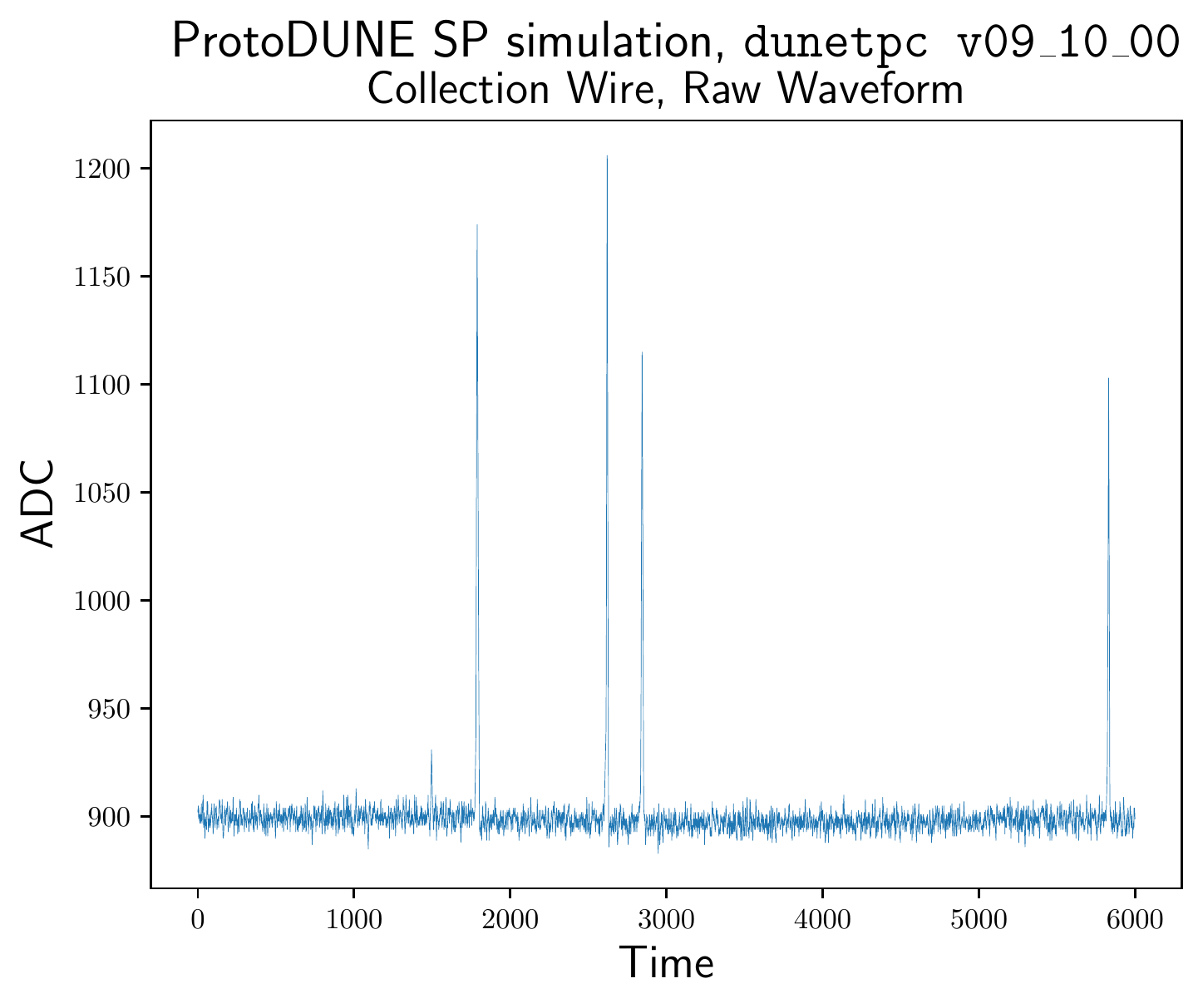}
  }
  \hfill
  \subfigure[Clear waveform]{
    \myincludegraphics[width=0.3\textwidth]{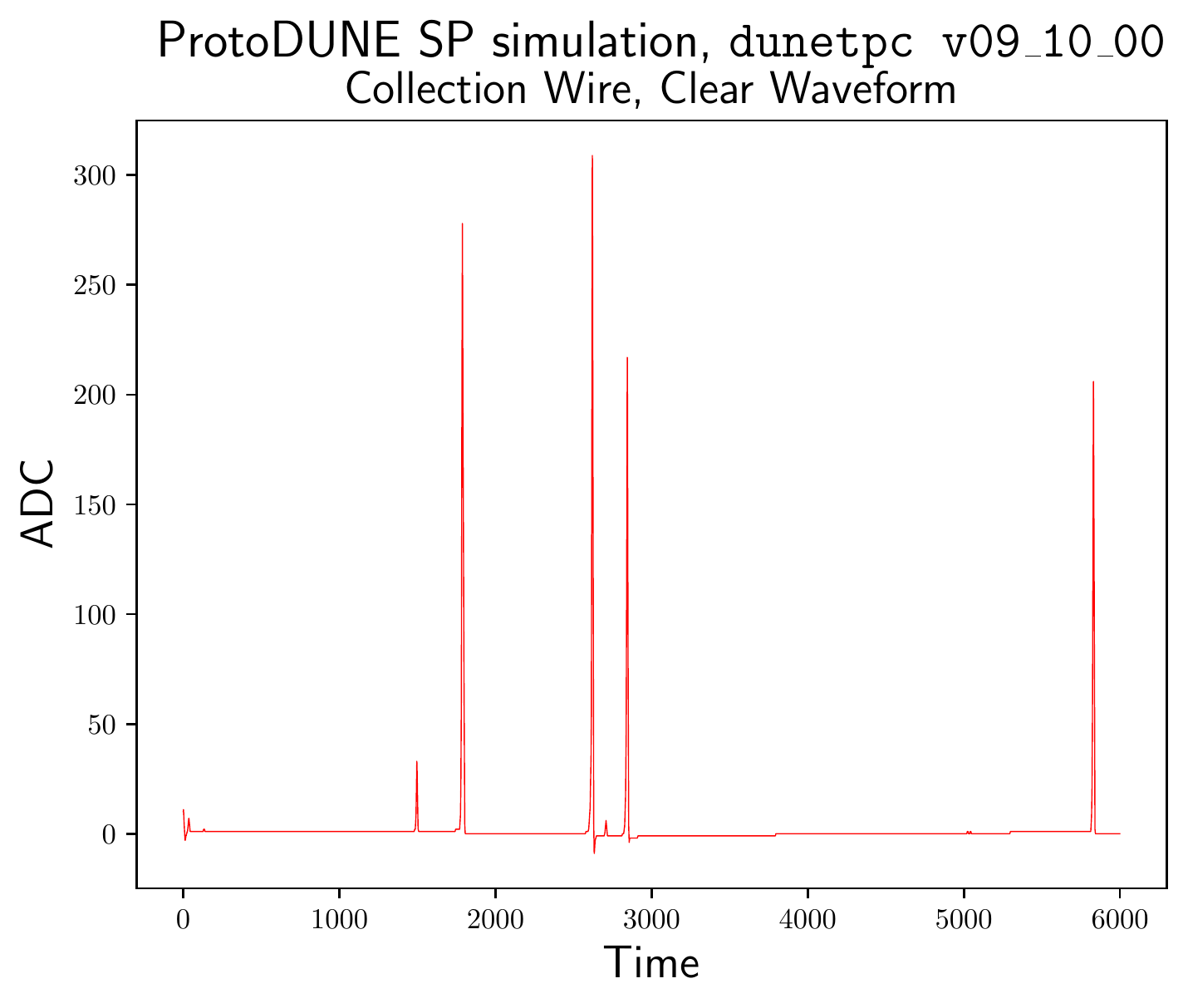}
  }
  \caption{Example taken from \dunetpc \vnine dataset. The horizontal orange
  dashed line in the first panel marks the channel to extract the waveforms from.}
  \label{fig:data v9}
\end{figure}

Table~\ref{tab:dunedn-datasets} lists the composition of the generated datasets.
In figure~\ref{fig:data v8}
and~\ref{fig:data v9} we show visual examples from \veight and \vnine datasets.
The plots contain simulated raw digits, and $1{-}$dimensional waveforms extracted
from the channel marked with the orange dashed line. We refer to waveforms with and
without noise contributions as noisy and clear waveforms, respectively. We consider
the \veight dataset a simplified version since it is smaller and contains easier
features to denoise and segment than the \vnine one. \vnine, indeed, is generated according
to a detector data-driven approach. In the latter sample, clear waveforms are non-zero
even in regions without signal, this reflects the impossibility to measure peaks
with infinite resolution. Moreover, low-frequency negative tails recorded just
after big spikes are visible in figure~\ref{fig:out v9}.

We split our datasets in $80/10/10\,\%$ slices for training, validation and testing.
The validation set is used to choose the best model, while the test set is employed
to present the final results given in this section. We aggregate the performance
results for each metric in a single measurement: that is the mean $\bar{x}$ and the standard
error of the mean $\sigma$ of the sampling distribution, representing the metric expected
value and its uncertainty. This criterion influences also the selection of the best model 
during training, namely, the network checkpoint at the epoch end that achieves the
lowest upper bound $\bar{x} + \sigma$ of the loss function interval. With this
choice, we aim to minimize both the bias and the variance of the model.

We remark that a single event at ProtoDUNE-SP is comprised of multiple views
of the activity in the detector. Therefore, one event does not count only as one
training point for our neural network. The GCNN network processes tiles
of the original images. Considering crops of size $32\times32$, each event in the
dataset contains $9\cdot10^4$ of those. The USCG-Net, instead, processes entire
views. The two datasets account for $18$ and $126$ of such views used to assess
the model performance and the associated uncertainty. Note that the dimensionality
of the output image is of the order of millions of pixels, each one contributing
to the evaluation of the final performance metric. 

\subsubsection{Network training}
\label{ssubsec:dunedn-train}

We employ the same pre-processing procedure for all our models, namely, we subtract
the median value from each ProtoDUNE-SP view independently. According to the approach
currently adopted by DUNE and based on the discussion in section~\ref{sec:deconvolution},
the median value gives a good estimate of the noise pedestal quantity. The median
subtraction step is fundamental to have waveform values centered around zero and
it is expected to provide accuracy enhancements and training stability. The inputs
are further normalized in the unit interval to prevent gradient divergence as it
is a technique commonly exploited in machine learning applications.

The GCNN data processing is constrained by the memory issues described above. As
already anticipated, we employ a data parallel approach, cropping the inputs into
$32\times32$ pixels images and processing every tile independently.
This solution requires some subtleties in the training process due to the nature
of the inputs.

Given the sparsity of the input images, it is likely that the majority of the crops
contain little to no signal. It is meaningless to feed the network with multiple
empty crops. To speed up the training, we decide to train on just a subset
of the available crops. This choice, in turn, triggers a second issue: sampling
randomly the subset of crops provides an extremely unbalanced dataset, where we
are very likely to miss crops containing interesting charge depositions. Hence,
we fix the percentage of crops containing signal to balance the signal-to-background
pixel ratio: in our experiments, we keep this quantity at $99\%$. The cropping
procedure works as follows. We mark as signal all the pixels in clear waveforms
that have non-zero ADC value count in a plane view. Then, we sample randomly the
desired number of crop centers, complying with the exact ratio of signal to
background percentage. Finally, within the plane views we specify a bounding box
around the drawn centers.

We pre-train the Region Of Interest (ROI) block of the GCNN network for image
segmentation, which is a classification task at the pixel level with labels usually
belonging to a finite set of exhaustive and mutually exclusive classes. The ROI
block, indeed, is trained to distinguish between pixels that contain or not
electron-induced current signals. The Roi Block is then attached as in~\ref{fig:gcnn}
to the GCNN network, which is trained for image denoising, which is a regression
aiming at subtracting the noise contributions from each pixel intensity.

The two parts of the network are optimized following an ad-hoc strategy. First,
we pre-train the ROI block alone on image segmentation for $100$ epochs, storing
the best weight configuration: at this stage, the block learns to distinguish between
signal and background, i.e. empty pixels. We employ the standard binary classification
loss function, namely binary cross-entropy. The optimizer follows the AMSGrad
variant~\citep{Reddi:2018} of the Adam algorithm~\citep{Kingma:2017} with a
starting learning rate of $10^{-3}$. After pre-training completion, we train
the full GCNN network on image denoising for $50$ epochs and save the
best-performing model on the validation dataset.

We choose again AMSGrad as the optimization algorithm for the full denoising network,
however, this time, we set the initial learning rate to a slightly higher value
$9\cdot10^{-3}$. We also build a custom loss function made of two contributions:
the mean squared error (MSE) $\mathcal{L}_{MSE}$ between
labels and outputs and a loss function $\mathcal{L}_{ssim}$ derived from the
statistical structural similarity index measure (stat-SSIM)~\cite{Channappayya:2008}.
stat-SSIM for two images $\xb$ and $\yb$, is obtained through the following equation:
\begin{equation}
  \mathrm{stat{\text-}SSIM} (\xb, \yb) =
  \frac{1}{n_p n_c} \sum_{pc}
  \braces{
    \frac{2\mu_x\mu_y + \epsilon_\mu}{\mu_x^2 + \mu_y^2 + \epsilon_\mu}
    \times
    \frac{2\mathrm{E}[(\xb-\mu_x)(\yb-\mu_y)] + \epsilon_{\sigma}}
      {\mathrm{E}[(\xb-\mu_x)^2]+ \mathrm{E}[(\yb-\mu_y)^2] + \epsilon_{\sigma}}
  }_{pc}
  \label{eqn:statssim}
\end{equation}
where $\mu_i$ is a shorthand for the image $\mathbf{i}$ expected value $\mathrm{E}[\mathbf{i}]$,
which is computed through convolution with an $11 \times 11$ Gaussian kernel of
standard deviation $\sigma=3$. All the other expectation values in the equation
are calculated by convoluting the quantity appearing in the argument with the same
Gaussian filter. $\epsilon_\mu$ and $\epsilon_\sigma$ are two regulators that
limit the maximum resolution at which the fractions in the equations are computed,
imposing a cut-off on the mean and variance expected values, respectively.

When both the numerator and the denominator of the fraction reach much smaller
values than the corresponding $\epsilon$, the output gets close to one. In the
experiments, $\epsilon_\mu$ and $\epsilon_\sigma$ are fixed to $0.5^2$. This
choice implies that, for the stat-SSIM computation, we estimate the means and
standard deviations of the distributions at scales larger than half of one ADC
value, namely the granularity of the recorded detector hits. The result is finally
averaged over the entire image containing $n_p$ pixels and $n_c$ channel dimensions.
The quantity in equation~\ref{eqn:statssim} takes values in the range $[-1,1]$ and
approaches $1$ only if $\xb=\yb$.

The associated loss function is then given by $\mathcal{L}_{ssim} = 1 - \mathrm{stat{\text-}SSIM}$:
it is a perceptual loss, in the sense that tries to assess the fidelity of the
image by focusing on structural information. It relies on the idea that pixels may
have strong correlations, especially when they are spatially close. In contrast,
MSE evaluates absolute differences of pixels, without taking into account any
dependence amongst them. More details on the interpretation of these quantities
can be found in~\cite{Wang:2009}. The two contributions in the loss function are
weighted as follows:
$\mathcal{L} = \alpha \cdot \mathcal{L}_{MSE} + (1-\alpha) \cdot w \cdot \mathcal{L}_{ssim}$.
We fine-tune the multiplicative parameter $w=10^{-3}$, to balance the gradients
with respect to the model's trainable parameters provided by the two terms in the
sum. The parameter $\alpha$ is fixed to $0.84$ as in~\cite{Zhao:2018}.

In the rest of this section, we will refer, with a slight abuse of notation, to a
CNN as a GCNN network with Graph Convolutional layers replaced by plain Convolutional
ones.

The training of the USCG-Net is straightforward: cropping is not used, so no
sampling method is needed. Although the entire model fits in a single GPU with
$\SI{16}{\giga\byte}$ of memory, we prefer to employ a sliding window mechanism
as in the original SCG-Net paper~\cite{Liu:2020scg}. We split the raw digits matrix
along the time dimension with a $2000$ pixel wide window and stride of $1000$
pixels. After the forward pass, the results are combined by averaging predictions
on overlapping regions. The USCG-Net is trained to minimize the MSE function
between model outputs and clear raw digits, with AMSGrad optimizer and a learning
rate of $10^{-3}$. We dropped the stat-SSIM contribution from the loss function
after we experienced training convergence problems including that term in the
experiment.

\subsubsection{Experimental results}

The metrics used to assess the goodness of our models and benchmark them against
the state-of-the-art approach are four: structural similarity index measure
(stat-SSIM), mean squared error (MSE) and peak signal-to-noise ratio (pSNR) are
the standard for image denoising tasks; the last is a custom one, dubbed integrated
mean absolute error (iMAE) and will be defined below.

The pSNR is a function of the MSE between a noise-free image $\xb$ and a denoised
one $\yb$:
\begin{equation}
  \mathrm{pSNR}(\xb,\yb) = 10\cdot \log_{10}
   \frac{\max^2(\xb)}{\mathrm{MSE}(\xb,\yb)}
\end{equation}
Note that the pSNR and stat-SSIM increase for better reconstruction of the signal
in the data, while the opposite happens for the MSE. We observe that these three
quantities struggle to grasp the goodness of the baseline model. Indeed, it aims
at fitting gaussian peaks in masked regions, rather than reconstructing the precise
shape of the spike contained in the raw digits. Hence, the baseline tool performs
inevitably poorly on the considered metrics and specifically, up to our knowledge,
there is no default metric in the literature to assess its performance.
Nonetheless, we define a custom quantity that tries to compare the different
approaches: we highlight that the deconvolution process does not preserve
waveform amplitudes, but their integrals, namely the charge on the ProtoDUNE-SP
wires. For such reason, we decide to evaluate the integrated Mean Absolute Error
(iMAE) on the wires integrated charge:
\begin{equation}
  \label{eq:imae}
  \mathrm{iMAE}(\xb, \yb) =
  \frac{1}{n_w}\sum_{w=1}^{n_{w}} \abs{\sum_t (\xb - \yb)_{wt}}
\end{equation}
where the sum inside the absolute value runs over time for the whole readout
window, while the outer sum aggregates the result over the wire dimension.

\begin{figure}
  \centering
      \myincludegraphics[width=\textwidth]{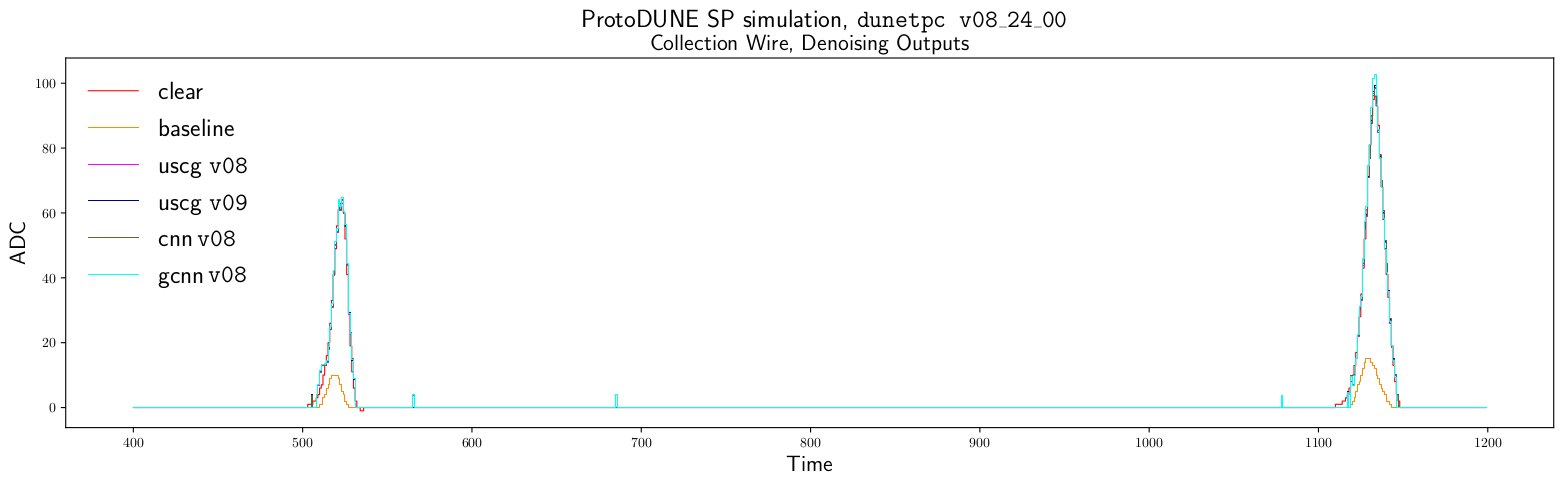}
  \caption{Detail of a raw waveform from  \dunetpc \veight dataset: label,
  traditional algorithm and neural network outputs. The version,
  \texttt{v08} or \texttt{v09}, next to the model name in the legend
  refers to which dataset the corresponding model was trained on.}
  \label{fig:out v8}
\end{figure}

\begin{figure}
  \centering
      \myincludegraphics[width=\textwidth]{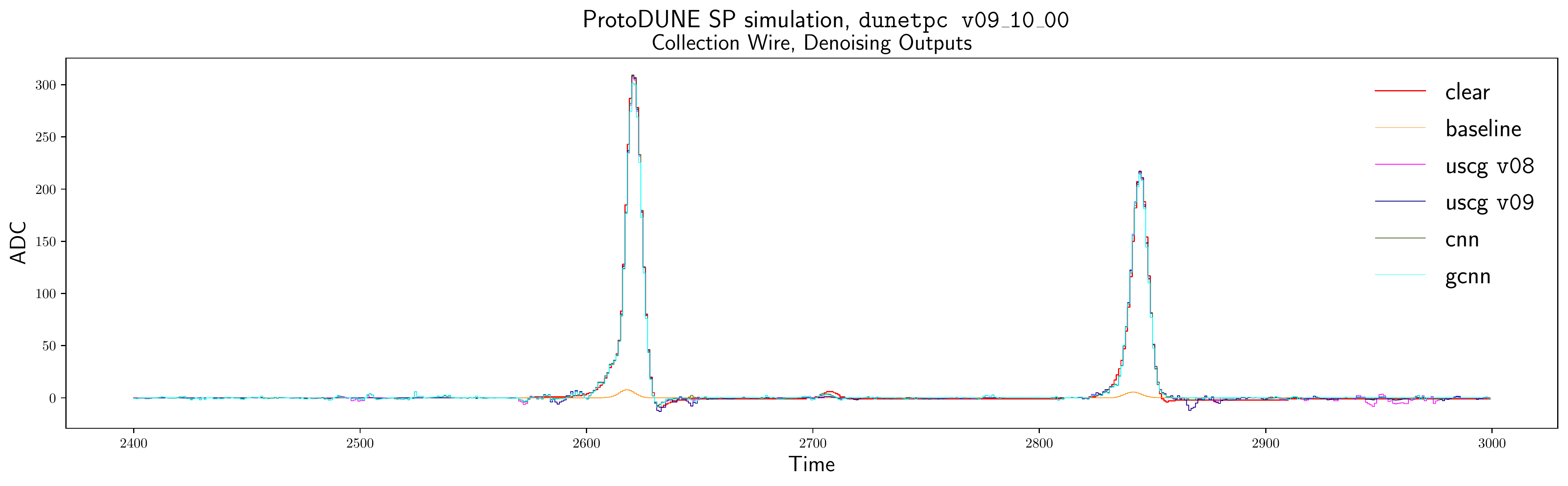}
  \caption{Detail of a raw waveform from  \dunetpc \vnine dataset: label,
  traditional algorithm and neural network outputs. The version,
  \texttt{v08} or \texttt{v09}, next to the model name in the legend
  refers to which dataset the corresponding model was trained on.}
  \label{fig:out v9}
\end{figure}

The deconvolution approach does not preserve amplitudes because of the filtering
function applied in Fourier space. If such a filter is normalized, then the
transformation preserves the integral. Furthermore, the deconvolution outputs
are known up to an overall normalization constant, which we fit on the datasets
to minimize the iMAE quantity. We show that, although we perform this
operation on the outputs of the baseline tool for a fair comparison against our
models, they nonetheless achieve a worse iMAE score. Table~\ref{tab:res-v08}
collects the metrics values evaluated on the \veight dataset. We gather \vnine
dataset results in table~\ref{tab:res-v09}. We present the scores only for the
events with $\SI{2}{\GeV}$ beam energy, since we find a flat distribution of
the considered metrics in the energy parameter. Figures~\ref{fig:out v8}
and~\ref{fig:out v9} show samples of labels and denoised waveforms.
\begingroup
\setlength\tabcolsep{2pt}
\begin{table}
  \scriptsize
  \centering
  \begin{tabular}{lllll}
    \hline\noalign{\smallskip}
    Model                      & stat-SSIM                 & PSNR                & MSE                & iMAE         \\
    \noalign{\smallskip}\hline\noalign{\smallskip}
    Baseline & \multicolumn{1}{c}{-}& \multicolumn{1}{c}{-} & \multicolumn{1}{c}{-} & $5391\pm 1622$    \\
    CNN {\tt v08}  & $0.471\pm 0.008$       & $67.3\pm 1.2$         & $0.57\pm 0.03$        & $287\pm 12$       \\
    GCNN {\tt v08} & $0.512\pm 0.011$     & $70.12\pm 1.4$        & $0.30\pm 0.01$        & $191.4\pm 2.6$    \\
    USCG {\tt v08} & $\bf 0.988\pm 0.005$ & $\bf 72.66\pm 1.54$   & $\bf 0.17\pm 0.02$    & $95.5\pm 8.5$     \\
    USCG {\tt v09} & $0.926\pm 0.007$     & $72.3\pm 1.5$         & $0.18\pm 0.02$        & $\bf 76.3\pm 8.2$ \\
    \hline\noalign{\smallskip}
  \end{tabular}
  \caption{
    Test metrics for denoising on \veight dataset. Results for collection plane
      and $\SI{2}{\GeV}$ beam energy only. \texttt{v08} or \texttt{v09} in the
      first column refer to which dataset the corresponding model was trained on.
  }
  \label{tab:res-v08}
\end{table}
\begin{table}
  \scriptsize
  \centering
  \begin{tabular}{lllll}
    \hline\noalign{\smallskip}
    Model                      & stat-SSIM            & PSNR              & MSE                & iMAE $[\times 10^3]$ \\
    \noalign{\smallskip}\hline\noalign{\smallskip}
    Baseline & \multicolumn{1}{c}{-}& \multicolumn{1}{c}{-} & \multicolumn{1}{c}{-} & $5.86\pm 0.52$       \\
    CNN {\tt v08}  & $0.37\pm 0.02$     & $57.3\pm 1.4$         & $5.79\pm 0.88$        & $4.16 \pm 0.36$      \\
    GCNN {\tt v08} & $0.40 \pm 0.02$    & $57.7\pm 1.5$         & $5.27\pm 0.69$        & $4.51 \pm 0.39$      \\
    USCG {\tt v08} & $0.65\pm 0.05$     & $61.1\pm 1.6$         & $2.3\pm 0.2$          & $\bf 2.18\pm 0.29$       \\
    USCG {\tt v09} & $\bf 0.81\pm 0.07$ & $\bf 61.8\pm 1.7$     & $\bf 1.99\pm 0.19$    & $2.25\pm 0.23$   \\
    \noalign{\smallskip}\hline
  \end{tabular}
  \caption{
      Test metrics for denoising on \vnine dataset. Results for collection plane
      and $\SI{2}{\GeV}$ beam energy only. \texttt{v08} or \texttt{v09} in the
      first column refer to which dataset the corresponding model was trained on.
  }
  \label{tab:res-v09}
\end{table}
\endgroup

USCG-Net-like networks exceed GCNN-like ones in all the collected metrics. In
order to have a first assessment of the quality of the neural network generalization power, we train
two versions of the USCG-Net: one on the \veight dataset and the other on the \vnine
dataset. We decide not to train the GCNN-like networks on the \vnine dataset after
we observed difficulties in training convergence as well as long training times
on such a big dataset. We evaluate these networks on both datasets.

Following expectations, the networks trained and applied on the same dataset lead
to better performance. The only exceptions are given by the iMAE columns,
where the USCG-Net trained on the dataset opposite to the testing one, achieves the best iMAE score.
The stat-SSIM index score drops significantly for GCNN-like networks.
All the networks, nonetheless, show hints of overall good generalization power when they are
applied to datasets not used for training. This fact is well supported by the PSNR
columns, which show that even the worst model achieves competitive results. We underline
that the USCG-Net is not trained according to the stat-SSIM quantity: adding an extra 
term in the loss function containing such a term could be considered a point of
further development of the present research.

As a final remark, we observe that the USCG-Net performs better than the other
architectures. Although a thorough investigation of the motivations behind this
performance gap has not been carried out yet, we advance two possible hypotheses
to try to explain this behavior. The first idea lies in the ability of the
USCG-Net to process entire planes at once, managing to connect pixels very far
apart. This might enhance the performance due to the particular nature of the
inputs: muons inside ProtoDUNE-SP are revealed by the detector as tracks that
extend for hundreds of centimeters, i.e. they span big portions of
the raw digit images as in figure~\ref{fig:data v8} and figure~\ref{fig:data v9}.
Therefore, the USCG-Net network is allowed to reconstruct entire track objects.
This is not possible with the cropping approach used for the GCNN network, which
inspects portions of inputs independently, inevitably splitting the signal.

The second insight is linked to the fact that we exploited a pre-trained
\texttt{ResNeXt-50} network in the early layers of the USCG-Net architecture.
Instead, the GCNN network is trained from scratch with randomly initialized
weights. It is a common technique in deep learning to exploit some layers of
a network pre-trained on a different task to help the training convergence.
This procedure is known as transfer learning. The key point is that the
pre-trained layers should already contain some basic domain knowledge that will
not have to be learned anymore. In our particular image-processing use case, it
could be like we were allowing the network to start the training with some
general knowledge of the geometrical objects that it might find within an input
image. This helped the whole training to converge to a more performant minimum
of the loss function search space.

%% file: chapters/partIII/chap6/chap6.tex
\chapter{Clustering at ProtoDUNE-SP with supervised learning}
\label{chap:slicing}
\thispagestyle{plain}

\minitoc

\renewcommand{\myincludegraphics}[2][width=\textwidth]{
    \includegraphics[#1]{chapters/partIII/chap6/plots/#2}
}

In this chapter, we present the implementation of a neural network trained to
solve the slicing problem. Slicing is an essential step in the reconstruction
workflow for neutrino detectors and, in particular, this work applies to
ProtoDUNE-SP detector. We set the stage by introducing the Pandora framework, a
dedicated software for event reconstruction in physics experiments. Pandora is
widely employed by many neutrino collaborations. In the second section of this
chapter, we detail a novel deep learning approach to the problem, namely the
Cluster Merging Network~\citep{slicing:2021}.

\section{The Pandora framework}

Event reconstruction at physics experiments is the process of extracting high-level
features from a collection of detector electronics signals to
determine the event particle content along with their properties. The output
of such workflow is a list of particle objects associated with their corresponding
detector signals. The Pandora framework~\cite{Marshall:2015} is a multi-pattern
recognition software dedicated to reconstruction, firstly built for linear
collider experiments, such as the International Linear Collider
(ILC)~\citep{Thomson:2009} and the Compact Linear Collider
(CLIC)~\citep{Marshall:2012}, and then extended to LArTPC detectors. In
particular, it has been tested in the context of the MicroBooNE
experiment~\citep{Acciarri:2018}. In the following we will focus on applications
to LArTPC detectors only, giving an overview of the Pandora reconstruction
framework.

The Pandora software development kit (SDK) is a tool implemented in \texttt{C++}
programming language and defines the data structures needed to run several
reconstruction algorithms. It is specifically designed to be interfaced with
external Event Data Models (EDMs), such as the one provided by
LArSoft~\citep{Church:2014} which implements the simulation software for
LArTPCs. The Pandora application can run in standalone mode providing an
algorithm to export the LArSoft simulated events into the Pandora format and
vice-versa. In this scenario, Pandora defines two kinds of
structures: Input and Algorithm objects. Products of the first kind are created by
the Pandora application before running the reconstruction algorithms and cannot
be changed by them. On the other hand, the Algorithm objects are created and
modified by reconstruction algorithms.

The Pandora SDK implements three kinds of Input Objects.
\emph{CaloHit}: represents a space point deposition in the detector and is the
fundamental building block of the Pandora event representation. Each CaloHit is
associated with the $x$, $y$, $z$ detector coordinates and the relative deposited
energy $E$. Within a LArTPC a CaloHit is built from processed raw signals of the
form discussed in chapter~\ref{chap:dunedn}: a single peak in an ROI region is
fitted with $N$ multiple Gaussians to allow the possibility to have
several single depositions within a single peak. Each fitted gaussian in the peak
is turned as a separate CaloHit. LArTPC raw data are described by a drift
coordinate ($x$ coordinate) and a readout channel number identifier, indeed,
Pandora CaloHits are initially given a null $y$ component and a $z$ component
referring to the position of the wire where the signal was first detected in the
wire pitch direction. The transformation from the wire plane representation to
the canonical detector $3{-}$dimensional cartesian coordinate system involves a
rotation-like operation that requires at least two wire plane coordinates to be
known for each CaloHit. An additional flag for each CaloHit signals which of the
$3$ different wire planes ($\mathrm{U}$, $\mathrm{V}$ or $\mathrm{W}$) recorded
the corresponding signal.

\emph{Track}: represents a collection of CaloHits objects arranged in a
continuous pattern. Beyond the CaloHits and their attributes belonging to the
Track object, this holds other global information like associated particle
momentum and trajectory direction. The Track object can also have methods to
retrieve the parent, daughter and sibling relationships with other Tracks with
the intent to build a particle tree to describe the interaction hierarchy
contained in the event.

\emph{MCParticle}: is an object collecting true information from simulation (of
course, detector data do not hold this kind of details) about an event particle.
This kind of data structure is carried within Pandora mostly for development
reasons, as it allows to cheat parts or even entire algorithms in the
reconstruction workflow. It can also be used to measure performance metrics for
the tools' implementations. The key idea is that the best method should reproduce
as much as possible the true information contained in the MCParticle objects and
the related MCParticleLists.

The Pandora framework implements also Algorithm objects: auxiliary products that
represent outputs at various steps of the reconstruction process. Below we
introduce the most important ones.

\emph{Cluster}: represents a collection of CaloHits carrying space and energy
information. It also usually contains estimations of Cluster global properties,
such as results of energy or geometrical fits. Being an Algorithm object, Pandora
implements also operations that allow the algorithms to create, read, split and
merge different Cluster objects.

\emph{Vertex}: this object might not correspond to a true CaloHit
deposition, but it is rather the output of a fitting procedure on the available
event CaloHits. It identifies the exact space point where an interaction took place
during the readout window. It naturally holds also information about the particle
involved in the interaction, highlighting the parent-daughter structure of the
physical scattering.

\emph{ParticleFlowObject} (PFO): this can be seen as the final reconstruction
output. It is a class containing all the information gathered through several
algorithms about an event particle. A PFO instance is linked with all the
information that a physics analysis might desire to investigate: as such, it is
associated with Clusters, Tracks and Vertices objects belonging to the particle
and it is conveniently put in the decay tree hierarchy representing the event
history wrap-up.

Pandora implements reconstruction in a modular way: the framework encompasses
\orderof{100} algorithms and is easily extended via sub-classing thanks
to the Object Oriented paradigm, it is built around. The modularity of Pandora
prefers to split the reconstruction process into several small steps with
dedicated algorithms, rather than a single end-to-end procedure: this ensures
more fine-grain control on the workflow and the ability to switch between
multiple implementations of the same method. The instructions and the list of
algorithms that should be run by the Pandora application are contained in XML
settings files.

The modularity of Pandora abstracts also the detector geometry to handle
different setups: specific geometry settings collect all the parameters needed
to design the detector where to run the reconstruction in. The main parameters
included in a geometry file for a LArTPC detector are the number of drift
chambers with their spatial extent in the cartesian frame of reference. The
orientation of the wires in each wire plane with respect to the vertical, which
fixes the wire pitch direction. The detector gaps simulate un-responsive
channels or regions populated with a low number of wires if any. The current
Pandora framework supports different geometries for LArTPC detectors, such as
ICARUS~\citep{Antonello:2012}, MicroBooNE, ProtoDUNE-SP, ProtoDUNE Dual Phase
(DP)~\citep{Scarpelli:2019} and the DUNE Far Detector. In the following, we will
focus on ProtoDUNE-SP only, but the discussion can be easily generalized to all
the different LArTPC realizations.

\begin{figure}
    \centering
    \myincludegraphics[width=0.9\textwidth]{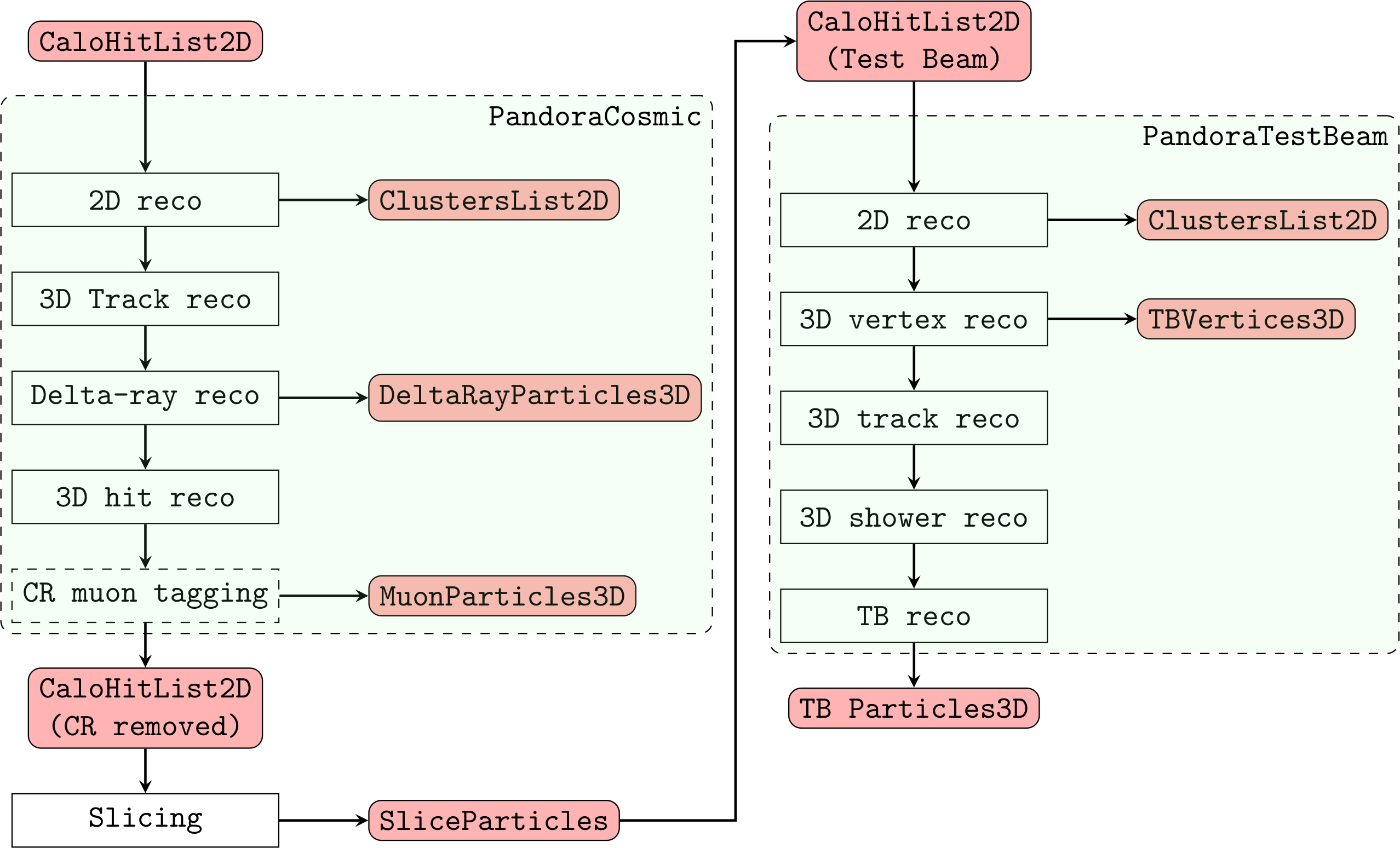}
    \caption{Scheme of the Pandora reconstruction workflow for a ProtoDUNE-SP
    event. The red blobs refer to data structures that serve as inputs for the
    Pandora algorithms or their products. White boxes identify high-level
    reconstruction step and are gathered into two main pipelines:
    \texttt{PandoraCosmic} and \texttt{PandoraTestBeam}. In the middle, the
    slicing algorithm aims at isolating the Test Beam related CaloHits. The CR
    muon tagging algorithm, labeled by a dashed box, is taken from the LArSoft
    implementation.}
    \label{fig:pandora-flowchart}
\end{figure}
Figure~\ref{fig:pandora-flowchart} represents the flowchart of a Pandora run for
a ProtoDUNE-SP reconstruction event. The inputs are the $2{-}$dimensional CaloHits
from either simulation or real detector data. The diagram, then, splits the
process into three main pipelines, namely \texttt{PandoraCosmic}, slicing and
\texttt{PandoraTestBeam}. The first step aims at reconstructing the CaloHits
belonging to cosmic rays (CRs): many LArTPC detectors are built underground (such
as the DUNE FD) and are shielded by hundreds of meters of rock, however, others
are surface detectors (like MicroBooNE and ProtoDUNE) and they record events
contaminated by CR muons interacting at high angles in the fiducial volume
producing long ionization tracks. This effect is even more emphasized by the
long readout windows of a few milli-seconds required to observe LArTPCs signals
due to the rather small electron drift speed, which includes many CR tracks in
each event. For example, we expect an $\orderof{50}$ of CR interactions for each
event at ProtoDUNE-SP (readout window of $\SI{3}{\milli\s}$). The
\texttt{PandoraCosmic} outputs a list of \texttt{Muon3DParticles} that
unambiguously identifies the CR muons CaloHits that will not be available for
the next steps.

The slicing algorithm runs just after the first main pipeline and its goal is to
group the CaloHits related to the same main interacting particle and its
subsequent decays. The output of such a procedure is comprised of multiple sets
of CaloHits named slices.

The \texttt{PandoraTestBeam} pipeline further analyzes the CaloHit remnants from
the previous steps and aims at reconstructing the data structures related to the
test beam (TB) depositions. In other detectors involving neutrino beams from
accelerator sources, such as MicroBooNE, this pipeline is replaced by the
analogous \texttt{PandoraNu}, described in detail by~\cite{Acciarri:2018}. The
final output of such groups of algorithms is a list of test beam or neutrino
$3{-}$dimensional particles that are inserted in the event interaction tree.

\paragraph{\texttt{PandoraCosmic} pipeline}

The \texttt{PandoraCosmic} pipeline is track-oriented, given that CRs
generate long ionization tracks in the detector rather than particle showers.
The inputs of this pipeline are all the $2{-}$dimensional CaloHits extracted from 
an event generated by LArSoft. At ProtoDUNE-SP, we identify three main CaloHitLists
associated with the three wire planes $\mathrm{U}$, $\mathrm{V}$ and
$\mathrm{W}$. Each CaloHitList undergoes a $2{-}$dimensional clustering procedure
that groups together the hits employing topological algorithms. The objective of
this first stage is to create high-purity small clusters to be merged afterward,
rather than complete ones. Purity is the fraction of hits in a cluster that
indeed belong to the true MCParticle associated with the cluster. On the other
hand, completeness is the percentage of the true MCParticle CaloHits that are
included in the Cluster2D.

After the first clusters are created, the Pandora framework tries to associate
and extend the clusters to increase completeness without affecting the
purity. This is done by extracting geometrical information about cluster pairs
considering geometrical information such as cluster proximity and orientation.
A further cluster refinement is done with the output of this merging step looking
for clusters that should be split, instead. Indeed, clusters that present
discontinuities or intersections in the overall track envelope are split into two
clusters to preserve the purity metric.

Once the three separate ClustersList2D are generated, the $3{-}$dimensional
reconstruction algorithms aim to match the different $2{-}$dimensional plane
views into a single detector view. The key idea is to link together the different
Cluster2D objects from the three views if they refer to the same Track3D. Several
algorithms run in this step to identify and resolve ambiguities among
the different plane views. Two main algorithms known as ThreeDTransverseTracks
and ThreeDLongitudinalTracks achieve these goals for ionization tracks with a long
or small extension in the $x$ drift coordinate, respectively.

The CR ray muons often generate small secondary particle interactions out of
their main ionization tracks. These are secondary electrons that trigger small
curly depositions known as delta rays. At this stage of the reconstruction the
CaloHits related to these particles are already grouped into clusters, but they
still un-assigned to any $3{-}$dimensional object. The DeltaRayMatching algorithm
aims at identifying these delta rays and linking them to their parent muon track
based on their proximity.

The input $2{-}$dimensional CaloHits are grouped into clusters and tracks and, 
at this stage, Pandora populates the list of $3{-}$dimensional CaloHits
reconstructed from the input ones. The CR reconstruction terminates with
identifying the track's end-points (the starting point is always at the highest
$y$ value since CRs come from above), creating a Muon3DParticle along with
Vertices objects at each delta ray emission. Before exiting the pipeline, the 
CaloHits associated with the reconstructed CR objects are removed from the
input lists.

\paragraph{Slicing}

After cosmic rays removal by the previous pipeline, the CaloHits should contain
only the TB or the neutrino-associated hits, depending on the specific LArTPC
application. However, since failures in CR identification might occur, whatever
downstream pipeline, namely \texttt{PandoraTestBeam} or \texttt{PandoraNu}, should
be able to handle CR remnants properly. Therefore, Pandora runs at this
intermediate stage the slicing algorithm: its primary objective is to separate
the CR remnants CaloHits from the TB or neutrino ones. As a plus, it proceeds to
assign the available CaloHits to separate clusters based on the main interacting
parent particle. This way, not only the TB or $\nu$ particles are kept divided
from the rest of the CaloHits, but also it should be easier to run once more
the reconstruction on the CR remnants.

The Pandora implementation of the slicing algorithm is based on topological
considerations and completely ignores the energy information contained by
the CaloHits. The core methods connected with this algorithm are the
\texttt{ThreeDSlidingFit} and the \texttt{ThreeDSlidingConeFit}, which
respectively consist in fitting a track and shower envelope around the available
clusters in the $3{-}$dimensional space after a first fit has been done on the
$2{-}$dimensional objects projections. The resulting tracks and shower clusters
are allowed to seed a new slice if they group more than $50$ CaloHits. Once a new
slice is created all the clusters associated with the seeding one are gathered
inside the same slice. The association of each candidate cluster is done through
pointing (mainly orientation and parent-daughter relationship) and proximity
(minimum inter-cluster distance) conditions with the main seeding cluster for
track-like products and percentage overlapping with the main shower cone envelope
for the shower-like ones.

The algorithm ensures that all the clusters get tested to be included into a
slice \linebreak
traversing all the possible clusters with a Breadth First Search approach.
This way, the CaloHits associated with each seeding cluster are collected inside
their relevant slice. However, it can be the case that small isolated clusters do
not get linked with any slice because they do not pass the conditional statements
listed above. The software, therefore, takes care of these remaining clusters
building a $3{-}$dimensional k-d tree graph for fast cluster querying:
the k-d tree nodes are fixed to the positions of the CaloHits already in the
slices and the centroid point of each remnant cluster is checked to be included
in the slice of the closest k-d tree node.

\begin{figure}
    \begin{minipage}[t]{0.47\textwidth}
        \begin{center}
        \myincludegraphics[scale=0.4]{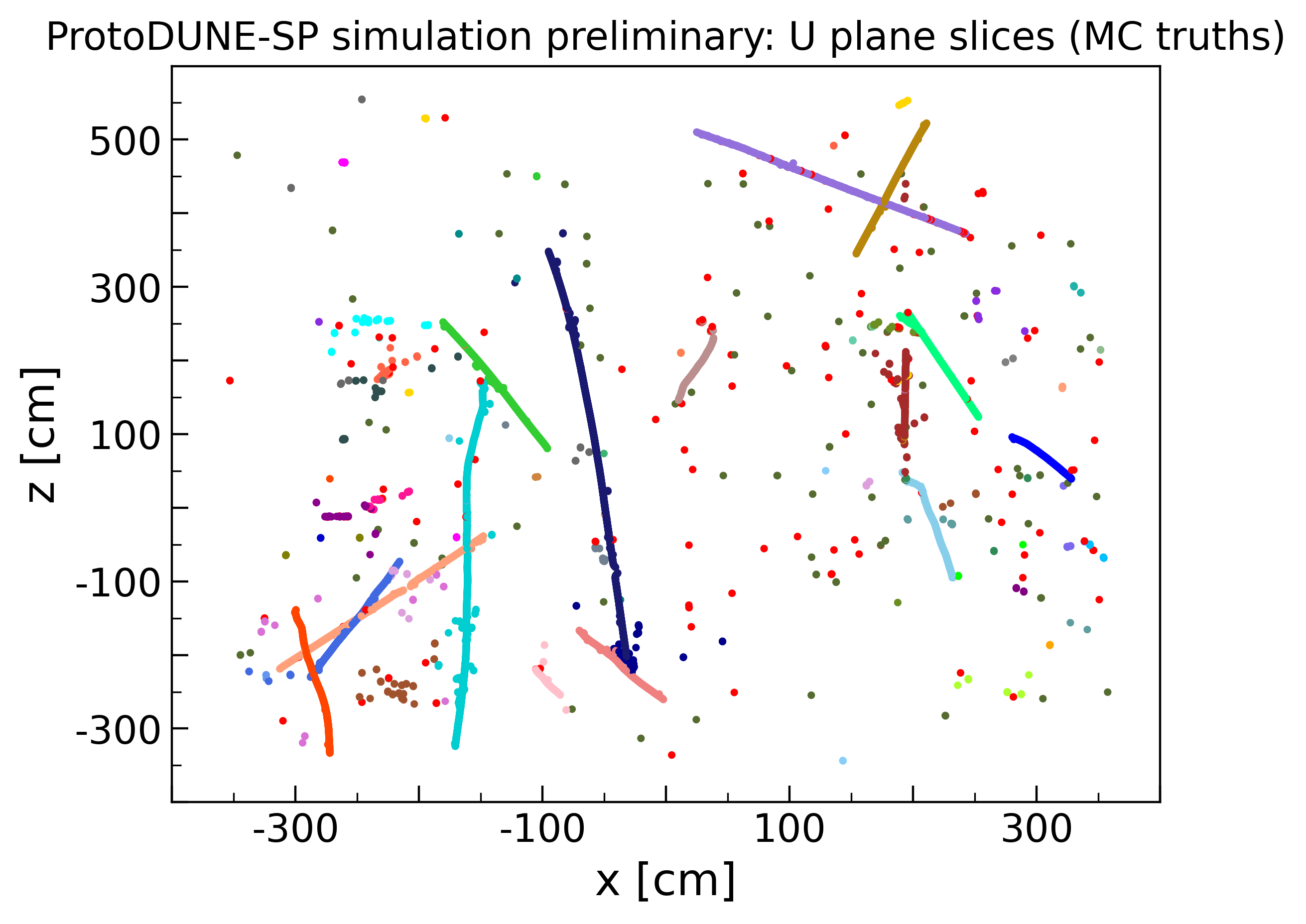}
        \caption{\footnotesize\label{fig:mc-slices}Cosmics rays and test beam
        event: U plane view. Hits with the same color belong to the same main
        Monte Carlo simulated particle.
        }
    \end{center}
\end{minipage}
\hfill
    \begin{minipage}[t]{0.47\textwidth}
        \begin{center}
            \myincludegraphics[scale=0.4]{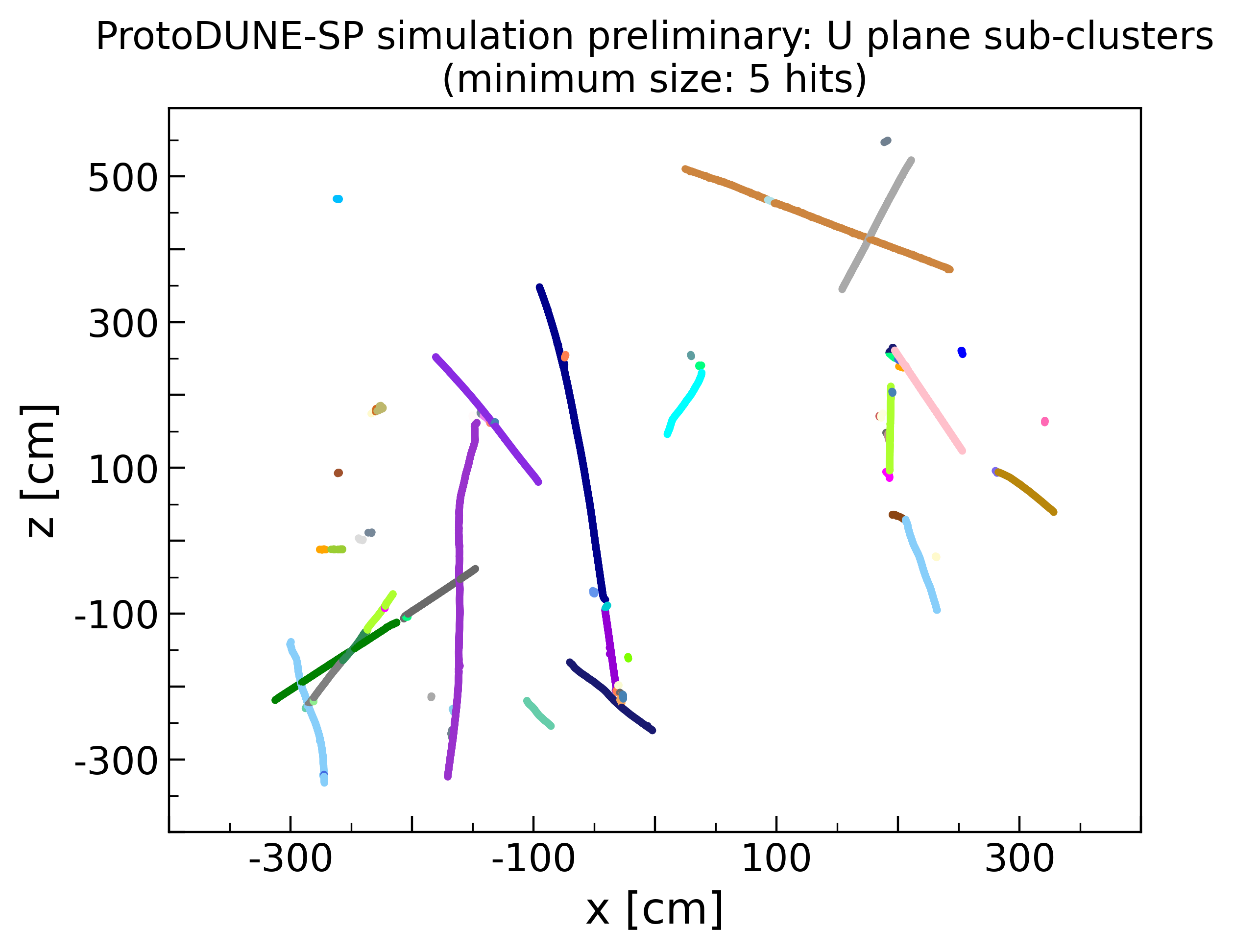}
        \caption{\footnotesize\label{fig:initial-clusters}U plane view: 2D initial
        clusters with more than $5$ hits computed by Pandora.}
    \end{center}
\end{minipage}
\end{figure}

Figure~\ref{fig:mc-slices} provides an example of the slicing tool output for an
event with cosmic rays plus a test beam. At ProtoDUNE-SP TB is a mixture of charged
particles with given energy (mainly $p$, $e^+$ and $\pi^+$). The hits in the image
are color-coded, depending on which final slice they belong to.

\paragraph{\texttt{PandoraTestBeam}}

The \texttt{PandoraTestBeam} pipeline is associated with the reconstruction of
the CaloHits due to the test beam interactions with the liquid Argon. The test
beam at ProtoDUNE is mainly comprised of hadrons, while neutrinos are the main
constituents of the beam flux to which MicroBooNE is exposed. Even if the
composition of the beams is different, Pandora treats the two problems almost
in the same way. We stress that the input CaloHits of this step should not
contain CR muon tracks anymore, even if remnants can still be addressed at this
level.

The TB pipeline differs from the CR one mainly by two aspects: the vertex
selection and the reconstruction of primary electromagnetic showers. In
particular, the vertex reconstruction starts considering all the possible vertex
candidates in the $3{-}$dimensional space as a result of the
CandidateVertexCreation algorithm. Pandora proceeds, then, to evaluate a score
metric for each vertex candidate to identify the fittest, which turns
out to be the one with the maximum score. The metric is built around three
different contributions that can be weighted to tunable parameters to tweak the
relative importance of each factor.

The first contribution is dubbed as \emph{energy kink} and depends on the sum
of each cluster total reconstructed energy projected on each $2{-}$dimensional
view, plus an impact parameter measuring the displacement of each cluster from
the vertex. The idea is that secondary interactions downstream of the interaction
vertex should be less energetic than primary particles, such as TB ones or CR.

The second contribution describes the \emph{asymmetry} of the vertex splitting:
it has been introduced to suppress vertex candidates that arise in the middle of
long tracks, splitting the depositions into approximately the same number of
downstream and upstream CaloHits. Pandora favors configurations with large
values of such asymmetry parameters, underlining that the vertex occurs near
the track end-points.

The last factor influencing the vertex metric is the \emph{beam deweighting
score}, which measures the $z$ coordinate placement of the vertex. The vertex
position is better placed at low values of the $z$ coordinate, which means near
the TB source. This quantity strongly depends on the normalized $z$ position of
the candidate vertex.

Once the TBVertices3D vector is computed, Pandora starts the track and shower
reconstruction. The former aims at finding the CR remnants, along with TB
particles (such as protons, pions and muons) that produce track-like depositions;
the latter identifies two kinds of objects: shower spines and shower branches.
The shower-like clusters are divided into two categories by the length of
the associated $2{-}$dimensional clusters: long clusters are spines and are
typically oriented towards the interaction vertex, while the shorter ones
represent branches and might also be displaced from the main spine. The
algorithm dedicated to recursively building up the $2{-}$dimensional showers is
named ShowerGrowing. After this stage, the shower projections are matched into
the $3{-}$dimensional objects by resolving all the possible ambiguities that may
arise.

Further algorithms refine and complete the high-level objects created during the
\texttt{PandoraTestBeam} pipeline so far. Therefore, a second
$3{-}$dimensional reconstruction is run with less stringent thresholds to ensure
that all the CaloHits are conveniently placed in their corresponding object.
The Pandora software is finally ready to build the particle hierarchy wrap-up,
recursively looking from the primary particles to their daughters while tagging 
all the primary and secondary interaction vertices in the event.

\section{The Cluster Merging Network}
\label{sec:cm-net}

\paragraph{Network design}

We introduce the novel slicing method based on the Cluster Merging Network
(CM-Net)~\citep{slicing:2021}: an agglomerative clustering algorithm, in which we train a neural
network to decide if sub-clusters from a 2D initial cluster pair should be merged
or not. A successful link of two 2D clusters should happen if they are
formed by hits generated by the same interaction. This information is
available in the simulation and it is used as the ground truth in the CM-Net
supervised training. The objective is to build an undirected graph, where the
nodes are the 2D initial clusters in an event plane view and the resulting
different disconnected parts of the graph are interpreted as slices.
Eventually, two specific 2D hits in a plane view will belong to the same slice if
there exists a path in the graph connecting their corresponding 2D initial
clusters.

\begin{figure}
    \begin{minipage}{0.47\textwidth}
        \begin{center}
        \myincludegraphics[scale=0.4, height=2in]{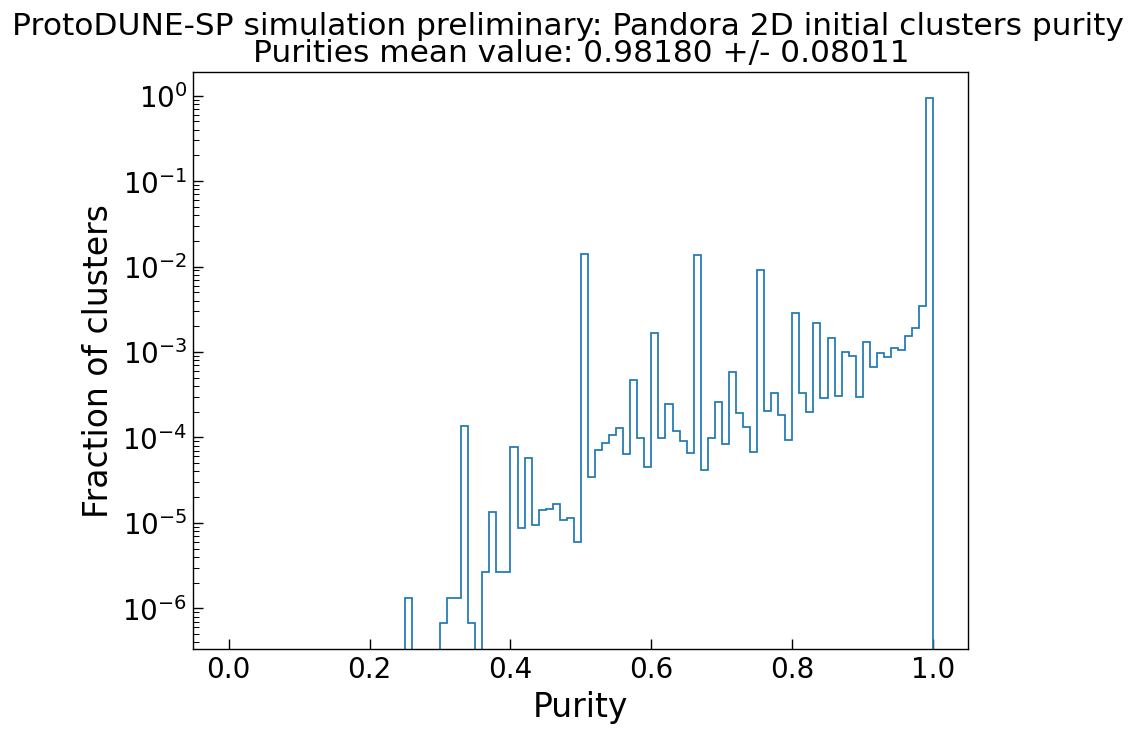}
        \caption{\footnotesize\label{fig:pur hist}2D initial clusters purity. The
        histogram is peaked towards high purity values.
        }
        \end{center}
    \end{minipage}
    \hfill
    \begin{minipage}{0.47\textwidth}
        \begin{center}
        \myincludegraphics[scale=0.4, height=1.8in]{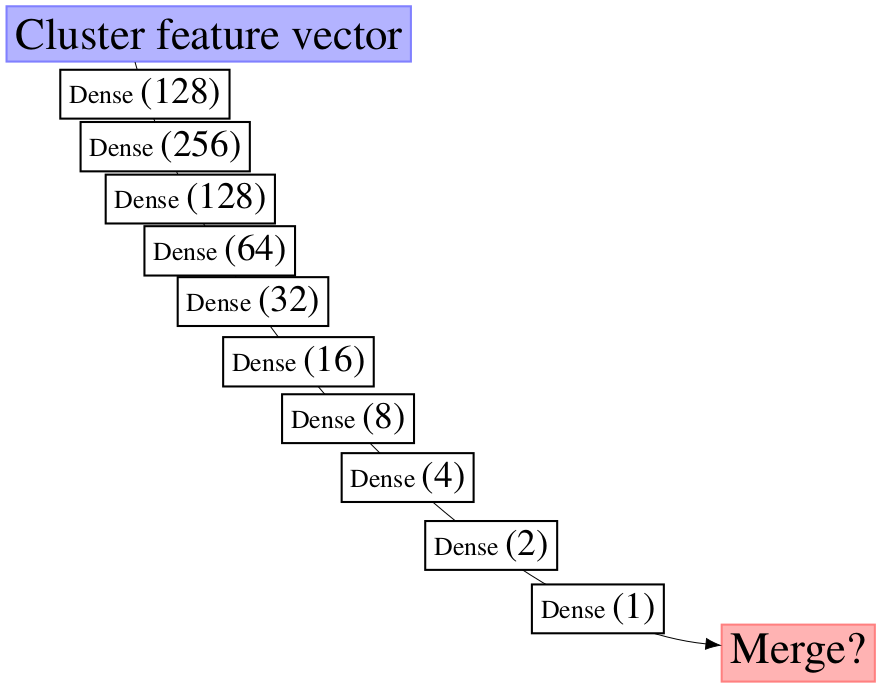}
        \caption{\footnotesize\label{fig:cmnet}The CM-Net network architecture:
        given a feature vector of the cluster pair, outputs the probability that
        the cluster pair should be merged.}
        \end{center}
    \end{minipage}
\end{figure}

We underline that this method is not able to split an initial 2D cluster into two
parts. Thus, amending reconstruction errors arising from a previous step of the
Pandora reconstruction is not possible. This approach is supported by the fact
that the Pandora reconstructed objects show high purity, that is the percentage
of hits in a 2D initial cluster falling in the same Monte Carlo true slice,
even before the slicing takes place. Figure~\ref{fig:pur hist} captures
this behavior: since a cluster merging procedure can only lower the purity of the
resulting set with respect to the initial cluster purities, this benchmark sets
an upper bound of the purity of the CM-Net approach. The mean of the purity
histogram exceeds $90\%$.

Figure~\ref{fig:cmnet} depicts our CM-Net, which is realized as a stack of fully
connected layers, with sigmoid final activation. The CM-Net takes as input a
vector of fixed size for each pair of sub-clusters in the plane view. The array
contains a set of $29$ intra-cluster features computed independently for each
sub-cluster. We add the same quantities extracted from the sample of hits that
would be obtained by merging the two candidate sub-clusters. Finally, we
concatenate two more inter-cluster scalar measures. The total length of the
network input vector is therefore of $89$ elements.

\begingroup
\renewcommand{\arraystretch}{1.3}
\begin{table}
    \scriptsize
    \centering
    \begin{tabular}{lll}
\hline\noalign{\smallskip}
\multicolumn{3}{c}{Intra-cluster features} \\
\noalign{\smallskip}\hline\noalign{\smallskip}
\multirow{3}*{Standard stats $(5)$} & Hits percentage & $|h_{\mathcal{C}}| / h_{\mathrm{tot}}$ \\
                           & Hits mean position & $\mathbf{\bar{h}} = (\bar{h}_x, \bar{h}_y)$ \\
                           & Cluster expected direction &  $\mathbf{v} = (v_x,v_y)$ \\[7pt]
\multirow{4}*{Covariance matrix $(10)$} & Elements & $Cov[h_x,h_y]_{ij}$ \\
                                        & Eigenvalues  & $\lambda_1, \lambda_2$ \\
                                        & Eigenvalues importance &$\lambda_1 / (\lambda_1 + \lambda_2)$  \\
                                        & Eigenvectors &$\mathbf{v}_{\lambda_1}, \mathbf{v}_{\lambda_2}$ \\[7pt]
\multirow{2}*{Post-PCA $(9)$}  & Cluster end-points & $\mathbf{h}^{\mathrm{t}}$, $\mathbf{h}^{\mathrm{l}}$, $\mathbf{h}^{\mathrm{r}}$, $\mathbf{h}^{\mathrm{b}}$\\
                               & Max end-points distance & ${\displaystyle \max_{i,j}} \|\mathbf{h}^i-\mathbf{h}^j\|$ \\[7pt]
\multirow{2}*{Calorimetry $(2)$} & Hit mean energy & $\bar{E}$ \\
                                     & Hit energy std & $\sigma_E$ \\[7pt]
TPC view $(3)$ & One-hot vector for U,V,W  & \\[15pt]
\hline\noalign{\smallskip}
\multicolumn{3}{c}{Inter-cluster features} \\
\noalign{\smallskip}\hline\noalign{\smallskip}
\multirow{2}*{Geometric stats $(2)$} & Minimum inter-cluster distance & ${\displaystyle\min_{\mathcal{C}_1,\mathcal{C}_2}} \|\mathbf{h}_{\mathcal{C}_1} - \mathbf{h}_{\mathcal{C}_2}\| $\\
                                   & Expected angular separation & $\abs{\mathbf{\hat{v}}_1\cdot\mathbf{\hat{v}}_2}$\\
\noalign{\smallskip}\hline   
    \end{tabular}
    \caption{List of the extracted intra and inter-cluster features. The first
    column groups the features into categories: the number in parentheses highlights
    how many components of the final feature vector the specific category gathers.
    }
    \label{tab:cluster-features}
\end{table}
\endgroup
Table~\ref{tab:cluster-features} collects the inter-cluster and intra-clusters
engineered features. We group the quantities into different categories. Five
standard values mainly come from the Pandora reconstruction and measure the
importance of the cluster in the event, in terms of the percentage of hits over
the total in the plane view, the average position in the plane and the expected
direction of the cluster. The latter is a quantity available within the Pandora
framework and gives information about the expected direction of the track
envelope in the plane.

The covariance matrix informs mainly about the shape of the point cloud formed
by the sub-cluster, namely how the hits are distributed in the plane. Hence, we
collect the elements of the covariance matrix as well as its eigenvalues
$\lambda_{1,2}$ and eigenvectors $\mathbf{v}_{\lambda_{1,2}}$. We further compute
the importance of the first eigenvalue normalizing it to the sum of the two:
$\lambda_1 / (\lambda_1 + \lambda_2)$.

The Principal Component Analysis (PCA) allows us to rotate and
rescale the position of each sub-cluster hit, identifying the directions of maximal
variance. Once the point cloud is transformed with the PCA algorithm, it is easy
to identify the hits at the boundaries of the point cloud $\mathbf{h}^{\mathrm{t,l,r,b}}$
(top, left, right, bottom) in the new $(x',z')$ plane, since they are now the
furthest ones away from the origin. Also, we consider the maximum euclidean pair
distance among the set of such end-point hits.

As opposed to the Pandora geometric approach, in this work, we include the \linebreak
calorimetry information, adding to the feature vector the mean and standard
deviation of the hit deposited energies.

Finally, we complete the list of the intra-cluster features with a one-hot vector
referring to which plane view the sub-cluster belongs to. This is a $3{-}$element
array corresponding to the different U, V and W planes.

The inter-cluster features read two scalar entries: the minimum inter-cluster
distance between hits and the expected angular separation, namely the cosine of
the angle between the two sub-cluster expected directions. In principle, the
former value should be inversely correlated to the probability that two
sub-clusters belong, in fact, to the same track. The latter quantity is
conveniently computed through the modulus of the inner product of the unit
vectors $\mathbf{\hat{v}}_{1,2}$ pointing towards the sub-cluster's expected
directions.

\paragraph{Dataset}

The dataset used to train and test our model collects only simulated data,
testing on detector data is out of the scope of the present work and can be
addressed in the future. We simulate $600$ events for training and $300$ for
testing purposes with the help of the LArSoft~\cite{Church:2014} framework.
Each event represents a test beam of charged particles with certain energy plus
cosmic rays interactions within ProtoDUNE-SP. We produce the same amount of
events for each of the available test beam energies: $0.3$, $0.5$, $1$, $2$, $6$,
$\SI{7}{\GeV}$. We remind the reader that each event is composed of three
different plane views and, in turn, it contains  $\mathcal{O}(\SI{10}{k})$ hits
and $\mathcal{O}(500)$ clusters of various size. We expect that each event
contains $\mathcal{O}(50)$ main interactions to be grouped into slices.

\paragraph{Training}

We train the CM-Net with the following setup. After collecting all the possible
2D input cluster pairs from the events in the training dataset, we filter out the
negative examples (those that should not be merged) to reach a $50\%$ balance
on the two classes. We reserve an array with $10\%$ of the examples as a
validation dataset. Validation is employed to choose the best model during
training, namely, the checkpoint on the epoch end that achieves the best validation
accuracy. The CM-Net is optimized using the Stochastic Gradient
Descent (SGD)~\cite{Ruder:2017} algorithm with a learning rate $\eta=5\cdot
10^{-3}$ and the binary cross-entropy as the loss function.

After training, we exploit the CM-Net to build the clusters graph: an edge
between a cluster pair is drawn if the output of the network exceeds a certain
threshold. In our experiment, we fix this threshold to $0.9$. We choose such
a high value to lower as much as possible the false positive rate: a single
positive mispredicted edge can cause a macroscopic effect at the graph level. 
It is possible, indeed, that false positive edges establish a bridge between
two large disconnected parts of the graph, linking them together when, in fact,
they should remain separate. The final CM-Net graph prediction is turned
into a set of slices containing 2D detector hits, that can be compared against
the state-of-the-art Pandora slicing method output.

\paragraph{Results}

We build a benchmark to compare the goodness of the CM-Net slicing reconstruction
against the current Pandora implementation. Since ProtoDUNE-SP is a surface
detector, it is crucial to correctly identify and keep separate slices associated
with test beam (TB) hits and cosmic rays (CR) hits. We remind the reader that in 
our experiments, we simulate a TB made of charged particles with a given energy.
The goal is to define metrics that allow us to tag the status of the
reconstruction on a plane view basis. Such figures of merit are the purity and
the completeness of a reconstructed TB slice: purity is the fraction of hits in
a reconstructed TB slice that is shared with the TB slice; completeness is the
fraction of hits from the true TB slice that is shared with the reconstructed
TB slice.

\begin{figure}
    \centering
    \begin{minipage}[t]{0.47\textwidth}
        \myincludegraphics[width=\textwidth]{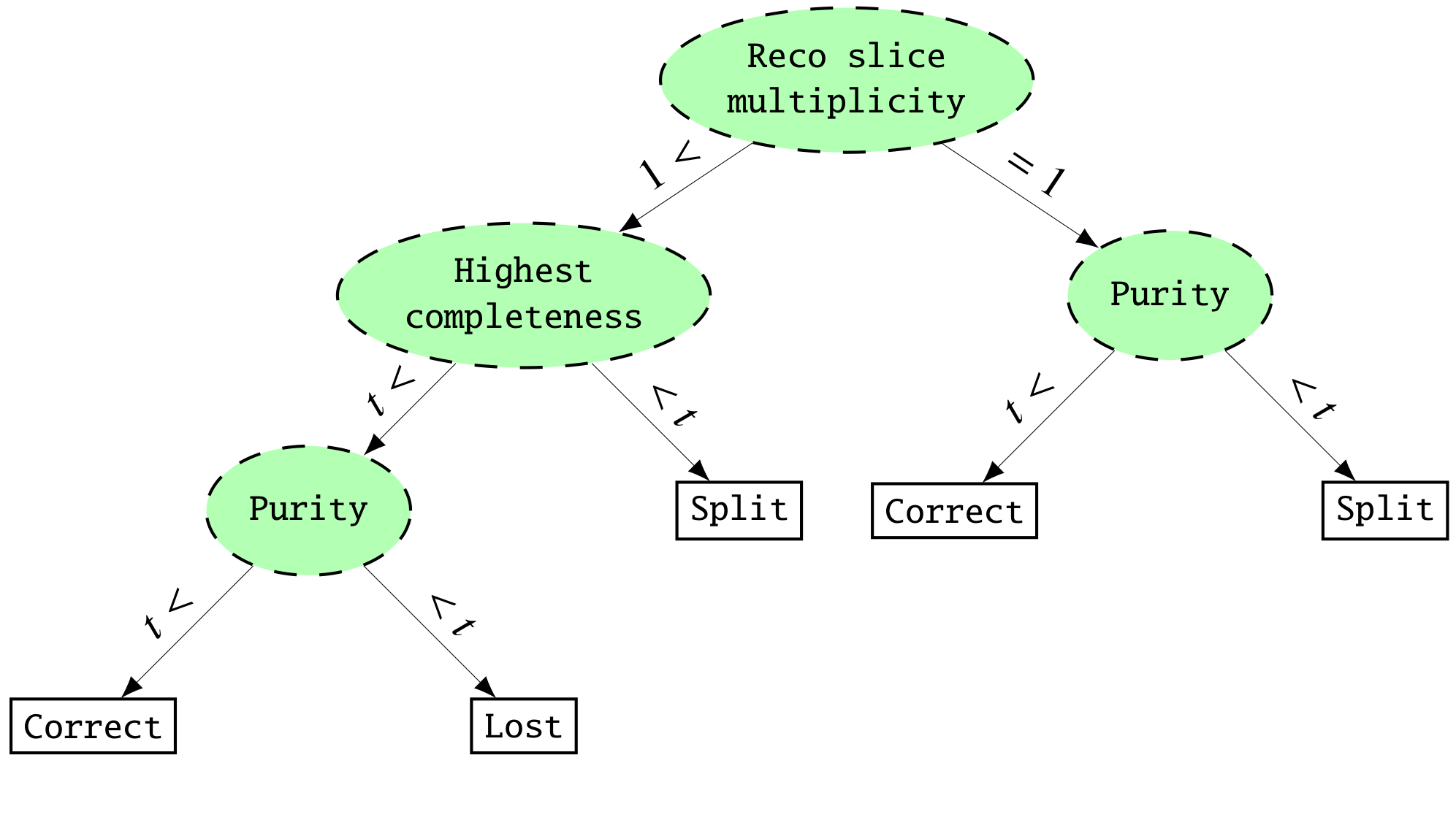}
        \caption{Test beam benchmark flowchart:
        marks each plane view as \texttt{correct}, \texttt{split} or
        \texttt{lost}. The threshold is $t=90\%$.}
        \label{fig:flowchart}
    \end{minipage}
    \hfill
    \begin{minipage}[t]{0.47\textwidth}
        \myincludegraphics[width=0.85\textwidth, height=1.8in]{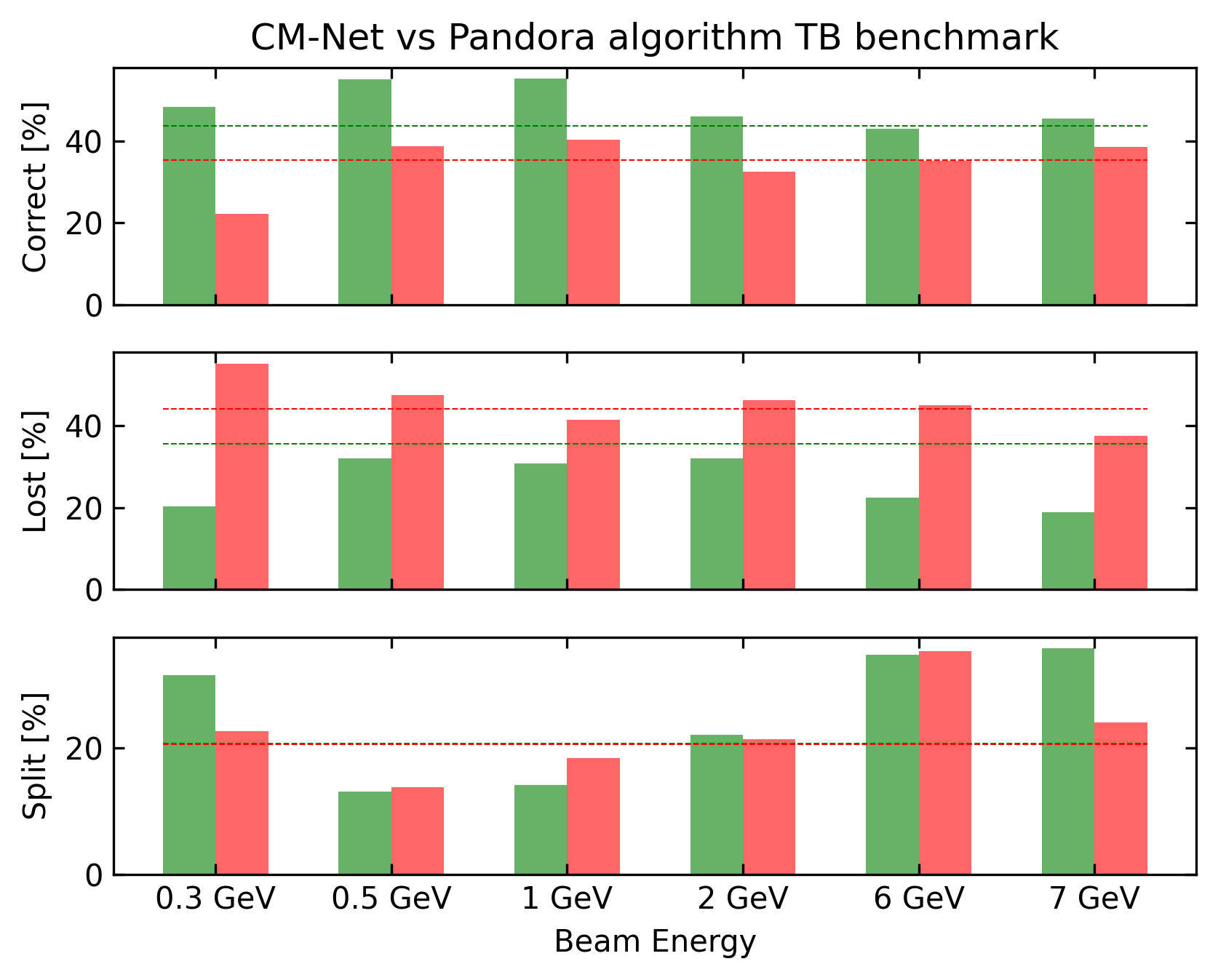}
        \caption{Reconstructing MC TB slices: green bars for CM-Net, red bars
        for Pandora. Dashed lines represent the overall TB energy score. Rows
        from top to bottom represent correct, lost and bottom percentages.}
        \label{fig:metrics}
    \end{minipage}
\end{figure}

Based on these definitions, we mark each plane in the test set with a flag within
the following set: \texttt{correct}, \texttt{split} or \texttt{lost}.
Figure~\ref{fig:flowchart} illustrates the tagging procedure. The first branching
depends on the multiplicity of the reconstructed slices containing true TB hits.
If all true TB hits fall into the same reconstructed slice, the only
discriminant on the reconstruction status is its purity: above $90\%$ purity
threshold, the plane view is said to be \texttt{correct}, otherwise is \texttt{lost}.
The optimal output should not involve multiple TB slices.
Nonetheless, we allow a small margin of error in our tagging procedure, when the
highest reconstructed TB slice has both completeness and purity above $90\%$.
Conversely, if completeness is below the threshold, the plane is naturally marked
as \texttt{split} or if the output, instead, falls short in purity, it provides
the evidence that the TB slice has inevitably been merged (\texttt{lost}) into a
CR larger slice.

Figure~\ref{fig:metrics} collects the results of the TB reconstruction benchmark.
The top row shows that CM-Net achieves better accuracy than the current Pandora
baseline for all the available beam energies. The improvement of our algorithm
has to be identified in the less number of clusters marked as \texttt{lost}.
The CM-Net algorithm presents similar performance for all available TB energies
for both the \texttt{correct} and \texttt{lost} kind of predictions, while it
achieves worse score for the \texttt{split} category for at the lowest
($\SI{0.3}{\GeV}$) and highest ($\SI{7}{\GeV}$) ends of the TB energy range.

We showed that the
proposed model outperforms the state-of-the-art implementation within
the Pandora framework. This approach is compatible with the Pandora interface
and can be appointed for integration into the software in the future.
Further work directions include the assessment of the relative importance of each
input feature in the CM-Net final prediction, as well as the optimization of the
network architecture itself, inspecting new and more flexible solutions, such as
the attention mechanism~\cite{Vaswani:2017} and its variants. Those
networks might exploit the data structure more naturally, avoiding manually
selecting and extracting feature arrays of fixed size from the clusters.

%% file: chapters/concl/Concl.tex
\chapter*{Future directions}
\addcontentsline{toc}{chapter}{Future directions}
\label{chap:concl}
\thispagestyle{plain}

In this section, we introduce the main ideas to further developments of the
research presented in this thesis work. We point out the main objectives for
each of the four main results delivered by this study.

\paragraph{\pdfflow}

The development of the \pdfflow software is considered complete.
The challenge is to provide code maintenance through time, understanding and
matching the needs of the users of the package both from the point of view of
theoretical and experimental physicists. The implementation is sufficiently
modular that should allow to easily include new algorithms to extend the
capabilities of the tool.

\paragraph{\madflow}

The fundamental end-point for the successful spread of the \madflow framework
among the HEP community is the need to promote the software to include
next-to-leading order calculations in an automated way, exploiting the full
potential of hardware accelerators. The present work marked an important
milestone showing the possibility to hold the computation of a high number of
events even for complicated processes with multiple final state partons. This
sheds light on the feasibility to go to higher order in perturbation theory.
However, such computations involve complex subtraction techniques to take care
of QCD infrared singularities that might reveal difficult to implement within the
TensorFlow library.

Other chances to improve the \madflow software within reach are the development
of a more efficient and GPU-compatible phase space sampler. The current
implementation of RAMBO for GPU employs a flat phase space that represents a
potential shortcoming in the whole pipeline: complex query processes might introduce
instabilities in the \vegasflow cross-section integration and reduce the overall
event generation efficiency, requiring sampling more events than the GPU might
hold, inevitably slowing down the entire run. Possible solutions can be
investigated from studying adaptive sampling methods also implemented by the
\mgamc package.

Continuous integration and continuous development can be interesting approaches
for the maintenance of the \madflow package in the future: the idea is to
design a fully automated test suite to check corner cases and identify potential
bugs in the \madflow \linebreak implementation to deploy the code in production
mode.

\paragraph{Denoising ProtoDUNE-SP raw data}

Interpretability is one of the main issues related to deep learning, which
represents a big question mark on the behavior of these algorithms in new
environments. The main future research goal for both denoising and slicing deep
learning applications should be to test the networks with larger datasets and
eventually on real data. This investigation might require the implementation of
automated solutions for the management of the deep learning resources: efficient
data storing and retrieval from collected databases, continuous training
techniques to inspect new models and automatically triggered hyperparameter
search strategies. The deployment of these methods requires advanced software
engineering knowledge and experience.

Specific objectives for further developments of the networks implemented in the
context of ProtoDUNE-SP raw data denoising are the integration within the LArSoft
framework to allow easier comparison with the methods currently used by the DUNE
collaboration. This means scanning the suitable solutions to interface the Python
code, in which the Graph Neural Networks are implemented, with the LarSoft
\texttt{C++} program.

Other interesting opportunities for further investigation of deep learning
methods in this study are model optimization and parameter pruning techniques
to provide faster inference with the presented neural networks
architectures.

\paragraph{Slicing algorithm at ProtoDUNE-SP}

A full test comparison with the Pandora implementation of the model shall be the
first phase of the future development of this research. The subsequent step would
be to integrate our solution within the Pandora framework and test it on real
detector data.

Further optimization of the current approach we introduced would be desirable to
provide faster inference performance.
The starting point would be to replace the sequential processing of the pairs of
$2{-}$dimensional sub-clusters in each plane view with more efficient operations.
The bottleneck might be represented by the fact that the number of sub-cluster
pairs to be inspected grows quadratically with the number of sub-clusters.
An interesting approach is given by the application of the attention
mechanism to the entire array of detector hits in each plane view.
This would allow to process entire plane views at the hit level with a single
network forward pass, rather than considering just sub-clusters pairs only.
The increased complexity of the algorithm would then require careful training
process management, although it might result in improved output performance.

%% file: chapters/pubs/pubs.tex
\chapter*{List of Publications}
\addcontentsline{toc}{chapter}{List of Publications} 
\thispagestyle{plain}
\hfill {\it As of \today}
\vspace{0.2cm}\\

\noindent{\bf Refereed publications}

\noindent PDFFlow: parton distribution functions on
GPU~\citep{Carrazza:2020pdfflow}.\\
\noindent MadFlow: automating Monte Carlo simulation on GPU for particle physics \linebreak
processes~\citep{Carrazza:2021madflow}.\\
\noindent Deep Learning Strategies for ProtoDUNE Raw Data
Denoising~\citep{Rossi:2022}.
\vspace{2cm}\\

\noindent {\bf Publications under review}

\noindent Slicing with deep learning models at ProtoDUNE-SP~\citep{slicing:2021}.
Proceedings of 20th International Workshop on Advanced Computing and Analysis
Techniques in Physics Research {\textemdash} ACAT 2021\footnote{Article already
accepted by the editor. The proceedings book has not been issued yet.}.
\vspace{2cm}\\


\noindent {\bf Publications in conference proceedings}

\noindent PDFflow: hardware accelerating parton density access~\citep{Rossi:2021}.
Proceedings of 40th International Conference on High Energy physics
{\textemdash} PoS(ICHEP2020).\\
\noindent MadFlow: towards the automation of Monte Carlo simulation on GPU for
particle \linebreak
physics processes~\citep{Carrazza:2021madflwpos}. Proceedings of 25th
International Conference on Computing in High Energy and Nuclear Physics
{\textemdash} CHEP 2021.

%% file: chapters/ackno/Ackno.tex
\chapter*{Acknowledgments}
\addcontentsline{toc}{chapter}{Acknowledgments}
\thispagestyle{plain}

\noindent
The thesis work has been done in cooperation with the University of Milan and
CERN openlab groups.

\noindent The research has been funded and supported by the IBM Power Development group,
that also provided hardware and intellectual contribution.

\noindent  Several people contributed to the realization of the present work. A huge thank
you to them who have made this possible.

\noindent To the supervisors Stefano Carrazza and Sofia Vallecorsa for their precious
guidance during the whole journey.

\noindent To Eric Aquaronne, that has always been present to organize and sponsor the
work throughout the IBM company.